\documentclass[12pt]{article}
\usepackage{preamble}
\usepackage{xr}
\usepackage{subfiles}

\usepackage[affil-it]{authblk} 

\title{\vspace{-5ex}Regression Discontinuity Designs Under Interference\vspace{-2ex}}
\author[1]{Elena Dal Torrione}
\author[2]{Tiziano Arduini}
\author[1]{Laura Forastiere}

\affil[1]{Department of Biostatistics, Yale University\vspace{-1ex}}
\affil[2]{Department of Economics and Finance, Tor Vergata University of Rome\vspace{-6ex}}

\begin{document}

\date{}

\maketitle

\begin{abstract} 
We extend the continuity-based framework to Regression Discontinuity Designs (RDDs) to identify and estimate causal effects under interference when units are connected through a network. Assignment to an “effective treatment,” combining the individual treatment and a summary of neighbors’ treatments, is determined by the unit’s score and those of interfering units, yielding a multiscore RDD with complex, multidimensional boundaries. We characterize these boundaries and derive assumptions to identify boundary causal effects. We develop a distance-based nonparametric estimator and establish its asymptotic properties under restrictions on the network degree distribution. We show that while direct effects converge at the standard rate, the rate for indirect effects depends on the number of scores fixed at the cutoff. Finally, we propose a variance estimator accounting for network correlation and apply our method to PROGRESA data to estimate the direct and indirect effects of cash transfers on school attendance.  
\newline
\newline
\textbf{Keywords:} Causal inference, regression discontinuity, interference, networks, local polynomials, statistical dependence.
\end{abstract}
\setcounter{page}{1}

\section{Introduction} \label{sec:introduction}

In a Regression Discontinuity Design (RDD), the treatment assignment is determined by whether an observed variable— known as \qq score" or \qq forcing" variable—exceeds a given cutoff. For instance, in conditional cash transfer (CCT) programs, eligibility for financial aid hinges on the household's poverty index relative to a predetermined threshold \citep[e.g.][]{battistinrettor2008,attanasio11}. 
Beyond CCT programs, RDDs have been applied across diverse domains, including education, the labour market, and medicine \citep{lefgren,battistinrettore2002, rddvaccine}. 
RDDs have gained renewed interest in recent years, leading to significant methodological advancements \citep{kolesar, optrdd, cct14}. However, the role of interference in RDDs remains underexplored. 

Interference arises whenever the \qq treatment" on one unit affects the outcome of another unit, as when units interact physically or socially. 
Interference can affect RDD settings. 
\citet{angelucci}, show that the Mexican PROGRESA/Oportunidades program increased consumption among ineligible households through informal loans from beneficiaries, while \citet{lalive} find spillover effects on school attendance among ineligible children living in treated villages 
due to social interactions.
\citet{londo2} find that a Colombian financial aid program designed to boost elite college enrollment among high-achieving, low-income students increased enrollment for ineligible students through demand and supply mechanisms.
Interference can also emerge in health, education, and {social interventions}. Vaccinating eligible individuals can protect ineligible individuals by reducing community viral load \citep{rddvaccine}; retaining students based on test scores may affect classmates through classroom dynamics \citep{matsudaira}; and cash assistance to displaced refugees may benefit ineligible households through resource sharing within kinship or social networks \citep{salti}.


Spillover effects are often of substantive interest. Positive spillover effects can be leveraged to achieve cost-effective interventions: designing optimal policies allocating financial aid to a targeted subset of families can increase overall school attendance in a village through beneficial spillover effects \citep[e.g.][]{viviano2024}. 
Conversely, negative spillovers, such as reduced employability among those excluded from job training programs \citep{crepon}, highlight the need for preventive measures to preserve policy effectiveness.
The literature on causal inference under interference has grown rapidly in recent years. Much of it assumes partial interference, allowing interference within, but not across, clusters \citep{sobel}. Under this framework, methods have been developed for both randomized experiments and observational studies \citep{hh, tchegen, perez, papadogeorgou19}. Other works consider network interference, where treatment effects propagate through more complex network structures.
\citep{aronowsamii17, leung20, vazquez, forastiere, ogburn22, leung2023unconfoundedness}.
Despite this growing literature, little attention has been paid to interference in RDDs.


This paper develops a framework for interference in the continuity-based approach to RDDs. We allow each unit's potential outcome to depend on an ``effective treatment'': the binary individual treatment and a spillover exposure given by an exposure mapping over specified interference sets, such as network neighborhoods or clusters. 
Our key insight is that the effective treatment assignment is a deterministic function of both the individual score and the scores of units within the interference set. Consequently, although each unit has a one-dimensional score, RDDs under interference can be analyzed as multivariate RDDs, building on the literature \citep{zajonc, keele_titiunik_2015}. Unlike standard multivariate RDDs, however, the multivariate treatment boundaries arise from the interference mechanism and the RDD individual treatment rule, rather than being given ex ante.
We characterize these boundaries and define novel causal estimands: boundary direct effects at a given spillover exposure and boundary indirect effects comparing different spillover exposures. We further define the overall direct effect and the overall indirect effect, {summarizing the marginal local effect of the individual treatment and of having one additional treated neighbor, respectively}. 
We establish identification by extending the standard RDD continuity assumptions.
In addition, we show that the traditional RDD approach identifies the overall direct effect, even if interference is ignored.

Building on the distance-based approach to multivariate RDDs \citep[e.g.][]{keele_titiunik_2015}, we propose an estimator that uses distance from the boundary
as the running variable in a local polynomial regression. We show that the convergence rate is determined by the codimension of the effective treatment boundary, i.e., the number of individual and neighbors' score components fixed at the cutoff. Our analysis reveals that direct effects attain the same rate as standard no-interference RDD estimators, whereas indirect effect estimators may converge more slowly. We also develop bias-corrected versions of these estimators. For the overall direct effect at the cutoff, we rely on the canonical local linear RDD estimator, while for the overall indirect effect, we introduce a local polynomial estimator. Both estimators converge at the standard RDD rate.
Our setting also features dependence in both outcome and score variables induced by network structures and by overlapping interference sets. We establish our asymptotic results under a weak dependence condition limiting the growth rate of network links, and we develop a novel variance estimator that accounts for network-induced correlation. 
To our knowledge, aside from \citet{bartalotti}-who analyze clustered settings where all units in a cluster lie on the same side of the cutoff- this is the first contribution to address network dependence in RDDs. 
Our analysis is complemented by a simulation study and an empirical application using data from PROGRESA/Oportunidades, where we estimate the direct and indirect effects of the program on children’s school attendance, defining interference sets by village, grade, and gender. We find an overall direct effect of about $0.07$, matching the direct effects reported by \citet{lalive} and \citet{apr}.
In contrast, the overall indirect effect is larger than the indirect effect of about $0.02$ found in these studies. 

Only recently has the multivariate RDD literature begun to formally analyze distance-based local polynomial estimators in settings without interference, drawing on tools from geometric measure theory \citep{cattaneo2025, cattaneoyu, cattaneoyuaggregate}. 
Our analysis uses related tools but differs in several fundamental respects. 
First, the treatment boundaries are not given; they arise from the interaction of the RD assignment rule and spillover exposure and must be characterized as part of the analysis. Second, conventional multivariate RDDs typically involve codimension-one boundaries, of dimension $d-1$ for a $d$-dimensional score, whereas our boundaries may have higher and heterogeneous codimensions depending on interference set size. Third, we allow non-identically distributed, network-dependent observations, 
requiring new asymptotic and variance-estimation methods that account for dependence in both outcomes and distance variables.
Finally, \citet{cattaneoyuaggregate} derive consistency results for two-dimensional designs with piecewise boundaries and two treatment regions. By contrast, we establish these results for general interference effects with multiple treatment regions. 
As in that literature, bias control requires stronger smoothness conditions. We provide primitive conditions that characterize the required smoothness in the general interference setting.


{Given the scarce literature on interference in RDDs, our work provides several contributions. As opposed to the applied literature that estimates spillover effects by relying on the presence of groups that are untreated by design as comparison groups \citep{angelucci, lalive, londo2},
our RDD approach leverages the eligibility cutoff to identify spillover effects on ineligible units through local comparisons around the threshold. \citet{castro} estimates spillovers effects from rural teacher incentives on vacancy fill rates in nearby non-rural schools using an additive specification and assuming that exposure to a nearby treated school is exogenous at the cutoff. 
However, this approach does not explicitly account for the RDD assignment mechanism governing neighbors’ treatment status, and does not condition on the score variable of neighboring schools. In contrast, we link spillover exposure to the RDD mechanism, identifying effects through discontinuities in neighbors’ treatment status. Our framework nests the specification in \citet{castro}, provides formal continuity conditions, and introduces a nonparametric estimator that accounts for neighborhood scores, with network-robust variance for valid inference.
}

Only a few papers examine interference in RDDs methodologically. \citet{aronow17} propose a difference-in-means estimator for the average direct effect near the cutoff under a local randomization assumption. In contrast, we estimate direct and indirect effects within a continuity-based framework. \citet{yongcai} analyzes a linear-in-means model in which interference occurs among units with similar score values and focuses on direct and spillover coefficients at the cutoff. Our framework instead defines interference sets based on a given network structure 
imposes no linear-in-means assumptions on potential outcomes, and estimates causal effects at boundaries where effective treatment changes discontinuously. Finally, \citet{borusyak2024} studies RDDs where units’ treatment status is an aggregate across multiple discontinuity events. Their method recovers in a simple way a local marginal effects of lower-level units' treatment on upper-level units' outcome. In contrast, our formalization of interference accommodates a richer set of causal effects, albeit with greater methodological complexity.

The paper is organized as follows: \red{Section \ref{sec:motivatingapp} presents our motivating application}; Section \ref{sec:framework} introduces the notation and methodological framework; Section \ref{sec:identification} illustrates our identifying assumptions and results; Section \ref{sec:estimation}
describes our estimation method; Section \ref{sec:mainoverall} considers overall direct and indirect effects; Section \ref{sec:simulation} includes the simulation study; Section \ref{sec:empirical} provides the empirical illustration; and Section \ref{sec:conclusion} concludes.
\red
{\section{Motivating Application}\label{sec:motivatingapp}}
The governmental program PROGRESA/Oportunidades, launched in 1997, sought to alleviate poverty in rural Mexico through cash transfers to poor households, conditional on regular school attendance for children and visits to health clinics. Among 506 selected rural villages, 320 were randomly assigned to the program. Within the 320 treatment villages, households qualified for the cash transfer according to a poverty index and an eligibility threshold, creating the basis for an RDD.
%
However, given potential interactions among children in the same village, a child's school enrollment may also depend on their peers' eligibility through mechanisms such as peer pressure, increased awareness among families, or shifting labor market incentives \citep{lalive, apr, attanasio11}. 

The program can therefore affect both eligible and ineligible children, raising two central causal questions. 
The first concerns the direct effect: how eligibility for the subsidy affects a child when peers' eligibility is held constant. The second concerns  spillover effects: how peers' eligibility affects a child when the child's own eligibility is held constant. The latter reveals whether the transfers generated benefits or harms beyond eligible children. PROGRESA also illustrates the methodological challenge of estimating these effects under an RDD: own treatment is determined by the child’s household poverty score, whereas peer exposure is jointly determined by the poverty scores of peers’ households. 
%
Our methodology addresses this problem and is applied to estimate these effects 
on children’s school enrollment in the 320 villages assigned to the program. 

\section{Framework}\label{sec:framework} 
\subsection{Interference}\label{subsec:interference} 
Consider $n$ units indexed by $i=1, \ldots, n$, connected through an undirected network $G=(\mathcal{N}, \mathcal{E})$ with edges, or links, $(i, j)=(j, i) \in \mathcal{E}$. The network is represented by an adjacency matrix $\boldsymbol{A}$, with $A_{i j}=1$ if $i$ and $j$ are linked and $A_{i j}=0$ otherwise.
Each unit is associated with 
a score variable $X_i \in \mathcal{X} \subseteq \mathbb{R}$, a binary treatment assignment $D_i \in\{0,1\}$, and an outcome variable $Y_i \in \mathbb{R}$, collected in the vectors $\bm{X}$, $\bm{D}$, and $\bm{Y}$ for all $n$ units.
Denote by $Y_i(\bm{d})$, with $\bm{d} \in \{0,1\}^{n}$, the potential outcome under $\bm d$, allowing $i'$s outcome to depend on the complete treatment vector in $\mathcal{N}$

We introduce an interference assumption that restricts interference within specified \emph{interference sets} and uses an \emph{exposure mapping function} to describe the interference mechanism. Specifically, for each unit $i \in \mathcal{N}$ we define an interference set $\mathcal{S}_i \subseteq (\mathcal{N}\setminus \{i\})$, \red{determined by the network $\bm A$}. 
For instance, $\mathcal{S}_i$ may contain all units that are connected to $i$ by a link in $\boldsymbol{A}$, i.e., the network neighborhood \citep{forastiere}. 
A special case is partial interference, where units are partitioned into clusters, such as villages or schools, and $\mathcal{S}_i$ contains only units in $i$ 's cluster \citep{sobel}. The interference set $\mathcal{S}_i$ can also be defined based on network (or cluster) features and covariates. 
We refer to elements of $\mathcal{S}_i$ as \emph{neighbors} of $i$, and we let $\pdelta$ and $\pscore$, with values $\bm{d}_{\smallN_i}$ and $\bm{x}_{\smallN_i}$, collect the treatments and the score variables of units in $\mathcal{S}_i$.
Additionally, we define the exposure mapping as a function that maps the treatment vector within $\mathcal{S}_i$ to an exposure value:  
$g : \{0,1\}^{|\smallN_i|} \to \mathcal{G}_i$, \red{where $\mathcal{G}_i$ is a discrete set with $\left|\mathcal{G}_i\right| \leq 2^{\left|S_i\right|}$}.
Let $G_i \in \mathcal{G}_i$ be the resulting random variable, with $G_i = g(\bm{D}_{\smallN_i})$. We refer to $G_i$ as the \emph{neighborhood treatment}.
Common examples include the number or proportion of treated neighbors, $G_i =  \sum_{j \in \mathcal{S}_i}D_j$ and $G_i = \tfrac{\sum_{j \in \mathcal{S}_i}D_j}{|\mathcal{S}_i|}$. Another type is the \emph{one-treated} exposure mapping, $G_i = \mathbbm{1}\{\sum_{j \in \mathcal{S}_i}D_j > 0\}$, which takes value one if there is at least one treated unit in $\mathcal{S}_i$
\footnote{Our definition accommodates vector-valued exposure mappings that classify neighbors into types \citep[e.g., parents][]{qu2021semiparametric}, but rules out continuously valued exposure mappings. 
}. 
%
%
%
We make the following assumptions.
\begin{assumption}[Consistency]\label{ass:consistency}For all $i\in \mathcal{N}$, if $\bm{D} = \bm{d}$ then $Y_i = Y_i(\bm{d})$.
\end{assumption}
\begin{assumption}[Interference]\label{ass:interfence}
For all $i \in \mathcal{N}$, and for all $\bm{D}$, $\bm{D}' \in \{0,1\}^n$ such that $D_i = D_i'$ and $g(\bm{D}_{\smallN_i}) = g(\bm{D}_{\smallN_i}')$, the following holds: 
 $   Y_i(\bm{D}) = Y_i(\bm{D}')$.
\end{assumption}
Assumptions \ref{ass:consistency} and \ref{ass:interfence} jointly define an extended version of SUTVA. Assumption \ref{ass:consistency} rules out multiple versions of treatment, while Assumption \ref{ass:interfence} allows unit $i$ 's outcome to depend on treatments within $\mathcal{S}_i$, but not outside it, and only through the exposure mapping. Potential outcomes can therefore be indexed by the \emph{effective treatment} $\left(D_i, G_i\right)$, where $D_i$ is the individual treatment and $G_i$ summarizes treatments within $\mathcal{S}_i$ \citep{manski}. Accordingly, $Y_i(d, g)$ denotes unit $i$'s potential outcome when $D_i=d$ and $G_i=g$\footnote{For units with empty interference sets, we leave potential outcomes undefined. Alternatively, one could define potential outcomes of the form $Y_i(d)$.}. Finally, Assumption \ref{ass:interfence} implies that the interference sets and the exposure mapping are correctly specified. 

{\color{black} In our PROGRESA application, units are children, $X_i$ is the household poverty index, $D_i$ indicates subsidy receipt, $Y_i$ is school enrollment, and $\bm A$ encodes village membership. Each child's interference set $\mathcal S_i$ consists of same-gender
children living in the same village and attending the same grade, since interactions in children friendship networks tend to be strong along these dimensions \citep{shrum}. We adopt the fraction of eligible neighbors as the exposure mapping, so that, for example, a child with two neighbors has $G_i\in\{0,1/2,1\}$, corresponding to zero,
one, or both neighbors receiving the subsidy. Finally, under Assumptions \ref{ass:consistency}-\ref{ass:interfence}, $Y_i(d,g)$ represents the child's potential enrollment under eligibility status $D_i = d$ when a fraction $g$ of their neighbors is eligible for the subsidy, regardless of which neighbors are eligible.}

\subsection{Effective Treatment Rule}

In RDDs, the treatment is assigned only to units whose individual score $X_i$ exceeds a known cutoff $c$. We formalize the RDD assignment through the following assumption.
\begin{assumption}[Individual treatment rule]\label{ass:indrule}
For all $i \in \mathcal{N}$,
\(
D_i = \mathbbm{1}(X_i \geq c) = t(X_i)
\).
\end{assumption}
%
\noindent The treatment assignment $D_i$ is then a deterministic function of the score variable.
This paper focuses on the causal effects of the treatment assignment, or intention-to-treat (ITT) effect, regardless of whether the treatment coincides (sharp RDD) or not (fuzzy RDD) with the assignment. For simplicity, we still refer to $D_i$ as the treatment.
Assumption \ref{ass:indrule} leads to the following property: 
\begin{property}[Non-probabilistic individual assignment]\label{property:nonprob}
    For all $i\in \mathcal{N}$, $\CPrr{D_i = 1}{X_i = x_i} = 0$ for $x_i < c$ and $\CPrr{D_i = 1}{X_i = x_i} = 1$ for $x_i \geq c$. 
\end{property}
%
Crucially, Property \ref{property:nonprob} induces a non-probabilistic assignment mechanism of the effective treatment $(D_i, G_i)$. 
For each unit $j \in \mathcal{S}_i$, the treatment assignment follows Assumption \ref{ass:indrule}, i.e., $D_{j} = t(X_{j})$. Therefore, $G_i$ can be rewritten as:
\(
G_i = g(\bm{D}_{\smallN_i}) = g(\{t(X_{j})\}_{j\in \smallN_i}) = e(\pscore)
\), the neighborhood treatment rule. Combining $D_i = t(X_i)$ and $G_i = e(\pscore)$, 
we obtain a deterministic \emph{effective treatment rule}:
\begin{equation}\label{eq:effective rule}
    (D_i, G_i) = (t(X_i), e(\pscore)) = \bm{F}(X_i, \pscore)
\end{equation}
From Eq. \eqref{eq:effective rule}, it is clear that the effective treatment is governed by multiple scores. Therefore, RDDs under interference can be reframed as a \qq multiscore design" \citep{rear}. 
As opposed to typical multiscore RDDs, where the individual treatment is assigned based on multiple individual scores (e.g., math and reading test scores), here
the first component, the individual treatment $D_i$, is determined by the individual one-dimensional score
$X_i$, and the second component, the neighborhood treatment $G_i$, is determined by the scores of neighbors $\bm{X}_{\mathcal{S}_i}$. 

Let us define the multiscore space $\mathcal{X}_{i\smallN_i} \subseteq \mathbb{R}^{|\mathcal{S}_i| +1}$ as the space of all elements $(x_i, \bm{x}_{\smallN_i})$.
The effective treatment rule defines a partition of $\scorespace$ into \emph{effective treatment regions}
\begin{equation*}
    \mathcal{X}_{i\smallN_i}(d,g) = \{(x_i, \bm{x}_{\smallN_{i}}) \in \mathcal{X}_{i\smallN_i} : \bm{F}(x_i, \bm{x}_{\smallN_i}) = (d,g)\}
\end{equation*} 
where the probability of receiving the effective treatment $(d,g)$ is equal to one. 
\red{In PROGRESA, for example, $\scorespace(0,0)$ contains the score configurations for which the child is ineligible and none of their peers are eligible. Under the assignment rule $D_i=\mathbbm{1}(X_i\geq c)$ and the fraction-treated exposure mapping, this region is given by $x_i<c$ and $x_j<c$ for every $j\in\mathcal{S}_i$. Similarly, $\scorespace(1,0)$ contains the configurations for which the child is eligible but none of their peers are eligible, corresponding to $x_i\geq c$ and $x_j<c$ for every $j\in\mathcal{S}_i$.}


\begin{property}[Non-probabilistic effective treatment assignment]\label{pr:intoverlap}
For all $i \in \mathcal{N}$, for all $d \in \{0,1\}$, and for all $g \in \mathcal{G}_i$
\begin{equation*}
    \CPrr{D_i = d, G_i = g}{(X_i, \pscore) = (x_i, \bm{x}_{\smallN_i})} = \begin{cases}
     1 & (x_i, \bm{x}_{\smallN_{i}}) \in  \mathcal{X}_{i\smallN_i}(d,g) \\
      0 & (x_i, \bm{x}_{\smallN_{i}}) \notin  \mathcal{X}_{i\smallN_i}(d,g)
     \end{cases}
 \end{equation*}
\end{property}
\noindent\sloppy
An immediate consequence of Property \ref{pr:intoverlap} is that conditional on the individual score $X_i$ and neighborhood score $\pscore$, the potential outcomes $Y_i(d,g)$ are independent of the effective treatment. Thus, conditional unconfoundedness is an inherent characteristic of the RDD with interference under Assumption \ref{ass:interfence}.



\subsection{Effective Treatment Boundaries}\label{sec:etb}

The non-probabilistic effective treatment assignment results in a structural lack of overlap, as units with the same value of $X_i$ and $ \pscore$ are never observed under different effective treatments. Therefore, unconfoundedness cannot be exploited directly as in other observational studies with interference \citep{papadogeorgou19, forastiere}. However, causal effects can be identified at common boundaries between effective treatment regions. 
The boundary of $\scorespace(d,g)$, denoted by $\bar{\mathcal{X}}_{i\smallN_i}(d,g)$, is the set of points $(x_i, \bm{x}_{\smallN_i})$ such that for any $\epsilon > 0$, a neighborhood of radius $\epsilon$ around $\ppoint$ contains elements in $\scorespace(d,g)$ and in its complement, i.e. $\scorespace(d,g)^c.$
We define the \emph{effective treatment boundary} $\Frontier$ as the intersection of two such boundaries, 
\[
\Frontier = \bar{\mathcal{X}}_{i\smallN_i}(d,g) \cap \bar{\mathcal{X}}_{i\smallN_i}(d',g'),
\]
the set of points in the multiscore space where the effective treatment changes from $(d,g) $ to $(d',g')$. Figure~\ref{fig:oneunit} illustrates the multiscore space for a unit with one neighbor,
$\mathcal S_i=\{j\}$ and $G_i=D_j$. This two-dimensional space is partitioned into four effective treatment regions, whose common boundaries is either a segment (e.g., top-left and bottom-left regions) or the point $(c,c)$ (bottom-left and top-right regions).


The geometry of $\Frontier$ depends on the interference set and the exposure mapping. 
%
Identifying it may be challenging, especially in high dimensions. 
We therefore provide a general explicit characterization.
Let $\treatspace(d,g) = \{(d_i,\bm{d}_{\smallN_i}) \in \treatspace\, s.t.\, (d_i, g(\bm{d}_{\smallN_i})) = (d,g)\}$, with elements 
$\bm{d}_{i\smallN_i}^{d,g}=(d_i, \bm{d}_{\smallN_i})^{d,g}$, the set of individual and neighbors' treatment that yield the effective treatment $(d,g)$. 
Then $\Frontier = \bigcup_{\treatspace(d,g) \times \treatspace(d',g')} \ell(\bm{d}_{i\smallN_i}^{d,g}, \bm{d}_{i\smallN_i}^{d',g'})$, with the set-valued function $\ell : \treatspace(d,g)\times\treatspace(d',g') \rightarrow \scorespace$ being 
\begin{equation}\label{eq:piecefunction}
\ell(\bm{d}_{i\smallN_i}^{d,g}, \bm{d}_{i\smallN_i}^{d',g'})
= \left\{ \ppoint \in \scorespace 
\; \middle|\; \forall z \in (\{i\} \cup \mathcal{S}_i),
    \begin{cases}
        x_z \geq c \quad d_z^{d,g}=d_z^{d',g'} = 1 \\
        x_z \leq c \quad d_z^{d,g} = d_z^{d',g'} = 0\\
        x_z = c \quad d_z^{d,g} \neq d_z^{d',g'}
    \end{cases} \right\}
\end{equation}
(see Supplementary Material \ref{appendix:mainresults} for the derivation). Then $\Frontier$ is identified by applying $\ell(\cdot)$ to each pair of elements 
$\bm{d}_{i\smallN_i}^{d,g}$ and $\bm{d}_{i\smallN_i}^{d',g'}$.
{
This characterization expresses $\Frontier$ as a finite union of linear pieces. Each piece has a \qq codimension", defined as the number of differing entries between treatment configurations, or equivalently, the number of score components fixed at the cutoff:
\begin{equation}\label{eq:codimensionpiece}
s_{\ell(\bm{d}_{i\smallN_i}^{d,g},\bm{d}_{i\smallN_i}^{d',g'})}
=
\sum_{z\in \{i\}\cup \smallN_i}
\mathbbm{1}\!\left(d_z^{d,g}\neq d_z^{d',g'}\right).
\end{equation}
Formally, each linear piece is a $(|\mathcal{S}_i| + 1 - s_{\ell(\cdot)})$-dimensional submanifold of $\scorespace$ with codimension $s_{\ell(\cdot)}$, which is the difference between the dimension of $\scorespace$, i.e., $|\mathcal{S}_i| + 1$, and that of the linear piece.
Consequently, $\Frontier$ is a rectifiable set (see Appendix \ref{appendix:boundformal} for the definition), with dimension $|\mathcal{S}_i| + 1 - \bar{s}_{i,d,g,d',g'}$ and codimension $\bar{s}_{i,d,g,d',g'} $, given by the smallest codimension among all boundary pieces:
\begin{equation}\label{eq:codimension}
\bar{s}_{i,d,g,d',g'}
= \min_{\treatspace(d,g) \times \treatspace(d',g')}
s_{\ell(\bm{d}_{i\smallN_i}^{d,g},\bm{d}_{i\smallN_i}^{d',g'})}.
\end{equation} 
As such, $\Frontier$ admits a well-defined surface measure, allowing us to define boundary average causal effects as integrals over $\Frontier$. 
The codimension $\bar{s}_{i,d,g,d',g'}$ is interpreted as the minimal number of treatment components required for the effective treatment to change from $(d,g)$ to $(d',g')$. By construction, $1 \leq \bar{s}_{i,d,g,d',g'} \leq |\mathcal{S}_i|+1$.
Larger interference sets imply a larger dimension of the score space, but not necessarily a larger codimension of the boundary. 

Figure~\ref{fig:twounits} illustrates the construction for a unit with two neighbors,
$\mathcal{S}_i=\{j,k\}$, so that $\scorespace\subseteq\mathbb{R}^3$ has coordinates
$(X_i,X_j,X_k)$. Under the one-treated exposure mapping, the boundary between
$\scorespace(0,0)$ and $\scorespace(0,1)$ is the union of
\(
\{X_i\leq c,\ X_j=c,\ X_k\leq c\}
\quad\text{and}\quad
\{X_i\leq c,\ X_j\leq c,\ X_k=c\}.
\)
Each piece fixes one score at the cutoff and therefore has dimension $3-1=2$ and
codimension one, whereas their intersection,
$\{X_i\leq c,\ X_j=X_k=c\}$, has dimension $3-2=1$ and codimension two.
For example, applying $\ell(\cdot)$ to
$\bm d_{i\smallN_i}^{0,0}=(0,0,0)$ and
$\bm d_{i\smallN_i}^{0,1}=(0,1,0)$ yields the first piece, with
$s_{\ell(\cdot)}=1$. Since the boundary contains pieces of codimension one, $\bar s_{i,0,1, 0,0}=1$.

\begin{figure}[t]
    \centering
    \includegraphics[width =0.4\textwidth]{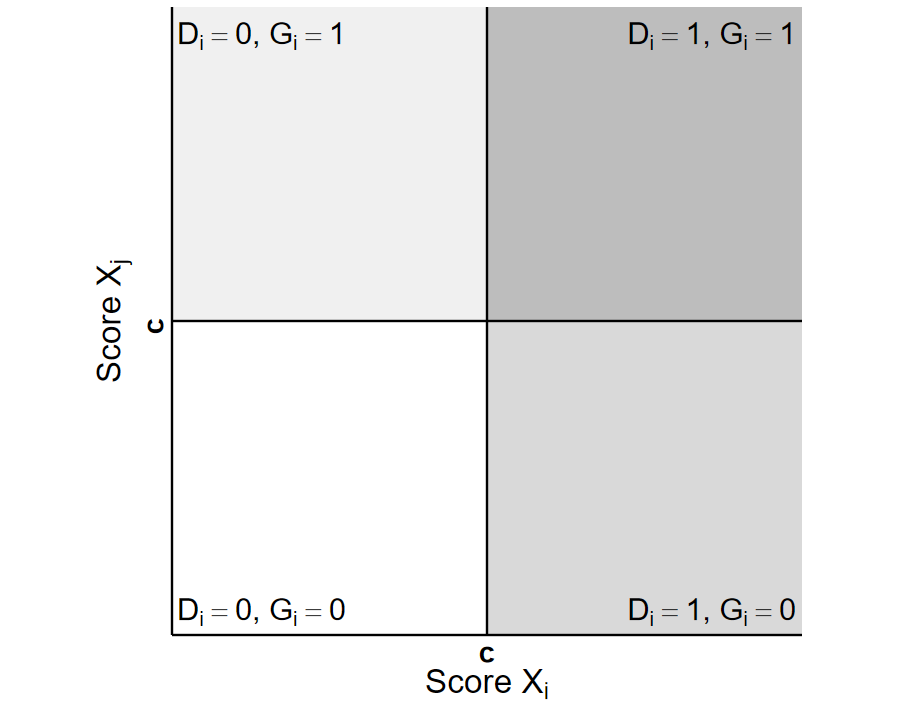}
    \vspace{-1mm}
    \caption{Score space for unit $i$ with interference set $\mathcal{S}_i = \{j\}$ and $G_i = D_j$.}
    \label{fig:oneunit}
\end{figure}
\begin{figure}[t]
    \centering
    \includegraphics[width =0.6\textwidth]{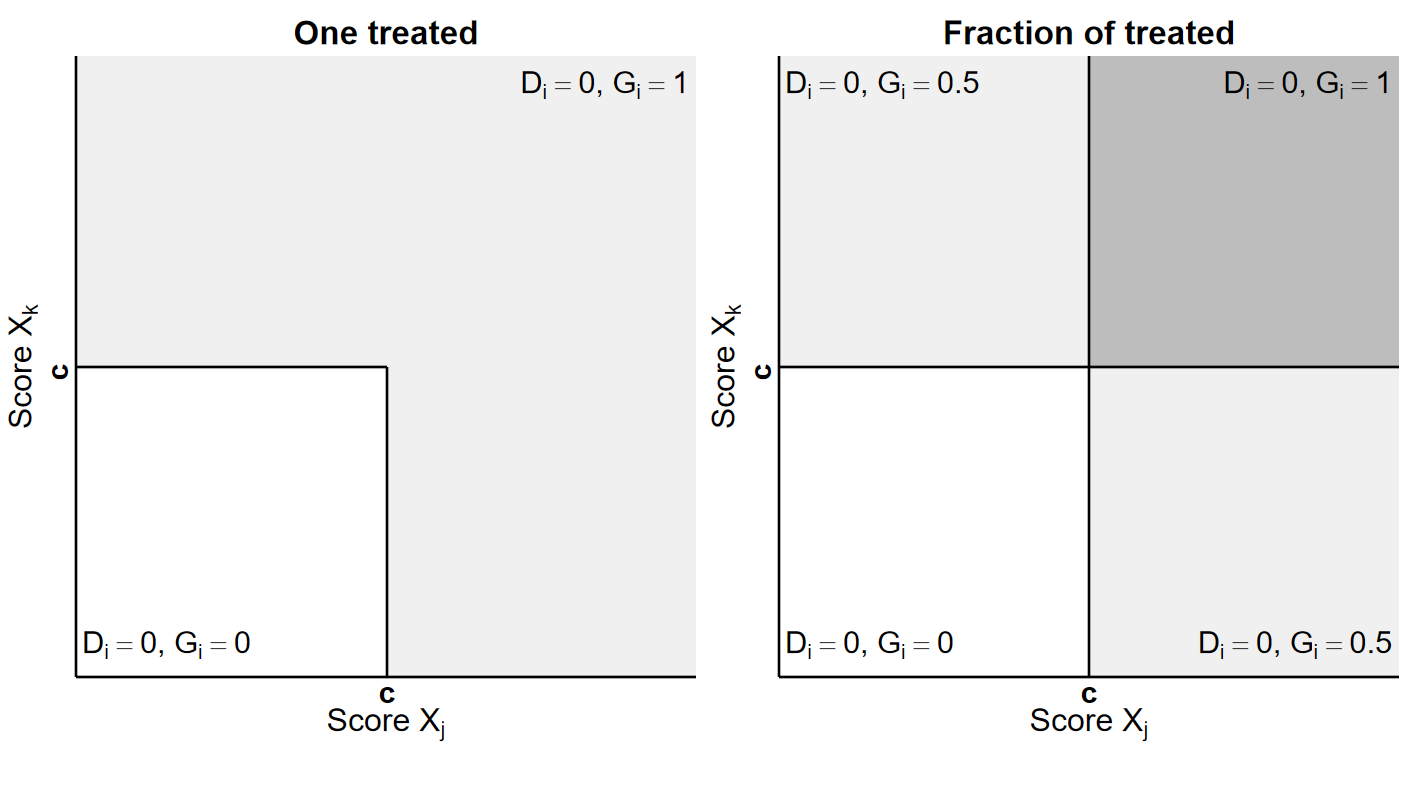}
    \vspace{-5mm}
    \caption{Score space for unit $i$ with interference set $\mathcal{S}_i = \{j, k\}$ fixing $X_i = x_i$ with $x_i < c$.}
\label{fig:twounits}
\end{figure}

{\color{black}
Under the fraction-of-treated mapping used in PROGRESA, the boundary between having
no eligible peers and all eligible peers for an ineligible child is
\[
\bar{\mathcal X}_{i\smallN_i}(0,1\mid0,0)
=
\{X_i\leq c,\;X_j=c\ \text{for every }j\in\mathcal S_i\}.
\]
This follows by applying $\ell(\cdot)$ to the unique treatment vectors associated
with $(D_i,G_i)=(0,1)$ and $(D_i,G_i)=(0,0)$, that is, $\bm d^{0,1}_{i\smallN_i} = (0,\bm 1_{|\mathcal S_i|})$ and $\bm d^{0,0}_{i\smallN_i} = (0,\bm 0_{|\mathcal S_i|})$.
Thus, all peers' scores must be at the eligibility cutoff. Consequently, \(
\bar s_{i, 0,1, 0,0}=|\mathcal S_i|
\), meaning all peers must cross the cutoff for the effective treatment to switch from $(0,0)$ to $(0,1)$. For a child with two peers, the right panel of Figure~\ref{fig:twounits} displays
this boundary, after fixing $X_i=x_i<c$, as the point $(X_j,X_k)=(c,c)$.
By contrast, 
\(
\bar{\mathcal X}_{i\smallN_i}(1,0\mid0,0)
=
\{X_i=c,\;X_j\leq c\ \text{for every }j\in\mathcal S_i\}
\)
has codimension one, as only the focal child $i$'s score $X_i$ must cross the cutoff for the effective treatment to change from $(0,0)$ to $(1,0)$. More generally, all boundaries of the form $\bar{\mathcal X}_{i\smallN_i}(1,g\mid0,g)$ have codimension one: $X_i$ is
fixed at $c$, while $\pscore$ remain in the region inducing $g$.
}

Although \eqref{eq:codimension} is a discrete minimization problem, $\bar{s}_{i,d,g,d'g'}$ can be easily computed for common exchangeable exposure mappings:
\begin{enumerate}[
    label=\alph*),
    ref=\theassumption.\alph*,
    leftmargin=*,
    topsep=0.2em,
    itemsep=0.5pt,
    parsep=0pt,
    partopsep=0pt
]
    \item if $G_i=\sum_{j\in\smallN_i}D_j$, then $\bar{s}_{i,d,g,d',g'}=|g-g'| + |d' - d|$ ;
    \item if $G_i=\frac{1}{|\mathcal{S}_i|}\sum_{j\in\smallN_i}D_j$, then
    $\bar{s}_{i,d,g,d',g'}=|\mathcal{S}_i|\cdot|g-g'| + |d' - d|$. 
    \item if $G_i=\mathbbm{1}(\sum_{j\in\smallN_i}D_j>0)$, then
    $ \bar{s}_{i,d,g,d',g'}
    = | g - g'| + |d - d'|$.
\end{enumerate}
Throughout we will denote the codimension simply by $\bar{s}_i$ to
simplify notation. This quantity will determine the convergence rate of the proposed estimator (Section
\ref{sec:estimation}).

\subsection{Data Generating Process}

We assume that the network \(\bm{A}\) is fixed, fully observed, and unaffected by treatment \(\bm{D}\). The data $\left(\bm{X}, \bm{D},\{Y_i(d, g)\}_{i \in \mathcal{N}, d \in\{0,1\}, g \in \mathcal{G}_i}\right)$ are jointly distributed according to a probability law conditional on the realized network $\bm{A}$, from which we observe one draw for a population of $n$ connected units.
This model-based approach treats potential outcomes as random variables, 
and it is similar to \citet{ogburn22} and \citet{leung2023unconfoundedness}, where the observed data are generated from a random process conditional on the network. 



In this setting, individuals may have different numbers of neighbors, implying a varying dimension of the multiscore spaces and effective treatment boundaries across units. Moreover, both outcomes and scores may not be identically distributed across units due to the network. 
Write potential outcomes as 
\begin{equation}\label{eq:obsinterference} 
Y_i(d,g) = m_{i,d,g}(X_i, \pscore) + \epsilon_{i,d,g}, \quad i = 1,\dots,n, 
\end{equation} 
for $d \in\{0,1\}$ and $g \in \mathcal{G}_i$, where
$
m_{i, d, g}\ppoint =\mathbb{E}_i\left[Y_i(d, g) \mid\left(X_i, \mathbf{X}_{\mathcal{S}_i}\right)=\ppoint, \mathbf{A}\right]
$
denotes the unit-specific conditional expectation function. Let $f_i(x_i,\bm x_{\smallN_i}| \bm A)$ denote the unit-specific joint density of $(X_i, \pscore)$.
{\color{black}
Define $\mathbf S:=(\mathcal S_1,\ldots,\mathcal S_n)$, the collection of interference sets. We summarize relevant interference set characteristics through the finite-dimensional vector
\(
\phi_i:=\phi(\mathbf S,i)\in\boldsymbol\Phi,
\)
where
\(
\phi: \mathfrak S_n\times\mathcal N
\rightarrow\boldsymbol\Phi,
\)
$\mathfrak S_n$ denotes the set of admissible collections of interference sets. We require $\phi_i$ to include the interference set size $|\mathcal S_i|$, and possibly other characteristics of the interference structure.
%
For each $\phi\in\boldsymbol\Phi$, define
$
\mathcal V_\phi:= \{i\in\mathcal N:\phi_i=\phi\}
$. 
%
%
All units $i\in\mathcal V_\phi$ share a common interference set size $|\mathcal S_i|=S_\phi$, and their effective treatment boundaries, indexed by  $\Frontierphi$, are geometrically identical up to a relabeling of the neighbors' score coordinates, with codimension $\bar s_i= \bar s_\phi$.


We next impose statistical homogeneity within structurally comparable strata by requiring units with the same $\phi_i$ to share the conditional outcome functions and joint score distributions.

\begin{assumption}[Network heterogeneity via stratification]\label{ass:networkheterogeneity}
For $i, j \in \mathcal{N}$ such that $\phi_i=\phi_j$, and for all $d, g$, the following holds:
\begin{enumerate}[
    label=\alph*),
    ref=\theassumption.\alph*,
    leftmargin=*,
    topsep=0.2em,
    itemsep=0.5pt,
    parsep=0pt,
    partopsep=0pt
]
\item Outcome homogeneity: $m_{i, d, g}\left(x_i, \boldsymbol{x}_{\mathcal{S}_i}\right)=m_{j, d, g}\left(x_j, \boldsymbol{x}_{\mathcal{S}_j}\right)$.
\item Density homogeneity: $f_i\left(x_i, \boldsymbol{x}_{\mathcal{S}_i}\right)=f_j\left(x_j, \boldsymbol{x}_{\mathcal{S}_j}\right)$.
\end{enumerate}
\end{assumption}


Assumption~\ref{ass:networkheterogeneity} imposes a coarsening of the interference structure: all relevant heterogeneity is summarized by the vector $\phi_i$. For any $i \in \mathcal{V}_\phi$, we can then define the common conditional outcome function, $m_{d, g}(\cdot ; \phi)$, and the common joint density $f(\cdot ; \phi)$.
When $\phi_i=|\mathcal S_i|$, units with equally sized interference sets share conditional outcome functions and score distributions.
More generally, $\phi_i$ may include community membership, centrality, or other features of the interference structure affecting outcomes and scores.
\red{
In our PROGRESA application, we define $\phi_i$ as the interference set size ($\phi_i=|\mathcal S_i|$). Therefore, a stratum $\mathcal{V}_\phi$ groups together children who share the same number of same-village, same-gender, same-grade neighbors. Under this definition of $\phi$, Assumption~\ref{ass:networkheterogeneity}  requires children in the same size-specific stratum to
share the same expected outcome conditional on own and neighbors' scores. 
This may fail if enrollment decisions depend also on specific characteristics of the interference set, 
such as village size or region membership. To mitigate this, we restrict the sample to three regions with similar geographic and economic characteristics; moreover, villages are small by program design, which limits the scope for such heterogeneity. In principle, $\phi_i$ could also include village-size bins or region membership, thereby weakening the assumption at the cost of fewer observations within each stratum.
}}

Assumption \ref{ass:networkheterogeneity} implies that the potential outcome conditional expectation over the effective treatment boundary, $\E_i[Y_i(d,g)\, |\, (X_i, \pscore) \in \Frontier ]$, is invariant within $ \mathcal{V}_\phi$, since its components depends on the unit only through $\phi$. In particular, for any $i\in\mathcal V_\phi$, it can be written as
\begin{equation}\label{eq:outcomeboundary}
\frac{
\displaystyle
\int_{\Frontierphi}
m_{d,g}\!\left(
x_i,
\bm{x}_{\smallN_i,i},
\bm{x}_{\smallN_i};
\phi
\right)
\densityphi\,
d\mathcal{H}^{S_\phi+1-\bar{s}_\phi}
\!\left(x_i,\bm{x}_{\mathcal{S}_i}\right)
}{
\displaystyle
\int_{\Frontierphi}
\densityphi\,
d\mathcal{H}^{S_\phi+1-\bar{s}_\phi}
\!\left(x_i,\bm{x}_{\smallN_i}\right)
}
\end{equation}
where the integral is taken with respect to the $(S_\phi+1-\bar s_\phi)$-dimensional Hausdorff
measure. We denote this common boundary conditional expectation by
\[
\CE{Y_i(d,g)}
{(X_i,\pscore)\in\Frontierphi,\phi_i=\phi}.
\]

\subsection{Stratum-specific Causal Estimands}
We define the stratum-specific \emph{boundary average causal effect} as the difference in the average potential outcomes under different effective treatments at the effective treatment boundary for units with network characteristics $\phi$:
\begin{equation}\label{def:boundeff}
    \boundeff =  \CE{Y_i(d,g) - Y_i(d',g')}{(X_i,\, \pscore) \in \Frontier, \phi} 
\end{equation}
with $d, d' \in \{0,1\}$, $g$,$g' \in \mathcal{G}_i$. This effect can be represented an integral over the effective treatment boundary of $\tau_{d,g | d',g'}(x_i, \bm x_{\smallN_i}; \phi) = \mplusphi - \mminusphi$, the average causal effect of the effective treatment $(d,g)$ compared to $(d',g')$ at a point $\ppoint$ for a unit in $\mathcal{V}_\phi$. 

%


We can further distinguish between boundary direct and indirect average effects. The stratum-specific \emph{boundary direct average effect} is defined as
\begin{equation}
\tau_{1,g\,|\,0,g}\big(\bar{\mathcal{X}}_{i\smallN_i}(1,g\,|\,0,g); \phi \big) = \CE{Y_i(1,g) - Y_i(0,g)}{(X_i\,,\pscore) \in  \bar{\mathcal{X}}_{i\smallN_i}(1,g\,|\,0,g), \phi}
\end{equation}
for $g \in \mathcal{G}_i$, isolates the (local) impact of the individual treatment for a fixed level of $G_i$ within $\mathcal{V}_\phi$. In contrast, the stratum-specific \emph{boundary indirect effect}, given by 
\begin{equation}
\tau_{d,g\,|\,d,g}(\bar{\mathcal{X}}_{i\smallN_i}(d,g\,|\,d,g'); \phi) = \CE{Y_i(d,g) - Y_i(d,g')}{(X_i\,,\pscore) \in  \bar{\mathcal{X}}_{i\smallN_i}(d,g\,|\,d,g'), \phi}
\end{equation}
for $d \in \{0,1\}$, $g$, $g'$ $\in \mathcal{G}_i$ captures the indirect effects of changing the neighborhood treatment $G_i$ for a fixed level of the individual treatment within $\mathcal{V}_\phi$.
For example, in our empirical application in Section \ref{sec:empirical},  if $\phi$ corresponds to units' degree, the direct boundary effect $\tau_{10|00}(\bar{\mathcal{X}}_{i\smallN_i}(10|00); \phi)$ captures the impact of being eligible for the PROGRESA/Oportunidades program for individuals with a given degree who are at the poverty threshold and  have friends that are below the threshold. This effect reveals whether the program directly benefits children whose peers are situated below the eligibility cutoff.
In contrast, the indirect boundary effect $\tau_{01|00}(\bar{\mathcal{X}}_{i\smallN_i}(01|00))$ measures the effect of having all eligible friends versus no eligible friends, for ineligible individuals whose peers are at the threshold. This estimand helps assess whether implementing such a program indirectly benefits those who are themselves ineligible but connected to eligible peers.

\subsection{Pooled Causal Estimands}\label{sec:pooledestimand}
We define a pooled boundary causal effect which aggregates stratum-specific effects across strata into a summary measure:
\begin{equation}\label{eq:pool}
\boundeffpooled = \frac{\sum_{\phi \in \Phi} \boundeff \, \cdot \limdensity \cdot \pi_{\phi}}
     {\sum_{\phi \in \Phi} \limdensity \cdot \pi_\phi}    
\end{equation}
where $\pi_\phi = |\mathcal{V}_\phi|/n$ is the proportion of units with network characteristic $\phi$, and $\limdensity = \int_{\Frontier} \densityphi  d\mathcal{H}^{S_\phi + 1 - \bar{s}_\phi}$ is the boundary density.
This estimand assigns greater weight to strata with higher boundary density, thereby accounting for the likelihood that units lie close to their corresponding boundaries. It can be interpreted as the boundary effect for a randomly drawn unit from the network, conditional on being near the effective treatment boundary. 
{\color{black}
Specifically, let $I$ be a random index drawn uniformly from $\{1,\dots,n\}$, then it can be shown that 
\(
\boundeffpooled = \mathbb{E}\big[Y_I(d,g) - Y_I(d',g') \,\big|\, (X_I, \bm X_{\smallN_I}) \in \mathcal{X}_{I\smallN_I}(d,g\, | d',g')\big]
\)
}
}
(Theorem \ref{th:randomsampling}).
In this sense, it provides a network analog of the standard RDD estimand studied in \citep{lee2008}, \red{and thus can be seen as the boundary effect for a representative node}. Similar weighted estimands have also been proposed in spatial settings \citep[e.g.][]{pollmann}.\footnote{An alternative aggregation is the simple average $\sum_{\phi}\boundeff\cdot \pi_\phi$, analogous to approaches used in difference-in-differences under interference \citep{xu2024differenceindifferencesinterference}. The two coincide when boundary densities are equal across strata. We discuss this alternative in Appendix \ref{appendix:heterogeneous}.}
Clearly, when units are homogeneous across $\phi$, the pooled estimand reduces to the common boundary effect $\boundeff$, which no longer depends on $\phi$. 
We also distinguish between pooled boundary direct effects $\tau_{1,g,|,0,g}(\mathcal{X}_{i\smallN_i}(1,g\,|\,0,g); \phi)$ and indirect effects $\tau_{d,g,|,d,g'}(\mathcal{X}_{i\smallN_i}(d,g\,|\,d,g'); \phi)$. 
\red{In PROGRESA, for instance, we consider the pooled direct effect with no eligible neighbors for units up to four neighbors.
This estimand represents the direct effect of eligibility for a randomly drawn unit with one to four neighbors, conditional on being near the relevant boundary and having no eligible neighbors. The pooled indirect effect is defined analogously and summarizes the effect of having all rather than no eligible neighbors among ineligible units with up to four neighbors.}

\section{Identification}\label{sec:identification}

Due to lack of overlap, at least one of the two potential outcome conditional expectations defining $\boundeff$  is not directly observed at the effective treatment boundary, and must therefore be recovered via extrapolation.
We assume that {within each stratum $\mathcal{V}_\phi$}, units near the boundary with similar own and neighbors' scores have similar average potential outcomes. We formalize this by imposing a continuity assumption on {$\mplusphi$}, which enables us to use observations from one side of the boundary to recover the unobserved conditional expectation.

Let $l(p,q)$ be the Euclidean distance between points $p$ and $q$. The minimum distance of a point $\ppoint \in \scorespace$ from the effective treatment boundary $\Frontierphi$ is $l_{d,g|d',g'}^{min}(x_i, \bm{x}_{\smallN_i}) = \min_{\Frontierphi} l\big(\ppoint, (\bar{x}_i, \bar{\bm{x}}_{\smallN_i})\big)$. Let
$$\eballboundphi = \{\ppoint \in \scorespace : l_{d,g|d',g'}^{min}(x_i, \bm{x}_{\smallN_i}) \leq \epsilon\}$$
be its $\epsilon$-neighborhood. Finally, let $\eballboundrightphi$ and $\eballboundleftphi$ be the intersections of $\eballboundphi$ with the effective treatment regions $\scorespace(d,g)$ and $\scorespace(d',g')$, respectively.
We impose the following assumptions:

\begin{assumption}[Identification]\label{ass:identification}
Assume that $X_i$ has bounded support.
For all {$i \in \mathcal{V}_\phi$}, for all $d$, $d' \in \{0,1\}$ and for all $g$, $g'\in \mathcal{G}_i$: 
\begin{enumerate}[
    label=\alph*),
    ref=\theassumption.\alph*,
    leftmargin=*,
    topsep=0.2em,
    itemsep=0.5pt,
    parsep=0pt,
    partopsep=0pt
]
\item\label{ass:pos} Score density positivity: ${\densityphi} >0$ for all $(x_i, \bm{x}_{\smallN_i}) \in \eballbound$.

\item\label{ass:cont} 
 Outcome continuity: {$\mplusphi$} is continuous at all $(x_i, \bm{x}_{\smallN_i}) \in \scorespace$.  
 
 \item\label{ass:densitycont} Score density continuity: {$\densityphi$} is continuous at all $(x_i, \bm{x}_{\smallN_i}) \in \scorespace$. 

\item\label{ass:bounded}Boundedness:
{$\big|\mplusphi\big|$} and {$\densityphi$} are bounded.
\end{enumerate}
\end{assumption}

\noindent Combined with Assumptions \ref{ass:consistency}-\ref{ass:indrule}, Assumption \ref{ass:identification} identifies the average causal effect at the effective treatment boundary\footnote{ A limitation of this continuity-based identification is that it can only identify causal estimands local to the boundary. This locality is a feature of RDDs, which cannot identify broader effects without stronger assumptions beyond continuity. }. 
\begin{theorem}[Identification]\label{th:cont}
 Suppose assumptions \ref{ass:consistency}-\ref{ass:identification} hold. Then for all $d$, $d'\in \{0,1\}$, $g$, $g'\in \mathcal{G}_i$
\begin{equation}
\begin{split}
\label{eq:limitboundaryeffects}
        \boundeff &= \lim_{\epsilon \to 0} \CE{Y_i}{(X_i,\pscore) \in \eballboundrightphi; \, \phi } \\
        & - \lim_{\epsilon \to 0}\CE{Y_i}{(X_i, \pscore) \in \eballboundleftphi; \, \phi } 
        \end{split}
\end{equation}
Proof. \textup{See Supplementary Material \ref{appendix:stratumid}.}
\end{theorem}

\noindent Theorem \ref{th:cont} identifies the boundary average causal effect {within stratum $\mathcal{V}_\phi$} by taking the difference between the limits of the conditional expectations of the observed outcomes on $\eballboundrightphi$ and $\eballboundleftphi$ as $\epsilon \to 0$. This result relies on Assumptions \ref{ass:identification}. Assumption \ref{ass:pos} ensures well-defined conditional expectations via a non-null score density near the boundary. Assumption \ref{ass:cont}) extends the continuity requirement from standard multivariate RDDs and enables the approximation of unobserved potential outcome expectations at the boundary using sufficiently close observations. It implies that individuals {with the same network characteristics}, and similar own and neighbor scores have similar expected outcomes. In our application, when estimating the effect of having all versus no neighbors eligible among the ineligible, it means that individuals with the same poverty index and neighbors just above or below the cutoff have comparable expected outcomes. 
Assumption \ref{ass:densitycont}) ensures continuous density weights at the boundary, preventing shifts in the baseline score variables distribution from spuriously driving differences in expected outcomes
Finally, Assumption \ref{ass:bounded}) is a technical condition allowing the interchange of the limit and integral operators.
Identification of the pooled causal effects follows as in Theorem \ref{th:cont} by extending Assumption \ref{ass:identification} to hold for all $\phi \in \Phi$. We refer to Appendix \ref{onlineappendix:irregular} for more details.   

\section{Estimation}\label{sec:estimation}

In standard RDDs, the average causal effect at the cutoff is typically estimated using local polynomial regression, which fits $Y_i$ on a low-order polynomial of the score on each side of the cutoff using observations within a bandwidth, denoted by $h_n$ \citep{hahn}. However, estimation in our setting is more complex due to multidimensional effective treatment boundaries.
A popular method is the distance-based approach, which replaces the multivariate score with each unit's shortest Euclidean distance to the treatment boundary \citep{dell10, keele_titiunik_2015}. Unlike frontier estimation, which subsets the data by boundary region \citep{matsudaira}, it uses the whole data along the boundary; unlike the binding approach, which computes a summary score \citep{rear}, it applies to general boundary shapes; Moreover, its local approach avoids the misspecification pitfalls of a global model.
We adapt the distance-based local polynomial estimator to our framework to estimate boundary average effects. We show that the convergence rate of distance-based local polynomial estimator for the stratum-specific boundary effects is determined by $\bar{s}_\phi$, which depends on the causal estimand.
{We extend our analysis to estimation of the pooled boundary estimands and introduce a pooled estimator that pools all observations based on their distance to the respective boundary without conditioning on network characteristics.}


\subsection{Distance-based Estimator}
We introduce the distance-based estimator for $\boundeff$.
Let $c = 0$ without loss of generality; let $\distvar = l_{d,g|d',g'}^{min}(X_i, \pscore)$ be minimum Euclidean distance from {its own} $\Frontier$ of a random point $(X_i, \bm{X}_{\smallN_i})$, with realization $\tilde{x}_{d,g,d',g'}$.
{Let $m(X_i, \pscore; \phi)$ denote the observed outcome conditional expectation on the own and neighborhood score for $i \in \mathcal{V}_\phi$, where $m(X_i, \pscore; \phi) =\sum_{d,g}\mathbbm{1}((X_i, \pscore) \in \scorespace(d,g))\cdot m_{d,g}(X_i, \pscore; \phi)$ under consistency, and define $\epsilon_i$ analogously from $\epsilon_{i,d,g}$
Estimation of $\boundeff$ will not rely on the direct estimation of {$m(X_i, \pscore; \phi)$}, but instead {$\mu(\distvar; \phi ) = \CE{Y_i}{\distvar, D_i, G_i; \phi}$}, the observed outcome expectation conditional on the distance variable and $(D_i, G_i)$.
{Within $\mathcal{V}_\phi$, the observed outcome is}
\begin{equation}
   {Y_i =  \mu(\distvar; \phi) + \error, \quad i \in \mathcal{V}_\phi} 
\end{equation}
where $\error =  m(X_i, \pscore; \phi) - \mu(\distvar; \phi) + \epsilon_i$.

For convenience, let us denote $\distvar$ by $\distvari$ when it is not ambiguous. The target distance-based outcome regression functions are 
$\muplus$ and
$\muminus$. 
Let {$\muplus[0] = \lim\limits_{\tilde{x}\to  0^+}\muplus[\tilde{x}]$ and $\muminus[0] = \lim\limits_{\tilde{x} \to  0^+}\muminus[\tilde{x}]$}, where $\tilde{x} = 0$ represents the effective treatment boundary. 
We estimate these limits by local polynomial regressions in $\distvari$ near zero.
Setting {$\beta_{d,g}(0; \phi) = [\muplus[0], \muplusderiv[(1)]]^T$ and $\beta_{d',g'}(0; \phi) = [\muminus[0], \muminusderiv[(1)]]^T$, the distance-based local linear estimators for $\beta_{d,g}(0, \phi)$ and $\beta_{d',g'}(0;\phi)$} are obtained as
\begin{equation}\label{eq:distrss}
    \begin{split}
   & {\hat{\beta}_{d,g}(h_n, \phi)} = \argmin\limits_{\bm{b} \in \mathbbm{R}^2}\sum_i{\mathbbm{1}(D_i = d, G_i = g, \phi_i = \phi)}(Y_i - {\bm r(\distvari)}^T\bm{b})^2\cdot K_{h_n}(\distvari) \\
    &{\hat{\beta}_{d',g'}(h_n; \phi)} = \argmin\limits_{\bm{b}\in \mathbbm{R}^2}\sum_i{\mathbbm{1}(D_i = d', G_i = g', \phi_i =  \phi)}(Y_i - {\bm r(\distvari)}^T\bm{b})^2\cdot K_{h_n}(\distvari)
    \end{split}
\end{equation}
\sloppy
\noindent where {$r(u) = [1 , u]$}, $h_n$ is a positive bandwidth and $K_{h_n}(u) = K(u/h_n)/h_n$ for a given kernel function $K(\cdot)$. 

\sloppy{Now let us set $\bm{Y} = [Y_1,..., Y_n]^T$ and $\tilde{\bm{X}}_{n} = [\tilde{X}_{1},..., \tilde{X}_{n}]^T$, and define $\tilde{X}(h_n) = [\bm{r}(\tilde{X}_{1}/h_n),..., \bm{r}(\tilde{X}_{n}/h_n)]$, i.e., the matrix containing the linear expansion of the distance variable, and the diagonal weighting matrices {$\Wplus$} with diagonal element {$w_{i,d,g}(h_n) = \mathbbm{1}(D_i = d, G_i = g, \phi_i = \phi)\cdot K_{h_n}(\distvari)$}. 
Finally, let {${\Gammaplusmain = \tilde{X}(h_n)^T\Wplus\tilde{X}(h_n)/n_\phi}$}
where $n_\phi$ is the number of units in $\mathcal{V}_\phi$.} Corresponding quantities for $(d',g')$ are defined analogously. 
Letting $H(h_n) = \text{diag}(1, h_n^{-1})$, for $h_n > 0$, 
the solutions to \eqref{eq:distrss} are
\begin{equation}\label{eq:betaestimators}
    \begin{split}
&{\hat{\beta}_{d,g}(h_n; \phi) = H(h_n)\Gammaplusmain[-1]\tilde{X}(h_n)^T\Wplus\bm{Y}/n_\phi}\\ 
&{\hat{\beta}_{d',g'}(h_n; \phi) = H(h_n)\Gammaminusmain[-1]\tilde{X}(h_n)^T\Wminus\bm{Y}/n_\phi}
\end{split}
\end{equation}
\noindent The final estimator for $\boundeff$ is the difference in the estimators of the intercepts of the distance-based outcome regression functions 
\begin{equation}\label{eq:estimatormain}
    {\estimator = \hat{\mu}_{d,g}(h_n; \phi) - \hat{\mu}_{d',g'}(h_n; \phi)}
\end{equation} 
where $\hat{\mu}_{d,g}(h_n; \phi) = e_1^T\hat{\beta}_{d,g}(h_n; \phi)$ and $\hat{\mu}_{d',g'}(h_n; \phi) = e_1^T\hat{\beta}_{d',g'}(h_n; \phi)$ with $e_1 = [1,0]^T$.

\subsection{Asymptotic Theory}\label{subsec:asy}

Statistical dependence within observations may arise from $G_i$ due to overlapping interference sets, and from network-induced correlation 
{We model this dependence for the collection of random variables $O_i = (\{Y_i(d,g)\}_{d,g}, X_i, \pscore, D_i, G_i)$, $i = 1,...n$, using \emph{dependency neighborhoods}, determined by the network $\bm{A}$}.
The dependency neighborhood of a unit $i$ is a collection of indices of observations potentially dependent on $O_i$. Formally, a collection of random variables $\{Z_i\}_{i = 1}^n$ has dependency neighborhoods $\bm{N}_{i} \subseteq \{1,...,n\}$, $i = 1,\ldots, n$, if $i \in \bm{N}_{i}$, and $Z_i$ is independent of $\{Z_j\}_{j \notin \bm{N}_{i}}$ \citep{ross11}. {We take $\bm N_i$ to be a function of the network $\bm A$, so that the entire dependence structure is induced by $\bm A$. This rules out additional sources of dependence that are not mediated by the network, such as global shocks affecting all units simultaneously.}

Dependency neighborhoods can be represented by a \emph{dependency graph} $\bm{W}$, a binary $n\times n$ matrix with entry $W_{ij} = 1$ if $j \in \bm{N}_{i}$ and $W_{ij} = 0$ otherwise. The dependency graph determines the amount of independent information available in the data as for any two disjoint subsets $I_1, I_2 \subseteq \{1,...,n\}$ where $W_{ij} = 0$ for all $i \in I_1$ and $j \in I_2$ we have $\{O_i : i \in I_1\} \indep \{O_j : j \in I_2\}$ by construction. Our asymptotic results are agnostic about the exact form of $\bm{W}$, and only rely on restrictions on its degree distribution. However, inference requires specifying $\bm{W}$ by mapping the assumed data-generating process and interference mechanism onto a dependency structure. 
Under network neighborhood interference, if $(X_i, \varepsilon_{i,d,g})$ are i.i.d. across units, two units are dependent whenever their neighborhoods overlap, since outcomes depend on $G_i$ and $\pscore$. Thus, $\bm N_i$ contains all units within two links of $i$, and $\bm W$ can be constructed by setting $W_{ij}=1$ if there exists a path of length at most two between $i$ and $j$, i.e., if $\big((\bm I + \bm A + \bm A^2)\big)_{ij} > 0$.
More generally, one can allow $(X_i,\varepsilon_{i,d,g})$ and $(X_j,\varepsilon_{j,d,g})$ to be dependent only between units that are neighbors or share a neighbor \citep{leung20, ogburn22}. 
Combined with overlapping interference sets, this dependence structure implies 
that $\bm W_{i j}=1$ if there exists a path of length at most four between $i$ and $j$, that is, if $\sum_{k=0}^4\left(\boldsymbol{A}^k\right)_{i j}>0$.
In clustered settings with partial interference, as in our PROGRESA application, one may allow arbitrary dependence within clusters and independence across clusters, leading to a block-diagonal $\bm W$.

\paragraph{Assumptions}
We consider an asymptotic framework where $n\to \infty$. Because all causal estimands and estimators are conditioned on $\bm{A}$, our asymptotic framework assumes a sequence of networks $\bm{A}_n$, preserving fundamental aspects of the network topology \citep{ogburn22}. Let $\bm W_{n}$ be the associated dependency graph sequence. We limit the asymptotic growth of data dependence via moment restrictions on $\bm W_n$.

\begin{assumption}[Local Dependence]\label{ass:localdependence}
$n^{-1}\sum_i |\bm{N}_{i,n}|^3$ and $n^{-1}\sum_i\sum_{j\neq i} (\bm{W}_n^3)_{ij}$ are bounded.
\end{assumption}
\noindent  The quantity $n^{-1}\sum_i |\bm{N}_{i,n}|^3$ is the empirical third moment of the degree distribution of $\bm{W}_n$ whereas $n^{-1}\sum_{j\neq i} (\bm{W}_n^3)_{ij}$ is the average number of walks of length 3 emanating from unit $i$. Bounding these quantities ensures that most units have a sufficiently small degree in the dependency graph with respect to the sample size, 
as $n$ becomes large. Assumption \ref{ass:localdependence} is analogous to Assumption 4 in \citet{leung20}, although here the dependency graphs are fixed. {It is satisfied under network neighborhood interference for networks with bounded degrees and can accommodate networks with unbounded degrees, such as those arising from the strategic network formation model in \citet{leung20}.}

{\color{black}
We next introduce some regularity assumptions.
For generic $(d,g)$, define the conditional variance
$
\sigmapluseps
=
\V\!\left[
Y_i(d,g)\mid (X_i,\pscore)=\ppoint;\phi
\right]$
Let $\widetilde{\mathcal S}_i=\mathcal S_i\cup\{i\}$, and define the joint score as
$\bm X_{ij}=\bm X_{\widetilde{\mathcal S}_i\cup\widetilde{\mathcal S}_j}$ the vector collecting the scores of units in $\tilde{\mathcal{S}}_i \cup \tilde{\mathcal{S}}_j$. Define its dimension as
\(
q_{ij}
:=
\left|
\widetilde{\mathcal S}_i\cup\widetilde{\mathcal S}_j
\right|
\),
where
\(
q_{ij}\in\mathcal Q_\phi
:=
\{S_\phi+1,\ldots,2S_\phi+2\}.
\)
Let $f_{ij}(\cdot;\phi)$ denote the density of $\bm X_{ij}$. For
$a,b \in\{(d,g),(d',g')\}$, define
\(
\kappa_{ij}^{a,b}(\bm x_{ij};\phi)
=
\C\!\left[
Y_i(a),Y_j(b)\mid \bm X_{ij}=\bm x_{ij};\phi
\right].
\)
Conditioning on \(\bm X_{ij}\) accounts for shared and non-shared score components when the interference sets overlap. Both $f_{ij}(\bm x_{ij}; \phi)$ and $\kappa_{ij}^{a,b}(\bm x_{ij};\phi)$ have thus domains of dimension $q_{ij}$. 
For $i,j\in\mathcal V_\phi$, define their \qq joint boundary" as
\[
\mathcal X_{ij}^{\phi}(d,g\,|\,d',g')
:=
\left\{
\bm x_{\widetilde{\mathcal S}_i\cup\widetilde{\mathcal S}_j}
\in\mathcal X_{\widetilde{\mathcal S}_i\cup\widetilde{\mathcal S}_j}:
\bm x_{\widetilde{\mathcal S}_i}\in\Frontierphi,
\;
\bm x_{\widetilde{\mathcal S}_j}\in\bar{\mathcal X}^{\phi}_{j\smallN_j}(d,g\,|\, d',g')
\right\},
\]
Let
\(
\mathcal K_i^\phi
=
\left\{
j\in(\bm N_i\setminus\{i\})\cap\mathcal V_\phi:
\operatorname{codim}(\mathcal X_{ij}^{\phi}(d,g\,|\,d',g')
)=\bar s_\phi
\right\}
\)
denote the set of dependent units with $i$ whose joint boundary has minimal codimension. Finally, let us define the approximation error as $\eta_a(x_i, \bm x_{\smallN_i};\phi)
:= m_{a}(x_i, \bm x_{\smallN_i}; \phi) 
-\E[ Y_i(a)\, |\, (X_i, \pscore) \in \Frontier; \phi]$,
and the covariance integral over the joint boundary, letting $\gamma_{a,b,ij}(x_{ij}, \phi) = \kappa_{ij}^{a,b}(x_{ij}; \phi) + \eta_a(x_i, \bm x_{\smallN_i};\phi) \eta_b(x_j, \bm x_{\smallN_j};\phi)$:
\[
\gamma_{a,b,n}(0;\phi)
=
\frac{1}{n_\phi}
\sum_{i\in\mathcal V_\phi}
\sum_{j\in\mathcal K_i^\phi}
\int_{\mathcal X_{ij}^{\phi}(d,g\,|\,d',g')
}
\gamma_{a,b,ij}(x_{ij}, \phi)f_{ij}(x_{ij};\phi)
\,d\mathcal H^{q_{ij}-\bar s_\phi}(x_{ij}).
\]

\begin{assumption}[Estimation regularity]\label{ass:regularitymain}

\begin{enumerate}[ label=(\roman*), ref=\theassumption.\roman*, leftmargin=*, topsep=0.2em, itemsep=0.2em ]

\item For some $\kappa>0$, the following holds for all
$\ppoint\in\eballboundphi[\kappa]$ and $i\in\mathcal V_\phi$:
\begin{enumerate}[ label=\alph*), ref=\theenumi.\alph*, leftmargin=1em, labelsep=0.4em, topsep=0.1em, itemsep=0pt, parsep=0pt, partopsep=0pt ]
    \item
    $\CE{Y_i^4}{X_i=x_i,\ \pscore=\bm x_{\smallN_i};\phi}$ is bounded.\label{ass:fourthmomentmain}
    \item
    $\sigmapluseps$ and $\sigmaminuseps$ are continuous and bounded away from zero.\label{ass:variancesmain}
    \item
    $\muplus$ and $\muminus$ are at least $3$-times continuously differentiable.
    \label{ass:smoothnessmain}
\end{enumerate} 
\item
For each $q\in\mathcal Q_\phi$, the collections
\(\{\kappa_{ij}^{a,b}(\cdot;\phi)\}_{i,j\in\mathcal V_\phi:\,q_{ij}=q}\)
 and
\(\{f_{ij}(\cdot;\phi)\}_{i,j\in\mathcal V_\phi:\,q_{ij}=q}\)
are uniformly bounded and equicontinuous, for
$a,b\in\{(d,g),(d',g')\}$, and  $\gamma_{a,b,n}(0;\phi) \to \gamma_{a,b}(0;\phi)$ for $n_\phi \to \infty$.\label{ass:covariancesmain}
\end{enumerate}
\end{assumption}
\noindent Part~\ref{ass:fourthmomentmain} imposes existence of moments. Part~\ref{ass:variancesmain} extends the usual
conditional variance regularity condition for local-polynomial RDD
estimation to hold near the boundary. Part~\ref{ass:covariancesmain} 
imposes equicontinuity of the joint score densities and conditional covariances within each joint score dimension, and requires the joint boundary covariance contribution to converge to a given limit.
It has an analogous role as Part~\ref{ass:variancesmain} for characterizing the asymptotic limit for the estimator's variance.}

{\color{black}
Part~\ref{ass:smoothnessmain} warrants further discussion. It assumes smoothness of the
distance-based outcome regressions, analogous to geographic RDDs \citep{keele_titiunik_2015},
and treats units similarly close to the boundary but on opposite sides as valid counterfactuals. However,
units with very different own- and neighbor-score compositions may have the same distance, so a
given distance aggregates heterogeneous boundary ``locations'' ---here configurations
$(X_i, \pscore)$ rather than spatial points. In PROGRESA, the distance for the indirect effect
$\tau_{01|00}(\bar{\mathcal{X}}_{i\smallN_i}(01|00);\phi)$ groups units with neighbors near $c$
but own scores anywhere below, while the distance for the direct effect
$\tau_{10|00}(\bar{\mathcal{X}}_{i\smallN_i}(10|00);\phi)$ groups units with $X_i$ near $c$ but
neighbors anywhere below. 

The assumption may fail if reporting or administrative practices induce sorting along particular portions of the boundary, as own and peer scores may correlate with socioeconomic factors affecting score manipulation across peer groups. For instance, with network homophily, better-off and better-informed households may help connected households strategically
report economic circumstances, while officials may exercise discretion based on household
connections. Children close to the boundary but under different effective treatments could then
differ systematically in own- and peer-score distributions (e.g. 
less-poor ineligible children clustering with peers just above the
cutoff), and in outcomes when these conditions affect enrollment.
The assumption is more plausible when households have limited scope to sort strategically around
the cutoff. PROGRESA's centralized census-based classification, undisclosed census purpose, and
limited community revision of eligibility reduced this risk. Plausibility also depends on the score range, since a narrower range implies more similar configurations at equal distance.
%
Smoothness can also fail absent any discontinuity: if the enrollment slope in own and neighbor
scores shifts abruptly (e.g.\ once liquidity constraints bind), the distance regressions stay
continuous but need not be differentiable. We verify Assumption~\ref{ass:smoothnessmain} from differentiability of $m_{d,g}(\cdot;\phi)$ and $f(\cdot;\phi)$ for a generic boundary
effect, in Supplementary Material \ref{appendix:differentiability}. Its plausibility can be assessed through covariate-balance tests across the boundary (Section~\ref{sec:empirical}) and by checking whether one-sided regressions show stable slopes and curvature across bandwidths, possibly localized to boundary segments (e.g. low- versus high-poverty peers).}

\paragraph{Central limit theorem}
Define {$\limdensity = \int_{\Frontierphi} \densityphi  d\mathcal{H}^{S_\phi + 1 - \bar{s}_\phi}$, where we recall $\bar{s}_\phi$ is the common codimension of units in $\mathcal{V}_{\phi}$}. Additionally, let $\omegaplus$ denote the integral over the boundary of the conditional variance of 
plus the squared approximation error, 
Eq. \eqref{eq:omegaexp} in Supplementary Material). Define $\omegaminus$ analogously but replacing $(d,g)$ with $(d',g')$. 
\begin{theorem}\label{th:asynormal}
\sloppy
\textcolor{black}{Suppose Assumptions~\ref{ass:consistency}--\ref{ass:regularitymain} and \ref{ass:kernel} (kernel regularity) hold.} If {$n_{\phi} h_n^{\bar{s}_{\phi}}$}$ \to \infty$, $h_n \to 0$, and 
${h_n = O\!\left(n_\phi^{-1/(4 + \bar{s}_\phi)}\right)}$
\[
\sqrt{n_\phi h_n^{\bar{s}_\phi}}\,
\frac{\big(\estimator - \boundeff\big) - h_n^{2}{B}_{d,g,d'g'}(h_n; \phi)}
{\sqrt{{V}_{d,g,d',g'}(h_n; \phi)}}
\xrightarrow{d} \mathcal{N}(0,1),
\]
where ${B}_{d,g,d'g'}(h_n; \phi)
= B^{\phi} + o(1)$, with 
\(
B^\phi = \Big(\tfrac{\muplusderiv[(2)]-\muminusderiv[(2)]}{2!}\Big)e_1^\top (\Gamma^{\bar{s}_\phi})^{-1}\theta^{\bar{s}_\phi}
\),
and ${V}_{d,g,d',g'}(h_n; \phi) = V^{\phi} + o(1)$, {\color{black}
\[
V^\phi = \Big(\frac{\omegaplus + \omegaminus + \bar \gamma_{d,g,d',g'}(0; \phi)}{\limdensity^2}\Big)e_1^\top (\Gamma^{\bar{s}_\phi})^{-1}\Psi^{\bar{s}_\phi}(\Gamma^{\bar{s}_\phi})^{-1} e_1 
\]
letting $\bar \gamma_{d,g,d',g'}(0; \phi) = \gamma_{d,g,d,g}(0;\phi) + \gamma_{d',g',d',g'}(0;\phi) - 2\gamma_{d,g, d',g'}(0;\phi)$.
}
The exact form of $\Gamma^{\bar{s}_\phi}$, $\theta^{\bar{s}_\phi}$ and $\Psi^{\bar{s}_\phi}$ is given in Supplementary Material \ref{appendix:estimator}, {page \pageref{eq:kernels}}.\\
Proof. \textup{See Supplementary Material~\ref{appendix:estimator}.} 
\end{theorem}


The asymptotic bias $h_n^{2}B^\phi$ depends on the second-order derivatives of the target outcome regression function, reflecting the linearization error; its contribution to the asymptotic distribution of $\estimator$ vanishes provided that $n_\phi h_n^{\bar{s}_\phi+4}\to 0$. 
The asymptotic variance $(n_{\phi}h_n^{\bar{s}_\phi})^{-1}V^\phi$ has convergence rate governed by the dimension of the effective treatment boundary through the parameter {$\bar{s}_\phi$}, i.e., the number of score components $\ppoint$ fixed at the cutoff in $\Frontierphi$. This arises because the volume of the boundary neighborhoods
 $\mathcal{B}_{h_n}^{d,g}(\Frontierphi)$ and $\mathcal{B}_{h_n}^{d',g'}(\Frontierphi)$ is proportional to {$h_n^{\bar{s}_\phi}$}, as shown in the proof of Theorem \ref{th:cont}. Consequently, the probability that an observation falls within such neighborhoods is proportional to {$h_n^{\bar{s}_\phi}$}, and thus, the estimator effectively uses approximately {$n_\phi h_n^{\bar{s}_\phi}$} observations.

As a consequence, the convergence rate of $\estimator$ exhibits a curse of dimensionality driven by the parameter $\bar{s}_\phi$. Importantly, this curse does not depend on the dimension of the multiscore space, i.e., $\mathcal{S}_\phi+1$, but rather on $\bar{s}_\phi$, which captures the number of components of the score vector $(X_i,\pscore)$ that must simultaneously fall within a bandwidth $h_n$ of the cutoff in order for an observation to contribute to the estimator.
To illustrate, consider the indirect effect comparing potential outcomes under zero versus all treated neighbors, for untreated units with two neighbors. The estimator assigns positive weight only to observations within  $\mathcal{B}_{h_n}^{d,g}(\Frontierphi)$ or $\mathcal{B}_{h_n}^{d',g'}(\Frontierphi)$), that is, with $\tilde{X}_i \leq h_n$ and under either effective treatment assignment. This condition requires that both neighbors' scores lie within the bandwidth—i.e., $||\pscore - c|| < h_n$—and that $X_i - c < 0$. The effective treatment boundary in this case has dimension $S_\phi + 1 - 2 = 1$, with $\bar{s}_\phi = 2$, and the convergence rate is $(n_\phi h_n^2)^{-1/2}$.
When instead only one component must be near the cutoff (i.e., $\bar{s}_\phi = 1$), the estimator averages over approximately $n_\phi h_n$ observations and achieves the standard univariate rate $(n_\phi h_n)^{-1/2}$ \citep{cct14}. This is the case for boundary direct effects $\tau_{1,g\,|\,0,g}\big(\bar{\mathcal{X}}_{i\smallN_i}(1,g\,|\,0,g)\big)$, which require only that $X_i$ be near the cutoff, 
and thus satisfy $\bar{s}_\phi = 1$, for all $\phi$. 
The codimension can also be one for some boundary indirect effects, e.g., 
$\tau_{0,1\,|\,0,0}\big(\bar{\mathcal{X}}_{i\smallN_i}(0,1\,|\,0,0); \phi\big)$ 
with sum-of-treated exposure mapping, which compares untreated units with zero versus one treated neighbor. However, boundary indirect effects may involve higher $\bar{s}_\phi$, depending on the exposure levels $g$ and $g'$ being compared, leading to a slower convergence rate.

{\color{black}
The asymptotic variance is generally affected by data dependence. In particular, $V^\phi$ depends on the boundary conditional variances and squared approximation errors, summarized by $\bar{\omega}_{d, g}(0 ; \phi)$ and $\bar{\omega}_{d^{\prime}, g^{\prime}}(0 ; \phi)$, while $\bar \gamma_{d, g,d',g'}(0 ; \phi)$ 
collects the cross-unit covariance contributions.
The latter arise because dependent units may have
overlapping boundary definitions induced by overlapping interference sets. A
dependent pair $(i,j)$ lies within distance $h_n$ of their respective boundaries with
probability of order
$h_n^{\operatorname{codim}(\bar{\mathcal{X}}^\phi_{ij}(d,g\mid d',g'))}$. When the joint boundary attains codimension $\bar s_\phi$ --- the case collected by
$\mathcal{K}^\phi_i$ --- this probability is of the same order as the marginal
localization probability that a single unit falls within $h_n$ of its own boundary, which governs the order of the individual variance terms. When this occurs, both units are localized by the same score coordinates near a common minimal-codimension boundary piece: their distance variables are determined by
the same leading score components, so localization of one unit within $h_n$ of
its boundary also localizes the other at the same order. The covariance contribution is non-negligible when this happens for a non-negligible fraction of dependent pairs.
%
In Figure~\ref{fig:twounits}, for instance, unit $i$ in region \(\scorespace(0,0)\) has distance
\(
\widetilde X_i= \min\{c - X_j,c - X_k\},
\)
set by whichever neighbor is closest to the cutoff; a dependent unit sharing that neighbor therefore shares the same distance. This contrasts with standard results for local-polynomial estimators under dependent data \citep[e.g.][]{jenish12}, where typically distinct scores induce a joint probability of two observations falling within distance $h_n$ of the cutoff of smaller order than the marginal, leading to  asymptotically negligible dependence.

\begin{remark}
\normalfont
Data dependence may vanish asymptotically in some special cases when the joint boundary of two dependent units has codimension higher than $\bar{s}_\phi$.  
\begin{assumption}\label{ass:commonpiecesmain}
$\overline{\mathcal{X}}_{ij}^\phi\left(d, g \mid d^{\prime}, g^{\prime}\right)$ has codimension greater than $\bar s_\phi$ for any $i,j \in \mathcal{V}_\phi$.
\end{assumption}

\noindent Under Assumption~\ref{ass:commonpiecesmain}, $\mathcal K_i^\phi=\varnothing$ for every $i$, and hence
$\bar \gamma_{d,g, d',g'}(0;\phi)=0$.
The condition holds automatically for direct effects: each unit's boundary requires its own score to equal the cutoff, so the joint boundary requires both $X_i=c$ and $X_j=c$ and therefore has codimension 2, greater than $\bar s_\phi=1$. In practice, Assumption \ref{ass:commonpiecesmain} can be assessed in a sample by computing the shares of pairs with the same distance. However, we usually expect overlapping interference sets to induce nonvanishing dependence in the asymptotic distribution of the estimator.
\end{remark}
}

\subsection{Variance Estimation}\label{subsec:varianceest}

We propose the following network-robust variance estimator for the variance $(n_\phi h_n^{\bar s_\phi})^{-1}V_{d,g,d',g'}(h_n;\phi)$ in Theorem \ref{th:asynormal}, building on \citet{leung20}: 
\begin{equation}\label{eq:var}
    \hat{{V}}_{d,g,d',g'}(h_n; \phi) = \frac{1}{n_\phi}e_1^T[\hat{V}_{d,g}(h_n, \phi) + \hat{V}_{d',g'}(h_n; \phi) - 2\hat{C}_{d',g',d,g}(h_n; \phi)]e_1
\end{equation}

\noindent where $\hat{V}_{d,g}(h_n; \phi) = \Gamma_{d,g}^{-1}(h_n; \phi)\hat{\bar{\Psi}}_{d,g}(h_n; \phi) \Gamma_{d,g}^{-1}(h_n; \phi)$ for a generic $(d,g)$ with 
\begin{equation}\label{eq:varianceest}   \quad \hat{\bar{\Psi}}_{d,g}(h_n; \phi) = \mathcal{M}_{d,g}(h_n; \phi)^T\bm{W}_n\mathcal{M}_{d,g}(h_n; \phi)/n_\phi \\  
\end{equation}
and $\hat{C}_{d,g,d',g'}(h_n, \phi) = \Gammaplusmain[-1]\hat{\bar{\Psi}}_{d,g,d',g'}(h_n) \Gammaminusmain[-1]$ with  
\begin{equation*}
  \quad \hat{\bar{\Psi}}_{d,g,d',g'}(h_n; \phi) = \mathcal{M}_{d,g}(h_n;\phi)^T\bm{W}_n\mathcal{M}_{d',g'}(h_n, \phi)/n_\phi  
\end{equation*}
Here, $\bm{W}_n$ is the dependency graph, and $\mathcal{M}_{d,g}(h_n; \phi)$ and $\mathcal{M}_{d',g'}(h_n; \phi)$ are $ n\times 2$ matrices with $i$-th row given by the weighted linear expansion of $\tilde{X}_{i}/h_n$ times the regression residual $\hat{u}_{i}$. 
{\color{black}
This estimator captures covariance within each region, \(\scorespace(d,g)\) and \(\scorespace(d',g')\), and across regions for observations within the bandwidth. It consistently estimates
\((n_\phi h^{\bar{s}_{\phi}})^{-1}V_{d,g,d',g'}(h_n;\phi)\), corresponding to a ``fixed-bandwidth'' perspective which treats \(\estimator\) as locally parametric under a fixed $h$. As \(h_n\to0\), it is consistent for the limiting variance, whether \(\bar \gamma_{d,g, d',g'}(0; \phi) \neq0\) or \(\bar \gamma_{d,g, d',g'}(0; \phi) = 0\) (Theorem~\ref{th:variance}). In the latter case, conventional variance estimators that ignore dependence may be asymptotically valid, but our estimator may improve inference by capturing finite-sample covariance arising from using a positive bandwidth in practice and is therefore recommended in general.


}

{\color{black}
\subsection{Bias Correction and Robust Confidence Intervals}
\label{sec:biascorrection}}

Our estimator requires a bandwidth selection. Standard RDD methods typically employ the asymptotic Mean Squared Error (MSE)-optimal bandwidth, which in our setting is
\[
h_{n}^{\mathrm{opt}}
=
\left(
\frac{\bar{s}_\phi V^\phi}
{4(B^\phi)^2}
\right)^{\frac{1}{\bar{s}_\phi+4}}
n_\phi^{-\frac{1}{\bar{s}_\phi+4}}.
\]
This bandwidth can be estimated using the plug-in approach of \citet{cct14}, based on preliminary estimates of the bias and variance components.
However, the MSE-optimal bandwidth does not satisfy the condition $n_\phi h_n^{\bar{s}_\phi + 4} \to 0$ needed for the bias to be asymptotically negligible. Following \citet{cct14}, we then recenter 
$\estimator$ by an estimate of the leading bias and rescale by its variance accounting for the added variability of the bias correction term, yielding confidence intervals robust to large bandwidths. Our bias-corrected estimator is given by:
\begin{equation}\label{eq:estimatorbc}
    \begin{split}
        \estimatorbc =& \estimator - \hat{{B}}_{d,g,d',g'}(h_n, b_n; \phi), \\
        \hat{{B}}_{d,g,d',g'}(h_n, b_n; \phi) = &h_n^{2}\Big[\frac{\hat{\mu}_{d,g}^{(2)}(0; \phi)}{2!} \hat{B}_{d',g'}(h_n; \phi) -\frac{\hat{\mu}_{d',g'}^{(2)}(0; \phi)}{2!}\hat{B}_{d,g}(h_n; \phi)\Big]
    \end{split}
\end{equation}
where $\frac{\hat{\mu}_{d,g}^{(2)}(0; \phi)}{2!}$ and $\frac{\hat{\mu}_{d',g'}^{(2)}(0; \phi)}{2!}$ are the distance-based local-quadratic estimators of the second-order derivatives with bandwidth $b_n$, and $\hat{B}_{d,g}(h_n) = e_1^T{\Gamma}^{-1}_{d,g}(h_n; \phi)\theta_{d,g}( h_n, \phi)$ for generic $(d,g)$.
The estimator is asymptotically normal under a similar asymptotic regime to \citet{cct14}.
\begin{theorem}\label{th:asybiascorretiontaumain}
Suppose Assumptions~\ref{ass:consistency}--
\ref{ass:regularitymain}, and~\ref{ass:kernel} hold. If $n_\phi\min\{h_n^{\bar{s}_\phi},b_n^{\bar{s}_\phi}\}\to \infty$, $n_\phi\min\{h_n^{4+\bar{s}_\phi},b_n^{4+\bar{s}_\phi}\}\max\{h_n^2,b_n^{2}\} \to 0$:
\begin{equation*}
    \frac{\estimatorbc - \boundeff}{\sqrt{V^{bc}_{d,g,d',g'}(h_n, b_n; \phi)}}\xrightarrow{d} \mathcal{N}(0, 1)
\end{equation*}
Proof. \textup{See Supplementary Material \ref{appendix3:biascorrection}.} 
\end{theorem}
\noindent The variance decomposes into the variances of the bias-corrected local-polynomial estimator in the two effective treatment regions and their covariance: $V_{d,g,d',g'}^{bc}(h_n, b_n; \phi) = V_{d,g}^{bc}(h_n, b_n; \phi) + V_{d',g'}^{bc}(h_n, b_n; \phi) - 2C^{bc}_{d,g,d', g'}(h_n, b_n; \phi)$. The first two terms include the variance of the original local linear estimator, of the bias-correction term, and their covariance within regions. The term $C^{bc}_{d,g,d', g'}(h_n, b_n; \phi)$ captures the covariance of the bias-corrected estimators across the two regions. 
We estimate these components using the dependency-graph correction introduced in Section~\ref{subsec:varianceest}, extended to account for the bias-correction variability. Since the expressions for the variance and its estimator involve cumbersome notation, we defer the complete formulas to Supplementary~\ref{appendix3:biascorrection}.

\subsection{Estimation of Pooled Boundary Causal Effects}\label{sec:idestpooled}
We extend our analysis to the estimation of the pooled estimand defined in Eq.~\eqref{eq:pool} and summarize key results. Additional details are provided in Online Appendix \ref{onlineappendix:irregular}.
Estimation proceeds by computing the distance of each unit from its effective treatment boundary—which may have varying dimension depending on the size of the interference set—and pooling all observations in a local linear regression\footnote{The alternative aggregate estimand which takes a simple average of stratum-specific boundary effects can be estimated by estimating $\boundeff$ separately conditioning on $\phi$, and then averaging the estimates. This procedure may lead to a limited sample size near the boundary. The pooled estimator instead aggregates observations across heterogeneous strata, thereby improving effective sample size.}.
We define the pooled local linear estimator as\begin{equation}\label{eq:pooled}
\estimatorpooled = \mupluspooled - \muminuspooled     
\end{equation}
where $\mupluspooled = e_1^T\betapluspooled$ and $\muminuspooled = e_1^T\betaminuspooled$, and $\betapluspooled$ and $\betaminuspooled$ are obtained as
\begin{equation}\label{eq:distrsspooled}
    \begin{split}
   & {\betapluspooled} = \argmin\limits_{\bm{b} \in \mathbbm{R}^2}\sum_i{\mathbbm{1}(D_i = d, G_i = g)}(Y_i - \bm r(\distvari)^T\bm{b})^2\cdot K_{h_n}(\distvari) \\
    &{\betaminuspooled} = \argmin\limits_{\bm{b}\in \mathbbm{R}^2}\sum_i{\mathbbm{1}(D_i = d', G_i = g')}(Y_i - \bm r(\distvari)^T\bm{b})^2\cdot K_{h_n}(\distvari)
    \end{split}
\end{equation}
Consistency and asymptotic normality of the pooled estimator can be shown similarly to Theorem \ref{th:asynormal}. Theorem \ref{lemma:consistencypooled} in Supplementary Material shows that the estimator is consistent for $\boundeffpooled$ when the codimension $\bar{s}_\phi$ is constant across strata. This result requires continuity of $\densityphi$, $\mplusphi$, and $\mminusphi$ to hold for each $\phi$, as well as a bounded number of strata $|\Phi_n|$ as $n \to \infty$ (Assumption \ref{ass:boundaryhet}). This latter assumption ensures that aggregation across strata preserves convergence of the sample matrices involved in the definition of the pooled estimator.
Asymptotic normality is established in Theorem \ref{th:heterogeneousconvergence}. The pooled estimator exhibits a second-order asymptotic bias depending on the second derivatives of the density-weighted distance-based outcome regressions, and an asymptotic variance of order $O(nh_n^{\bar{s}})$, that depends on the conditional boundary variance terms averaged across strata. The bias result requires smoothness of the distance-based outcome regression functions $\muplus$ and $\muminus$ for each $\phi$, as well as smoothness of the stratum-specific density integral averaged over the set of points at a given distance from the boundary.

Importantly, when the codimension varies across strata, the estimator is not consistent for $\boundeffpooled$, Instead, it converges to the weighted average of boundary effects over strata with the minimal codimension. 
\begin{theorem}
\label{thm:teo3}
Suppose Assumptions~\ref{ass:consistency}--\ref{ass:networkheterogeneity}, \ref{ass:localdependence}, and \ref{ass:kernel}, \ref{ass:boundaryhet} in Supplementary material hold. Let Assumption~\ref{ass:identification} and $\ref{ass:variancesmain}$ hold for all $\phi$, and let $\bar{s} = \min_\phi \bar{s}_\phi$.
If $n h_n^{\bar{s}} \to \infty$ and $h_n \to 0 $
then
\[
\estimatorpooled \xrightarrow{p} \frac{\sum_{\phi : \bar{s}_\phi = \bar{s}} \boundeff \, \cdot \limdensity \cdot \bar{\pi}_{\phi}}
     {\sum_{\phi : \bar{s}_\phi = \bar{s}} \limdensity \cdot \bar{\pi}_\phi}
\]
Proof. \textup{See Supplementary Material~\ref{appendix:heterogeneous}.} 
\end{theorem}
%
\noindent where $\bar{\pi}_\phi$ is the limiting fraction of units in $\mathcal{V}_\phi$. The probability limit in Theorem \ref{thm:teo3} arises because the probability that an observation in a specific stratum lies near the stratum-specific boundary is proportional to $h_n^{\bar{s}_\phi}$, so the contribution to the probability limits from observations in strata with codimension $\bar{s}_\phi > \bar{s}$ vanish asymptotically relative to those with minimal codimension.
\red{
This limiting estimand remains meaningful: it aggregates the boundary effects among strata with minimal codimension and can be interpreted as the boundary effect for a randomly drawn unit from this subset of the network. For example, in the PROGRESA application, the pooled indirect effect $\tau_{0,1|0,0}(\bar{X}_{i\smallN_i}(0,1|0,0))$, is defined over units with one to four neighbors. Because units' codimension here equals the interference-set size, the pooled estimator asymptotically targets the effect for units with one neighbor. When the objective is instead to recover an effect over all degree strata, or higher-codimension strata, researchers should use the stratum-specific estimators and aggregate them using the corresponding density weights. This approach may, however, be less precise or infeasible when some strata contain few observations near the boundary.}

{\color{black}
\section{Boundary Overall Effects}\label{sec:mainoverall}}

It is often useful to define a summary measure of direct and indirect effects under interference. We first consider estimation of the \emph{boundary overall direct effects}, which provides a causal interpretation for a conventional RDD analysis under interference. We then introduce an overall indirect effect summarizing the marginal effect of adding one treated neighbor.

\red{\subsection{Boundary overall direct effects}}

We define the stratum-specific \emph{boundary overall direct effect}:
\begin{equation}\label{chapter1def:overalleffect}
    \overall(\phi) = \sum_{g \in \mathcal{G}_i}\tau_{1,g\,|\,0,g}(\bar{\mathcal{X}}^\phi_{i\smallN_i}(1,g\,|\,0,g), \phi )\cdot \CPrr{G_i = g}{X_i = c, \phi}
\end{equation}  
This effect averages boundary direct effects over the probability distribution of $G_i$ conditional on $X_i = c$ and $\phi_i = \phi$ \footnote{This effect corresponds to a weighted average of point direct effects $\tau_{1,g|0,g}(x_i, \bm x_{\smallN_i}; \phi)= m_{1, g}(x_i, \bm x_{\smallN_i}; \phi) - m_{0,g}(x_i, \bm x_{\smallN_i}; \phi)$ across the union of all boundaries of the type $\bar{\mathcal{X}}_{i\smallN_i}(1,g|0,g)$. Particularly, it holds that $ \bigcup_{g}\bar{\mathcal{X}}^\phi_{i\smallN_i}(1,g|0,g) = \{\ppoint \in \scorespace\, : \, x_i = c\}$. 
%
}, and thus summarizes the effect of the individual treatment at the cutoff within $\mathcal{V}_\phi$ \citep[see e.g.][for related definitions]{forastiere}.
In the example in Figure \ref{fig:oneunit}, $\overall(\phi)$ averages the boundary point effects along the line $X_i = c$, the union of the boundaries $\bar{\mathcal{X}}^\phi_{i\smallN_i}(1,1\,|\,0,1)$ and $\bar{\mathcal{X}}^\phi_{i\smallN_i}(1,0\,|\,0,0)$. 
We further define the pooled boundary overall direct effects:
\begin{equation}\label{eq:pooledoverall}
\overall = 
\frac{\sum_{\phi} \overall(\phi)\cdot f_{X_i}(c; \phi) \cdot \pi_\phi}
{\sum_{\phi} f_{X_i}(c; \phi)\cdot \pi_\phi},
\end{equation}
%
where $f_{X_i}(c; \phi)$ denotes the marginal density of $X_i$ at the cutoff $x=c$ within stratum $\mathcal{V}_\phi$. This estimand can be interpreted as the overall boundary effect for a randomly drawn unit from the network at the cutoff.   


In a no-interference RDD, the discontinuity in the observed outcome
regression identifies the average causal effect at the cutoff. We show that, in our framework, this strategy precisely identifies $\overall(\phi)$ under Assumptions \ref{ass:consistency}-\ref{ass:identification}, and within the stratum $\mathcal{V}_\phi$. We then show that aggregating the stratum-specific outcome regressions using the stratum-specific density of the individual score near the cutoff identifies $\overall$.
\begin{theorem}[Identification of $\tau_{1|0}(\phi)$]\label{th:overall}
    Under assumptions \ref{ass:consistency}-\ref{ass:identification}
    \begin{equation*}
        \overall(\phi)  = \lim_{x_i \downarrow c}\CE{Y_i}{X_i = x_i; \phi} - \lim_{x_i \uparrow c}\CE{Y_i}{X_i = x_i; \phi}
    \end{equation*}
Proof. \textup{See Supplementary Material \ref{appendix:overall}.}
\end{theorem}
\begin{corollary}[Identification of $\tau_{1|0}$]\label{corollary:overall}
Suppose the conditions of Theorem~\ref{th:overall} hold for every $\phi$. Then
\[
\overall = \lim_{x \downarrow c}\frac{\sum_\phi \mathbb{E}\left[Y_i \mid X_i=x, \phi\right] \cdot f_{X_i}(x ; \phi) \cdot \pi_\phi}{\sum_\phi f_{X_i}(x ; \phi) \cdot \pi_\phi} - \lim_{x \uparrow c}\frac{\sum_\phi \mathbb{E}\left[Y_i \mid X_i=x, \phi\right] \cdot f_{X_i}(x ; \phi) \cdot \pi_\phi}{\sum_\phi f_{X_i}(x ; \phi) \cdot \pi_\phi}
\]
\end{corollary}
%
%
Theorem \ref{th:overall} crucially relies on Assumption \ref{ass:densitycont} (score density continuity), which ensures that the distribution of $\pscore$ conditional on $X_i$, and hence of $G_i$, is balanced on each side of the cutoff. The corollary further shows that the difference in the right and left limit of the density-weighted pooled outcome regression identifies $\overall$.

\sloppy
Theorem \ref{th:overall} yields $\overall = \lim_{x \downarrow 0}\muoverall[x] - \lim_{x \uparrow 0}\muoverall[x]$, with $\muoverall[x] = \CE{Y_i}{X_i = x;; \, \phi}$ with $c = 0$.
A natural choice is to estimate these limits by local linear regression in $X_i$, without considering $\pscore$, as in no-interference RDDs. For $d\in\{0,1\}$, define
\begin{align*}
\hat{\beta}_{d}(h_n;\phi)
&=
\argmin_{b_0,b_1\in\mathbbm{R}}
\sum_i
\mathbbm{1}(D_i = d)\mathbbm{1}(\phi_i=\phi)(Y_i-b_0-b_1X_i)^2K_{h_n}(X_i),\\
\hat{\beta}_{d}(h_n)
&=
\argmin_{b_0,b_1\in\mathbbm{R}}
\sum_i
\mathbbm{1}(D_i = d)
(Y_i-b_0-b_1X_i)^2K_{h_n}(X_i).
\end{align*}
Here, $\hat{\beta}_{d}(h_n;\phi)$ uses observation within stratum $\mathcal{V}_\phi$ whereas $\hat{\beta}_{d}(h_n)$ aggregates observations across strata. The stratum-specific and pooled estimators are, respectively,
\[
\hat{\tau}_{1|0}(h_n;\phi)
=
e_1^\top\!\left [
\hat{\beta}_{1}(h_n;\phi)-\hat{\beta}_{0}(h_n;\phi)
\right ],
\qquad
\hat{\tau}_{1|0}(h_n)
=
e_1^\top\!\left [
\hat{\beta}_{1}(h_n)-\hat{\beta}_{0}(h_n)
\right].
\]
The pooled estimator $\hat{\tau}_{1 \mid 0}\left(h_n\right)$ is exactly the conventional local-linear RDD estimator that would be used if network strata and interference were ignored. 
Define $\tilde{\mu}_1\left(x_i ; \phi\right)=\mathbb{E}\left[Y_i \mid X_i=x_i ; \phi\right]$ for $x_i \geq 0$, and $\tilde{\mu}_0\left(x_i ; \phi\right)=\mathbb{E}\left[Y_i \mid X_i=x_i ; \phi\right]$ for $x_i<0$. The next theorem establishes asymptotic normality for $\overallest$ under the same assumptions as section \ref{sec:estimation} but replacing the smoothnes assumption on \ref{ass:smoothnessmain} with smoothness of $\muoverallplus$ and $\muoverallminus$.

\begin{theorem}\label{th:overallnormal}
Suppose Assumptions~\ref{ass:consistency}--\ref{ass:networkheterogeneity}, \ref{ass:localdependence}, and~\ref{ass:kernel} hold. 
\begin{enumerate}[(i)]
\item 
For given $\phi$, suppose Assumption  \ref{ass:identification}, \ref{ass:fourthmomentmain}---\ref{ass:variancesmain} hold and that
$\widetilde{\mu}_d(\cdot;\phi)$ is $3$-times continuously differentiable. If $n_{\phi}h_n\to\infty$, $h_n\to0$, and
$h_n=O(n_\phi^{-1/5})$, 
\[
\sqrt{n_\phi h_n}
\frac{
\{\hat\tau_{1|0}(h_n;\phi)-\overall(\phi)\}
-h_n^2B_{1,0}(h_n;\phi)
}{
\sqrt{V_{1,0}(h_n;\phi)}
}
\overset{d}{\longrightarrow}\mathcal N(0,1).
\]

\item Suppose \ref{ass:boundaryhet} holds and, for all $\phi$, the
conditions in part~(i) hold and $f_{X_i}(\cdot;\phi)$ is 3-times
continuously differentiable. If $nh_n\to\infty$, $h_n\to0$, and
$h_n=O(n^{-1/5})$, 
\[
\sqrt{nh_n}
\frac{
\{\hat\tau_{1|0}(h_n)-\overall\}
-h_n^2B_{1,0}(h_n)
}{
\sqrt{V_{1,0}(h_n)}
}
\overset{d}{\longrightarrow}\mathcal N(0,1).
\]
\end{enumerate}
The bias and variance terms are defined in Supplementary
Material~\ref{appendix:estimatoroverall}, page~\pageref{eq:bias-overall}. \\
Proof. \textup{See Supplementary Material~\ref{appendix:estimatoroverall}.}
\end{theorem}

Both estimators' asymptotic bias depend on the second derivatives of
the relevant outcome regressions at the cutoff. The variances are
of order $(n_\phi h_n)^{-1}$ and $(nh_n)^{-1}$ for the stratum-specific and pooled estimators, respectively, and depend on the conditional outcome variance and score density at the cutoff. Indeed, because both effects are defined on subsets of the multiscore space with codimension one, their estimators attain the canonical one-dimensional RDD convergence rate. Dependence-robust variance estimators and local-quadratic bias corrections can be constructed analogously to sections \ref{subsec:varianceest} and \ref{sec:biascorrection}. 
An important implication of part~(ii) is that the conventional RDD
estimator remains consistent for a meaningful causal estimand, $\overall$, even under network heterogeneity and interference. Moreover, the
contribution of data dependence to the asymptotic variance $V_{1|0}(h_n)$ vanishes, so inference based on an $i.i.d.$ variance estimator may remain asymptotically valid. 
As discussed in Section~\ref{subsec:varianceest}, however, data dependence may still affect the estimator's variance in finite samples, so our dependence-robust variance estimator remains preferable for inference.

\red{\subsection{Boundary overall indirect effects}}\label{sec:indirectoverall}
We define the \textit{boundary overall indirect effect} as a measure of the marginal effect of having an additional treated neighbor: 
\begin{equation}
\label{eq:overallindirect}
    \begin{aligned}
 \tau_{indirect}(\phi) =  
 \sum_{d_i, \bm d_{\smallN_i}}
\!\!\!\! \;\hspace{-1em}\sum_{\substack{ \: \:\bm{d}_{\smallN_i}^{\prime}:\\\, d_j'\ge d_j \, \forall j\in \smallN_i, \\ 
\sum_{j\in \smallN_i} |d_j^{\prime} - d_j| = 1}}
&\hspace{-1.5em}\CE{Y_i(d_i, g(\bm{d}^{\prime}_{\smallN_i}))-Y_i(d_i, g(\bm{d}_{\smallN_i}))}{(X_i, \pscore) \in \ell\big((d_i,\bm{d}_{\smallN_i}'),(d_i, \bm{d}_{\smallN_i})\big); \phi}\\[-2em]
&\quad \cdot \CPrr{D_i=d_i, \bm{D}_{\smallN_i}'=\bm{d}_{\smallN_i}'}{\exists j \in \mathcal{S}_i: X_j=c; \, \phi }
    \end{aligned}
\end{equation}
Here, $\ell\big((d_i,\bm{d}_{\smallN_i}'),(d_i, \bm{d}_{\smallN_i})\big)$, with $\ell(\cdot)$ defined in Eq. \eqref{eq:piecefunction}, fixes at the cutoff the score of the neighbor whose treatment switches from 0 to 1 between $\boldsymbol{d}_{\smallN_i}$ and $\boldsymbol{d}_{\smallN_i}^{\prime}$. This estimand averages the boundary indirect effects comparing neighborhood treatment vectors that differ by one treated neighbor, over the distribution of the individual and neighbors' treatment, conditional on having at least one neighbor at the cutoff. 
%
While its general definition appears cumbersome, it simplifies once we specify the exposure mapping. For instance, under \emph{sum-of-treated} $G_i = \sum_{j \in \mathcal{S}_i} D_j$, this estimand is
%
%
the weighted average of the boundary indirect effects of increasing the neighborhood treatment level from $g$ to $g+1$,
weighted by the probability of being at the corresponding boundary conditional on having at least one neighbor at the cutoff (see Online Appendix \ref{onlineappendix:indirectoverall}). 
Finally, we can define a pooled counterparts of this type of effect:
\begin{equation}
    \tau_{indirect} = \frac{\sum_{\phi} \tau_{\text{indirect}}(\phi), f_{\text{indirect}}(c; \phi)}{\sum_{\phi} f_{\text{indirect}(c, \phi}}
\end{equation}
where $f_{\text{indirect}}(c; \phi)$ denotes the density over the multiscore subspace of in which at least one neighbor’s score equals the cutoff.

We defer identification and estimation details to the Online Appendix \ref{onlineappendix:indirectoverall}. 
Here we highlight that $\tau_{\text{indirect}}(\psi)$ and $\tau_{\text{indirect}}$ are identified by difference in limits of the observed outcome conditional on the score of the neighbor closest to the cutoff, as it approaches the cutoff.  Both are estimated by local polynomial regression on the score of the neighbor closest to the cutoff. We denote these estimators by
$\hat{\tau}_{\text{indirect}}(h_n; \phi)$ and $\hat{\tau}_{\text{indirect}}$. Notably, because these estimands are defined on boundary subsets with codimension one, the estimators achieve the fast $(nh_n)^{-1/2}$ convergence rate, in contrast to  estimators of boundary indirect effects comparing specific values of the neighborhood treatment $g$ and $g'$, which  may converge more slowly. 

{\color{black}
\section{Simulation Study}\label{sec:simulation}}
We present a simulation study calibrated to the PROGRESA dataset to evaluate the performance of our method in a realistic setting that incorporates the observed network structure and empirically grounded distributions of the score variables. We consider a sample of $n = 4{,}900$ individuals from the treatment villages. Each unit $i$ is characterized by a set of covariates $\mathbf{Z}_i$, including demographic and household characteristics, a poverty index $X_i$, and a village identifier $C_i$. Peer groups are defined as individuals within the same village, grade, and gender.

The score $X_i$ is generated according to a linear model with both village- and peer group- level dependence:
\begin{equation}
X_i = \alpha + \mathbf{Z}_i'\beta + u_{C_i} + v_{g(i)} + \varepsilon_i,
\end{equation}
where $u_{C_i} \sim \mathcal{N}(0,\sigma^2_{\text{village}})$ is a village-level random effect, $v_{g(i)} \sim \mathcal{N}(0,\sigma^2_{\text{group}})$ is a peer-group effect, and $\varepsilon_i \sim \mathcal{N}(0,\sigma^2_{\text{id}})$ is an idiosyncratic error. The index $g(i)$ identifies units with identical peer groups. The covariates $\mathbf{Z}_i$ include enrollment status in 1997, housing characteristics, and the mother’s education level. Treatment is given by $D_i = \mathbbm{1}(X_i \geq 0)$. Under this specification, a typical simulation yields approximately $63\%$ share of treated units, closely matching the empirical proportion in the data. 
Consistent with our empirical application, we adopt a fraction-of-treated exposure mapping, $G_i = \tfrac{1}{|\smallN_i|}\sum_{j \in \mathcal{S}_i} D_j$.
Outcomes are generated as a nonlinear function of own and neighborhood scores:
\begin{equation*}
Y_i = 0.5 + 1.5D_i + G_i + 0.15 |\mathcal{S}_i|\cdot G_i + 0.5X_i + 4.3D_iX_i^2 - 1.5(1 - D_i)X_i^2 + 0.3\bar{X}{\smallN_i} + \varepsilon_i^Y,
\end{equation*}
where $\bar{X}_{\smallN_i} = |\mathcal{S}i|^{-1} \sum_{j \in \mathcal{S}_i} X_j$ denotes the average neighbor score and $|\mathcal{S}_i|$ is the degree of unit $i$. 
The outcome error term incorporates network dependence through neighbors’ shocks: $\varepsilon_i^Y = D_i\big(2\frac{\sum_{j\in\smallN_i}u_j}{|\mathcal{S}_i|} + u_i\big) - 1.4(1 -D_i)\cdot \big(2\frac{\sum_{j\in\smallN_i}u_j}{|\mathcal{S}_i|} + u_i\big)$ with $u_i \overset{i.i.d.}{\sim} \mathcal{N}(0,1)$.

We estimate the pooled overall direct and indirect effect, as well as the pooled direct effect $\tau_{10|00}(\bar{\mathcal{X}}_{i\smallN_i}(10|00))$ and indirect effect $\tau_{01|00}(\bar{\mathcal{X}}_{i\smallN_i}(01|00))$ for individuals with peer group sizes ranging from 1 to 4, using our bias-corrected local linear \emph{pooled} estimators. 
%
The results in Table~\ref{tab:realistic} show that our method performs well under a realistic distribution of the score variables and interference structure.
The estimates are virtually unbiased for their respective asymptotic targets, including those for $\tau_{10|00}(\bar{\mathcal{X}}_{i\smallN_i}(10|00))$ and $\tau_{01|00}(\bar{\mathcal{X}}_{i\smallN_i}(01|00))$, which are based on relatively small samples near the boundary. These sample sizes are comparable in magnitude to those observed in the empirical application. \red{In particular, $\hat{\tau}^{bc}_{10|00}(h_n, b_n)$ targets the pooled effect over degrees 1--4, whereas $\hat{\tau}^{bc}_{01|00}(h_n, b_n)$ is consistent for the minimum-codimension (degree-one) effect; its reported bias is relative to this asymptotic target.}
%
The dependence-robust variance estimator achieves good coverage despite the limited effective sample size. It nearly attains the nominal coverage rate for the overall direct and indirect effects, outperforming the $iid$ variance estimator, which tends to undercover. For $\tau_{10|00}(\bar{\mathcal{X}}_{i\smallN_i}(10|00))$ and $\tau_{01|00}(\bar{\mathcal{X}}_{i\smallN_i}(01|00))$, the two variance estimators perform similarly in coverage.

\textcolor{black}{
Appendix \ref{appendix:simulations} provides additional simulations with clustered data, varying cluster sizes, and networked data, further assessing the finite-sample performance of our estimators and inference procedures across different dependence structures and network configurations.}

\begin{table}[H]
\centering
\footnotesize
\begin{tabular}{lcccccccc}
  \hline
 & Bias & S.D& S.E & S.E$_{i.i.d.}$ & C.R. & C.R.$_{i.i.d.}$& $N_{d',g'}$ & $N_{d,g}$ \\ 
  \hline
$\hat{\tau}_{1|0}^{bc}(h_n, b_n)$ & 0.0079 & 0.7704 & 0.7407 & 0.6375 & 0.9430 & 0.8994 & 563.4356 & 463.9132 \\ 
  $\hat{\tau}^{bc}_{indirect}(h_n, b_n)$ & -0.0067 & 1.8331 & 1.7072 & 1.2533 & 0.9380 & 0.8302 & 885.4838 & 744.0430 \\ 
  $\hat{\tau}^{bc}_{10|00}(h_n, b_n)$ & -0.0014 & 2.0045 & 1.8262 & 1.7949 & 0.9284 & 0.9240 & 111.3436 & 111.7864 \\ 
  $\hat{\tau}^{bc}_{01|00}(h_n, b_n)$ & 0.0468 & 3.0962 & 2.7910 & 2.7504 & 0.9288 & 0.9266 & 121.3244 & 108.6190 \\ 
   \hline
\end{tabular}
\caption{\footnotesize Simulation results over 5,000 replications. 
 S.D. = empirical standard error. S.E. = estimated standard error.
 S.E.$_{i.i.d.}$ = estimated standard error with $i.i.d.$ variance estimator.
 C.R. = coverage rate. C.R.$_{i.i.d.}$ = coverage rate corresponding to the $i.i.d.$ variance estimator.
 $N_{d,g}$ and $N_{d',g'}$ = effective sample size on each side of the treatment boundary.}
\label{tab:realistic}
\end{table}

\section{Empirical Illustration}\label{sec:empirical}

We apply the proposed framework to the PROGRESA application described in Section \ref{sec:motivatingapp}. Following \citet{lalive}, we restrict the sample to children who lived with their mothers, had complete enrollment information for 1997 and 1998, and had completed grades 3–6 by October 1997. 
 We focus on three rural regions with similar eligibility thresholds and on children enrolled in 1997–1998, who therefore had an opportunity to interact before treatment.
Our final sample consists of 4,928 children. 
The household poverty index determines eligibility, and the outcome is enrollment in 1999–2000.
For each child, $\mathcal{S}_i$ comprises same-gender and same-grade children living in the same village at baseline. We treat these interference sets as fixed over the study period, supported by the program's timing and the absence of evidence for inter-village relocation.
The exposure mapping is the fraction of treated neighbors:
$G_i = \frac{\sum_{j \in \mathcal{S}_{i}} D_j}{|\mathcal{S}_i|}$.
%

\red{We focus on pooled effects for children with one to four peers, retaining 2,776 observations.
The remaining 43.67\% of the full sample are excluded because configurations with no eligible peers are extremely rare in larger interference sets.} We estimate (i) the pooled boundary direct effects of the cash transfer on school enrollment in the academic year 1999-2000 when having all treated peers ($G_i = 1$) and no treated peers ($G_i = 0$); (ii) the pooled boundary spillover effects of having all treated peers ($G_i=1$) versus no treated peers ($G_i=0$), both for treated and untreated units. We estimate these pooled effects for children with at most four neighbors, as for larger groups the number of units with $G_i=0$ is extremely small. We also estimate the pooled boundary overall direct effect of receiving the cash transfer, and the boundary overall indirect effect of having an additional peer receiving the subsidies using \red{observations with at least one neighbor.}  

\red{
Table \ref{table:boundaries} describes the corresponding effective-treatment boundaries, the range of the minimum-distance measure, and the number of observations in each effective-treatment region. The boundaries associated with the direct effects have codimension one in all four degree strata. Consequently, the pooled direct-effect estimators target the pooled effects across these strata. By contrast, for the all-versus-none indirect effects, codimension equals interference-set size and therefore varies across strata. Although units with one to four peers contribute in finite samples, the pooled indirect effect estimators asymptotically target the indirect effect for units with one peer, the unique minimum-codimension stratum, as discussed in Section~\ref{sec:idestpooled}. Units with interference-set size one constitute approximately $15.18 \%$ of the full sample. Hence, asymptotically, this estimator targets the pooled effect for this percentage of the network, discarding the remaining 84.82\%.
}


\begin{table}[t]
\centering
\resizebox{0.95\textwidth}{!}{
\begin{tabular}{lccccccc}
    \midrule
    & & \multicolumn{2}{c}{All observations} & \multicolumn{2}{c}{Treatment} & \multicolumn{2}{c}{Control} \\ 
    \cmidrule(lr){3-4}  \cmidrule(lr){5-6}  \cmidrule(lr){7-8} 
      & Boundary & $N$ & Range & $N$ & Range & $N$ & Range \\
     \midrule
   $\tau_{1|0}$
& $X_i = 0,\, X_j \in \mathbbm{R}\ \forall j \in \mathcal{S}_i$
& 3789 & $[-542.50,467.00]$
& 2578 & $[0.00,467.00]$
& 1211 & $[-542.50,-0.10]$ \\

$\tau_{\mathrm{indirect}}$
& $X_i \in \mathbbm{R},\, \exists j \in \mathcal{S}_i:X_j=0$
& 3789 & $[-259.00,288.50]$
& 2353 & $[0.00,288.50]$
& 1436 & $[-259.00,-0.10]$ \\
  $\tau_{10|00}(\bar{\mathcal{X}}_{i\smallN_i}(10|00))$ &  $X_i = 0, \, X_j < 0\: \forall j \in \mathcal{S}_i$ & 450 & $[-527.75, 378.00]$ & 229 & $[0.04, 378.00]$ & 221 & $[-527.75, -0.50]$ \\
 $\tau_{11|01}(\bar{\mathcal{X}}_{i\smallN_i}(11|01))$ &  $X_i = 0,\, X_j \geq 0 \:\forall j \in \mathcal{S}_i$  & 1271 & $[-508.00, 391.50]$ & 963 & $[1.50, 391.50]$ & 308 & $[-508.00, -0.50]$  \\
 $\tau_{01|00}(\bar{\mathcal{X}}_{i\smallN_i}(01|00))$ &  $X_i < 0,\, X_j = 0 \:\forall j \in \mathcal{S}_i$ & 529 & $[-533.12, 542.07]$ & 308 & $[0.04, 542.07]$ & 221 & $[-533.12, -0.50]$ \\
 $\tau_{11|10}(\bar{\mathcal{X}}_{i\smallN_i}(11|10))$ &  $X_i \geq 0,\, X_j = 0 \: \forall j \in \mathcal{S}_i$ &  1192 & $[-509.80, 599.46]$ & 963 & $[1.50, 599.46]$ & 229 & $[-509.80, -0.50]$ \\
 \midrule
    \end{tabular}}
\caption{Boundaries and sample sizes for each effect for PROGRESA data. The range refers to the individual score variable in the first line and the minimum distance score in the remaining lines.}
    \label{table:boundaries}
\end{table}

Estimates are reported in Table~\ref{tab:emp_results}. 
The bandwidth is the MSE-optimal bandwidth selected for each effect as described in Section~\ref{sec:biascorrection}. We employ a triangular kernel, the standard choice in RDDs \citep{cattaneo_idrobo_titiunik_2020}, the Euclidean distance metric, and our local linear estimator ($p = 1$), which is more stable than higher-order polynomials.
Standard errors and confidence intervals are computed using the variance estimator described in Subsection \ref{subsec:varianceest}. 
Our estimates indicate positive and significant boundary direct effects of the subsidy receipt for children with ineligible and eligible neighbors, \red{with $\hat{\tau}_{10 \mid 00}^{b c}\left(h_n, b_n\right)$ significant at the $5 \%$ level and  $\hat{\tau}_{11 \mid 01}^{b c}\left(h_n, b_n\right)$ at the $10 \%$ level $(p=0.089)$.} The boundary indirect effect estimates suggest a larger impact of having all eligible neighbors versus all ineligible neighbors for eligible children than for ineligible children. Both estimates, however, are not significant. The varying effect sizes in direct effects may reflect local factors. A lower poverty index correlates with better living conditions and more educated parents \citep{lalive}. Children with non-poor friends, who are more likely to attend school, may be more encouraged to attend school upon receiving a subsidy, explaining the larger estimate of $\hat{\tau}^{bc}_{10|00}(h_n, b_n)$ compared to $\hat{\tau}^{bc}_{11|01}(h_n, b_n)$. 
The estimators $\hat{\tau}^{bc}_{01|00}(h_n, b_n)$ and $\hat{\tau}^{bc}_{11|10}(h_n, b_n)$ estimate a spillover effect on ineligible (non-poor) and eligible (poor) children, respectively, whose friends have a poverty index near the cutoff. The larger estimate for eligible children may suggest that peer subsidy receipt reinforces schooling decisions when households receive the subsidy, whereas this channel may be weaker for ineligible children.

Finally, our estimates indicate a positive boundary overall direct effect,\red{ significant at the 10\% level}, suggesting an overall impact of eligibility to PROGRESA on children's school enrollment. The estimate for $\tau_{indirect}$ is also positive and \red{significant at the 10\% level}. It thus provides suggestive evidence of a (local) spillover effect.
Overall, these results imply that the program directly increased school attendance among children—especially those with untreated peers—and offer suggestive evidence of positive spillover effects for children with at least one peer at the eligibility cutoff. From a policy perspective, policymakers wishing to improve the effectiveness of the program could prioritize eligible (poor) children whose peers are ineligible (non-poor), as the individual return to the subsidy is higher in such environments.
Note, however, that we focus on estimating individual-level estimands 
and do not address village-level “overall effects” of assignment strategies (e.g., comparing village-level mean outcomes if all eligible households in a village were treated versus none). 
One could express such village-level effects as a combination of individual-level total and spillover effects \citep{hh}. 
Establishing this link 
remains a direction for future research.

\begin{table}[t]
\centering
\resizebox{0.9\textwidth}{!}{
\footnotesize{
\begin{tabular}{lcccccccc}
  \hline
 & Estimate & S.E. & C.I. & p-value & $N_{d',g'}$ & $N_{d,g}$ \\ 
  \hline
  $\hat{\tau}^{bc}_{10|00}(h_n, b_n)$ & 0.277 & 0.122 & (0.038, 0.517) & 0.023 & 114 & 115\\ 
   $\hat{\tau}^{bc}_{11|01}(h_n, b_n)$ & 0.122 & 0.072 & (-0.018, 0.262) & 0.089 & 219 & 426 \\ 
   $\hat{\tau}^{bc}_{01|00}(h_n, b_n)$ & 0.065 & 0.129 & (-0.187, 0.317) & 0.612 & 143 & 196 \\ 
   $\hat{\tau}^{bc}_{11|10}(h_n, b_n)$ & 0.144 & 0.103 & (-0.058, 0.346) & 0.163 & 154 & 282\\ 
\hline
 $\hat{\tau}^{bc}_{1|0}(h_n, b_n)$  & 0.067 & 0.040 & (-0.010, 0.145) & 0.089 & 751 & 1203\\ 
  $\hat{\tau}^{bc}_{indirect}(h_n, b_n)$ & 0.058 & 0.035 & (-0.010, 0.126) & 0.095 & 817 & 1079\\
   \hline
\end{tabular}}}
\caption{Estimates for PROGRESA data with robust confidence intervals, MSE-optimal bandwidths. S.E. = estimated standard errors, C.I. = Normal approximation confidence intervals. p-value = Normal approximation two-sided p-value. $\text{N}_{d,g}$, $\text{N}_{d',g'}$ = Effective sample sizes in each side of the boundary.}
\label{tab:emp_results}
\end{table}

\paragraph{Diagnostics and robustness}
Our framework relies on continuity of the outcome regression functions and the score density, and on correct specification of the interference set and exposure mapping. The continuity assumptions plausibility depends on the inability of individuals or their peers to manipulate scores to sort across the boundaries. In PROGRESA, manipulation could arise if households strategically misreported assets or other inputs to the poverty score.
In practice, a range of diagnostics can be used to assess these conditions. 

We implement a set of diagnostic checks, with details reported in Online Appendix~\ref{onlineappendix:empirical}. 
First, we conduct covariate balance tests at the boundary, including both individual and peer-aggregated covariates, using our distance-based estimator. We find no systematic discontinuities, with only mild imbalances in parental education.
Second, we implement a density discontinuity test based on the marginal score distribution, complemented with dependence-robust inference, and find no strong evidence of discontinuity. Additional approaches include placebo-boundary tests where no treatment discontinuity is present. Here, however, such tests are not straightforward, as the effective treatment boundary is multidimensional and a placebo boundary must be chosen to preserve the geometry of the original problem. We leave this extension to future work.
Third, given the estimator's local nature, we assess sensitivity to bandwidth choice. Qualitative conclusions remain robust across alternative bandwidths.
%
%
Fourth we examine robustness to the specification of the exposure mapping. An alternative specification yields qualitatively similar direct effects while highlighting the sensitivity of spillovers to the definition of the exposure mapping. 

Additional checks could involve varying the interference set (e.g., defining peers by age rather than gender). Here, our choice is motivated by previous literature utilizing village- and grade-level peer groups \citep[e.g.,][]{lalive}—and the empirical tendency for children’s friendship networks to exhibit gender-based homophily. Ultimately, such choices are application-specific, should be guided by domain knowledge, and depend on the availability of relational data, such as friendship or kinship networks.

\section{Conclusions} \label{sec:conclusion}

This paper developed an analytical framework for interference in continuity-based RDDs. We showed that the RDD assignment rule induces a multiscore RDD in which effective treatment depends on individuals' and their neighbors' scores. We characterized the resulting multiscore space and effective treatment boundaries, introduced novel causal estimands, and established their identification under continuity assumptions, leveraging treatment unconfoundedness implied by the assignment rule. One could instead impose unconfoundedness conditional on summary statistics of neighbors' scores rather than the full vector, but its validity would depend on choosing those summaries correctly. We leave this extension to future work.

For estimation, we proposed a distance-based local polynomial estimator for boundary effects, derived its asymptotic properties under local dependence, and developed a variance estimator robust to network dependence. We showed that its convergence rate depends on the codimension of the effective treatment boundary, with the conventional RDD rate recovered at codimension one. These results extend beyond our interference framework and are of more broad interest for multivariate RDDs with complex multiscore assignment rules. To our knowledge, this is also the first study of local polynomial estimation with network-dependent data.

Several limitations of our framework suggest directions for future work. First, we assume a fixed network, as is plausible over short horizons such as in PROGRESA. Extending the framework to jointly model stochastic network formation and outcomes when networks evolve or respond to treatment would be valuable \citep{moon}. Second, while we assume a correctly specified exposure mapping, the true mapping is often unknown. In RDDs, misspecified exposure mappings may lead to interpretable estimands and valid inference \citep{vazquez, savje}, while potentially simplifying the effective treatment boundaries, a possibility worth investigating. Third, in fuzzy RDDs our method focuses on intention-to-treat effects. While these capture overall spillovers, they cannot disentangle channels operating through treatment receipt from those operating directly through treatment assignment. For instance, in PROGRESA, having eligible peers may encourage enrollment through greater household awareness, regardless of whether those peers receive the transfer. Under interference, noncompliance introduces further complexities \citep{imai_noncompliance}, and extending monotonicity from univariate to multivariate RDDs has proven challenging \citep{choi}. Our framework may provide a foundation for studying these settings.

The convergence rate of the distance-based estimator deteriorates when the effective treatment boundary has low dimension relative to the multiscore space. Furthermore, reducing the multiscore space to distance from the boundary discards location information and may be suboptimal \citep{optrdd}. As an alternative, \citet{cattaneoyu} propose location-based estimators using multivariate local polynomial methods to aggregate heterogeneous effects along a boundary. Extending these approaches to our setting would require explicit localization over boundary positions and exposure mappings, which is likely to induce substantial data sparsity in practice. More generally, interference in large groups or networks can generate high-dimensional effective treatment boundaries with heterogeneous codimensions. Adapting flexible nonparametric methods designed for high-dimensional, graph-structured data, such as tree-based methods or GNNs, is therefore an important direction for future research.

\bibliography{biblio}

\begin{bibunit}[apalike]

\emptythanks
\title{Online Appendix to \qq Regression Discontinuity Designs Under Interference"}

\date{}
\maketitle

\begin{appendices}
\numberwithin{table}{section}
\numberwithin{theorem}{section}
\numberwithin{equation}{section}
\setcounter{page}{1}

\section*{Appendix Roadmap}
This Online Appendix collects results and diagnostics referenced in the main text. Appendix~\ref{onlineappendix:irregular} treats identification and estimation of the pooled estimand of Section~\ref{sec:pooledestimand}, including asymptotic normality of the pooled estimator in Section~\ref{sec:idestpooled}. Appendix~\ref{onlineappendix:indirectoverall} gives identification and estimation for the boundary overall indirect effects of Section~\ref{sec:indirectoverall}. Appendix~\ref{appendix:simulations} reports additional simulation evidence: a clustered design with block-diagonal dependence, with and without bias correction (Appendix \ref{onlineappendix:mainsim}); small-world networks, for both pooled and degree-conditional estimators (Appendix \ref{onlineappendix:networksim}); and designs with increasing group size, hence higher-dimensional score spaces and boundaries (Appendix \ref{onlineappendix:bigclustersim}). Appendix~\ref{onlineappendix:empirical} contains additional details and diagnostics related to the empirical application of Section~\ref{sec:empirical}: Appendix \ref{onlineappendix:boundary} contains the construction of the effective treatment boundaries and the minimum-distance computation for the exchangeable exposure mappings; Appendix \ref{onlineappendix:diagnostics} develops and implements the covariate-balance and score-density tests, reports a bandwidth sensitivity analysis, and re-estimates the PROGRESA effects under an alternative exposure mapping. Proofs of all results in the main text and the Online Appendix are in the Supplementary Material.

\section{\red{Pooled Causal Estimand}}\label{onlineappendix:irregular}
This section discusses identification and estimation of the pooled estimand of Section.~\ref{sec:pooledestimand}. 
Identification is achieved by imposing Assumption \ref{ass:identification} to hold for all $\phi$, namely $\densityphi$ is positive in a neighborhood of the boundary, and continuous and bounded, and that $m_{d,g}(x_i, \bm x_{\smallN_i}; \phi)$ is continuous and bounded. 

\begin{theorem}\label{th:idpooled}
Suppose Assumptions \ref{ass:consistency}-\ref{ass:identification} hold, with \ref{ass:identification} holding for all $\phi \in \Phi$, and that $\bar{s}_\phi = \bar{s}$ for all $\phi$. Then for random index $I$ drawn uniformly from $\{1, ..., n\}$:     
    \begin{align*}
        \boundeffpooled =& \lim_{\epsilon \to 0^+} \CE{Y_I}{ (X_I, \bm X_{\smallN_I}) \in  \mathcal{B}_{\epsilon}^{d,g}\big(\Frontier\big)}\\ 
        - & \lim_{\epsilon \to 0^+} \CE{Y_I}{ (X_I, \bm X_{\smallN_I}) \in  \mathcal{B}_{\epsilon}^{d',g'}\big(\Frontier\big)}
    \end{align*}
Proof. \textup{See Supplementary Material~\ref{appendix:pooledproof} }.

\end{theorem}
%
\noindent Note that here $ \CE{Y_I}{ (X_I, \bm X_{\smallN_I}) \in  \mathcal{B}_{\epsilon}^{d,g}\big(\Frontier\big); \, \phi }$ corresponds to 
\[
\frac{\sum_{\phi} \CE{Y_i}{(X_i,\pscore) \in \mathcal{B}_{\epsilon}^{d,g}\big(\Frontier\big); \, \phi }\Prr((X_i,\pscore) \in \mathcal{B}_{\epsilon}^{d,g}\big(\Frontier\big); \, \phi )}{\sum_{\phi}\Prr((X_i,\pscore) \in \mathcal{B}_{\epsilon}^{d,g}\big(\Frontier\big); \, \phi )\cdot \pi_\phi}
\]
for generic effective treatment level $(d,g)$.
%
As described in Section \ref{sec:idestpooled}, the pooled estimand can be estimated by a distance-based local linear pooled estimator $\estimatorpooled$ that aggregates all observations based on their distance to their effective treatment boundary. The estimator is asymptotically normal, 

\begin{theorem}\label{th:heterogeneousconvergence}
{Suppose Assumptions \ref{ass:consistency}--\ref{ass:networkheterogeneity}, \ref{ass:boundaryhet}, \ref{ass:pooledcovariance}, \ref{ass:kernel} hold, and let \ref{ass:identification}, \ref{ass:regularity}.i and \ref{ass:smoothnesshet} hold for all $\phi$.} Let $\bar{s}_{\phi}=\bar{s}$ for all $\phi$.
If $n h_n^{\bar{s}} \to \infty$, $h_n \to 0$, and 
$h_n = O\!\left(n^{-1/(4+ \bar{s})}\right)$, then
\[
\sqrt{n h_n^{\bar{s}}}\,
\frac{(\estimatorpooled - \boundeffpooled) - h_n^{2}B_{d,g,d',g'}(h_n)}
{\sqrt{V_{d,g,d',g'}(h_n)}}
\xrightarrow{d} \mathcal{N}(0,1),
\]
where $B_{d,g,d',g'}(h_n) = B_{d,g,d',g'} + o(1)$ and $V_{d,g,d',g'}(h_n) = V_{d,g,d',g'} + o(1)$,
\begin{equation*}
B_{d,g,d',g'} 
= \Big(\frac{\bar{\mu}^{2}_{d,g}(0)-\bar{\mu}^{2}_{d',g'}(0)}{2!}\Big)e_1^\top (\Gamma^{\bar{s}})^{-1}\theta^{\bar{s}},
\end{equation*}
and 
\begin{equation*}
    V_{d,g,d',g'}
= \Big(\frac{\omegaplushet + \omegaminushet + \red{\bar \gamma_{d,g, d',g'}(0)}}{\densitypooledbar[0]^2}\Big)e_1^\top (\Gamma^{\bar{s}})^{-1}\Psi^{\bar{s}}(\Gamma^{\bar{s}})^{-1}
e_1.
\end{equation*}
Proof. \textup{See Supplementary Material \ref{appendix:heterogeneous}} 
\end{theorem}
\noindent The bias depends on the second derivatives of the pooled outcome regression functions $\bar{\mu}_{d,g}(\cdot)$ and $\bar{\mu}_{d',g'}(\cdot)$, defined on the distance variable $\tilde{x}$, defined as density-weighted averages across strata:
\begin{equation*}
\bar{\mu}_{d,g}(\tilde{x}) = \frac{\sum_{\phi} \mu_{d,g}(\tilde{x};\phi)\, f_{d,g|d',g'}(\tilde{x};\phi)\, \pi_\phi}{\sum_\phi f_{d,g|d',g'}(\tilde{x};\phi)\, \pi_\phi}, 
\qquad 
\bar{\mu}_{d',g'}(\tilde{x}) = \frac{\sum_{\phi} \mu_{d',g'}(\tilde{x};\phi)\, f_{d,g|d',g'}(\tilde{x};\phi)\, \pi_\phi}{\sum_\phi f_{d,g|d',g'}(\tilde{x};\phi)\, \pi_\phi},
\end{equation*}
where $f_{d,g|d',g'}(\tilde{x};\phi)$ is the integral of the stratum-specific density over the points distance $\tilde{x}$ from the boundary 
\begin{equation*}
f_{d,g|d',g'}(\tilde{x};\phi) = \int_{\partial \mathcal{B}_{\tilde{x}}(\Frontierphi)} 
f(x_i, \bm{x}_{\smallN_i};\phi)\, d\mathcal{H}^{\mathcal{S}_\phi + 1 - \bar{s}_\phi},
\end{equation*}
with $\partial\mathcal{B}_{\tilde{x}}(\Frontierphi) = \{\ppoint \in \scorespace : l_{d,g|d',g'}^{min}(x_i, \bm{x}_{\smallN_i}) = \tilde{x}\}$.
The variance depends on an aggregate boundary variance components, $\bar{\omega}_{d,g}(0) = \sum_\phi \omega_{d,g}(0; \phi)\, \pi_\phi$ and $\bar{\omega}_{d',g'}(0) = \sum_\phi \omega_{d',g'}(0; \phi)\, \pi_\phi$, and on the aggregate boundary density $\densitypooledbar[0] = \sum_\phi f_{d,g|d',g'}(0;\phi)\, \pi_\phi$.
{\color{black}
It further depends on the covariance components
\(
\gamma_{d,g, d',g'}(0)
=
\sum_{\phi,\phi'}\bar\pi_\phi \cdot \bar \gamma_{d,g,d',g'}(0;\phi,\phi'),
\), where $\bar \gamma_{d,g,d',g'}(0;\phi,\phi') = \gamma_{d,g}(0;\phi,\phi') + \gamma_{d',g'} - 2 \gamma_{d,g,d',g'}(0;\phi,\phi')$, where, for $a,b \in \{(d,g), (d',g')\}$ each generic term  is the limit of
\[
\gamma_{a,b,n}(0;\phi,\phi')
:=
\frac{1}{n_\phi}
\sum_{i\in\mathcal V_\phi}
\sum_{j\in\mathcal K_i^{\phi,\phi'}}
\int_{\mathcal X_{ij}^{\phi,\phi'}(d,g\mid d',g')}
\gamma_{a,b,ij}(x_{ij};\phi,\phi')
f_{ij}(x_{ij};\phi,\phi')
,d\mathcal H^{q_{ij}-\bar s}(x_{ij}).
\]
Here, $\gamma_{a,b,ij}(x_{ij};\phi,\phi') = \kappa_{ij}^{a,b}(\bm x_{ij};\phi,\phi') +  \eta^{a}(x_i, \bm x_{\smallN_i};\phi)\eta^{b}(x_j, \bm x_{\smallN_j};\phi')$ and 
\(
\kappa_{ij}^{a,b}(\bm x_{ij};\phi,\phi')
=
\C\left[
Y_i(a),Y_j(b)
\mid
\bm X_{ij}=\bm x_{ij},\phi,\phi'
\right],
\)
is the conditional outcome covariance across strata, \(\mathcal X_{ij}^{\phi,\phi'}(d,g\mid d',g')\) is the boundary intersection for units in strata \(\phi\) and \(\phi'\), and \(\mathcal K_i^{\phi,\phi'}\) contains the dependent units whose boundary intersection has codimension \(\bar s\).
}
Inference can be conducted using the same variance estimator as in Section \ref{subsec:varianceest}, applied to the full sample rather than within each stratum. Bias correction can similarly be implemented using the procedure described in Section \ref{sec:biascorrection}, estimating second derivatives via a quadratic pooled estimator.

We note that alternative aggregate estimands can be defined. For example, the literature on causal inference under network interference considers simple averages of stratum-specific effects \citep[e.g.][]{xu2024differenceindifferencesinterference}. In our setting, this corresponds to
\[
\tau_{d,g| d',g'}^{\text{simple}}(\Frontier) = \sum_\phi \tau_{d,g|d',g'}(\Frontier;\phi)\, \pi_{\phi},
\]
which assigns equal weight to each stratum regardless of the likelihood of being near the boundary. By contrast, the pooled estimand $\boundeffpooled$ weights strata proportionally to their boundary density. The two estimands coincide when $f_{d,g|d',g'}(0;\phi)$ is constant across strata.
The simple estimand $\tau_{d,g| d',g'}^{\text{simple}}(\Frontier)$ can be estimated by computing stratum-specific estimators and averaging across strata. However, this approach may result in high variance due to the sparsity of observations near the stratum-specific boundaries. In contrast, the pooled estimator aggregates all observations within a given distance from their boundary, leading to greater estimation precision.

\section{Boundary Overall Indirect Effects}\label{onlineappendix:indirectoverall}

This section discusses interpretation, identification and estimation results for the the stratum-specific and pooled boundary overall indirect effects defined in section \ref{sec:indirectoverall}, denoted be $\tau_{\text {indirect }}(\phi)$ and $\tau_{\text {indirect }}$.
The effect $\tau_{\text {indirect }}(\phi)$ represents the marginal effect of having an additional treated neighbor at the relevant boundary. As highlighted in Section \ref{sec:indirectoverall} this effect appears cumbersome in its general form, but it simplifies once we specify the interference set and exposure mapping.
For the sum-of-treated exposure mapping $G_i=\sum_{j \in \mathcal{S}_i} D_j$, this estimand becomes:
\begin{equation*}
\begin{aligned}
\tau_{\text{indirect}}(\phi)
={}&
\sum_{\substack{
d\in\{0,1\}\\[-0.3ex]
g\in\{0,\ldots,|\mathcal{S}_i|-1\}
}}
\tau_{d,g+1,d',g'}
\left(
\overline{\mathcal{X}}_{i\mathcal{S}_i}^{\phi}(d,g+1\mid d,g);\phi
\right)
\\[-0.7ex]
&\quad{}\cdot
\mathbb{P}\left(
(X_i,\boldsymbol{X}_{\mathcal{S}_i})
\in
\overline{\mathcal{X}}_{i\mathcal{S}_i}(d,g+1\mid d,g)
\,\middle|\,
\exists j\in\mathcal{S}_i:X_j=c;\phi
\right).
\end{aligned}
\end{equation*}
which represents the weighted average of the boundary indirect effect of increasing the neighborhood treatment level from $g$ to $g+1$ $\tau_{d, g+1, d^{\prime}, g^{\prime}}\left(\overline{\mathcal{X}}_{i s_i}^\phi(d, g+1 \mid d, g) ; \phi\right)$, weighted by the probability of being at $\overline{\mathcal{X}}_{i s_i}^\phi(d, g+1 \mid d, g)$ conditional on having at least one neighbor at the cutoff. Under one-treated exposure mapping:
\begin{equation*}
\tau_{\text {indirect }}(\phi)=\sum_{d \in\{0,1\}} \tau_{d, 1 \mid d, 0}\left(\overline{\mathcal{X}}_{i \mathcal{S}_i}^\phi(d, 1 \mid d, 0)\right) \cdot \mathbb{P}\left(\left(X_i, \boldsymbol{X}_{\mathcal{S}_i}\right) \in \overline{\mathcal{X}}_{i \mathcal{S}_i}^\phi(d, 1 \mid d, 0) \mid \exists j \in \mathcal{S}_i: X_j=c\right)
\end{equation*}
i.e., the weighted average of the boundary indirect effects of having at least one treated neighbor versus none, for both treated and untreated units.

Let $X_{\smallN_i}^{min} = X_{j^*}$, where $j^* = \argmin_{j \in \smallN_i} |X_j - c|$ denote the score of the neighbor closest to the cutoff. We denote its generic value by $x_{\smallN_i}^{\min}$.
Geometrically, $|X_{\smallN_i}^{\min} - c|$ represents the orthogonal distance from the point $\bm{X}_{\smallN_i}$ to the subspace of $\scorespace$ where at least one component of $\bm{x}_{\smallN_i}$ equals $c$. Then, under Assumptions~\ref{ass:consistency}–\ref{ass:identification}, 
\[
\tau_{indirect}(\phi) = \lim_{x_{\smallN_i}^{min} \downarrow c}\CE{Y_i}{X_{\smallN_i}^{min} = x_{\smallN_i}^{min}; \phi} - \lim_{x_{\smallN_i}^{min} \uparrow c}\CE{Y_i}{X_{\smallN_i}^{min} = x_{\smallN_i}^{min}; \phi}
\]
The proof follows easily from similar arguments as in Theorem~\ref{th:cont} and Theorem~\ref{th:overall}, and is thus omitted.  

Based on this identification, $\tau_{indirect}(\phi)$ can be estimated via univariate local linear regression using $X_{\smallN_i}^{min}$ as the one-dimensional regressor, letting $c = 0$:
\begin{equation}
\begin{split}
    \hat{\tau}_{indirect}(h_n; \phi) = &e_1^T\hat{\beta}_{1,indirect}(h_n; \phi) - e_1^T\hat{\beta}_{0,indirect}(h_n; \phi), \\
    \hat{\beta}_{1,indirect}(h_n; \phi) =& \argmin\limits_{b0,b1 \in \mathbbm{R}}\sum_i\mathbbm{1}(X^{min}_{\smallN_i} \ge 0, \phi_i = \phi)(Y_i - b_0 - b_1X^{min}_{\smallN_i})^2K_{h_n}(X^{min}_{\smallN_i}), \\
    \hat{\beta}_{0,indirect}(h_n; \phi) = & \argmin\limits_{b0,b1 \in \mathbbm{R}}\sum_i\mathbbm{1}(X^{min}_{\smallN_i} < 0; \phi_i = \phi)(Y_i - b_0 - b_1X^{min}_{\smallN_i})^2K_{h_n}(X^{min}_{\smallN_i}),
    \end{split}
\end{equation}
Let $\mu_{1, indirect}(x^{min}_{\smallN_i}; \phi) = \CE{Y_i}{X_{\smallN_i}^{min} = x^{min}_{\smallN_i}; \phi}$ for $x^{min}_{\smallN_i} \ge 0$, and $\mu_{0, indirect}(x^{min}_{\smallN_i}; \phi) = \CE{Y_i}{X_{\smallN_i}^{min} = x^{min}_{\smallN_i}; \phi}$ for $x^{min}_{\smallN_i} \le 0$. 
\begin{theorem}
Suppose Assumptions~\ref{ass:consistency}--\ref{ass:localdependence} hold. Further, suppose that Assumptions 
\ref{ass:regularity}, and~\ref{ass:kernel} hold, but replacing Assumption~\ref{ass:smoothness} with: $\mu_1(x^{min}_{\smallN_i}; \phi)$ and $\mu_0(x^{min}_{\smallN_i}; \phi)$ are $3$-times continuously differentiable.  
If $n_\phi h_n \to \infty$, $h_n \to 0$, and $h_n = O(n_\phi^{-1/5})$, then
\[
\sqrt{n_\phi h_n}
\frac{\big(\hat{\tau}_{indirect}(h_n; \phi) - \tau_{indirect}(\phi)\big) - h_n^{2}B_{indirect}(h_n; \phi)}{\sqrt{V_{indirect}(h_n; \phi)}}
\xrightarrow{d} 
\mathcal{N}(0,1),
\]
where $B_{indirect}(h_n) = O(1)$ and $V_{indirect}(h_n) = O(1)$.\\
Proof. \textup{The proof is analogous to Theorem \ref{th:asynormal} and \ref{th:overallnormal}, and thus, is omitted for brevity.}
\end{theorem}

\noindent This estimator achieves the $n_\phi h_n^5$ rate. Specifically, $\hat{\tau}_{indirect}(h_n; \phi)$ averages observations in a neighborhood of the subset of $\scorespace$ where at least one component of $\pscore$ equals $c$, which is itself a boundary formed by a union of linear pieces.
The codimension of this subset is one, implying an asymptotic variance of order $O\big((n_\phi h_n)^{-1}\big)$ and the usual RDD convergence rate. Thus, while estimators of boundary indirect effects $\boundeff$ comparing specific values $g$ and $g'$ may converge more slowly, this aggregate measure attains the canonical fast rate of standard RDDs.
An estimator of the pooled version of this effect, denoted by $\hat{\tau}^{indirect}(h_n)$, is obtained by pooling all observations with $X_{\smallN_i}^{min}$ within a given distance from the cutoff. Asymptotic properties can then be derived analogously to Theorem \ref{th:heterogeneousconvergence}.

\section{Additional Simulation Results}\label{appendix:simulations}

\subsection{Clustered Data with Equal Cluster Size}\label{onlineappendix:mainsim}
We report the results of a simulation study demonstrating the finite-sample properties of our estimators , and their bias-corrected version.
In this exercise, units are clustered in groups of size 3 ($\bm{A}$ is a block diagonal matrix). 
Interference occurs within clusters, so that $|S_i| = 2$ for each $i$. We use the sum-treated exposure mapping, common in the literature \citep[see, e.g.,][]{aronowsamii17}, resulting in $G_i\in \{0,1,2\}$ and a partition of the multiscore space into six treatment regions. Outcomes are simulated from the model 
\begin{equation}\label{mod:simul}
        Y_i = 0.5 + 1.5D_i + G_i + 0.5X_i + 4.3D_iX_i^2 - 1.5(1 - D_i)X_i^2 + 0.3\bar{\bm{X}}_{\smallN_i}  + \epsilon_i
\end{equation}
\noindent
where $\bar{\bm{X}}_{\smallN_i}$ is the average score within $\mathcal{S}_i$. Here, $X_i$ is independently and identically distributed as a truncated normal with mean -0.3 and variance 1, and upper and lower bound -5 and 5, while the error terms are given by $\epsilon_i = D_i\big(2\frac{\sum_{j\in\smallN_i}u_j}{|\mathcal{S}_i|} + u_i\big) - 1.4(1 -D_i)\cdot \big(2\frac{\sum_{j\in\smallN_i}u_j}{|\mathcal{S}_i|} + u_i\big)$ with $u_i \overset{i.i.d.}{\sim} \mathcal{N}(0,1)$, so that the correlation between the error terms of two neighbors depends on their treatment. Observations are then dependent only within each cluster with a block-diagonal dependency graph. In addition, Model \eqref{mod:simul} is non-linear in the individual score, which enables us to evaluate how the estimators differ with and without bias correction.
{Here, all units in the sample have the same neighborhood degree, and the same conditional outcome expectation and score distribution, and thus the same boundary effects. We therefore drop the index $\phi$ from the estimands and estimators notation.}

We simulate 1000 datasets with increasing sample size $n=\{750, 1500, 3000, 6000\}$ and increasing number of clusters $Z = \{250, 300, 1000, 2000\}$. For each, we estimate the boundary overall direct effect $\overall$, the boundary direct effects $\tau_{10|00}(\mathcal{X}_{i\smallN_i}(10|00))$ and, the boundary indirect effects $\tau_{01|00}(\mathcal{X}_{i\smallN_i}(01|00))$ and $\tau_{02|00}(\mathcal{X}_{i\smallN_i}(02|00))$, using the estimator in Eq.\eqref{eq:estimatormain} (Table \ref{tab_mc_one}). 
For the first two effects, the effective treatment boundaries has dimension $3 - \bar{s}$ with $\bar{s}=1$. In contrast, the effective treatment boundary for $\tau_{02|00}(\mathcal{X}_{i\smallN_i}(02|00))$ has $\bar{s}=2$, implying a slower convergence rate for the corresponding estimator.
We compute the standard error and confidence intervals using the variance estimator in Eq. \eqref{eq:var}. For comparison, we include the heteroskedasticity-robust plug-in residuals variance estimator without weights assuming i.i.d. data, and the cluster-robust variance estimator of \citet{bartalotti19} accounting for dependence among observations on the same side of the cutoff, both implemented in \texttt{rdrobust} R package \citep{rdrobust}. 
We estimate the optimal bandwidth following the plug-in procedure in \citet{cct14}. In addition, for sample size $n = 3000$, we estimate the same causal effects using the bias-corrected estimators described Section~\ref{sec:biascorrection}, denoted by $\hat{\tau}^{bc}_{d,g|d',g'}(h_n, b_n)$ (Table \ref{tab:biascorrection}).

\paragraph{Results without bias correction.}Table \ref{tab_mc_one} presents the results for the estimators without bias correction. 
The estimators $\hat{\tau}_{1|0}(h_n)$ and $\hat{\tau}_{10|00}(h_n)$ exhibit some bias. In contrast, the estimator $\hat{\tau}_{01|00}(h_n)$ exhibit small bias. This discrepancy arises from the different degrees of non-linearity of the target outcome regression functions. In addition, the standard deviation of $\hat{\tau}_{01|00}(h_n)$ is consistently higher than that of $\hat{\tau}_{10|00}(h_n)$, due to clusterization at the distance level. As expected, $\hat{\tau}_{02|00}(h_n)$ has the highest variance among all estimators.
\begin{table}[t]
\centering
\resizebox{0.95\textwidth}{!}{
\footnotesize{
\begin{tabular}[T]{llccccccccccc}           
            \hline

    & & Bias & S.D. & S.E. & S.E.$_{i.i.d.}$ & S.E.$_{cl}$& C.R. & C.R.$_{i.i.d.}$ & C.R.$_{cl}$& $\bar{N}_{d,g}$ & $\bar{N}_{d',g'}$  \\ 

  \hline 
 
 \multirow{4}{*}{n = 750}& $ \hat{\tau}_{1|0}(h_n)$  &  -0.184  &   0.643  &   0.577  &   0.536  &   0.561  &   0.912  &   0.886  &   0.899  & 148.341  & 177.908  \\
~&$ \hat{\tau}_{10|00}(h_n)$  &  -0.189  &   0.939  &   0.812  &   0.780  &   0.823  &   0.902  &   0.890  &   0.906  &  57.658  &  69.806  \\
~&$ \hat{\tau}_{01|00}(h_n)$  &  -0.031  &   1.489  &   1.340  &   1.117  &   1.357  &   0.933  &   0.872  &   0.940  &  98.904  &  96.580  \\
~&$ \hat{\tau}_{02|00}(h_n)$  &  -0.064  &   4.294  &   2.982  &   2.932  &   3.060  &   0.855  &   0.847  &   0.862  &  22.952  &  34.422  \\
\hline 
\multirow{4}{*}{n = 1500}& $ \hat{\tau}_{1|0}(h_n)$  &  -0.156  &   0.446  &   0.423  &   0.395  &   0.411  &   0.918  &   0.891  &   0.907  & 279.640  & 330.839  \\
~&$ \hat{\tau}_{10|00}(h_n)$  &  -0.141  &   0.644  &   0.588  &   0.565  &   0.592  &   0.911  &   0.897  &   0.912  & 113.016  & 135.519  \\
~&$ \hat{\tau}_{01|00}(h_n)$  &  -0.030  &   1.052  &   0.943  &   0.781  &   0.949  &   0.928  &   0.851  &   0.928  & 203.788  & 198.410  \\
~&$ \hat{\tau}_{02|00}(h_n)$  &   0.146  &   2.840  &   2.259  &   2.187  &   2.285  &   0.920  &   0.912  &   0.921  &  48.221  &  72.686  \\
\hline 
\multirow{4}{*}{n = 3000}& $ \hat{\tau}_{1|0}(h_n)$  &  -0.137  &   0.337  &   0.314  &   0.295  &   0.306  &   0.909  &   0.883  &   0.896  & 512.603  & 596.299  \\
~&$ \hat{\tau}_{10|00}(h_n)$  &  -0.166  &   0.463  &   0.433  &   0.415  &   0.435  &   0.907  &   0.895  &   0.907  & 216.641  & 255.929  \\
~&$ \hat{\tau}_{01|00}(h_n)$  &  -0.042  &   0.737  &   0.669  &   0.549  &   0.671  &   0.935  &   0.864  &   0.935  & 418.464  & 404.431  \\
~&$ \hat{\tau}_{02|00}(h_n)$  &  -0.016  &   2.070  &   1.642  &   1.574  &   1.651  &   0.913  &   0.896  &   0.915  &  97.101  & 146.636  \\
\hline 
\multirow{4}{*}{n = 6000}& $ \hat{\tau}_{1|0}(h_n)$  &   -0.106  &    0.245  &    0.235  &    0.222  &    0.229  &    0.908  &    0.893  &    0.901  &  920.610  & 1051.498  \\
~&$ \hat{\tau}_{10|00}(h_n)$  &   -0.127  &    0.364  &    0.322  &    0.310  &    0.323  &    0.890  &    0.872  &    0.890  &  399.292  &  464.010  \\
~&$ \hat{\tau}_{01|00}(h_n)$  &    0.015  &    0.511  &    0.475  &    0.389  &    0.475  &    0.942  &    0.873  &    0.943  &  840.688  &  811.067  \\
~&$ \hat{\tau}_{02|00}(h_n)$  &    0.046  &    1.384  &    1.184  &    1.127  &    1.187  &    0.930  &    0.920  &    0.931  &  197.973  &  299.199  \\
 \hline 
 \end{tabular}}}
 \caption{Simulation results. One thousand replications. 
 S.D. = empirical standard error. S.E. = estimated standard error. S.E.$_{i.i.d.}$ = estimated standard error with $i.i.d.$ variance estimator. S.E.$_{cl}$ = estimated standard error with the cluster-robust variance estimator. 
 C.R. = coverage rate. C.R.$_{i.i.d.}$ = coverage rate corresponding to the $i.i.d.$ variance estimator. C.R.$_{cl}$ = coverage rate corresponding to the cluster-robust variance estimator. 
 $N_{d,g}$ and $N_{d',g'}$ = effective sample size on each side of the treatment boundary.}
 \label{tab_mc_one}
\end{table}
Our variance estimator generally provides more accurate estimates of the standard deviation than the $i.i.d.$ estimator, with the magnitude of improvement depending on the degree of clustering induced by the treatment assignment. \red{The improvement is particularly substantial for $\hat{\tau}_{01|00}(h_n)$. With clusters of size three, consider any pair of dependent units and label the cluster ${i,j,k}$, where $k$ denotes the remaining unit. For the generic dependent pair $i$ and $j$, the joint boundary is $\{X_i\leq c,X_j\leq c,X_k=c\}$ and therefore has codimension one, equal to the individual-boundary codimension. Hence, Assumption~\ref{ass:commonpiecesmain} fails and covariance contributes to the asymptotic variance, as discussed in Section~\ref{sec:estimation}. Consistently, i.i.d. standard errors understate the empirical variability and coverage remains around $0.85$--$0.87$ as $n$ increases, while dependence-robust coverage approaches the nominal level. By contrast, for $\hat{\tau}_{02|00}(h_n)$, the joint boundary is $\{X_i=X_j=X_k=c\}$, with codimension three, exceeding the individual-boundary codimension of two. Assumption~\ref{ass:commonpiecesmain} therefore holds, so the covariance contribution vanishes asymptotically. Correspondingly, i.i.d. coverage improves with sample size, indicating that the remaining performance gap is due to finite-sample dependence.}
The performance of the cluster-robust variance estimator is nearly identical to that of our variance estimator for $\hat{\tau}_{10|00}(h_n)$, $\hat{\tau}_{01|00}(h_n)$  and $\hat{\tau}_{02|00}(h_n)$. This is expected: under cluster dependence, these two estimators tend to coincide when the dependent observations are located on the same side of the zero-distance cutoff, as is the case here. In contrast, for $\hat{\tau}_{1|0}(h_n)$, our variance estimator outperforms the cluster-robust alternative, as it properly accounts for the covariance between units on opposite sides of the cutoff.
The coverage of $\hat{\tau}_{02|00}(h_n)$ improves with larger sample sizes, while the coverage of $\hat{\tau}_{01|00}(h_n)$ approaches the nominal level in smaller samples, reflecting the slower convergence of $\hat{\tau}_{02|00}(h_n)$. 
Coverage also stabilizes for $\hat{\tau}_{1|0}(h_n)$ and $\hat{\tau}_{10|00}(h_n)$ in smaller samples due to their faster convergence.  
However, for these two estimators, coverage settles below the nominal level because the selected bandwidths do not fully eliminate the leading bias in their distributions.

\paragraph{Results with bias correction.}Table \ref{tab:biascorrection} presents the results for the bias-corrected estimators. The estimators for $\overall$ and $\tau_{10|00}(\bar{\mathcal{X}}_{i\smallN_i}(10|00))$ effectively remove the leading bias from the estimate. In addition, the estimated confidence intervals also provide better coverage of the true values across effects, reaching a rate close to the nominal level. Here, our standard error estimator and the i.i.d. estimator perform similarly, with the largest difference observed in the estimation of the standard error of $\hat{\tau}^{bc}_{01|00}(h_n, b_n)$ and $\hat{\tau}^{bc}_{02|00}(h_n, b_n)$.

\begin{table}[H]
\centering
\resizebox{0.95\textwidth}{!}{
\footnotesize{
\begin{tabular}[T]{llcccccccccccc}

            \hline

    & & Bias & S.D. & S.E. & S.E.$_{i.i.d.}$ & S.E.$_{cl}$& C.R. & C.R.$_{i.i.d.}$ & C.R.$_{cl}$& $\bar{N}_{d,g}$ & $\bar{N}_{d',g'}$  \\ 

  \midrule 
 
 \multirow{4}{*}{n = 3000}& $ \hat{\tau}^{bc}_{1|0}(h_n, b_n)$  &   0.007  &   0.349  &   0.338  &   0.320  &   0.331  &   0.940  &   0.926  &   0.934  & 512.878  & 596.780  \\
~&$ \hat{\tau}^{bc}_{10|00}(h_n, b_n)$  &   0.035  &   0.505  &   0.483  &   0.467  &   0.484  &   0.949  &   0.939  &   0.950  & 215.830  & 255.597  \\
~&$ \hat{\tau}^{bc}_{01|00}(h_n, b_n)$  &   0.005  &   0.845  &   0.788  &   0.651  &   0.791  &   0.938  &   0.875  &   0.938  & 417.212  & 404.337  \\
~&$ \hat{\tau}^{bc}_{02|00}(h_n, b_n)$  &   0.101  &   2.553  &   2.161  &   2.090  &   2.173  &   0.931  &   0.918  &   0.931  &  97.388  & 147.831  \\
 \midrule 
 \end{tabular}}}
 \caption{Simulation results using bias correction. One thousand replications. 
 S.D. = empirical standard error. S.E. = estimated standard error.
 S.E.$_{i.i.d.}$ = estimated standard error with $i.i.d.$ variance estimator. S.E.$_{cl}$ = estimated standard error with the cluster-robust variance estimator. 
 C.R. = coverage rate. C.R.$_{i.i.d.}$ = coverage rate corresponding to the $i.i.d.$ variance estimator. C.R.$_{cl}$ = coverage rate corresponding to the cluster-robust variance estimator. $N_{d,g}$ and $N_{d',g'}$ = effective sample size on each side of the treatment boundary.}
 \label{tab:biascorrection}
\end{table}


\subsection{Network Data}\label{onlineappendix:networksim}

The following simulation study evaluates the finite-sample performance of the pooled and conditional estimators, $\hat{\tau}_{d,g|d',g'}(h_n, b_n)$ and $\estimatorbc$
under network interference, where units' interference set coincides with their network neighborhood— the set of nodes directly linked to them. The neighborhood treatment $G_i$ is defined via the one-treated exposure mapping.
For sample sizes $n = \{750, 1500, 3000\}$, we simulate small-world networks \citep{wattstrogatz} using the \texttt{sample\_smallworld()} function from the R package \texttt{igraph}. Small-world networks are characterized by high clustering, meaning that two nodes sharing a neighbor tend to be connected. We fix the average node degree at 4 by setting the argument \texttt{nei = 2}, and vary the rewiring probability to generate networks with different levels of clustering, as measured by the 
average local clustering coefficients, denoted by $\bar{C}$ \citep{wattstrogatz}.
%
We also allow the potential outcome model to depend on unit degree, thereby inducing heterogeneity in boundary effects across degree strata. 
This simulation design allows us to assess the performance of the proposed estimators in network settings for both the pooled estimator and the conditional estimator on the degree, and to investigate how network clustering affects performance—particularly through the induced clustering in the distance variable.

We consider two scenarios: \textbf{Scenario A}, with rewiring probability $p = 0.15$, resulting in $\bar{C} = 0.20$; \textbf{Scenario B} with rewiring probability $p = 0.05$, resulting in $\bar{C} = 0.37$. Outcomes are simulated from the model 
\begin{equation}
        Y_i = 0.5 + 1.5D_i + G_i + 0.15 |\mathcal{S}_i|\cdot G_i + 0.5X_i + 4.3D_iX_i^2 - 1.5(1 - D_i)X_i^2 + 0.3\bar{\bm{X}}_{\smallN_i}  + \varepsilon_i
\end{equation}

\noindent
where $X_i$ is i.i.d. a truncated normal with mean -0.5 and variance 1, and upper and lower bound -5 and 5, while the error terms are given by $\varepsilon_i = D_i\big(2\frac{\sum_{j\in\smallN_i}u_j}{|\mathcal{S}_i|} + u_i\big) - 1.4(1 -D_i)\cdot \big(2\frac{\sum_{j\in\smallN_i}u_j}{|\mathcal{S}_i|} + u_i\big)$ with $u_i \overset{i.i.d.}{\sim} \mathcal{N}(0,1)$. Observations are thus dependent up to second-order network neighbors (second-order dependence), and the boundary average effects are homogeneous across units with the same degree.

Simulation results are reported in Tables \ref{tab:net_clus} 
for the pooled estimators and the conditional estimators for units with degree equal to 4. The table reports the empirical bias, empirical standard deviation, estimated standard error (using both our proposed variance estimator and the standard i.i.d. estimator), and the associated confidence interval coverage rates. We also report the average effective sample sizes ($N_{d,g}$ and $N_{d',g'}$) and the proportion of units with unique score variables and minimum distance values (column \qq\% Unique"). Values less than one in this column indicate overlapping distance variables among units.
The proposed pooled estimators are unbiased across all sample sizes and scenarios. The standard deviation decreases substantially with sample size in both settings. The overall effect estimator $\hat{\tau}^{bc}_{1|0}(h_n, b_n)$ consistently has the lowest standard deviation, due to arger effective sample sizes. In contrast, $\hat{\tau}_{10|00}(h_n, b_n)$ exhibits the highest standard deviation, driven by a smaller fraction of units with $(D_i = 1, G_i = 0)$. 
The estimator $\hat{\tau}_{01|00}(h_n, b_n)$ has a relatively large effective sample size but still exhibits high variance. This is explained by the clustering structure of the distance variable: in Scenario A, approximately 80\% of units have unique distance values, while in Scenario B only 74\% do. This increase in overlap leads to higher standard deviation in Scenario B.
The conditional estimators $\hat{\tau}^{bc}_{1|0}(h_n, b_n; \phi)$ and $\hat{\tau}^{bc}_{d,g|d',g'}(h_n, b_n; \phi)$ for units with degree 4 are also approximately unbiased, except for the indirect effect at $n = 750$ due to very small effective sample sizes. They exhibit a variance pattern similar to the pooled estimators, but with systematically larger variance due to the reduced sample within the degree stratum.

For the pooled estimators, our variance estimator performs well, approaching the empirical standard deviation in large samples. Coverage rates using this estimator converge to the nominal level, whereas the i.i.d. confidence intervals tend to undercover the true value, although their coverage improves with larger sample sizes. For $\hat{\tau}^{bc}_{01|00}(h_n, b_n)$, however, the coverage does not improve in large samples and it is substantially worse in Scenario B, with higher clustering in the network.
For the conditional estimators, differences between our variance estimator and the i.i.d. estimator—and the associated confidence intervals—are less pronounced in Scenario A, except for the overall direct effect. In contrast, in Scenario B, our variance estimator achieves improved coverage.
Overall, our variance estimator is more accurate than the $i.i.d$ one, especially in Scenario B with higher clustering.  

In conclusion, this simulation study demonstrates an overall good performance of the proposed distance-based estimators and dependence-robust variance estimator. It also highlights how increased clustering in the network can induce higher clustering in the distance variable. When this occurs, relying on the i.i.d. variance estimator leads to consistently poor coverage.


\begin{table}[H]
\centering                        
\footnotesize

\caption*{\normalsize \textbf{Scenario A}}  

\resizebox{0.96\textwidth}{!}{%
\begin{tabular}{llccccccccc}
\hline
& & Bias & S.D. & S.E. & S.E.$_{i.i.d.}$ & C.R. & C.R.$_{i.i.d.}$ 
& $\bar{N}_{d,g}$ & $\bar{N}_{d',g'}$ & \% Unique \\ 
\hline

\multirow{3}{*}{n = 750} 
& $\hat{\tau}^{bc}_{1|0}(h_n,b_n)$     
& 0.026 & 1.178 & 1.067 & 0.831 & 0.925 & 0.837 
& 134.541 & 183.627 & 1.000 \\ 

& $\hat{\tau}^{bc}_{10|00}(h_n,b_n)$   
& 0.027 & 2.539 & 2.060 & 2.054 & 0.880 & 0.877 
& 30.739 & 41.192 & 1.000 \\ 

& $\hat{\tau}^{bc}_{01|00}(h_n,b_n)$   
& -0.011 & 2.313 & 1.974 & 1.893 & 0.900 & 0.899 
& 113.684 & 76.886 & 0.795 \\ 

\midrule

\multirow{3}{*}{n = 750} 
& $\hat{\tau}^{bc}_{1|0}(h_n,b_n;\phi)$     
& -0.036 & 1.779 & 1.534 & 1.259 & 0.899 & 0.838 
& 53.869 & 73.030 & 1.000 \\

& $\hat{\tau}^{bc}_{10|00}(h_n,b_n;\phi)$   
& -1.161 & 31.930 & 3.778 & 3.815 & 0.775 & 0.785 
& 10.394 & 13.457 & 1.000 \\

& $\hat{\tau}^{bc}_{01|00}(h_n,b_n;\phi)$   
& -0.099 & 4.121 & 3.033 & 3.086 & 0.859 & 0.868 
& 37.103 & 26.700 & 0.885 \\

\midrule

\multirow{3}{*}{n = 1500} 
& $\hat{\tau}^{bc}_{1|0}(h_n,b_n)$   
& 0.050 & 0.798 & 0.765 & 0.598 & 0.936 & 0.859 
& 263.750 & 358.313 & 1.000 \\ 

& $\hat{\tau}^{bc}_{10|00}(h_n,b_n)$ 
& 0.038 & 1.674 & 1.474 & 1.441 & 0.920 & 0.913 
& 64.333 & 87.464 & 1.000 \\ 

& $\hat{\tau}^{bc}_{01|00}(h_n,b_n)$
& -0.025 & 1.589 & 1.414 & 1.317 & 0.929 & 0.915 
& 234.421 & 159.958 & 0.798 \\ 

\midrule

\multirow{3}{*}{n = 1500} 
& $\hat{\tau}^{bc}_{1|0}(h_n,b_n;\phi)$     
& 0.06 & 1.14 & 1.09 & 0.89 & 0.94 & 0.88 
& 108.30 & 147.42 & 1.00 \\ 

& $\hat{\tau}^{bc}_{10|00}(h_n,b_n;\phi)$   
& 0.03 & 2.77 & 2.15 & 2.15 & 0.86 & 0.86 
& 22.92 & 30.38 & 1.00 \\ 

& $\hat{\tau}^{bc}_{01|00}(h_n,b_n;\phi)$   
& -0.10 & 2.66 & 2.23 & 2.19 & 0.90 & 0.90 
& 78.02 & 56.26 & 0.89 \\ 

\midrule

\multirow{3}{*}{n = 3000} 
& $\hat{\tau}^{bc}_{1|0}(h_n,b_n)$   
& 0.036 & 0.576 & 0.548 & 0.438 & 0.940 & 0.874 
& 501.658 & 664.884 & 1.000 \\ 

& $\hat{\tau}^{bc}_{10|00}(h_n,b_n)$ 
& 0.032 & 1.149 & 1.050 & 1.024 & 0.934 & 0.929 
& 129.772 & 175.393 & 1.000 \\ 

& $\hat{\tau}^{bc}_{01|00}(h_n,b_n)$
& 0.053 & 1.059 & 0.982 & 0.900 & 0.948 & 0.921 
& 504.585 & 334.135 & 0.797 \\ 

\midrule

\multirow{3}{*}{n = 3000} 
& $\hat{\tau}^{bc}_{1|0}(h_n,b_n;\phi)$     
& 0.03 & 0.85 & 0.78 & 0.64 & 0.94 & 0.87 
& 214.77 & 289.29 & 1.00 \\ 

& $\hat{\tau}^{bc}_{10|00}(h_n,b_n;\phi)$   
& 0.01 & 1.79 & 1.53 & 1.52 & 0.91 & 0.91 
& 48.06 & 64.13 & 1.00 \\ 

& $\hat{\tau}^{bc}_{01|00}(h_n,b_n;\phi)$   
& 0.01 & 1.78 & 1.58 & 1.52 & 0.93 & 0.92 
& 163.52 & 115.59 & 0.89 \\

\hline
\end{tabular}%
}

\end{table}


\begin{table}[H]
\centering
\footnotesize

\caption*{\normalsize \textbf{Scenario B}}  

\resizebox{0.96\textwidth}{!}{%
\begin{tabular}{llccccccccc}
\hline
& & Bias & S.D. & S.E. & S.E.$_{i.i.d.}$ & C.R. & C.R.$_{i.i.d.}$ 
& $\bar{N}_{d,g}$ & $\bar{N}_{d',g'}$ & \% Unique \\ 
\hline

\multirow{3}{*}{n = 750} 
& $\hat{\tau}^{bc}_{1|0}(h_n,b_n)$ 
& -0.020 & 1.090 & 1.050 & 0.814 & 0.932 & 0.859 
& 134.251 & 184.451 & 1.000 \\ 

& $\hat{\tau}^{bc}_{10|00}(h_n,b_n)$ 
& 0.036 & 2.461 & 2.005 & 1.992 & 0.901 & 0.904 
& 29.272 & 39.542 & 1.000 \\ 

& $\hat{\tau}^{bc}_{01|00}(h_n,b_n)$ 
& 0.069 & 2.607 & 2.133 & 1.942 & 0.897 & 0.877 
& 102.776 & 73.006 & 0.741 \\ 

\midrule

\multirow{3}{*}{n = 750} 
& $\hat{\tau}^{bc}_{1|0}(h_n,b_n;\phi)$ 
& 0.01 & 1.29 & 1.23 & 0.98 & 0.94 & 0.87 
& 90.92 & 124.49 & 1.00 \\ 

& $\hat{\tau}^{bc}_{10|00}(h_n,b_n;\phi)$ 
& -0.05 & 3.21 & 2.37 & 2.38 & 0.86 & 0.86 
& 18.51 & 24.83 & 1.00 \\ 

& $\hat{\tau}^{bc}_{01|00}(h_n,b_n;\phi)$ 
& 0.01 & 3.30 & 2.55 & 2.42 & 0.88 & 0.86 
& 64.39 & 46.46 & 0.79 \\ 

\midrule

\multirow{3}{*}{n = 1500} 
& $\hat{\tau}^{bc}_{1|0}(h_n,b_n)$ 
& -0.031 & 0.800 & 0.748 & 0.586 & 0.932 & 0.854 
& 264.120 & 355.919 & 1.000 \\ 

& $\hat{\tau}^{bc}_{10|00}(h_n,b_n)$ 
& -0.103 & 1.627 & 1.427 & 1.399 & 0.903 & 0.897 
& 61.228 & 83.298 & 1.000 \\ 

& $\hat{\tau}^{bc}_{01|00}(h_n,b_n)$ 
& -0.045 & 1.722 & 1.531 & 1.355 & 0.925 & 0.889 
& 213.629 & 149.108 & 0.741 \\ 

\midrule

\multirow{3}{*}{n = 1500} 
& $\hat{\tau}^{bc}_{1|0}(h_n,b_n;\phi)$ 
& -0.03 & 0.96 & 0.87 & 0.69 & 0.92 & 0.85 
& 183.96 & 248.77 & 1.00 \\ 

& $\hat{\tau}^{bc}_{10|00}(h_n,b_n;\phi)$ 
& -0.10 & 2.05 & 1.69 & 1.68 & 0.89 & 0.89 
& 39.92 & 53.61 & 1.00 \\ 

& $\hat{\tau}^{bc}_{01|00}(h_n,b_n;\phi)$ 
& -0.04 & 2.07 & 1.82 & 1.67 & 0.93 & 0.90 
& 135.54 & 96.86 & 0.78 \\ 

\midrule

\multirow{3}{*}{n = 3000} 
& $\hat{\tau}^{bc}_{1|0}(h_n,b_n)$ 
& 0.002 & 0.560 & 0.538 & 0.428 & 0.940 & 0.880 
& 502.501 & 665.696 & 1.000 \\ 

& $\hat{\tau}^{bc}_{10|00}(h_n,b_n)$ 
& -0.028 & 1.068 & 1.015 & 0.985 & 0.941 & 0.934 
& 123.655 & 168.285 & 1.000 \\ 

& $\hat{\tau}^{bc}_{01|00}(h_n,b_n)$ 
& 0.040 & 1.198 & 1.083 & 0.942 & 0.935 & 0.883 
& 445.985 & 305.454 & 0.738 \\ 

\midrule

\multirow{3}{*}{n = 3000} 
& $\hat{\tau}^{bc}_{1|0}(h_n,b_n;\phi)$ 
& 0.01 & 0.63 & 0.61 & 0.50 & 0.95 & 0.88 
& 362.96 & 485.56 & 1.00 \\ 

& $\hat{\tau}^{bc}_{10|00}(h_n,b_n;\phi)$ 
& -0.01 & 1.26 & 1.18 & 1.16 & 0.94 & 0.93 
& 84.38 & 114.68 & 1.00 \\ 

& $\hat{\tau}^{bc}_{01|00}(h_n,b_n;\phi)$ 
& 0.03 & 1.45 & 1.29 & 1.15 & 0.94 & 0.89 
& 288.80 & 202.20 & 0.78 \\ 

\hline
\end{tabular}%
}


\vspace{1em}
\caption{\footnotesize
Simulation results with small-world network.
S.D. = empirical standard error.
S.E. = estimated standard error.
S.E.$_{i.i.d.}$ = estimated standard error with $i.i.d.$ 
heteroskedasticity-robust HC0 variance estimator.
C.R. = coverage rate.
C.R.$_{i.i.d.}$ = $i.i.d.$ coverage rate.
$N_{d,g}$ and $N_{d',g'}$ = effective sample size.
\% Unique = percentage of unique distance variables.
}
\label{tab:net_clus}

\end{table}

\subsection{Varying Group Size}\label{onlineappendix:bigclustersim}

We consider a scenario with equally sized groups. 
We keep the total sample size fixed at $n = 3000$ and vary the cluster size, denoted by $M$. Data are generated as in Section \ref{sec:simulation}, but using the one-treated exposure mapping.
The objective is to assess the performance of our proposed (pooled) bias-corrected estimators in settings with larger group sizes, which correspond to higher-dimensional multiscore spaces and effective treatment boundaries than those considered in Section~\ref{sec:simulation}. The dimension of the multiscore space for each unit is given by $M$. Here, although the total sample size remains constant, the probability of observing certain neighborhood-treatment configurations (e.g., $G_i=0$) decreases with $M$, reducing the effective sample size available for estimation near each boundary.

We simulate 1,000 datasets with group sizes $M \in \{3,4,5,6\}$ and total sample size $n = 3000$, corresponding to $\{1000, 750, 600, 500\}$ groups, respectively. For brevity, we only consider the pooled estimators $\hat{\tau}^{bc}_{1|0}(h_n, b_n)$, $\hat{\tau}^{bc}_{10|00}(h_n, b_n)$, and $\hat{\tau}^{bc}_{01|00}(h_n, b_n)$, and report the results in Table~\ref{tab:mc_one_fixednumber}. 
All estimators are approximately unbiased across different group sizes. 
The estimators $\hat{\tau}_{10|00}^{bc}(h_n, b_n)$ and $\hat{\tau}^{bc}_{01|00}(h_n, b_n)$ remain approximately unbiased for $M = \{3,4,5,6\}$ but exhibit increasing standard deviations as the group size grows. 
The increase in the empirical standard deviation is more pronounced for $\hat{\tau}_{01|00}^{bc}(h_n, b_n)$, due to the high percentage of observations sharing the same distance measure, as indicated in column ``\% Unique.'' 
Our variance estimator for $\hat{\tau}_{10|00}^{bc}(h_n, b_n)$ and $\hat{\tau}^{bc}_{01|00}(h_n, b_n)$ is accurate for smaller group sizes ($M = \{3,4\}$) but worsens in performance for larger groups, resulting in lower coverage rates. 
In contrast, the variance estimator for $\hat{\tau}^{bc}_{1|0}(h_n, b_n)$ remains relatively accurate across group sizes.

The high standard deviation of $\hat{\tau}_{10|00}^{bc}(h_n, b_n)$ and $\hat{\tau}^{bc}_{01|00}(h_n, b_n)$, together with the worsening performance of their variance estimators, arises from the reduced effective sample size. 
In our data generating process, larger groups lead to lower observed probabilities for some neighborhood treatment configurations. 
In particular, the observed probabilities $\Pr(G_i = 0)$ equal approximately $\{0.38, 0.23, 0.14, 0.09\}$ as group size increases. To illustrate the impact of the effective sample size, we repeat the simulation focusing on $\hat{\tau}^{bc}_{01|00}(h_n, b_n)$ but choose the cutoff $c = -0.3 + \Phi^{-1}\!\big(0.35^{1/(M-1)}\big)$,
which ensures that the observed probability $\Pr(G_i = 0) \approx 0.35$ for each group size $M$. 
This adjustment keeps a larger effective sample size for estimation. 
As shown in Table~\ref{tab:mc_one_fixedn}, this modification substantially reduces the estimator's standard error relative to Table~\ref{tab:mc_one_fixednumber}. 
Moreover, our variance estimator provides more accurate variance estimates and improved coverage, even in the presence of substantial clustering at the distance-measure level.

Overall, the results show that while our estimation method yields accurate inference when the effective sample size is sufficient, large groups may lead to sparse data within each effective treatment arm and thus less precise inference.
Nevertheless, our results indicate that even if specific boundary effects cannot be precisely estimated due to limited observations, the boundary overall direct effect can still be estimated accurately when the total sample size is large, as in this setting.

\begin{table}[H]
\centering
\resizebox{0.8\textwidth}{!}{
\begin{tabular}{llcccccccc}

            \hline

  & & Bias & S.D. & S.E. & C.R. & $\bar{N}_{d,g}$ & $\bar{N}_{d',g'}$ & \% Unique  \\ 

  \midrule 
 
   \multirow{3}{*}{M = 3}& $ \hat{\tau}^{bc}_{1|0}(h_n, b_n)$  &  -0.013  &   0.651  &   0.639  &   0.950  & 603.549  & 723.666  &   1.000  \\
~&$ \hat{\tau}^{bc}_{10|00}(h_n, b_n)$  &   0.009  &   1.032  &   0.987  &   0.950  & 243.326  & 294.503  &   1.000  \\
~&$ \hat{\tau}^{bc}_{01|00}(h_n, b_n)$  &   0.039  &   1.402  &   1.280  &   0.929  & 481.672  & 411.999  &   0.637  \\
\midrule 
\multirow{3}{*}{M = 4}& $ \hat{\tau}^{bc}_{1|0}(h_n, b_n)$  &   0.010  &   0.552  &   0.552  &   0.944  & 585.236  & 695.398  &   1.000  \\
~&$ \hat{\tau}^{bc}_{10|00}(h_n, b_n)$  &   0.036  &   1.104  &   1.069  &   0.946  & 150.242  & 181.642  &   1.000  \\
~&$ \hat{\tau}^{bc}_{01|00}(h_n, b_n)$  &  -0.065  &   1.638  &   1.568  &   0.931  & 393.081  & 271.343  &   0.455  \\
\midrule 
\multirow{3}{*}{M = 5}& $ \hat{\tau}^{bc}_{1|0}(h_n, b_n)$  &  -0.026  &   0.509  &   0.504  &   0.958  & 575.230  & 681.489  &   1.000  \\
~&$ \hat{\tau}^{bc}_{10|00}(h_n, b_n)$  &  -0.034  &   1.324  &   1.221  &   0.928  &  92.785  & 111.590  &   1.000  \\
~&$ \hat{\tau}^{bc}_{01|00}(h_n, b_n)$  &   0.094  &   2.375  &   1.952  &   0.885  & 281.102  & 169.292  &   0.350  \\
\midrule 
\multirow{3}{*}{M = 6}& $ \hat{\tau}^{bc}_{1|0}(h_n, b_n)$  &  -0.018  &   0.482  &   0.473  &   0.938  & 565.136  & 668.042  &   1.000  \\
~&$ \hat{\tau}^{bc}_{10|00}(h_n, b_n)$  &  -0.020  &   1.507  &   1.408  &   0.929  &  56.543  &  68.610  &   1.000  \\
~&$ \hat{\tau}^{bc}_{01|00}(h_n, b_n)$  &  -0.016  &   2.939  &   2.388  &   0.888  & 188.998  & 105.040  &   0.284  \\
 \midrule 
 \end{tabular}  }
 \caption{\footnotesize Simulation results over one thousand replications.
 S.D. = empirical standard error. S.E. = estimated standard error.
 C.R. = coverage rate.  $N_{d,g}$ and $N_{d',g'}$ = effective sample size on each side of the treatment boundary.}
\label{tab:mc_one_fixednumber}
 \end{table}

 \begin{table}[H]
\centering
\resizebox{0.8\textwidth}{!}{
\begin{tabular}{llcccccccc}
\hline
  & & Bias & S.D. & S.E. & C.R. & $\bar{N}_{d,g}$ & $\bar{N}_{d',g'}$ & \% Unique  \\ 
  \midrule 
   \multirow{1}{*}{M = 3}& $ \hat{\tau}^{bc}_{01|00}(h_n, b_n)$  &   0.018  &   1.438  &   1.344  &   0.942  & 448.701  & 368.789  &   0.642  \\
\midrule 
\multirow{1}{*}{M = 4}& $ \hat{\tau}^{bc}_{01|00}(h_n, b_n)$  &   0.041  &   1.382  &   1.280  &   0.942  & 526.895  & 436.620  &   0.439  \\
\midrule 
\multirow{1}{*}{M = 5}& $ \hat{\tau}^{bc}_{01|00}(h_n, b_n)$  &  -0.015  &   1.327  &   1.252  &   0.941  & 562.791  & 474.355  &   0.330  \\
\midrule 
\multirow{1}{*}{M = 6}& $ \hat{\tau}^{bc}_{01|00}(h_n, b_n)$  &   0.010  &   1.369  &   1.243  &   0.935  & 580.149  & 498.828  &   0.264  \\
 \midrule 
 \end{tabular}}
 \caption{\footnotesize Simulation results over one thousand replications. S.D. = empirical standard error. S.E. = estimated standard error.
 C.R. = coverage rate.  $N_{d,g}$ and $N_{d',g'}$ = effective sample size on each side of the treatment boundary.}
\label{tab:mc_one_fixedn}
 \end{table}

\section{Diagnostics and Additional Empirical Results}\label{onlineappendix:empirical}

\subsection{Construction}\label{onlineappendix:boundary}

The empirical application in Section \ref{sec:empirical} uses the fraction of treated neighbors,
$ 
G_i = \frac{1}{|\smallN_i|}\sum_{j \in \mathcal{S}_i} D_j,
$
and considers the following effects: (i) the direct effect of own treatment when all neighbors are treated ($G_i = 1$) versus when no neighbors are treated ($G_i = 0$); and (ii) the indirect effect of having all neighbors treated ($G_i = 1$) versus none treated ($G_i = 0$), both when the individual is untreated and when treated.

Based on the procedure described in Section~\ref{sec:etb} 
the corresponding effective treatment boundaries are constructed as follows. First, we characterize the treatment vector configurations that correspond to each effective treatment assignment:
\begin{equation*}
\begin{split}
\treatspace(1,1) &= \{(d_i^{1,1} = 1, \bm d_{\smallN_i}^{1,1} = \bm 1)\}, \quad 
\treatspace(1,0) = \{(d_i^{1,0} = 1, \bm d_{\smallN_i}^{1,0} = \bm 0)\}, \\
\treatspace(0,1) &= \{(d_i^{0,1} = 0, \bm d_{\smallN_i}^{0,1} = \bm 1)\}, \quad 
\treatspace(0,0) = \{(d_i^{0,0} = 0, \bm d_{\smallN_i}^{0,0} = \bm 0)\}.
\end{split}
\end{equation*}
In this setting, each set contains a single treatment configuration.

Second, recalling that 
$
\Frontier = \bigcup_{\treatspace(d,g) \times \treatspace(d',g')} 
\ell\big(\bm d_{i\smallN_i}^{d,g}, \bm d_{i\smallN_i}^{d',g'}\big),
$
where $\ell(\cdot)$ is defined in Equation \eqref{eq:piecefunction}, we construct the boundary corresponding to each effect. For the direct effect comparing $(d,g) = (1,0)$ to $(d',g') = (0,0)$, we obtain
\begin{equation*}
\begin{split}
\mathcal{X}_{i\smallN_i}(1,0 \mid 0,0) 
=&\ \ell\big((d_i^{1,0} = 1, \bm d_{\smallN_i}^{1,0} = \bm 0), (d_i^{0,0} = 0, \bm d_{\smallN_i}^{0,0} = \bm 0)\big)\\
=&\ \{ \ppoint \in \scorespace \,:\, x_i = c,\ \bm x_{\smallN_i} < c \},
\end{split}
\end{equation*}
since $d_i^{1,0} \neq d_i^{0,0}$ while $\bm d^{1,0}_{\smallN_i} = \bm d^{0,0}_{\smallN_i} = \bm 0$. Hence, the discontinuity from $(0,0)$ to $(1,0)$ occurs when the own score crosses the cutoff, while all neighbors' scores remain below the cutoff.
For the indirect effect comparing $(d,g) = (0,1)$ to $(d',g') = (0,0)$, we obtain
\begin{equation*}
\begin{split}
\mathcal{X}_{i\smallN_i}(0,1 \mid 0,0) 
=&\ \ell\big((d_i^{0,1} = 0, \bm d_{\smallN_i}^{0,1} = \bm 1), (d_i^{0,0} = 0, \bm d_{\smallN_i}^{0,0} = \bm 0)\big) \\
=&\ \{ \ppoint \in \scorespace \,:\, x_i < c,\ \bm x_{\smallN_i} = \bm c \},
\end{split}
\end{equation*}
since $d_i^{0,1} = d_i^{0,0}$ while $\bm d^{0,1}_{\smallN_i} \neq \bm d^{0,0}_{\smallN_i}$ in all coordinates. Hence, the discontinuity occurs when the own score is below the cutoff and all neighbors' scores are at the cutoff.
The boundaries corresponding to the remaining effects are constructed analogously.

Finally, the overall direct effect averages boundary point effects over configurations with $x_i = c$, so that the corresponding boundary is the subspace
\(
\{\ppoint \in \scorespace \,:\, x_i = c\}.
\)
In contrast, the overall indirect effect is defined over configurations where at least one neighbor is at the cutoff, that is,
\(
\{\ppoint \in \scorespace \,:\, \exists j \in \mathcal{S}_i \text{ such that } x_j = c\}.
\)

\paragraph{Practical simplification.}
The construction of the effective treatment boundary described in Section \ref{sec:etb} and Eq. \eqref{eq:piecefunction} involves a union over treatment configurations and is therefore combinatorial in full generality. However, it simplifies substantially for many commonly used exchangeable exposure mappings.
To illustrate, consider the sum-of-treated exposure mapping $G_i = \sum_{j \in \mathcal{S}_i} D_j$
where $|\mathcal{S}_i|$ denotes the size of the interference set. The number of neighbor treatment configurations yielding exposure level $g$ is equal to $\binom{|\mathcal{S}_i|}{g}$, and similarly for $g'$. A direct application of the general construction would therefore require evaluating
$
\binom{|\mathcal{S}_i|}{g}\binom{|\mathcal{S}_i|}{g'}
$
pairs of treatment vectors. However, due to exchangeability, any two such pairs that differ only by a permutation of neighbor indices generate geometrically equivalent boundary pieces, and the boundary admits a simple representation  
%
Without loss of generality, suppose $g > g'$. Then, the boundary is characterized by:
\begin{enumerate}[
    label=$\bullet$,
    ref=\theassumption.\alph*,
    leftmargin=*,
    topsep=0.2em,
    itemsep=0.5pt,
    parsep=0pt,
    partopsep=0pt
]
    \item $g'$ neighbors' score components satisfying $x_j \ge c$;
    \item $g - g'$ neighbors' score components satisfying $x_j = c$;
    \item $|\mathcal{S}_i| - g$ neighbors' score components satisfying $x_j < c$.
    \item $x_i = c$ if $d_i \neq d_i'$;
\end{enumerate}
The same logic applies to other exposure mappings—such as fraction of treated, one-treated, or majority —which can be expressed as functions of the number of treated neighbors. In these cases as well, the boundary admits a simple representation.

For instance, consider the one-treated exposure mapping $G_i = \mathbbm{1}\left\{\sum_{j \in \mathcal{S}_i} D_j \ge 1\right\}$.
Let
\(
S_i=\sum_{j\in\mathcal S_i}D_j,
\) so that $G_i=\mathbf 1\{S_i\ge 1\}.$
Then the effective treatment region $\scorespace^{\mathrm{one}}(d,0)$ under one-treated exposure mapping coincides with $\scorespace^{\mathrm{sum}}(d,0)$ under sum-of-treated exposure mapping, while
\[
\scorespace^{\mathrm{one}}(d,1)
=
\bigcup_{k=1}^{|\mathcal{S}_i|} \scorespace^{\mathrm{sum}}(d,k).
\]
Accordingly, the boundary between $(d,1)$ and $(d',0)$ under the one-treated mapping can be written as
\[
\mathcal X_{i\smallN_i}^{\mathrm{one}}(d,1 \mid d',0)
=
\bigcup_{k=1}^{|\mathcal{S}_i|}
\mathcal X_{i\smallN_i}^{\mathrm{sum}}(d,k \mid d',0),
\]
which reduces to
\[
\left\{
\ppoint \in \scorespace :
\begin{array}{l}
x_i = c \text{ if } d \neq d', \\
\exists j \in \mathcal{S}_i \text{ such that } x_j = c, \\
x_z \leq c \text{ for all } z \in \mathcal{S}_i 
\end{array}
\right\}.
\]

\paragraph{Distance computation}
The minimum distance of $(X_i, \pscore)$ from the boundary can be computed efficiently for exchangeable exposure mappings by exploiting the permutation invariance of the boundary. In the case of the \emph{sum-treated} exposure mapping, suppose $g > g'$ and let $m = |\mathcal S_i|$. After centering scores at the cutoff, $\bar X_j = X_j - c$ for $j \in \{i\} \cup \mathcal S_i$, let $\bar X_{j(1)} \ge \cdots \ge \bar X_{j(m)}$, $j \in \mathcal S_i$, denote the ordered peer scores. The boundary is characterized by $g'$ neighbors' score components above the cutoff, $g-g'$ equal to the cutoff, and $|\mathcal{S}_i|-g$ below it, together with $\bar X_i = 0$ when $d \neq d'$. The closest boundary point is obtained by coordinate-wise projection onto these constraints, yielding
\[
\tilde{X}_i
=
\mathbf{1}\{d \neq d'\}\bar X_i^2
+ \sum_{k=1}^{g'} \bigl(\min\{\bar X_{j(k)}, 0\}\bigr)^2
+ \sum_{k=g'+1}^{g} \bar X_{j(k)}^2
+ \sum_{k=g+1}^{|\mathcal{S}_i|} \bigl(\max\{\bar X_{j(k)}, 0\}\bigr)^2.
\]

These expressions follow from permutation invariance, which allows the minimization to be restricted to ordered boundary points. The closest boundary point is then obtained by projecting the ordered score vector onto the boundary, implemented component-wise. Similar formulas can be derived for other exchangeable exposure mappings, such as one-treated, using a similar logic. However, the resulting expressions depend on the specific geometry of the boundary, and we therefore do not provide a unified formula.

\subsection{Specification Testing}\label{onlineappendix:diagnostics}

Standard RDD practice relies on a set of specification and robustness tests, including tests for discontinuities in baseline covariates and placebo outcomes, density tests of the score variable, and checks for sensitivity to bandwidth choices \citep{imblem}. We develop a covariate test and a score density test tailored to our framework and apply both to our PROGRESA application in Section \ref{sec:empirical}. Additionally, we conduct a bandwidth sensitivity analysis and discuss potential directions for further testing.
Unlike in a standard RDD, our framework also relies on the specification of the interference set and the exposure mapping. Consequently, robustness checks must also assess sensitivity to these modeling choices by examining alternative specifications. To illustrate this, we repeat our empirical analysis using a different exposure mapping. 

\paragraph{Covariate Test.}
Our identification strategy relies on continuity of the conditional mean outcome at the effective treatment boundary. As in standard RDD, this assumption can be partially assessed by testing for discontinuities in pretreatment covariates. Evidence of such discontinuities may indicate sorting of individuals or their neighbors around the boundary, potentially invalidating such continuity assumption.

Our test is implemented by adapting our distance-based local polynomial estimator to treat covariates as outcomes for each estimand of interest, which automatically allows us to account for the network dependence structure. Importantly, in our interference framework, such discontinuities should be assessed not only for individual covariates but also for neighbors' characteristics. This is because neighbors may also self-select around the cutoff. Consequently, individuals whose neighbors are just above versus just below the cutoff may have systematically different neighborhood characteristics, which could be correlated with the individuals' outcomes.

We illustrate this procedure in the PROGRESA application using a set of baseline covariates, including children’s age, grade, and gender, housing characteristics (e.g., roof and floor materials), and parental education, and the corresponding averages for neighbors. Results are reported in Tables \ref{tab:covtest} and \ref{tab:covtest2}. Overall, covariates are well balanced across the effective treatment boundary, with no systematic differences across most individual and peer characteristics, with only a few significant or marginally significant discontinuities in parental education and housing conditions among peer averages for specific direct and indirect effects.
%
%
%
While these differences are moderate in magnitude and isolated to specific configurations, their detection underscores the value of our testing framework. By explicitly accounting for peer characteristics in settings with interference, our approach uncovers neighborhood-level compositional differences that standard approaches ignoring interference would fail to detect.

\paragraph{Score density test.}
In RDDs, density tests of the score variable are used to detect score manipulation that may challenge the outcome continuity assumption. In our framework, by contrast, score density continuity is itself an identifying assumption (Assumption \ref{ass:densitycont}). Testing for density discontinuities therefore directly assesses this assumption, while also providing evidence of manipulation that could invalidate outcome continuity (\ref{ass:cont}).

Directly testing the continuity of $\densityphi$ would require complex, multidimensional methods. However, a necessary implication of the continuity of $\densityphi$ is the continuity of the marginal density of the individual score, $f(x_i; \phi)$, which follows from the Dominated Convergence Theorem. Thus, focusing on this marginal density, or the average marginal density $\sum_\phi f(x_i; \phi)\cdot \pi_\phi$ when considering on the pooled estimands, provides a tractable yet rigorous test for our assumption.
Therefore, we implement a univariate density test for the average marginal density based on local polynomial method implemented in the \texttt{rddensity} package. While the estimator itself is identical to the one in the package, we complement the approach with a dependence-robust variance estimator based on the dependency graph to account for network correlation.
In our PROGRESA application, under dependence-robust inference, the test yields a t-statistics $T = 1.8378$ and a p-value $p = 0.0661$, so that the null hypothesis of continuity cannot be rejected at the conventional $5\%$ level. Results are similar under iid inference ($ T = 1.7941$, $p = 0.0728 $). Given the similarity, we report the density plot with iid confidence bands. Overall, these findings provide no strong evidence of density discontinuity.

This testing approach offers a simple method to diagnosing score manipulation in networks. An alternative approach would be to test the joint score density using the distance to the boundary in a way that directly reflects the boundary density $\limdensity$. However, such an estimator would need to account for the boundary’s geometry and codimension in both its construction and bandwidth selection to be valid. We leave the formal development of this alternative density test to future work.

\begin{figure}[H]
\centering
\includegraphics[width=0.5\linewidth]{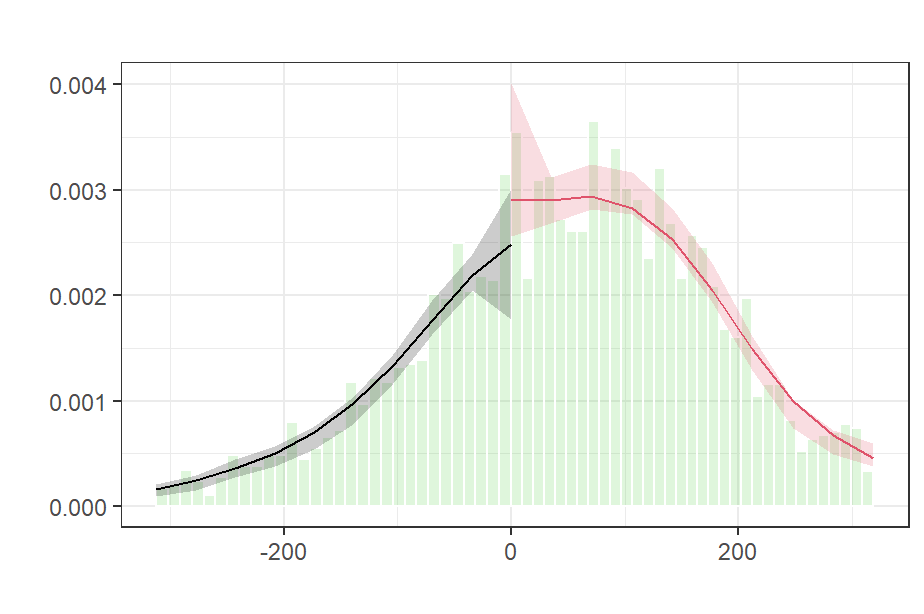}
\caption{ \footnotesize Estimated density of the running variable at the cutoff using local polynomial smoothing. Shaded areas denote confidence intervals.}
\label{fig:density}
\end{figure}

\begin{table}[!htbp]\centering
\caption{Covariate balance around the effective treatment boundary for direct effects}
\label{tab:covtest}
\scriptsize
\begin{tabular}{lcccc}
\toprule
Covariate & Mean$_{d',g'}$ & Mean$_{d,g}$ & Diff (SE) & p-value \\
\midrule
\multicolumn{5}{l}{\textbf{Panel A: Effect $\tau_{10|00}\big(\mathcal{X}_{i\smallN_i}(10|00)\big)$ --- Own covariates}} \\
edu97          & 4.448 & 4.279 & -0.169 (0.322) & 0.600 \\
age97          & 11.229 & 11.130 & -0.100 (0.630) & 0.875 \\
gender97       & 0.375 & 0.687 & 0.312* (0.176) & 0.077 \\
roof97         & 4.500 & 4.496 & -0.004 (0.663) & 0.995 \\
tin97          & 0.292 & 0.479 & 0.186 (0.149) & 0.211 \\
asbest97       & 0.095 & 0.053 & -0.042 (0.090) & 0.641 \\
tiles97        & 0.159 & 0.042 & -0.118 (0.118) & 0.318 \\
cementblock97  & 0.035 & 0.209 & 0.174 (0.117) & 0.136 \\
floor97        & 1.323 & 1.471 & 0.148 (0.164) & 0.367 \\
cement97       & 0.232 & 0.476 & 0.243* (0.143) & 0.089 \\
headhhrel97    & 3.888 & 3.615 & -0.273 (0.586) & 0.641 \\
mother\_edu97  & 2.473 & 2.389 & -0.084 (0.733) & 0.909 \\
father\_edu97  & 2.996 & 2.497 & -0.499 (0.809) & 0.537 \\
\addlinespace
\multicolumn{5}{l}{\textit{Neighbors (average covariates)}} \\
peermean\_edu97         & 4.448 & 4.279 & -0.169 (0.322) & 0.600 \\
peermean\_age97         & 11.019 & 11.238 & 0.218 (0.532) & 0.682 \\
peermean\_gender97      & 0.375 & 0.687 & 0.312* (0.176) & 0.077 \\
peermean\_roof97        & 5.094 & 4.991 & -0.103 (0.608) & 0.866 \\
peermean\_tin97         & 0.353 & 0.313 & -0.040 (0.141) & 0.775 \\
peermean\_asbest97      & 0.091 & 0.024 & -0.067 (0.069) & 0.333 \\
peermean\_tiles97       & 0.117 & 0.054 & -0.062 (0.106) & 0.555 \\
peermean\_cementblock97 & 0.082 & 0.342 & 0.260** (0.122) & 0.033 \\
peermean\_floor97       & 1.493 & 1.648 & 0.154 (0.161) & 0.338 \\
peermean\_cement97      & 0.417 & 0.601 & 0.185 (0.141) & 0.191 \\
peermean\_headhhrel97   & 3.567 & 3.302 & -0.265 (0.410) & 0.518 \\
peermean\_mother\_edu97 & 2.067 & 2.868 & 0.801 (0.690) & 0.246 \\
peermean\_father\_edu97 & 4.241 & 2.827 & -1.414 (0.899) & 0.116 \\
\midrule

\multicolumn{5}{l}{\textbf{Panel B: Effect $\tau_{11|01}\big(\mathcal{X}_{i\smallN_i}(11|01)\big)$ --- Own covariates}} \\
edu97          & 4.544 & 4.521 & -0.023 (0.239) & 0.922 \\
age97          & 11.610 & 11.114 & -0.497 (0.401) & 0.216 \\
gender97       & 0.430 & 0.394 & -0.036 (0.113) & 0.747 \\
roof97         & 5.081 & 4.742 & -0.339 (0.436) & 0.437 \\
tin97          & 0.466 & 0.416 & -0.050 (0.102) & 0.622 \\
asbest97       & 0.114 & 0.114 & -0.000 (0.061) & 0.994 \\
tiles97        & 0.069 & 0.080 & 0.011 (0.049) & 0.822 \\
cementblock97  & 0.159 & 0.123 & -0.036 (0.073) & 0.619 \\
floor97        & 1.329 & 1.380 & 0.051 (0.108) & 0.636 \\
cement97       & 0.339 & 0.381 & 0.042 (0.098) & 0.670 \\
headhhrel97    & 3.452 & 3.114 & -0.338 (0.293) & 0.248 \\
mother\_edu97  & 1.876 & 2.199 & 0.323 (0.508) & 0.525 \\
father\_edu97  & 2.682 & 3.013 & 0.331 (0.638) & 0.604 \\
\addlinespace
\multicolumn{5}{l}{\textit{Neighbors (average covariates)}} \\
peermean\_edu97         & 4.544 & 4.521 & -0.023 (0.239) & 0.922 \\
peermean\_age97         & 11.201 & 11.254 & 0.053 (0.314) & 0.866 \\
peermean\_gender97      & 0.430 & 0.394 & -0.036 (0.113) & 0.747 \\
peermean\_roof97        & 4.437 & 4.237 & -0.199 (0.404) & 0.622 \\
peermean\_tin97         & 0.348 & 0.360 & 0.012 (0.087) & 0.888 \\
peermean\_asbest97      & 0.112 & 0.117 & 0.004 (0.059) & 0.943 \\
peermean\_tiles97       & 0.121 & 0.057 & -0.065 (0.051) & 0.205 \\
peermean\_cementblock97 & 0.057 & 0.103 & 0.046 (0.055) & 0.403 \\
peermean\_floor97       & 1.257 & 1.388 & 0.131 (0.082) & 0.113 \\
peermean\_cement97      & 0.233 & 0.280 & 0.047 (0.079) & 0.556 \\
peermean\_headhhrel97   & 3.093 & 3.214 & 0.121 (0.283) & 0.669 \\
peermean\_mother\_edu97 & 2.168 & 2.711 & 0.543 (0.388) & 0.161 \\
peermean\_father\_edu97 & 2.554 & 3.321 & 0.767 (0.519) & 0.139 \\
\bottomrule
\end{tabular}

\begin{minipage}{0.95\textwidth}
\footnotesize
\textit{Notes:} Mean$_{d',g'}$ and Mean$_{d,g}$ denote the estimated average covariate values on the left and
right of the relevant effective treatment boundary. Diff reports the estimated discontinuity, with
standard errors in parentheses. Significance markers are: * $p<0.10$, ** $p<0.05$, and *** $p<0.01$.
\end{minipage}
\end{table}
\begin{table}[!htbp]\centering
\caption{Covariate balance around the effective treatment boundary for indirect effects}
\label{tab:covtest2}
\scriptsize
\begin{tabular}{lcccc}
\toprule
Covariate & Mean$_{d',g'}$ & Mean$_{d,g}$ & Diff (SE) & p-value \\
\midrule
\multicolumn{5}{l}{\textbf{Panel A: Effect $\tau_{01|00}\big(\mathcal{X}_{i\smallN_i}(01|00)\big)$ --- Own covariates}} \\
edu97          & 4.370 & 4.427 & 0.057 (0.315) & 0.856 \\
age97          & 10.854 & 11.538 & 0.685 (0.588) & 0.244 \\
gender97       & 0.301 & 0.559 & 0.258 (0.167) & 0.123 \\
roof97         & 5.183 & 5.093 & -0.090 (0.744) & 0.904 \\
tin97          & 0.386 & 0.364 & -0.021 (0.155) & 0.890 \\
asbest97       & 0.042 & 0.000 & -0.041 (0.065) & 0.521 \\
tiles97        & 0.123 & 0.067 & -0.056 (0.106) & 0.595 \\
cementblock97  & 0.078 & 0.346 & 0.267* (0.139) & 0.054 \\
floor97        & 1.535 & 1.745 & 0.210 (0.187) & 0.261 \\
cement97       & 0.437 & 0.621 & 0.184 (0.160) & 0.249 \\
headhhrel97    & 3.432 & 3.120 & -0.313 (0.419) & 0.456 \\
mother\_edu97  & 2.062 & 2.725 & 0.663 (0.847) & 0.434 \\
father\_edu97  & 4.510 & 2.841 & -1.670* (0.993) & 0.093 \\
\addlinespace
\multicolumn{5}{l}{\textit{Neighbors (average covariates)}} \\
peermean\_edu97         & 4.370 & 4.427 & 0.057 (0.315) & 0.856 \\
peermean\_age97         & 11.295 & 11.581 & 0.286 (0.561) & 0.611 \\
peermean\_gender97      & 0.301 & 0.559 & 0.258 (0.167) & 0.123 \\
peermean\_roof97        & 4.322 & 4.600 & 0.278 (0.604) & 0.645 \\
peermean\_tin97         & 0.363 & 0.514 & 0.151 (0.156) & 0.334 \\
peermean\_asbest97      & 0.058 & 0.152 & 0.094 (0.090) & 0.295 \\
peermean\_tiles97       & 0.118 & 0.043 & -0.075 (0.105) & 0.473 \\
peermean\_cementblock97 & 0.040 & 0.136 & 0.096 (0.099) & 0.334 \\
peermean\_floor97       & 1.366 & 1.401 & 0.035 (0.176) & 0.842 \\
peermean\_cement97      & 0.255 & 0.379 & 0.125 (0.151) & 0.409 \\
peermean\_headhhrel97   & 3.990 & 3.156 & -0.834 (0.595) & 0.162 \\
peermean\_mother\_edu97 & 2.460 & 2.114 & -0.347 (0.847) & 0.682 \\
peermean\_father\_edu97 & 3.285 & 2.496 & -0.789 (0.846) & 0.351 \\
\midrule

\multicolumn{5}{l}{\textbf{Panel B: Effect $\tau_{11|10}\big(\mathcal{X}_{i\smallN_i}(11|10)\big)$ --- Own covariates}} \\
edu97          & 4.639 & 4.479 & -0.160 (0.326) & 0.623 \\
age97          & 11.315 & 11.409 & 0.094 (0.461) & 0.839 \\
gender97       & 0.338 & 0.354 & 0.016 (0.152) & 0.917 \\
roof97         & 4.546 & 4.305 & -0.241 (0.596) & 0.685 \\
tin97          & 0.373 & 0.322 & -0.051 (0.143) & 0.720 \\
asbest97       & 0.139 & 0.030 & -0.109 (0.102) & 0.288 \\
tiles97        & 0.106 & 0.133 & 0.027 (0.097) & 0.783 \\
cementblock97  & 0.082 & 0.114 & 0.032 (0.086) & 0.711 \\
floor97        & 1.340 & 1.397 & 0.057 (0.138) & 0.678 \\
cement97       & 0.265 & 0.201 & -0.064 (0.112) & 0.569 \\
headhhrel97    & 3.021 & 3.375 & 0.354 (0.432) & 0.413 \\
mother\_edu97  & 2.208 & 2.822 & 0.614 (0.673) & 0.362 \\
father\_edu97  & 2.719 & 3.717 & 0.998 (0.882) & 0.258 \\
\addlinespace
\multicolumn{5}{l}{\textit{Neighbors (average covariates)}} \\
peermean\_edu97         & 4.639 & 4.479 & -0.160 (0.326) & 0.623 \\
peermean\_age97         & 11.474 & 10.832 & -0.642 (0.531) & 0.227 \\
peermean\_gender97      & 0.338 & 0.354 & 0.016 (0.152) & 0.917 \\
peermean\_roof97        & 5.489 & 4.960 & -0.529 (0.744) & 0.477 \\
peermean\_tin97         & 0.288 & 0.374 & 0.086 (0.160) & 0.590 \\
peermean\_asbest97      & 0.083 & 0.060 & -0.024 (0.097) & 0.806 \\
peermean\_tiles97       & 0.120 & 0.170 & 0.050 (0.093) & 0.592 \\
peermean\_cementblock97 & 0.166 & 0.136 & -0.030 (0.092) & 0.743 \\
peermean\_floor97       & 1.497 & 1.521 & 0.024 (0.142) & 0.866 \\
peermean\_cement97      & 0.486 & 0.387 & -0.100 (0.115) & 0.385 \\
peermean\_headhhrel97   & 3.330 & 3.434 & 0.104 (0.405) & 0.798 \\
peermean\_mother\_edu97 & 2.682 & 2.499 & -0.183 (0.824) & 0.824 \\
peermean\_father\_edu97 & 3.551 & 1.599 & -1.952** (0.944) & 0.039 \\
\bottomrule
\end{tabular}

\begin{minipage}{0.95\textwidth}
\footnotesize
\textit{Notes:} Mean$_{d',g'}$ and Mean$_{d,g}$ denote the estimated average covariate values on the left and
right of the relevant effective treatment boundary. Diff reports the estimates discontinuity, with
standard errors in parentheses. Significance markers are: * $p<0.10$, ** $p<0.05$, and *** $p<0.01$.
\end{minipage}
\end{table}
\paragraph{Bandwidth sensitivity.}
To assess the robustness of our baseline estimates, we conduct a bandwidth sensitivity analysis. Table \ref{tab:bandwidth} reports estimates obtained using half (0.5) and one-and-a-half (1.5) times the MSE-optimal bandwidth used in the main specification. 
As expected, reducing the bandwidth decreases the effective sample size and leads to larger standard errors across all estimands. Conversely, increasing the bandwidth yields smaller standard errors.
Across bandwidth choices, our main qualitative findings remain broadly unchanged. The direct effect $\hat{\tau}^{bc}_{10|00}(h_n, b_n)$ is consistently positive and statistically significant, or close to conventional significance levels. The direct effect $\hat{\tau}^{bc}_{11|01}(h_n, b_n)$ decreases under the smaller bandwidth and is not statistically significant, but remains stable under the larger bandwidth \red{and becomes significant}.
The indirect effects remain statistically indistinguishable from zero across all specifications, but they are more sensitive to the choice of a smaller bandwidth than the direct effects. The estimates shrink toward zero under the narrower bandwidth, which may reflect the more limited local sample size available for these estimands. However, they remain stable in magnitude for a larger bandwidth choice.
Overall, the results support the qualitative robustness of the main conclusions, while highlighting sensitivity in magnitude at smaller bandwidths, especially for the indirect effects. Additional sensitivity analyses, such as varying the polynomial order or kernel, could also be informative; we omit them here for brevity but recommend their use in practice.
 
\begin{table}[H]
\centering
\footnotesize
\begin{tabular}{lcc}
\toprule
 & $\tfrac{1}{2}h_n$ & $\tfrac{3}{2}h_n$ \\
\midrule
$\hat{\tau}^{bc}_{10|00}(h_n, b_n)$
& 0.338 (0.155) [0.034, 0.642]
& 0.230 (0.107) [0.020, 0.440] \\

$\hat{\tau}^{bc}_{11|01}(h_n, b_n)$
& 0.009 (0.097) [$-0.182$, 0.199]
& 0.142 (0.063) [0.019, 0.265] \\

$\hat{\tau}^{bc}_{01|00}(h_n, b_n)$
& $-0.041$ (0.162) [$-0.358$, 0.276]
& 0.041 (0.113) [$-0.180$, 0.262] \\

$\hat{\tau}^{bc}_{11|10}(h_n, b_n)$
& $-0.064$ (0.158) [$-0.373$, 0.246]
& 0.126 (0.085) [$-0.040$, 0.292] \\
\bottomrule
\end{tabular}
\caption{Bandwidth sensitivity using robust bias-corrected estimation. Standard errors are reported in parentheses and 95\% confidence intervals in square brackets.}
\label{tab:bandwidth}
\end{table}

\paragraph{Alternative Exposure Mapping.}
We repeat the analysis of Section \ref{sec:empirical} but employing an exposure mapping which takes value one if half or more of the neighborhood of an individual is treated: 
\[
G_i = \sum_j \mathbbm{1}\Big(D_j \geq \frac{|\mathcal{S}_i|}{2}\Big)
\]
Peer groups are defined as before, by village, grade, and gender. Table \ref{tab:majority} reports the pooled estimates for individuals with one up to four neighbors.. Boundary overall effect estimates are omitted because they are invariant to the exposure mapping definition.
Several interesting patterns emerge from these results. First, the direct effect estimates appear stable, and are slightly attenuated. \red{$\hat{\tau}_{10|00}(h_n, b_n)$ is significant at the 5\% level, as in Section \ref{sec:empirical}, and $\hat{\tau}^{bc}_{11|01}(h_n, b_n)$ is still significant at the 10\% level.}
On the other hand, the indirect effect estimate \red{$\hat{\tau}^{bc}_{01|00}(h_n, b_n)$}—which captures the impact of having half or more of the neighborhood treated compared to less than half when the focal individual is untreated—is larger and statistically significant at a 10\% significance level compared to the previous analysis. 
Here, the change in the exposure mapping improves statistical power and convergence. The comparison under the previous exposure mapping, the fraction of treated neighbors, required all neighbors to be treated or untreated, which involves high codimension, resulting in sparse data around the boundaries. By utilizing a mapping characterized by codimension $1$, we only require a single marginal neighbor's score to cross the cutoff. This admits significantly more data around the boundary and yields a fast convergence rate, allowing us to detect effects more precisely.

A different definition of the exposure mapping inherently changes the definition of the effective treatment boundary, meaning that the effects are evaluated on different subpopulations. Under the previous analysis, indirect effects were evaluated for individuals having all neighbors close to the cutoff, and the estimate was biased towards low-degree units, as high-degree units have a relatively smaller probability of having all neighbors at the cutoff. Here, instead, the indirect effect is for individuals whose first or second largest neighborhood score is at the cutoff, which represents a larger subpopulation and includes higher-degree units. The fact that the indirect effect estimate becomes larger and more precise suggests that individuals with more extensive networks may drive spillovers, and that the program could be beneficial for the untreated when a large share of neighbors is treated.
%
It would be highly valuable to derive what happens under a misspecified exposure mapping—such as when the assumed exposure is coarser than the true behavioral model—which might identify a weighted summary measure of the true effects, as explored in \cite{vazquez}. We leave this extension to future work.
\begin{table}[H]
\centering
\footnotesize
\begin{tabular}{lccccccc}
  \hline
 & Estimate & S.E. & 95\% C.I. & p-value & $N_{d,g}$ & $N_{d',g'}$\\ 
  \hline
  $\hat{\tau}^{bc}_{10|00}(h_n, b_n)$ & 0.210 & 0.106 & (0.003, 0.418) & 0.047 & 158& 168\\ 
  $\hat{\tau}^{bc}_{11|01}(h_n, b_n)$ & 0.090 & 0.051 & (-0.010, 0.190) & 0.078 & 407 & 698\\ 
  $\hat{\tau}^{bc}_{01|00}(h_n, b_n)$ & 0.134 & 0.078 & (-0.019, 0.288) & 0.086 & 239 & 370 \\ 
  $\hat{\tau}^{bc}_{11|10}(h_n, b_n)$ & 0.089 & 0.059 & (-0.026, 0.205) & 0.130 & 269 & 590 \\ 
  \hline
\end{tabular}
\caption{\footnotesize Estimates for PROGRESA data with robust confidence intervals, MSE-optimal bandwidths. S.E. = estimated standard errors, C.I. = Normal approximation confidence intervals. p-value = Normal approximation two-sided p-value. $\text{N}_{d,g}$, $\text{N}_{d',g'}$ = Effective sample sizes.}
\label{tab:majority}
\end{table}

\end{appendices}

\putbib[biblio]
\end{bibunit}

\clearpage

\begin{bibunit}[apalike]

\appendix 
\renewcommand\thesection{S.\arabic{section}} 
\setcounter{theorem}{0}
\setcounter{assumption}{0}
\setcounter{lemma}{0}
\setcounter{corollary}{0}

\renewcommand{\thetheorem}{\thesection.\arabic{theorem}}
\renewcommand{\theassumption}{\thesection.\arabic{assumption}}
\renewcommand{\thelemma}{\thesection.\arabic{lemma}}
\renewcommand{\thecorollary}{\thesection.\arabic{corollary}}

\title{
\vspace{-3em}
Supplementary Material to \qq Regression Discontinuity Designs Under Interference"}
\date{}

\maketitle

\thispagestyle{empty}
\startcontents[supp]
\printcontents[supp]{}{1}[3]{}

\setcounter{page}{1}
\section{Appendix Roadmap}

This ``Supplementary Material'' appendix collects the proofs of the formal results stated in the main text and in the Online Appendix, together with the regularity conditions and key notation.

Section \ref{appendix:mainresults} derives the representation of each effective treatment
boundary as a finite union of linear pieces and characterizes its codimension. Section
\ref{appendix:boundformal} parametrizes the boundary average causal effects and proves the
identification results: Theorem \ref{th:cont} for stratum-specific effects, Theorem
\ref{th:idpooled} for pooled effects, and Theorem \ref{th:overall} with Corollary
\ref{corollary:overall} for the boundary overall direct effect.
Section \ref{appendix:estimator} develops the asymptotic theory for the stratum-specific
distance-based estimator at a generic polynomial order $p$. Lemmas
\ref{lemma:convergencematricesgamma}--\ref{lemma:convergencematricestheta} establish the
probability limits of the sample matrices entering the estimator; Lemma
\ref{lemma:consistency} establishes consistency; Lemma \ref{lemma:variances} derives the
limiting variance; and Theorem \ref{th:asynormaltau} combines these with Stein's method to
establish asymptotic normality, yielding Theorem~\ref{th:asynormal} at
$p = 1$. Lemma \ref{lemma:psiconsistency} and Theorem \ref{th:variance} establish consistency of the proposed variance estimator, and Section
\ref{appendix:differentiability} provides primitive conditions for the smoothness required of the distance-based regression functions. Section \ref{appendix3:biascorrection} extends
these arguments to robust bias correction, establishing asymptotic normality of the
bias-corrected estimator in Theorem \ref{th:asybiascorretiontau}, which corresponds to
Theorem \ref{th:asybiascorretiontaumain} at $p = 1$ with bias correction by local quadratic
estimation.
Section \ref{appendix:heterogeneous} studies the pooled distance-based estimator: Lemma
\ref{lemma:convergencebiashet} and Theorem \ref{lemma:consistencypooled} establish
sample-matrix convergence and consistency when the codimension is common across strata;
Theorem \ref{lemma:convergencelowcod} characterizes the probability limit when it varies
across strata, yielding Theorem~\ref{thm:teo3} in the main text; Lemma
\ref{lemma:varianceshe} derives the limiting variance; and Theorem
\ref{th:heterogeneousconvergence} establishes asymptotic normality, which is
Theorem~\ref{th:heterogeneousconvergence}. Section \ref{appendix:estimatoroverall} contains the asymptotic
distribution for the boundary overall direct effect estimators in Theorem \ref{th:asynormaltauov} (Theorem~\ref{th:overallnormal} in the main
text).

\section{Effective Treatment Boundary Derivation}\label{appendix:mainresults}

We provide the derivation of the equality $$\Frontier = 
\bigcup_{\treatspace(d,g) \times \treatspace(d',g')} \ell(\bm{d}_{i\smallN_i}^{d,g}, \bm{d}_{i\smallN_i}^{d',g'})$$ in Subsection~\ref{sec:etb}.  
The derivation uses the fact that given the individual treatment rule in Assumption~\eqref{ass:indrule}, each element $\ddpoint[d,g] \in \treatspace(d,g)$ can be mapped back to the set $\scorespace(\ddpoint[d,g]) = \{\ppoint \in \scorespace : \forall z \in (\{i\}\cup \mathcal{S}_i),\; x_z \geq c \; \text{if}\; d_z^{d,g} = 1,\; x_z < c\; \text{if}\; d_z^{d,g} = 0\}$, and that each treatment region is the union of such sets over $\treatspace(d,g)$, i.e.,
\begin{equation*}
\scorespace(d,g) = \bigcup_{\treatspace(d,g)}\scorespace(\ddpoint[d,g])    
\end{equation*}

\begin{proof}
Recall that the closure of a set is the union of its interior with its boundary. In addition, the closure of a union of sets is equal to the union of their closures. Therefore, denoting the closure of a set $A$ as $Cl(A)$, we have 
\begin{equation*}
    Cl\big(\scorespace(d,g)\big) = \bigcup_{\treatspace(d,g)}Cl\big(\scorespace(\ddpoint[d,g])\big)    
\end{equation*}
where clearly $Cl\big(\scorespace(\ddpoint[d,g])\big) = \{\ppoint \in \scorespace : \forall z \in (\{i\}\cup \mathcal{S}_i), x_z \geq c \; \text{if}\; d_z^{d,g} = 1,\; x_z \leq c\; \text{if}\; d_z^{d,g} = 0\}$ .

Now by definition $\Frontier = \bar{\mathcal{X}}_{i\smallN_i}(d,g) \cap \bar{\mathcal{X}}_{i\smallN_i}(d',g')$, where $\bar{\mathcal{X}}_{i\smallN_i}(d,g)$ and $\bar{\mathcal{X}}_{i\smallN_i}(d',g')$ denote the boundary of $\scorespace(d,g)$ and $\scorespace(d',g')$. Moreover, because $\scorespace(d,g)$ and $\scorespace(d',g')$ are disjoint, the intersection of their closure is equal to the intersection of their boundaries, $Cl\big(\scorespace(d,g)\big) \cap Cl\big(\scorespace(d',g')\big) = \bar{\mathcal{X}}_{i\smallN_i}(d,g) \cap \bar{\mathcal{X}}_{i\smallN_i}(d',g')$.
Note that $Cl\big(\scorespace(d,g)\big) \cap Cl\big(\scorespace(d',g')\big)$ is equal to 
\begin{equation*}
    \Big(\bigcup_{\treatspace(d,g)} Cl\big(\scorespace(\ddpoint[d,g])\big)\Big) \bigcap\Big( \bigcup_{\treatspace(d',g')} Cl\big(\scorespace(\ddpoint[d',g'])\big)\Big)
\end{equation*}
which in turn is equal to 
\begin{equation*}
    \bigcup_{\treatspace(d,g) \times \treatspace(d',g')} \Big(Cl\big(\scorespace(\ddpoint[d,g])\big)\bigcap Cl\big(\scorespace(\ddpoint[d',g'])\big)\Big)
\end{equation*}
\sloppy
Finally, we note that the intersection $Cl(\scorespace(\ddpoint[d,g])) \cap Cl(\scorespace(\ddpoint[d',g']))$ is equal to the set $ \{\ppoint \in \scorespace : \forall z \in (\{i\}\cup \mathcal{S}_i),\; x_z \geq c \; \text{if}\; d_z^{d,g} = d_z^{d',g'} = 1,\; x_z \leq c\; \text{if}\; d_z^{d,g} = d_z^{d',g'} = 0,\; x_z = c\; \text{if}\; d_z^{d,g} \neq d_z^{d',g'}\}$ which is the definition of $\ell(\ddpoint[d,g], \ddpoint[d',g'])$. 
\end{proof}

\section{Identification}\label{appendix:boundformal}

\subsection{Boundary parametrization}

\paragraph{Preliminaries}
Let us formally define the notion of $n-$rectifiable set. Let $\mathcal{H}^n$ denote the $n$-dimensional Hausdorff measure.
\begin{definition}[n-rectifiable set \citep{simon}]
A set $M \subset \mathbb{R}^{n+k}$ is said to be countably $n$-rectifiable if
$$
M \subset M_0 \cup\left(\cup_{j=1}^{\infty} F_j\left(\mathbb{R}^n\right)\right)
$$
where $\mathcal{H}^n\left(M_0\right)=0$ and $F_j: \mathbb{R}^n \rightarrow \mathbb{R}^{n+k}$ are Lipschitz functions for $j=1,2, \ldots $.
\end{definition}
\noindent An equivalent characterization is the following:  
a set $M$ is countably $n$-rectifiable if and only if $M \subset \bigcup_{j=0}^{\infty} N_j$, where $\mathcal{H}^n(N_0)=0$ and each $N_j$, $j \ge 1$, is an $n$-dimensional embedded $C^1$ submanifold of $\mathbb{R}^{n+k}$ \citep[Lemma~11.1]{simon}.

The effective treatment boundary $\Frontier$ is the union of finitely many linear pieces in the multiscore. Therefore, It is $(|\mathcal{S}_i| + 1 - \bar{s}_i)-$rectifiable set, with dimension $|\mathcal{S}|_i + 1 - \bar{s}_i$, and codimension $\bar s_i$, where $\bar{s}_i$ is the minimum codimension across the linear pieces.

Under Assumption \ref{ass:networkheterogeneity}, the stratum-specific boundary causal effect is 
\begin{equation}\label{eq:boundformal}
    \boundeff = \frac{\sum\limits_{\ddpoint[d,g],\ddpoint[d',g']}\int\limits_{\ell(\ddpoint[d,g], \ddpoint[d',g'])}\tau_{d,g|d',g'}(x_i, \bm x_{\smallN_i}; \phi) \densityphi d\mathcal{H}^{\mathcal{S}_\phi + 1 - \bar{s}_\phi}}{\sum\limits_{\ddpoint[d,g],\ddpoint[d',g']}\int\limits_{\ell(\ddpoint[d,g], \ddpoint[d',g'])}\densityphi\mathcal{H}^{\mathcal{S}_\phi + 1 - \bar{s}_\phi}}
\end{equation}
where the summation ranges over all pairs $\ddpoint[d,g], \ddpoint[d',g']$ in $\treatspace(d,g)\times\treatspace(d',g')$. Here the pieces $\ell(\cdot)$ should in principle be indexed by $\phi$, but we omit this index for simplicity.
Note that integrals over pieces $\ell(\cdot)$ with codimension $s_{\ell(\cdot)}$ larger than $\bar{s}_i$ vanish since their $\mathcal{H}^{|\mathcal{S}_i| + 1 - \bar{s}_i}$-dimensional Hausdorff measure is zero. The above expression is equivalent to summing over only those pieces $\ell(\cdot)$ with codimension $\bar{s}_i$.

For each $\hcurve$, define the following index sets:
\begin{itemize}
    \item $F_{\hcurve}$: the set of indices $z \in (\mathcal{S}_i \cup \{i\})$ such that $x_z = c$ on $\hcurve$ (i.e., the constrained components);
    \item $R_{\hcurve}$: the complement set of indices where $x_z$ varies within some interval.
\end{itemize}
%

\noindent We henceforth drop the arguments $\bm d _{i\smallN_i}^{d,g}$ and $\bm d _{i\smallN_i}^{d',g'}$ and simply write $\curve$ for each piece. We also use $\sum_{\curve}$ and $\bigcup_{\curve}$ to denote summation and union over the relevant set of pieces $\hcurve$ indexed by $\bm d _{i\smallN_i}^{d,g}$ and $\bm d _{i\smallN_i}^{d',g'}$.

We partition each $\ppoint$ into $\bm{x}_{\cindex}$, where each component is fixed at $c$ on $\curve$, and $\bm{x}_{\findex}$, which ranges over the rectangular domain 
\begin{equation}\label{eq:rectangulardomain}
\mathcal{R}_{\curve} := \prod_{z \in R_{\curve}} I_z(\curve)   
\end{equation}
where $I_z(\curve)  = [a, c]$, if $x_z \le c$ on $\curve$, and $I_z(\curve)  = [c, b]$ if $x_z \geq c$ on $\curve$. 

The structure of each $\curve$ leads to the following simple parametrization of \eqref{eq:boundformal}:
\begin{equation}\label{eq:piecewiseintegral}
 \eqref{eq:boundformal} = \frac{\sum\limits_{\curve \, : \,\ccard = \bar{s}_\phi}\int\limits_{\freeregion}\tau_{d,g|d',g'}(\bm{c}_{\cindex},\bm{x}_{\findex}; \phi) f(\bm{c}_{\cindex},\bm{x}_{\findex}; \phi) d\bm{x}_{\findex}}
    {\sum\limits_{\curve \, : \, \ccard = \bar{s}_\phi}\int\limits_{\freeregion}f(\bm{c}_{\cindex},\bm{x}_{\findex}; \phi) d\bm{x}_{\findex}}
\end{equation}
%



\subsection{Stratum-specific Effects — Theorem \ref{th:cont}}
\begin{proof}[Proof of Theorem \ref{th:cont}]\label{appendix:stratumid}
We start by proving convergence of 
$$\CE{Y_i(d,g)}{(X_i,\pscore) \in \eballboundrightphi; \phi}$$
%
%
We can write $\CE{Y_i(d,g)}{(X_i,\pscore) \in \Frontier; \phi}$ as 
\begin{equation}
    \frac{\sum\limits_{\curve \, : \,\ccard = \bar{s}_\phi}\int\limits_{\freeregion}m_{d,g}(\bm{c}_{\cindex},\bm{x}_{\findex}, \phi)f(\bm{c}_{\cindex},\bm{x}_{\findex}; \phi)d\bm{x}_{\findex}}
    {\sum\limits_{\curve \, : \, \ccard = \bar{s}_\phi}\int\limits_{\freeregion}f(\bm{c}_{\cindex},\bm{x}_{\findex}; \phi)d\bm{x}_{\findex}}
\end{equation}
Each piece is a function $\hcurve$, and, as shown in Section S.1, it is the intersection of the closure of $\scorespace(\bm d_{i\smallN_i}^{d,g})$ and $\scorespace(\bm d_{i\smallN_i}^{d',g'})$, where
$$\scorespace(\bm d_{i\smallN_i}^{d,g}) = \big\{\ppoint \in \scorespace : \forall z \in (\{i\}\cup \mathcal{S}_i),\; x_z \geq c \; \text{if}\; d_z^{d,g} = 1,\; x_z < c\; \text{if}\; d_z^{d,g} = 0\big\}$$ 
Denote by $\curveball$ the $\varepsilon$-neighborhood around $\curve$ restricted to $\scorespace(\bm d_{i\smallN_i}^{d,g})$:
\begin{equation}\label{eq:curveball}
\curveball = \big\{ \ppoint \in \scorespace : \bm{x}_{\cindex} \in \disc, \; \bm{x}_{\findex} \in \freeregion \big\},
\end{equation}
where
$$
\disc := \Big\{ \bm{x}_{\cindex} : \|\bm{x}_{\cindex} - \bm{c}_{\cindex}\| \leq \varepsilon, \; x_z \geq c \text{ if } d_z^{d,g} = 1, \; x_z < c \text{ if } d_z^{d,g} = 0 \text{ for } z \in F_{\curve} \Big\}.
$$
Geometrically, $\disc$ is the intersection of a $\ccard$-dimensional ball of radius $\varepsilon$ centered at $\bm{c}_{\cindex}$ with an orthant in the space of elements $\bm{x}_{\cindex}$ with origin $\bm{c}_{\cindex}$. Its volume is given by 
\begin{equation}\label{eq:vol}
\operatorname{Vol}(\disc) = \ballvolume,
\qquad 
\ballvolume = \frac{1}{2^{\ccard}} \cdot \frac{\pi^{\ccard / 2}}{\Gamma\left(\frac{\ccard}{2}+1\right)} \cdot \varepsilon^{\ccard},
\end{equation}
where $\Gamma$ is the Euler gamma function.

Note that $\eballboundright = \bigcup_{\curve} \curveball$. We can write  the expectation $\CE{Y_i(d,g)}{(X_i,\pscore) \in \eballboundright}$ as   
\begin{equation}\label{eq:fibers}
    \frac{\sum\limits_{\curve }\int\limits_{\freeregion}\int\limits_{\disc}\hspace{-1em}m_{d,g}(\bm{x}_{\cindex}, \bm{x}_{\findex}; \phi) f(\bm{x}_{\cindex}, \bm{x}_{\findex}; \phi)d\bm{x}_{\cindex}d\bm{x}_{\findex} + I_\varepsilon}
{\sum\limits_{\curve }\int\limits_{\freeregion}\int\limits_{\disc}\hspace{-1em} f(\bm{x}_{\cindex}, \bm{x}_{\findex}; \phi) d\bm{x}_{\cindex}d\bm{x}_{\findex} + I'_\varepsilon} 
\end{equation}
and the terms $I_\varepsilon$ and $I_{\varepsilon}'$ include the integrals over the subspace $\Big(\bigcup_{\curve} \curveball\Big) \setminus \eballbound$.
We first establish 
\begin{equation}\label{eq:avgint}
\frac{1}{\ballvolume}\int_{\disc} f(\bm{x}_{\cindex}, \bm{x}_{\findex}; \phi) \, d\bm{x}_{\cindex} \to  f(\bm{c}_{\cindex}, \bm{x}_{\findex}; \phi)
\end{equation}
for fixed $\bm{x}_{\findex}$. By Assumption~\ref{ass:densitycont}, the density $f$ is continuous. Given $\bm{x}_{\findex}$ and any $\delta > 0$, there exists $r > 0$ such that whenever $\|\bm{x}_{\cindex} - \bm{c}_{\cindex}\| \leq r$,  
\[
\big| f(\bm{x}_{\cindex}, \bm{x}_{\findex}; \phi) - f(\bm{c}_{\cindex}, \bm{x}_{\findex}; \phi) \big| < \delta. 
\]
Therefore, for $\varepsilon < r$ by integral triangle inequality 
\begin{equation*}
\begin{split}
\frac{1}{\ballvolume}\Bigg|\int_{\disc} 
   \big(f(\bm{x}_{\cindex}, \bm{x}_{\findex}; \phi) - f(\bm{c}_{\cindex}, \bm{x}_{\findex}; \phi)\big)\, d\bm{x}_{\cindex}\Bigg| 
< \, \delta.
\end{split}
\end{equation*}
Thus, \eqref{eq:avgint} converges to $f(\bm{c}_{\cindex}, \bm{x}_{\findex};\phi)$ as $\varepsilon \to 0$. Moreover,
\begin{equation*}
\begin{split}
\lim_{\varepsilon \to 0} \frac{1}{\ballvolume}
   \int\limits_{\freeregion} \int\limits_{\disc}
   \kern-1em f(\bm{x}_{\cindex}, \bm{x}_{\findex}; \phi) \, d\bm{x}_{\cindex}\, d\bm{x}_{\findex}
&= \frac{1}{\ballvolume} \int\limits_{\freeregion} 
   \lim_{\varepsilon \to 0}\int\limits_{\disc}
   \kern-1em f(\bm{x}_{\cindex}, \bm{x}_{\findex}; \phi) \, d\bm{x}_{\cindex}\, d\bm{x}_{\findex} \\
&= \int\limits_{\freeregion} f(\bm{c}_{\cindex}, \bm{x}_{\findex}; \phi)\, d\bm{x}_{\findex},
\end{split}
\end{equation*}
by the Dominated Convergence Theorem, since $f$ is bounded by Assumption~\ref{ass:bounded}. By the same argument, we obtain convergence of
$$
\frac{1}{\ballvolume}
   \int\limits_{\freeregion}\int\limits_{\disc}
   m_{d,g}(\bm{x}_{\cindex}, \bm{x}_{\findex}; \phi) f(\bm{x}_{\cindex}, \bm{x}_{\findex}; \phi) \,
   d\bm{x}_{\cindex}\, d\bm{x}_{\findex}
$$
to
$$
\int_{\freeregion} m_{d,g}(\bm{c}_{\cindex}, \bm{x}_{\findex}; \phi) \,
   f(\bm{c}_{\cindex}, \bm{x}_{\findex}; \phi)\, d\bm{x}_{\findex}.
$$
because $m_{d,g}$ and $f$ are continuous and bounded by Assumptions \ref{ass:cont}-\ref{ass:densitycont} and \ref{ass:bounded}.

Hence, each element of the numerator and denominator in \eqref{eq:fibers} is of order 
$O(\varepsilon^{\ccard})$. Moreover, since both the score density and the conditional mean of potential outcomes are bounded by Assumption \ref{ass:bounded}, and the region of integration of $I_\varepsilon$ and $I_\varepsilon'$ has volume $O(\varepsilon^{\bar{s}_i+1})$, we obtain 
$|I_\varepsilon| \lesssim \varepsilon^{\bar{s}_i+1}$ and 
$|I_\varepsilon'| \lesssim \varepsilon^{\bar{s}_i+1}$.
\footnote{To see this, note that each tubular neighborhood $\curveball$ is contained in a rectangular box 
$\cube = \{\ppoint \in \scorespace : \bm{x}_{\findex} \in \freeregion, \; |x_z - c|\leq \varepsilon \text{ for all } z \in \cindex\}$, with volume $O(\varepsilon^{\ccard})$. 
The intersection of tubes $\{\curveball : \curve \in \tilde{C}\}$ for any collection $\tilde{C}$ of pieces is then contained in the intersection of the corresponding boxes $\{\cube : \curve \in \tilde{C}\}$. 
This set is described as follows: $x_z \in [a,c]$ if $z \in \cap_{\tilde{C}}\negindex$, 
$x_z \in [c,b]$ if $z \in \cap_{\tilde{C}}\posindex$, where $\negindex$ and $\posindex$ are index sets collecting the indices of the components varying in $[a,c]$ and $[c,b]$, respectively, in each $\curve$, and $|x_z - c|\leq \varepsilon$ if $z \in \cap_{\tilde{C}}\cindex$ or if $z \notin \cap_{\tilde{C}}\posindex$ and $z \notin \cap_{\tilde{C}}\negindex$. Hence, the intersection set is rectangular with volume $O(\varepsilon^{\bar{s}_i+1})$.}

By dividing numerator and denominator of \eqref{eq:fibers} by 
$V_{\varepsilon}^{\bar{s}_\phi}$ we obtain
\begin{equation*}
\begin{split}
\eqref{eq:fibers} &\to 
\frac{\sum\limits_{\curve: s_{\curve} = \bar{s}_\phi}
\int_{\freeregion} m_{d,g}(\bm{c}_{\cindex},\bm{x}_{\findex}; \phi) 
f(\bm{c}_{\cindex},\bm{x}_{\findex}; \phi) \, d\bm{x}_{\findex}}
{\sum\limits_{\curve: s_{\curve} = \bar{s}_\phi}
\int_{\freeregion} f(\bm{c}_{\cindex},\bm{x}_{\findex}; \phi) \, d\bm{x}_{\findex}},
\end{split}
\end{equation*}
by the Dominated Convergence Theorem and Assumptions 
\ref{ass:cont}--\ref{ass:densitycont}, \ref{ass:bounded}, where only the integral terms with $\ccard = \bar{s}_\phi$ contribute in the limit. Hence,
%
\begin{equation*}
\eqref{eq:fibers} \to
\CE{Y_i(d,g)}{(X_i, \pscore) \in \Frontier; \phi},
\end{equation*}
Analogously, one can show
$\CE{Y_i(d',g')}{(X_i, \pscore) \in \eballboundleft, \phi}$ 
converges to 
$\CE{Y_i(d',g')}{(X_i, \pscore) \in \Frontier; \phi}$.

Therefore,
\begin{equation*}
\begin{split}
\boundeff 
&= \CE{Y_i(d,g) - Y_i(d',g')}{(X_i,\pscore) \in \Frontier; \phi} \\
&= \lim_{\varepsilon \to 0}\CE{Y_i(d,g)}{(X_i,\pscore) \in \eballboundright; \phi} \\
& -\; \lim_{\varepsilon \to 0}\CE{Y_i(d',g')}{(X_i,\pscore) \in \eballboundleft; \phi} \\
&= \lim_{\varepsilon \to 0}\CE{Y_i}{(X_i,\pscore) \in \eballboundright; \phi} \\
&-\; \lim_{\varepsilon \to 0}\CE{Y_i}{(X_i,\pscore) \in \eballboundleft; \phi},
\end{split}
\end{equation*}
where the last equality follows from Assumptions \ref{ass:consistency}-\ref{ass:indrule}.
\end{proof}


\subsection{Pooled Effects — Theorem \ref{th:idpooled}}\label{appendix:pooledproof}

We show preliminarily that $\boundeffpooled$ admits a natural representation as the expected boundary effect for a randomly sampled unit from the network.
\begin{theorem}\label{th:randomsampling}
Let $I$ be a random index drawn uniformly from $\{1,\dots,n\}$. Then
\[
\boundeffpooled = \mathbb{E}\big[Y_I(d,g) - Y_I(d',g') \,\big|\, (X_I, \bm X_{\smallN_I}) \in \mathcal{X}_{I\smallN_I}(d,g\, | d',g')\big]
\]
\end{theorem}
\begin{proof}

By the Law of Iterated Expectation,
\begin{equation}\label{eq:boundpoolLIE}
\begin{aligned}
&\mathbb{E}\Big[
    Y_I(d,g) - Y_I(d',g')
    \,\Big|\,
    (X_I,\bm X_{\mathcal N_I})
    \in \mathcal X_{I\mathcal N_I}(d,g \mid d',g')
\Big]
\\
&\quad =
\sum_{\phi \in \Phi}
\underbrace{
\mathbb{E}\Big[
    Y_I(d,g) - Y_I(d',g')
    \,\Big|\,
    (X_I,\bm X_{\mathcal N_I})
    \in \mathcal X_{I\mathcal N_I}(d,g \mid d',g'),
    \phi_I = \phi
\Big]
}_{\boundeff}
\\
&\qquad\quad {}\times
\mathbb{P}\Big(
    \phi_I = \phi
    \,\Big|\,
    (X_I,\bm X_{\mathcal N_I})
    \in \mathcal X_{I\mathcal N_I}(d,g \mid d',g')
\Big).
\end{aligned}
\end{equation}

By Bayes’ theorem,
\[
\mathbb{P}\big(\phi_I = \phi \,\big|\, (X_I, X_{\smallN_I}) \in \mathcal{X}_{I\smallN_I}(d,g \mid d',g')\big)
=
\frac{\mathbb{P}\big((X_I, X_{\smallN_I}) \in \mathcal{X}_{I \smallN_I}(d,g \mid d',g'), \phi_I=\phi\big)}
{\mathbb{P}\big((X_I, X_{\smallN_I}) \in \mathcal{X}_{I\smallN_I}(d,g \mid d',g')\big)}.
\]
Using the definition of $\limdensity$ and $\pi_\phi$, which is the fraction of units in $\mathcal{V}_\phi$, so that $\pi_\phi = \mathbb{P}(\phi_I=\phi)$, we obtain\footnote{With a slight abuse of notation, we write quantities such as $\mathbb{P}\big((X_I, X_{\smallN_I}) \in \mathcal{X}_{I \smallN_I}(d,g \mid d',g')\big)$ to denote the corresponding boundary density integrals.}
\begin{equation}\label{eq:pooledprobLIE}
\begin{split}
&\mathbb{P}\big((X_I, X_{\smallN_I}) \in \mathcal{X}_{I\smallN_I}(d,g \mid d',g')\big)
= \sum_\phi \limdensity \cdot \pi_\phi\\
& \mathbb{P}\big((X_I, X_{\smallN_I}) \in \mathcal{X}_{I \smallN_I}(d,g \mid d',g'), \phi_I=\phi\big)
= \limdensity \cdot \pi_\phi.
\end{split}
\end{equation}
Combining \eqref{eq:pooledprobLIE} with \eqref{eq:boundpoolLIE} yields the expression for $\boundeffpooled$.
\end{proof}%

\begin{proof}{Proof of Theorem \ref{th:idpooled}}
\sloppy
By the Law of Iterated Expectations and analogous argument as Theorem~\ref{th:randomsampling} the quantity $\CE{Y_I}{ (X_I, \bm X_{\smallN_I}) \in  \mathcal{B}_{\epsilon}^{d,g}\big(\Frontier\big)}$ equals to
    \begin{align*}
    \frac{\sum_{\phi} \CE{Y_i}{(X_i,\pscore) \in \mathcal{B}_{\epsilon}^{d,g}\big(\Frontierphi\big); \, \phi }\Prr((X_i,\pscore) \in \mathcal{B}_{\epsilon}^{d,g}\big(\Frontierphi\big); \, \phi )}{\sum_{\phi}\Prr((X_i,\pscore) \in \mathcal{B}_{\epsilon}^{d,g}\big(\Frontierphi\big); \, \phi )\cdot \pi_\phi}
    \end{align*}
The proof follows by dividing numerator and denominator by the volume $V_{\varepsilon}^{\bar{s}}$, as defined in the proof of Theorem \ref{th:cont} in \eqref{eq:vol}. Then, following the proof of  Theorem \ref{th:cont} for generic $(d,g)$ applied to all $\phi$ we have 
     \[
     \CE{Y_i}{(X_i,\pscore) \in \mathcal{B}_{\epsilon}^{d,g}\big(\Frontier\big); \, \phi } \to \CE{Y(d,g)}{(X_i, \pscore) \in \Frontier, \phi}
     \]
     and
     \[
     \frac{1}{V_\varepsilon^{\bar{s}}}\Prr((X_i,\pscore) \in \mathcal{B}_{\epsilon}^{d,g}\big(\Frontier\big); \, \phi ) \to \limdensity
     \]
     which yields the conclusion.     
 \end{proof}

\subsection{Boundary Overall Direct Effects — Theorem \ref{th:overall} and Corollary \ref{corollary:overall}}\label{appendix:overall}

\begin{proof}[Proof of Theorem \ref{th:overall}]
Let us define $\mathcal{X}^\phi_{\smallN_i} \subseteq \mathbbm{R}^{|\smallN_i|}$ as the set of elements $ \bm{x}_{\smallN_i}$ for units in stratum $\mathcal{V}_\phi$. Further, for all $g \in \mathcal{G}_i$ define $ \mathcal{X}^{\phi}_{\smallN_i}(g) = \{\bm{x}_{\smallN_i} \in \mathcal{X}_{\smallN_i}^\phi : e(\bm{x}_{\smallN_i}) = g\}$. 

Note that $\bar{\mathcal{X}}^{\phi}_{i\smallN_i}(1,g\,|\,0,g) = 
 \{(x_i, \bm{x}_{\smallN_i}) \in \scorespace : x_i = c,\, \bm{x}_{\smallN_i}\in \mathcal{X}^{\phi}_{\smallN_i}(g)\}$ for all $g \in \mathcal{G}_i$.
 By Assumption \ref{ass:densitycont} we have that $\lim_{x_i \to c} f(x_i,\bm{x}_{\smallN_i}; \phi) =  f(c,\bm{x}_{\smallN_i}; \phi)$. 
 Moreover, $$ \lim_{x_i \to c} \int\limits_{\mathcal{X}^\phi_{\smallN_i}} f(x_i,\bm{x}_{\smallN_i}; \phi)d\bm{x}_{\smallN_i} = \int\limits_{\mathcal{X}^\phi_{\smallN_i}} f(c,\bm{x}_{\smallN_i}; \phi)d\bm{x}_{\smallN_i}$$ by the Dominated Convergence Theorem following from Assumption \ref{ass:densitycont} and \ref{ass:bounded}. Because $f(x_i; \phi) = \int_{\mathcal{X}_{\smallN_i}}f(x_i,\bm{x}_{\smallN_i}; \phi)d\bm{x}_{\smallN_i} $ it follows that  $\lim_{x_i \to c}f(x_i; \phi) = f(c, \phi)$.

We have 
\begin{equation*}\label{eq:bo}
\begin{split}
    \lim_{x_i \downarrow c}\CE{Y_i}{X_i = x_i; \phi} &=  \lim_{x_i \downarrow c}\sum_{g \in \mathcal{G}_i} \int_{\mathcal{X}^\phi_{\smallN_i}(g)}\CE{Y_i}{X_i = x_i, \pscore = \bm{x}_{\smallN_i}; \phi}\cdot\frac{f(x_i, \bm{x}_{\smallN_i}; \phi)}{f(x_i; \phi)}d\bm{x}_{\smallN_i}\\
    & = \lim_{x_i \downarrow c}\sum_{g \in \mathcal{G}_i} \int_{\mathcal{X}^\phi_{\smallN_i}(g)}\CE{Y_i(1,g)}{X_i = x_i, \pscore = \bm{x}_{\smallN_i}; \phi}\cdot\frac{f(x_i, \bm{x}_{\smallN_i}; \phi)}{f(x_i; \phi)}d\bm{x}_{\smallN_i} \\
    & = \sum_{g \in \mathcal{G}_i} \int_{\mathcal{X}^\phi_{\smallN_i}(g)}\CE{Y_i(1,g)}{X_i = c, \pscore = \bm{x}_{\smallN_i}; \phi}\cdot\frac{f(c, \bm{x}_{\smallN_i}; \phi)}{f(c; \phi)}d\bm{x}_{\smallN_i} \\
\end{split}
\end{equation*}
where the first equality come from the Law of Iterated expectations, the second equality from Assumptions \ref{ass:consistency}-\ref{ass:interfence}, and the third equality follows from the Dominated Convergence Theorem, since $\frac{f(x_i, \bm{x}_{\smallN_i})}{f(x_i)}$ is continuous, and $\CE{Y_i(1,g)}{X_i = x_i, \pscore = \bm{x}_{\smallN_i}}$ is continuous and bounded by Assumption \ref{ass:cont}  and \ref{ass:bounded}.

Note $\CPrr{G_i = g}{X_i = c; \phi} =  \int\limits_{\mathcal{X}^\phi_{\smallN_i}(g)}\frac{f(c, \bm{x}_{\smallN_i}; \phi)}{f(c; \phi)}d\bm{x}_{\smallN_i}$, and $\CE{Y_i(1,g)}{(X_i, \bm{x}_{\smallN_i}) \in  \bar{\mathcal{X}}_{i\smallN_i}(1,g,|0,g); \phi}$ equals
\begin{equation*}
\int\limits_{\mathcal{X}^\phi_{\smallN_i}(g)}\CE{Y_i(1,g)}{X_i = c, \pscore = \bm{x}_{\smallN_i}}\cdot\frac{f(c, \bm{x}_{\smallN_i}; \phi)}{\int\limits_{\mathcal{X}^\phi_{\smallN_i}(g)}f(c, \bm{x}_{\smallN_i}; \phi)}d\bm{x}_{\smallN_i}
\end{equation*} 
By multiplying and dividing each element of the summation in the last line of \eqref{eq:bo} by $\int\limits_{\mathcal{X}^\phi_{\smallN_i}(g)}f(c, \bm{x}_{\smallN_i}; \phi)d\bm{x}_{\smallN_i}$, we obtain
\begin{equation*}
  \lim_{x_i\downarrow c}\CE{Y_i}{X_i = x_i; \phi} =   \CE{Y_i(1,g)}{(X_i, \bm{x}_{\smallN_i}) \in  \bar{\mathcal{X}}_{i\smallN_i}(1,g,|0,g); \phi}\cdot \CPrr{G_i = g}{X_i = c; \phi}
\end{equation*}
 By analogous arguments 
\begin{equation*}
    \begin{split}
    \lim_{x_i\uparrow c}\CE{Y_i}{X_i = x_i; \phi} & = \sum_{g \in \mathcal{G}_i}\CE{Y_i(0,g)}{(X_i, \bm{x}_{\smallN_i}) \in  \bar{\mathcal{X}}_{i\smallN_i}(1,g,|0,g); \phi}\cdot \CPrr{G_i = g}{X_i = c; \phi} \\
\end{split}
\end{equation*}
from which the conclusion follows.
\end{proof}

\begin{proof}[Proof of Corollary~\ref{corollary:overall}]
The result follows from Theorem~\ref{th:overall} and continuity of $f_{X_i}(x; \phi)$, established in the proof of Theorem~\ref{th:overall}. Specifically,
\[
\lim_{x \downarrow c}
\frac{\sum_\phi \mathbb{E}[Y_i\mid X_i=x,\phi] \cdot f_{X_i}(x;\phi)\cdot \pi_\phi}
{\sum_\phi f_{X_i}(x;\phi)\cdot \pi_\phi}
=
\frac{\sum_\phi \left(\lim_{x\downarrow c}\mathbb{E}[Y_i\mid X_i=x,\phi]\right)f_{X_i}(c;\phi)\cdot \pi_\phi}
{\sum_\phi f_{X_i}(c;\phi)\cdot\pi_\phi},
\]
and analogously for $x\uparrow c$. Hence,
\[
\begin{aligned}
& \lim_{x \downarrow c}
\frac{\sum_\phi \mathbb{E}[Y_i\mid X_i=x,\phi]\cdot f_{X_i}(x;\phi)\cdot \pi_\phi}
{\sum_\phi f_{X_i}(x;\phi)\cdot \pi_\phi}
- \lim_{x \uparrow c}
\frac{\sum_\phi \mathbb{E}[Y_i\mid X_i=x,\phi]\cdot f_{X_i}(x;\phi)\cdot \pi_\phi}
{\sum_\phi f_{X_i}(x;\phi)\cdot \pi_\phi}
\\
& =
\frac{
\sum_\phi
\left(
\lim_{x\downarrow c}\mathbb{E}[Y_i\mid X_i=x,\phi]
-
\lim_{x\uparrow c}\mathbb{E}[Y_i\mid X_i=x,\phi]
\right)\cdot 
f_{X_i}(c;\phi) \cdot \pi_\phi
}{
\sum_\phi f_{X_i}(c;\phi)\cdot \pi_\phi
}\\
& =
\frac{\sum_\phi \overall(\phi)\cdot f_{X_i}(c;\phi)\cdot \pi_\phi}
{\sum_\phi f_{X_i}(c;\phi)\cdot \pi_\phi},
\end{aligned}
\]
where the last equality follows from Theorem~\ref{th:overall}.
\end{proof}

\section{Stratum-Specific Estimation}\label{appendix:estimator}

This section provides the asymptotic theory for $\estimator$. 
%
Recall the definition of the distance-based outcome regression functions:
\begin{equation*}
    \muplus = \CE{Y_i}{\distvari = \tilde{x}_i, D_i = d, G_i = g; \phi}, \quad
    \muminus  = \CE{Y_i}{\distvari = \tilde{x}_i, D_i = d', G_i = g', \phi}
\end{equation*} 
In what follows, results are stated for $\estimatorapp$, the distance-based $p-$order local polynomial estimator for $\boundeff$. 

\subsection{Notation}\label{subsec:notation}

\noindent Let $p,q \in \mathbb{Z}_{+}$. Define $\rp[u] = [1 , u, u^2, ...,u^p]$ and let $K_{h^{s}}(u) = K(u/h)/h^{s}$, where $K(\cdot)$ is a kernel function. The estimator $\estimatorapp$ is given by : 
\begin{align*}
    & \estimatorapp = e_1^T\betaplus\: - e_1^T\betaminus\\
    & \betaplus = \argmin\limits_{\bm{b} \in \mathbbm{R}^{p+1}}\sum_i\mathbbm{1}(D_i = d, G_i = g, \phi_i = \phi)(Y_i - \rp[\tilde{X}_{i}]^T\bm{b})^2\cdot K_{h_n}(\tilde{X}_{i}) \\
    &\betaminus = \argmin\limits_{\bm{b}\in \mathbbm{R}^{p+1}}\sum_i\mathbbm{1}(D_i = d', G_i = g', \phi_i = \phi)(Y_i - \rp[\tilde{X}_{i}]^T\bm{b})^2\cdot K_{h_n}(\tilde{X}_{i})
\end{align*}
where $h_n$ is a positive bandwidth and $e_1$ is the $(p+1)$-dimensional vector with the first entry equal to 1 and 0 elsewhere. We set $\Iplus = \mathbbm{1}(D_i = d, G_i = g, \phi_i = \phi)$ and $\Iminus = \mathbbm{1}(D_i = d', G_i = g', \phi_i = \phi)$, and $\bm{Y} = [Y_1,..., Y_n]^T$. In addition, we define 
\begin{align*}
        & \tilde{X}_p(h) = [\rp[\tilde{X}_1/h],..., \rp[\tilde{X}_n/h]]^T, \\
        & W_{d,g}(h; \phi)= \text{diag}(\Iplus[1]\cdot K_{h^{\bar s}}(\distvari[1]),...,\Iplus[n] \cdot K_{h^{\bar s}}(\distvari[n])),\\
        &W_{d',g'}(h, \phi)= \text{diag}(\Iminus[1]\cdot K_{h^{\bar s}}(\distvari[1]),..., \Iminus[n] \cdot K_{h^{\bar s}}(\distvari[n])).
\end{align*}
and the following sample quantities: 
\[
\begin{aligned}
  \begin{aligned}[t]
  &\Gamma_{d,g,p}(h; \phi) = \tilde{X}_p(h)^T W_{d,g}(h)\tilde{X}_p(h)/n_\phi,\\
  &t_{d,g,p}(h; \phi) = \tilde{X}_p(h)^T W_{d,g}(h; \phi)\bm{Y}/n_\phi,\\
  &\theta_{d,g,p,q}(h, \phi) = \tilde{X}_p(h)^T W_{d,g}(h_n; \phi) S_q(h)/n_\phi,
  \end{aligned}
  \qquad
  \begin{aligned}[t]
  &\Gamma_{d',g',p}(h; \phi) = \tilde{X}_p(h)^T W_{d',g'}(h; \phi)\tilde{X}_p(h)/n_\phi,\\
  &t_{d',g',p}(h; \phi) = \tilde{X}_p(h)^T W_{d',g'}(h; \phi)\bm{Y}/n_\phi,\\
  &\theta_{d',g',p,q}(h; \phi) = \tilde{X}_p(h)^T W_{d',g'}(h; \phi) S_q(h)/n_\phi,
  \end{aligned}
\end{aligned}
\]
with $S_q(h) = [(\distvari[1]/h)^q,...,(\distvari[n]/h)^q]$.
Denote by $\Gammaplusbc{,p}{h}{}$, $\Gammaminusbc{,p}{h}{}$, $\bar{t}_{d,g,p}(h)$, $\bar t_{d',g',p}(h)$, $\bar{\theta}_{d,g,p,q}(h_n)$, $\bar{\theta}_{d',g',p,q}(h_n)$ their respective expectations. Let $H(h_n) = \text{diag}(1, h^{-1}, ..., h^{-p})$. Then:
\begin{align*}
&\betaplus = H(h_n)\Gammaplus[-1]\tplus, \quad 
\betaminus = H(h_n)\Gammaminus[-1]\tminus
\end{align*}
Next, we define the population linear projection coefficients:
\begin{equation}\label{eq:linearprojection}
\begin{split}
\beta_{d,g,p}(h_n; \phi) =& H(h_n)\Gammaplusbc{,p}{h_n}{-1}\bar{t}_{d,g,p}(h_n; \phi), \quad
\beta_{d',g',p}(h_n; \phi) = H(h_n)\Gammaminusbc{,p}{h_n}{-1}\bar{t}_{d',g',p}(h_n; \phi)
\end{split}
\end{equation}
where 
\begin{align*}
\beta_{d,g,p}(h_n; \phi) =& \argmin\limits_{\bm{b} \in \mathbbm{R}^{p+1}}\E\Big[\sum_i\mathbbm{1}(D_i = d, G_i = g, \phi_i = \phi)(Y_i - \rp[\distvari]^T\bm{b})^2\cdot K_{h_n}(\distvari)\Big] \\
\beta_{d',g',p}(h_n; \phi) =& \argmin\limits_{\bm{b} \in \mathbbm{R}^{p+1}}\E\Big[\sum_i\mathbbm{1}(D_i = d', G_i = g', \phi_i = \phi)(Y_i - \rp[\distvari]^T\bm{b})^2\cdot K_{h_n}(\distvari)\Big]
\end{align*}
We introduce the following centered quantities:
\begin{equation}\label{eq:centerest}
\begin{aligned}
\begin{aligned}[t]
        &\tplus[*] = \Xp^T \Wplus \bm {\nu}_{d,g} /n, \\    &\betaplus[*] = H(h_n)\Gammaplus[-1] \tplus[*], \\
        &t^*_{d,g,d',g',p}(h_n; \phi) =  \tplus[*] - \tminus[*], \\
        &\hat{\tau}^{*}_{d,g,d',g',p}(h_n; \phi) = e_1^T\betaplus[*] - e_1^T\betaminus[*].
    \end{aligned}
    \quad
    \begin{aligned}[t]
        &\tminus[*] = \Xp^T \Wminus \bm \nu_{d',g'}/n \\
        &\betaminus[*] = H(h_n)\Gammaminus[-1] \tminus[*] \\
    \end{aligned}
\end{aligned}
\end{equation}
with $\bm\nu_{d,g} = [\nu_{1,d,g}, ..., \nu_{n,d,g}]$ and $\bm\nu_{d',g'} = [\nu_{1,d',g'}, ..., \nu_{n,d',g'}]$ where for generic $i$ and $(d,g)$: 
\begin{equation*}
    \nu_{i,d,g} = Y_i - \rp[\distvari]^T\beta_{d,g,p}(h_n; \phi)
\end{equation*}
We denote the scaled variance and covariance of $\tplus[*]$ and $\tminus[*]$ by
\[
\begin{aligned}
\Psiplusbc{p}{}{h_n} &= n h_n^{\bar{s}} \V[\tplus[*]], \quad
\Psiminusbc{p}{}{h_n} = n h_n^{\bar{s}} \V[\tminus[*]], \\[-2pt]
\Psiplusminusbc{p}{}{h_n} &= n h_n^{\bar{s}} \C[\tplus[*], \tminus[*]]
\end{aligned}
\]

\subsection{Notation for Bias and Variance}\label{sec:asyquantities}

\noindent\textbf{Kernel matrices}:
Let $\unitdisc$ be the intersection of a $s$-dimensional unit ball centered at zero $\bm{0}_s$ with a generic orthant in $\mathbbm{R}^{s}$. We define 
\begin{equation}\label{eq:kernels}
\begin{aligned}
    &\Gamma_p^s = \int_{\unitdisc}K(||\bm{u}_s||)\rp[||\bm{u}_s||]\rp[||\bm{u}_s||]^Td\bm{u}_s \\
    &\theta_{p,q}^s =  \int_{\unitdisc}K(||\bm{u}_s||)(||\bm{u}_s||)^q\rp[||\bm{u}_s||]d\bm{u}_s \\
    &\Psi_p^s =  \int_{\unitdisc}K(||\bm{u}_s||)^2\rp[||\bm{u}_s||]\rp[||\bm{u}_s||]^Td\bm{u}_s
\end{aligned}
\end{equation}
\noindent\textbf{Boundary density.} 
\[
\limdensity = \int_{\Frontierphi} \densityphi \, d\mathcal{H}^{S_\phi + 1 - \bar{s}_\phi},
\]

\noindent\textbf{Conditional variances.} 
For generic $(d,g)$ let
\[
\sigmapluseps = \V[Y_i(d,g) \mid X_i = x_i, \pscore = \bm{x}_{\smallN_i}; \phi], 
\]

Define the boundary variance components for generic $(d,g)$ as
\begin{equation}\label{eq:omegaexp}
\begin{split}
\bar{\omega}_{d,g}(0) = \int_{\Frontierphi}[\eta_{d,g}(x_i, \bm x_{\smallN_i}; \phi) + \sigmapluseps \Big]^2\densityphi \, d\ppoint,
\end{split}
\end{equation}
where 
\[
\eta_{d,g}(x_i, \bm x_{\smallN_i}; \phi) = m_{d,g}(x_i, \bm x_{\smallN_i}; \phi)  -  \tfrac{\int_{\Frontierphi} m_{d,g}(x_i, \bm x_{\smallN_i}; \phi) \densityphi}{\int_{\Frontierphi} \densityphi} 
\]
the squared deviation of the conditional mean of $Y_i(d,g)$ on a point from its boundary average.

{\color{black}
\noindent\textbf{Conditional covariances.}
Let $\widetilde{\mathcal S}_i=\mathcal S_i\cup\{i\}$, and let
$\bm X_{ij}=\bm X_{\widetilde{\mathcal S}_i\cup\widetilde{\mathcal S}_j}$ with density $f_{ij}(\cdot;\phi)$. For
$a,b\in\{(d,g),(d',g')\}$, define
\[
\kappa_{ij}^{a,b}(\bm x_{ij};\phi)
=
\C\!\left[
Y_i(a),Y_j(b)\mid \bm X_{ij}=\bm x_{ij};\phi
\right].
\]
Further define the joint boundary for two units $i$ and $j$ as:
\[
\mathcal X_{ij}^{\phi}(d,g\,|\,d',g')
:=
\left\{
\bm x_{\widetilde{\mathcal S}_i\cup\widetilde{\mathcal S}_j}
\in\mathcal X_{\widetilde{\mathcal S}_i\cup\widetilde{\mathcal S}_j}:
\bm x_{\widetilde{\mathcal S}_i}\in\Frontierphi,
\;
\bm x_{\widetilde{\mathcal S}_j}\in\bar{\mathcal X}_{j\smallN_j}(d,g\,|\, d',g')^{\phi}
\right\},
\]
The boundary covariance component is
\[
\gamma_{a,b}(0;\phi)
= \lim_{n_\phi \to \infty}
\frac{1}{n_\phi}
\sum_{i\in\mathcal V_\phi}
\sum_{j\in\mathcal K_i^\phi}
\int_{\mathcal X_{ij}^{\phi}(d,g\,|\,d',g')
}
\gamma_{a,b,ij}(x_{ij}, \phi)f_{ij}(x_{ij};\phi)
\, d \mathcal{H}^{q_{ij} - \bar{s}_\phi}(x_{ij}).
\]
where $\mathcal K_i^\phi
=
\left\{
j\in(\bm N_i\setminus\{i\})\cap\mathcal V_\phi:
\operatorname{codim}(\mathcal X_{ij}^{\phi}(d,g\,|\,d',g')
)=\bar s_\phi
\right\}
$, $\gamma_{a,b,ij}(x_{ij}, \phi) = \kappa_{ij}^{a,b}(x_{ij}; \phi) + \eta_a(x_i, \bm x_{\smallN_i};\phi) \eta_b(x_j, \bm x_{\smallN_j};\phi)$, and $q_{ij}\in\mathcal Q_\phi
:=\{S_\phi+1,\ldots,2S_\phi+2\}.$
}

\noindent\textbf{Fixed-bandwidth bias and variance.} The fixed-bandwidth bias component is given by
\begin{equation}\label{eq:biasfixed}
\begin{split}
B_{d,g,d',g',p}(h_n; \phi) =& \frac{\muplusderiv[p+1]}{(p+1)!}\Gammaplusbc{,p}{h_n}{-1}\thetaplusbc{p}{p+1}{h_n} \\
-& \frac{\muminusderiv[p+1]}{(p+1)!}\Gammaminusbc{,p}{h_n}{-1}\thetaminusbc{p}{p+1}{h_n}
\end{split}
\end{equation}
where we recall $\muplusderiv[p+1]$ and $\muminusderiv[p+1]$ are the $p+1$-order derivatives of $\muplus$ and $\muminus$ in $\tilde{x}=0$.

For the variance component first define:
\begin{equation}\label{eq:varcomponenest}
\begin{aligned}
    V_{d,g,p}(h_n; \phi) =& \Gammaplusbc{,p}{h_n}{-1}\Psiplusbc{p}{}{h_n}\Gammaplusbc{,p}{h_n}{-1}\\
    V_{d',g',p}(h_n; \phi) =& \Gammaminusbc{,p}{h_n}{-1}\Psiminusbc{p}{}{h_n}\Gammaminusbc{,p}{h_n}{-1}\\
    C_{d,g,d',g',p}(h_n; \phi) =& \Gammaplusbc{,p}{h_n}{-1}\Psiplusminusbc{p}{}{h_n}\Gammaminusbc{,p}{h_n}{-1}
    \end{aligned}
\end{equation}

The fixed-bandwidth variance component is given by
\begin{equation}\label{eq:variancefix}
    V_{d,g,d',g',p}(h_n; \phi) = V_{d,g,p}(h_n; \phi) + V_{d',g',p}(h_n; \phi) - 2C_{d',g',d,g,p}(h_n; \phi)
\end{equation}

where for generic $(d,g)$:
\begin{equation}\label{eq:vardecomp}
\begin{split}
\bar{\Psi}_{d,g, p}(h_n, \phi) &= \underbrace{\frac{1}{n_\phi h_n^{\bar{s}_\phi}}\sum_i
\E\!\big[I_{i,d,g, \phi} \Kh^2\, \rr\rr^T\, \nu_{i,d,g}^2\big]}_{\text{individual variance component}}\\
&\quad+ \underbrace{\frac{1}{n_\phi h_n^{\bar{s}_\phi}}\sum_i \sum_{j\in\bm N_i\setminus\{i\}}
\E\!\big[I_{i,d,g, \phi} I_{j,d,g, \phi}\, \Kh \Kh[j]\, \rr\, r_p\left(\Xh[j]\right)^T\, \nu_{i,d,g} \nu_{j,d,g}\big]}_{\text{covariance component}}.
\end{split}
\end{equation}
and $\bar{\Psi}_{d,g,,d',g', p}(h_n, \phi)$ is
\begin{equation}\label{eq:covdecomp}
\begin{split}
\frac{1}{n_\phi h_n^{\bar{s}_\phi}}\sum_i \sum_{j\in\bm N_i\setminus\{i\}}
\E\!\big[I_{i,d,g, \phi} I_{j,d',g', \phi}\, \Kh \Kh[j]\, \rr\, r_p\left(\Xh[j]\right)^T \nu_{i,d,g} \nu_{j,d',g'}\big]
\end{split}
\end{equation}
\noindent\textbf{Asymptotic bias and variance terms} 
The leading bias and variance terms are given by
\begin{equation}\label{eq:asybias}
B^\phi_{d,g,d',g',p, p+1} 
= \left(\frac{\muplusderiv[p+1]-\muminusderiv[p+1]}{(p+1)!}\right)
e_1^\top (\Gamma_p^{\bar{s}_\phi})^{-1}\theta_{p,p+1}^{\bar{s}_\phi},
\end{equation}
{\color{black}
\begin{equation}\label{eq:asyvariance}
V^\phi_{d,g,d',g',p}
= \left(\frac{\omegaplus + \omegaminus + \bar \gamma_{d,g,d',g'}(0; \phi)}{\limdensity^2}\right)
e_1^\top (\Gamma_p^{\bar{s}_\phi})^{-1}\Psi_p^{\bar{s}_\phi}(\Gamma_p^{\bar{s}_\phi})^{-1} e_1.
\end{equation}
with $\bar{\gamma}_{d,g,d',g'}(0; \phi) = {\gamma}_{d,g,d,g}(0; \phi) + {\gamma}_{d',g',d,'g,'}(0; \phi) - 2\gamma_{d,g,d',g'}(0; \phi)$.}

\noindent\textbf{Note.} 
In the main text, we report results for $p = 1$ and drop the subscript $p$ to simplify notation (e.g., $\Gamma^{\bar{s}_\phi}$ instead of $\Gamma^{\bar{s}_\phi}_{p}$)). We also denote the asymptotic bias and variance terms by $B^{\phi}$ and $V^{\phi}$ for notational convenience. In this appendix, we retain the full notation with $p$ and the treatment indices.

\subsection{Assumptions}
In addition to Assumptions \ref{ass:consistency}-\ref{ass:localdependence}, we will use the following assumptions. Here Assumption \ref{ass:regularity} restates Assumption \ref{ass:regularitymain} extending Part \ref{ass:smoothnessmain} to generic $p$-degree local polynomial estimators. 

\begin{assumption}\label{ass:regularity}

\begin{enumerate}[ label=(\roman*), ref=\theassumption.\roman*, leftmargin=*, topsep=0.2em, itemsep=0.2em ]

\item For some $\kappa_\phi>0$, the following holds for all
$\ppoint\in\eballboundphi[\kappa_\phi]$ and $i\in\mathcal V_\phi$:
\begin{enumerate}[ label=\alph*), ref=\theenumi.\alph*, leftmargin=1em, labelsep=0.4em, topsep=0.1em, itemsep=0pt, parsep=0pt, partopsep=0pt ]
    \item
    $\CE{Y_i^4}{X_i=x_i,\ \pscore=\bm x_{\smallN_i};\phi}$ is bounded.\label{ass:fourthmoment}
    \item
    $\sigmapluseps$ and $\sigmaminuseps$ are continuous and bounded away from zero.\label{ass:variances}
    \item
    $\muplus$ and $\muminus$ are at least $(p + 2)$-times continuously differentiable.
    \label{ass:smoothness}
\end{enumerate}
\vspace{-1em}
\item 
For each $q\in\mathcal Q_\phi$, the collections
\(\{\kappa_{ij}^{a,b}(\cdot;\phi)\}_{i,j\in\mathcal V_\phi:\,q_{ij}=q}\)
 and
\(\{f_{ij}(\cdot;\phi)\}_{i,j\in\mathcal V_\phi:\,q_{ij}=q}\)
are uniformly bounded and equicontinuous, for
$a,b\in\{(d,g),(d',g')\}$, and  $\gamma_{a,b,n}(0;\phi) \to \gamma_{a,b}(0;\phi)$ for $n_\phi \to \infty$.\label{ass:covariances}
\end{enumerate}
\end{assumption}


\begin{assumption}\label{ass:kernel}
The kernel function $K:\mathbbm{R} \to \mathbbm{R}$ has support on $ [-1,1]$, is bounded and non-negative on $[-1,1]$, and is positive, symmetric and continuous on $(-1,1)$. 
\end{assumption}


\subsection{Asymptotic Theory}\label{subsec:asytheory}

We establish the asymptotic distribution of our estimator.  
Recall that the boundary $\Frontier$ is the union of pieces $\curve$ (Appendix~\ref{appendix:mainresults}), and that each neighborhood $\eballboundrightphi$ is the union of tubular neighborhoods $\curveball$, with
\[
\operatorname{Vol}(\curveball)=O(\varepsilon^{\ccard}), 
\qquad 
\operatorname{Vol}(\eballboundrightphi)=O(\varepsilon^{\bar s_\phi}),
\]
(see proof of Theorem~\ref{th:cont}). We decompose $\betaplus$ as follows:
\begin{equation}\label{eq:expansionbeta}
    \begin{split}
        \betaplus = & \beta_{d,g,p}(h_n; \phi) + H(h_n)\Gammaplus[-1]\Xp\Wplus\bm \nu_{d,g} \\
        = &  \beta_{d,g,p}(h_n; \phi) + \betaplus[*]
    \end{split}
\end{equation}
%
%
Under Assumption \ref{ass:smoothness} a $(p+1)$-order Taylor expansion of $\muplus[\distvari]$ to the right of the cutoff yields
\begin{equation*}
\begin{split}
    \beta_{d,g,p}(h_n; \phi) = & \betaestimandplus + h_n^{p+1}\frac{\mu_{d,g}^{(p+1)}(0; \phi)}{(p+1)!}H(h_n)\Gammaplusbc{,p}{h_n}{-1}\thetaplusbc{p}{p+1}{h_n} + T_{d,g}(h_n; \phi)    
\end{split}
\end{equation*}
where $\betaestimandplus = [\mu_{d,g}(0; \phi), \;\mu^{(1)}_{d,g}(0; \phi),\,...,\, \mu_{d,g}^{(p)}(0; \phi)/p!]^T$ and 
\[
T_{d,g}(h_n; \phi) = H(h_n)\Gammaplusbc{,p}{h_n}{-1}\E\big[\Xp\Wplus T_{d,g,p}/n_\phi\big], \quad T_{d,g,p}=[T_{d,g,1},\dots,T_{d,g,n}]
\]
with each Taylor reminder component satisfying $|T_{d,g,i}|\leq \sup_{\tilde x}|\mu_{d,g}^{(p+2)}(\tilde x; \phi)|\cdot|\tilde X_i^{p+2}|/(p+2)!$ An analogous expansion holds for $\beta_{d',g',p}(h_n; \phi)$. Combining the two, we obtain
\begin{equation}\label{eq:decompositiontau}
    \begin{split}
        \estimatorapp = & \boundeff + h_n^{p+1}B_{d,g,d',g',p}(h_n; \phi) + T_{d,g,d'g'}(h_n)\\
        + & \estimatorapp[*]
    \end{split}
\end{equation}
where $B_{d,g,d',g',p}(h_n; \phi)$ is given in Eq. \eqref{eq:biasfixed} 
and $T_{d,g,d'g'}(h_n) = e_1^T\big(T_{d,g}(h_n) - T_{d',g'}(h_n)\big)$. 


\phantomsection
\addcontentsline{toc}{subsubsection}
  {Lemmas~\ref{lemma:convergencematricesgamma}--\ref{lemma:convergencematricestheta}: Convergence of sample matrices}

\begin{lemma}[Convergence of $\Gammaplus$ and $\Gammaminus$]\label{lemma:convergencematricesgamma}

Suppose Assumptions \ref{ass:consistency}--\ref{ass:networkheterogeneity}, \ref{ass:pos}, \ref{ass:densitycont}, \ref{ass:localdependence} and \ref{ass:kernel} hold. 
If $n_\phi h_n^{\bar{s}_\phi} \to \infty$ then
\[
\Gammaplus = \Gammaplusbc{,p}{h_n}{} + o_p(1), \quad  \Gammaminus = \Gammaminusbc{,p}{h_n}{} + o_p(1)
\]
where $\Gammaplusbc{,p}{h_n}{}$ and $\Gammaminusbc{,p}{h_n}{}$ are bounded and bounded away from zero.
Moreover, if $h_n \to 0$ 
    \[
    \Gammaplusbc{}{h_n}{}  = \limdensity \Gamma_p^{\bar{s}_\phi} + o(1),
    \qquad 
    \Gammaminusbc{}{h_n}{} = \limdensity \Gamma_p^{\bar{s}_\phi} + o(1).
    \]
\end{lemma}

\begin{proof}

\noindent\textbf{Expectation of $\Gammaplus$}. The generic element of $\Gammaplus$ is 
\[
F_{d,g,k}(h_n; \phi) = \frac{1}{n_\phi h_n^{\bar{s}^\phi}}\sum_i \Iplus\, K\!\left(\Xh\right)\left(\Xh\right)^k, \quad k = 0,...,2p
\]
We expand $\Gammaplusbc{,p}{h_n}{} = \E[\Gammaplus]$
\begin{equation*}
\begin{split}
&= \frac{1}{h_n^{\bar{s}_\phi}} \int\limits_{\eballboundrightphi[h_n]} 
   K\left(\XXh\right)r_p\left(\XXh\right)r_p\left(\XXh\right)^T \densityphi\, d\ppoint \\
&= \frac{1}{h_n^{\bar{s}_\phi}} \sum_{\curve}\int_{\curveball[h_n]} 
   K\left(\XXh\right)r_p\left(\XXh\right)r_p\left(\XXh\right)^T \densityphi\, d\ppoint 
   + \frac{1}{h_n^{\bar{s}_\phi}} I_{ih_n} \\
&= \frac{1}{h_n^{\bar{s}_\phi}} \sum_{\curve}
   \int_{\freeregion}\int_{\discest}
   K\!\left(\tfrac{\|\bm{x}_{\cindex}\|}{h_n}\right)
   \!r_p\left(\tfrac{\|\bm{x}_{\cindex}\|}{h_n}\right)r_p\left(\tfrac{\|\bm{x}_{\cindex}\|}{h_n}\right)^T
   f(\bm{x}_{\cindex},\bm{x}_{\findex}; \phi)\,
   d\bm{x}_{\cindex}\, d\bm{x}_{\findex}
   + \frac{1}{h_n^{\bar{s}_\phi}}I_{ih_n} \\
&=  \sum_{\curve}  h_n^{\ccard - \bar{s}_\phi}\int_{\freeregion}\int_{N^{\ccard}}
   K(\|\bm u\|)r_p(\|\bm u\|)r_p(\|\bm u\|)^T f(h_n\bm u,\bm x_{\findex})\,
   d\bm u\, d\bm x_{\findex}
   + \frac{1}{h_n^{\bar{s}^\phi}}I_{ih_n}
\end{split}
\end{equation*}
where the second equality comes from decomposing $\eballboundrightphi[h_n]$ as the union of $\curveball[h_n]$, with $I_{ih_n}$ collecting overlap terms (as in proof of Theorem \ref{th:cont}); the third equality follows from: (i) $\curveball = \freeregion \times \discest$, as defined in Eq. \ref{eq:curveball}, where we recall $\freeregion$ is the rectangular domain of the coordinates that are not fixed at $c$ on $\curve$ (Eq. \eqref{eq:rectangulardomain}), and $\cindex$ is the set of coordinates fixed at $c$ on $\curve$; (ii) $\distvari=\|\bm X_{\cindex}\|$ on $\curveball[h_n]$; the last equality follows by the change of variables $\bm{x}_{\cindex}=h_n\bm u$, which maps $\discest$ to the normalized domain $N^{\ccard}$, i.e., the unit ball in $\mathbbm{R}^{s_{\curve}}$ intersected with a generic orthant.

$\Gammaplusbc{,p}{h_n}{}$ is bounded and bounded away from zero by Assumptions \ref{ass:pos}, \ref{ass:densitycont} and \ref{ass:kernel}.

\medskip
\noindent\textbf{Variance concentration}
We show that the variance of $F_{k,d,g}(h_n; \phi)$ converges to zero:
\begin{equation*}
\begin{split}
\V[F_{k,d,g}(h_n; \phi)] 
&= \frac{1}{n_\phi^2h_n^{2\bar{s}_\phi}}\sum_i 
   \V\!\left[ \Iplus K\!\left(\Xh\right)\left(\Xh\right)^k\right] \\
&\quad + \frac{1}{n_\phi^2h_n^{2\bar{s}_\phi}}\sum_i\sum_{j \in \bm{N}_i\setminus\{i\}}
   \C\!\left[ \Iplus K\!\left(\Xh\right)\!\left(\Xh\right)^k,\,
     \Iplus[j]K\!\left(\Xh\right)\left(\Xh\right)^k \right].
\end{split}
\end{equation*}
Expanding,
\begin{equation*}
\begin{split}
\V[F_{k,d,g}(h_n; \phi)] 
&= \frac{1}{n_\phi^2h_n^{2\bar{s}_\phi}}\sum_i 
    \E\!\left[ \Iplus K\!\left(\Xh\right)^2\left(\Xh\right)^{2k}\right] \\
&\quad + \frac{1}{n_\phi^2h_n^{2\bar{s}_\phi}}\sum_i\sum_{j \in \bm{N}_i\setminus\{i\}} 
   \E\!\left[ \Iplus K\!\left(\Xh\right)\!\left(\Xh\right)^k\;
      \Iplus[j]K\!\left(\Xh\right)\left(\Xh\right)^k\right] \\
&\quad - \frac{1}{n_\phi^2h_n^{2\bar{s}_\phi}}\sum_i 
   \E\!\left[ \Iplus K\!\left(\Xh\right)\left(\Xh\right)^k\right]^2 \\
&\quad - \frac{1}{n_\phi^2h_n^{2\bar{s}_\phi}}\sum_i\sum_{j \in \bm{N}_i\setminus\{i\}} 
   \E\!\left[ \Iplus K\!\left(\Xh\right)\left(\Xh\right)^k\right] \,
   \E\!\left[  \Iplus[j] K\!\left(\Xh[k]\right)\left(\Xh[j]\right)^k\right].
\end{split}
\end{equation*}
The first term is $O((n_\phi h_n^{\bar{s}_\phi})^{-1}) = o(1)$, by the same arguments used in computing the expectation.
The second term is also $O((n_\phi h_n^{\bar{s}_\phi})^{-1})$: by the arithmetic–geometric inequality it is bounded by
\[
\frac{1}{n_\phi h_n^{2\bar{s}_\phi}} \max_i \E\!\left[\Iplus K\!\left(\frac{\tilde{X}_i}{h_n}\right)\left(\Xh\right)^k\right] 
\cdot \frac{\sum_i |\bm{N}_i|/n}{n_\phi/n} 
= O((n_\phi h_n^{\bar{s}_\phi})^{-1}),
\]
since $\frac{1}{n}\sum_i |\bm{N}_i| = O(1)$ by Assumption \ref{ass:localdependence}. The last two terms are $O(n^{-1})$ by Assumption \ref{ass:localdependence}, since each expectation term is $O(h_n^{\bar{s}_\phi})$. Hence, the component-wise variance of $\Gammaplus$ converges to zero, which yields the first conclusion.  

\medskip
\noindent\textbf{Limiting expectation computation} 
For $h_n \to 0$:%
\begin{equation*}
\begin{split}
\Gammaplusbc{,p}{h_n}{} =& \sum_{\curve}  h_n^{\ccard - \bar{s}_\phi}\int_{\freeregion}\int_{N^{\ccard}}
   K(\|\bm u\|)r_p(\|\bm u\|)r_p(\|\bm u\|)^T f(h_n\bm u,\bm x_{\findex})\,
   d\bm u\, d\bm x_{\findex}
   + \frac{1}{h_n^{\bar{s}^\phi}}I_{ih_n}  \\
   =& \limdensity \Gamma_p^{\bar{s}_\phi} + o(1)  
\end{split}
\end{equation*}
The second equality follows because all integral terms are bounded, so the contributions from each term in the summation with $\ccard>\bar{s}_\phi$ are $O(h_n)$, as is the overlap term since $I_{ih_n}=O(h_n^{\bar{s}_\phi+1})$ as shown in Theorem~\ref{th:cont}. For terms with $\ccard=\bar{s}_\phi$, Assumption~\ref{ass:densitycont} and the Dominated Convergence Theorem yield the limit.
Hence, $\Gammaplus = \limdensity \Gamma_p^{\bar{s}_\phi} + o_p(1)$.
The conclusions for $\Gammaminus$ follows by analogous arguments.
\end{proof}

\begin{lemma}[Convergence of $\tplus$ and $\tminus$]\label{lemma:convergencematricest}
Suppose Assumptions \ref{ass:consistency}--\ref{ass:localdependence}, \ref{ass:variances}, and \ref{ass:kernel} hold. 
If $n_\phi h_n^{\bar{s}_\phi} \to \infty$ and $h_n < \kappa$
then
\[
\tplus = \bar{t}_{d,g,p}(h_n; \phi) + o_p(1), \quad  \tminus = \bar{t}_{d',g',p}(h_n; \phi) + o_p(1)
\] 
where $\bar{t}_{d,g,p}(h_n; \phi)$ and $\bar{t}_{d',g',p}(h_n; \phi)$ are bounded. Moreover, if $h_n \to 0$ for generic $(d,g)$
    \[
    \bar{t}_{d,g,p}(h_n; \phi)  = \theta_{p,0}^{\bar{s}_\phi} \hspace{-1em} \int\limits_{\Frontierphi} \hspace{-1em}m_{d,g}(x_i, \bm x_{\smallN_i}; \phi) \densityphi + o(1).
    \]
\end{lemma}

\begin{proof}
\noindent Since the procedure is analogous to Lemma \ref{lemma:convergencematricesgamma} we only sketch the proof. 

We first compute the limiting expectation of $\tplus$: 
\begin{equation*}
\begin{split}
    \bar{t}_{d,g,p}(h_n; \phi) = & \sum_{\curve}  h_n^{\ccard - \bar{s}_\phi}\int_{\freeregion}\int_{N^{\ccard}}
   K(\|\bm u\|)r_p(\|\bm u\|) m_{d,g}(h_n\bm u,\bm x_{\findex}) f(h_n\bm u,\bm x_{\findex}; \phi)\,
   d\bm u\, d\bm x_{\findex}
   + \frac{1}{h_n^{\bar{s}}}I_{ih_n} \\
   = & \theta^{\bar{s}}_{p,0} \int\limits_{\Frontierphi}m_{d,g}(x_i, \bm x_{\smallN_i}; \phi) \densityphi + o(1)
\end{split} 
\end{equation*}
by Assumption \ref{ass:cont}, \ref{ass:densitycont} and \ref{ass:kernel} and the Dominated Convergence Theorem. Similarly, one obtains the limiting expectation for $\tminus$.
Convergence to zero of the component-wise variance of $\tplus$ and $\tminus$ can also be shown similarly to Lemma~\ref{lemma:convergencematricesgamma} using boundedness of the conditional variance (Assumption \ref{ass:variances}).
\end{proof}

\begin{lemma}[Convergence of $\theta_{d,g,p,q}(h_n)$ and $\theta_{d',g', p, q}({h_n})$]\label{lemma:convergencematricestheta}
Suppose Assumptions \ref{ass:consistency}--\ref{ass:networkheterogeneity}, \ref{ass:pos}, \ref{ass:densitycont}, \ref{ass:localdependence} and \ref{ass:kernel} hold. 
If $n_\phi h_n^{\bar{s}_\phi} \to \infty$ then
\[
\theta_{d,g,p,q}(h_n) = \thetaplusbc{p}{q}{h_n} + o_p(1), \quad  \theta_{d',g', p, q}({h_n}) = \thetaminusbc{p}{q}{h_n} + o_p(1)
\]
where $\thetaplusbc{p}{q}{h_n}$ and $\thetaminusbc{p}{q}{h_n}$ are bounded and bounded away from zero.
Moreover, if $h_n \to 0$ 
    \[
    \thetaplusbc{p}{q}{h_n}  = \limdensity \theta_{p,q}^{\bar{s}} + o(1),
    \qquad 
    \thetaminusbc{p}{q}{h_n} = \limdensity \theta_{p,q}^{\bar{s}} + o(1).
    \]
\end{lemma}

\begin{proof}
Since the procedure is again analogous to Lemma \ref{lemma:convergencematricesgamma} we only sketch the proof. We first compute the limiting expectation of $\theta_{d,g,p,q}(h_n)$ by a similar decomposition as in Lemma~\ref{lemma:convergencematricesgamma}:
\begin{equation*}
\begin{split}
\thetaplusbc{p}{q}{h_n} =& \sum_{\curve}h_n^{s_{\curve} - \bar{s}_\phi}
   \int_{\freeregion}\int_{N^{\bar{s}_{\curve}}}
   K(\|\bm u\|)r_p(\|\bm u\|)\|\bm u \|^q f(\bm u h_n,\bm x_{\findex}; \phi)\,
   d\bm u\, d\bm x_{\findex} + \frac{1}{n h_n^{\bar s}}I_{ih_n} \\
   =& \limdensity \theta_{p,q}^{\bar{s}} + o(1)
\end{split}
\end{equation*}
by the Dominated Convergence Theorem.
Hence, $\thetaplusbc{p}{q}{h_n} = \limdensity \theta_{p,q}^{\bar{s}} +  o(1)$. Similarly, one obtains the limiting expectation of $\theta_{d',g',p,q}(h_n; \phi)$.
Convergence to zero of the component-wise variance of $\theta_{d,g,p,q}(h_n; \phi)$ and $\theta_{d,g',p,q}(h_n; \phi)$ is shown as Lemma~\ref{lemma:convergencematricesgamma}.
\end{proof}

\phantomsection
\addcontentsline{toc}{subsubsection}
  {Lemma~\ref{lemma:consistency}: Consistency}
\begin{lemma}\label{lemma:consistency}
Suppose Assumptions \ref{ass:consistency}--\ref{ass:localdependence}, \ref{ass:variances}, \ref{ass:kernel} hold. 
If $n_\phi h_n^{\bar{s}_\phi} \to \infty$ and $h_n \to 0$
then
\[
\estimatorapp \xrightarrow{p} \boundeff
\]

\end{lemma}
\begin{proof}
The proof follows by noting that
\[
e_1^T\betaplus = e_1^T\Gamma_{d,g,p}^{-1}(h_n, \phi) \tplus \xrightarrow{p} \frac{\int_{\Frontierphi} m_{d,g}(x_i, \bm x_{\smallN_i}; \phi)\densityphi}{\int_{\Frontierphi}\densityphi}
\]
and, analogously,
\[
e_1^T\betaminus \xrightarrow{p} \frac{\int_{\Frontierphi} m_{d',g'}(x_i, \bm x_{\smallN_i}; \phi)\densityphi}{\int_{\Frontierphi}\densityphi}
\]
by Lemma ~\ref{lemma:convergencematricesgamma} and \ref{lemma:convergencematricest}, continuous mapping and the identity $e_1^T(\Gamma_p^{\bar{s}_\phi})^{-1}\theta_{p,0}^{\bar{s}_\phi} = 1$.
\end{proof}

\phantomsection
\addcontentsline{toc}{subsubsection}
  {Lemma~\ref{lemma:variances}: Limiting variance}

\begin{lemma}[Variance of $t_{d,g,d',g'}^*(h_n; \phi)$]\label{lemma:variances}
Suppose Assumptions \ref{ass:consistency}--\ref{ass:localdependence},
\ref{ass:variances}, \ref{ass:covariances} and \ref{ass:kernel} hold. If $h_n < \kappa_\phi$,
\[
n_\phi h_n^{\bar{s}_\phi}\V[t_{d,g,d',g'}^*(h_n, \phi)]
= \Psiplusbc{p}{}{h_n} + 2\Psiminusbc{p}{}{h_n} - \Psiplusminusbc{p}{}{h_n},
\]
is bounded and bounded away from zero. Moreover, if $nh_n^{\bar{s}}\to \infty$ and $h_n \to 0$
\[
\Psiplusbc{p}{}{h_n} + \Psiminusbc{p}{}{h_n} - \Psiplusminusbc{p}{}{h_n}
= (\omegaplus+\omegaminus + \bar \gamma_{d,g,d',g'}(0; \phi))\Psi_p^{\bar{s}_\phi} + o(1).
\]
with $\bar \gamma_{d,g,d',g'}(0; \phi) = \bar \gamma_{d,g,d,g}(0; \phi) + \gamma_{d',g',d',g'}(0; \phi) - 2 \gamma_{d,g,d',g'}(0; \phi) $

\end{lemma}

\begin{proof}
We start from
\[
\V[t_{d,g,d',g'}^*(h_n)]
= \V[\tplus[*]] + \V[\tminus[*]] - 2\C[\tplus[*], \tminus[*]]
\]

\medskip
\noindent\textbf{Decomposition of the variance}
Expanding the variance yields
\begin{equation*}
\begin{split}
n h_n^{\bar{s}} \V[\tplus[*]]
&= \underbrace{\frac{1}{n_\phi h_n^{\bar{s}_\phi}}\sum_i
\E\!\big[\Iplus \Kh^2\, \rr\rr^T\, \nu_{i,d,g}^2\big]}_{T_1}\\
&\quad+ \underbrace{\frac{1}{n_\phi h_n^{\bar{s}_\phi}}\sum_i \sum_{j\in\bm N_i\setminus\{i\}}
\E\!\big[\Iplus \Iplus[j]\, \Kh \Kh[j]\, \rr\, r_p(\Xh[j])^T\, \nu_{i,d,g} \nu_{j,d,'g'}\big]}_{T_2}.
\end{split}
\end{equation*}

\medskip
\noindent\textbf{Limit of $T_1$.}
Expanding $\nu_{i,d,g}^2$ gives
\begin{align*}
T_1
&= \sum_{h_i} h_n^{s_{\curve}-\bar{s}_\phi}
\!\int_{\freeregion}\!\int_{\unitdisc[s_{\curve}]}
K(\|\bm u\|)^2 r_p(\|\bm u\|) r_p(\|\bm u\|)^T
\big[
m_{d,g}(\bm u h_n, \bm x_{\findex}; \phi)^2
+ \sigmaplus[\bm u h_n, \bm x_{\findex}; \phi]\\
&\quad
+ (r_p(\|\bm u\|h_n)^T\beta_{d,g,p}(h_n; \phi))^2
- 2\, m_{d,g}(\bm u h_n, \bm x_{\findex}; \phi)\, r_p(\|\bm u\|h_n)^T \beta_{d,g,p}(h_n; \phi)
\big]
f(\bm u h_n, \bm x_{\findex}; \phi)\, d\bm u\, d\bm x_{\findex}.
\end{align*}
As $h_n \to 0$
\[
r_p(\|\bm u\| h_n)\beta_{d,g,p}(h_n; \phi)
\;\to\;
\frac{\int_{\Frontierphi} m_{d,g}(x_i, \bm x_{\smallN_i}; \phi) \densityphi}
     {\int_{\Frontierphi} \densityphi},
\]
using $r_p(\|\bm u\| h_n)\beta_{d,g,p}(h_n; \phi) = r_p(\|\bm u\|)\Gammaplusbc{}{h_n}{}\bar{t}_{d,g,p}(h_n; \phi)$,
Lemmas~\ref{lemma:convergencematricesgamma} and \ref{lemma:convergencematricest}, and $(\Gamma_p^{\bar s})^{-1}\theta_{p,0}^{\bar{s}} = e_1$.
\sloppy
\noindent Then by continuity of $m_{d,g}(\cdot; \phi)$, $f(\cdot; \phi)$, $\sigmaplus[\cdot; \phi]$, 
(Assumptions~\ref{ass:densitycont}-\ref{ass:cont} and \ref{ass:variances}), and by the Dominated Convergence Theorem, $T_1 = \omegaplus \Psi_p^{\bar{s}_\phi} + o(1)$. 

{\color{black}
\medskip
\noindent\textbf{Limit of $T_2$.}
Consider the term 
\[
\E\!\big[\Iplus \Iplus[j]\, \Kh \Kh[j]\, \rr\, r_p\left(\Xh[j]\right)^T\, \nu_{i,d,g} \nu_{j,d,'g'}\big], \quad i \neq j, \; j \in \bm N_i
\]

The domain of integration of this expectation is supported on the intersection of the
$h_n$-neighborhoods of the two boundaries $\Frontierphi$ and $\mathcal{\bar{X}_{j\smallN_j}}(d,g\,|\,d',g')$. We claim that this intersection
can be replaced, up to a set of volume $O(h_n^{\bar s_\phi+1})$, by the
$h_n$-neighborhood of their intersection
$\mathcal{\bar X}_{ij}^\phi(d,g\mid d',g')$. To prove this, we show that 
\[
\operatorname{Vol}\!\left(
[\eballboundright[h_n] \cap \mathcal{B}_{h_n}^{d,g}(\mathcal{X}^{\phi}_{j\smallN_j}(d,g|d',g') )]
\setminus \mathcal{B}_{h_n}^{d,g}(\mathcal{X}^{\phi}_{ij}(d,g|d',g'))
\right)
=O(h^{\bar s_\phi+1}).
\]

For a coordinate-aligned piece $\ell$, let
$\mathcal B_{h_n,\infty}(\ell;\phi)$ denote its $h_n$-neighborhood under
the Chebyshev distance. For any two pieces $\ell_i \in \Frontierphi$ and $\ell_j \in \mathcal{X}_{j\smallN_j}^{\phi}(d,g,\, | \, d', g')$ with nonempty intersection,
\[
\mathcal B_{h_n,\infty}(\ell_i;\phi)
\cap
\mathcal B_{h_n,\infty}(\ell_j;\phi)
=
\mathcal B_{h_n,\infty}(\ell_i\cap\ell_j;\phi),
\qquad
\mathcal B_{h_n}(\ell;\phi)
\subseteq
\mathcal B_{h_n,\infty}(\ell;\phi).
\]  
Hence, if the codimension of 
$(\ell_i\cap\ell_j)$ is greater than $\bar s_\phi$, then
\(
\operatorname{Vol}\!\left(
\mathcal B_{h_n}(\ell_i;\phi)
\cap
\mathcal B_{h_n}(\ell_j;\phi)
\right)
=
O(h_n^{\bar s_\phi+1}).
\)
If the codimension of 
$(\ell_i\cap\ell_j)$ is $\bar s_\phi$, then the two pieces fix the same $\bar s_\phi$ coordinates at the cutoff.
Let
\(
F_{\ell_{ij}}=F_{\curve}\cup F_{\ell_j},
\)
and, for any $\bm x_{ij}$, let $\bm x_{ij}^*$ be obtained by replacing $\bm x_{F_{\ell_{ij}}}$ by $ 0\mathbf 1_{F_{\ell_{ij}}}$ and leaving all other coordinates
unchanged. Consider
\[
E_{ij}
=
\left[
\mathcal B_{h_n}(\ell_i;\phi)
\cap
\mathcal B_{h_n}(\ell_j;\phi)
\right]
\setminus
\mathcal B_{h_n}(\ell_i\cap\ell_j;\phi).
\]
If $\bm x_{ij}\in E_{ij}$, then
\(
\|\bm x_{F_{\ell_{ij}}}\|_2\le h_n.
\)
Moreover, $\bm x_{ij}^*\notin\ell_i\cap\ell_j$, since otherwise
\(
d_2(\bm x,\ell_i\cap\ell_j)
\le
\|\bm x-\bm x_{ij}^*\|_2
=
\|\bm x_{F_{\ell_{ij}}}\|_2
\le h_n,
\)
contradicting $\bm x_{ij}\in E_{ij}$.
On the other hand,
\(
\bm x_{ij} \in
\mathcal B_{h_n,\infty}(\ell_i\cap\ell_j;\phi),
\)
Since $\bm x_{ij}^*$ differs from $\bm x_{ij}$ only by setting the coordinates in
$F_{\ell_{ij}}$ exactly equal to the cutoff, it follows that
\[
\bm x_{ij}^*
\in
\mathcal B_{h_n,\infty}(\ell_i\cap\ell_j;\phi)
\setminus
(\ell_i\cap\ell_j).
\]
For a coordinate-aligned piece, the set of such values of the remaining
coordinates has volume $O(h_n)$. Together with
$\|\bm x_{F_{\ell_{ij}}}\|_2\le h_n$, whose volume is
$O(h_n^{\bar s_\phi})$, this gives
\(
\operatorname{Vol}(E_{ij})
=
O(h_n^{\bar s_\phi+1})
\).
Since there are finitely many pairs of pieces, the same conclusion holds
for the corresponding full boundary intersections.
Consequently, we can write %
\[ \begin{split} T_2 =&\; \frac{1}{n_\phi h_n^{\bar s_\phi}} \sum_{i\in V_\phi} \sum_{j\in\bm N_i\setminus \{i\}} \int_{\mathcal B_{h_n}^{d,g} (\bar{\mathcal X}_{ij}^{\phi}(d,g\mid d',g'))} K\!\left( \frac{\widetilde X(x_i,\bm x_{\smallN_i})}{h_n} \right) K\!\left( \frac{\widetilde X(x_j,\bm x_{\smallN_j})}{h_n} \right) \\ &\qquad\times r_p\!\left( \frac{\widetilde X(x_i,\bm x_{\smallN_i})}{h_n} \right) r_p\!\left( \frac{\widetilde X(x_j,\bm x_{\smallN_j})}{h_n} \right)^T \gamma_{d,g,ij}(\bm x_{ij};\phi) f_{ij}(\bm x_{ij};\phi) \,d\bm x_{ij} +O(h_n). \end{split} \] 
Moreover, pairs satisfying \(\operatorname{codim}\!\left( \bar{\mathcal X}_{ij}^{\phi}(d,g\mid d',g') \right) > \bar s_\phi \) contribute $O(h_n)$ after normalization. Since $n_\phi^{-1}\sum_i|\bm N_i|=O(1)$ by Assumption~\ref{ass:localdependence}, their aggregate contribution is also $O(h_n)$. Therefore we can further write, 
\[ \begin{split} T_2 =&\; \frac{1}{n_\phi h_n^{\bar s_\phi}} \sum_{i\in V_\phi} \sum_{j\in\mathcal{K}^\phi _i } \int_{\mathcal B_{h_n}^{d,g} (\bar{\mathcal X}_{ij}^{\phi}(d,g\mid d',g'))} K\!\left( \frac{\widetilde X(x_i,\bm x_{\smallN_i})}{h_n} \right) K\!\left( \frac{\widetilde X(x_j,\bm x_{\smallN_j})}{h_n} \right) \\ &\qquad\times r_p\!\left( \frac{\widetilde X(x_i,\bm x_{\smallN_i})}{h_n} \right) r_p\!\left( \frac{\widetilde X(x_j,\bm x_{\smallN_j})}{h_n} \right)^T \gamma_{d,g,d,g,ij}(\bm x_{ij};\phi) f_{ij}(\bm x_{ij};\phi)\,d\bm x_{ij} +O(h_n). \end{split} 
\] 
For each $j\in\mathcal K_i^\phi$ we have $F_{\curve}=F_{\ell_j}=F_{\ell_{ij}}$, where $\ell_{ij}:=\ell_i\cap\ell_j$, so that $\widetilde X(x_i,\bm x_{\smallN_i})=\|\bm x_{F_{\ell_{ij}}}\|_2
=\widetilde X(x_j,\bm x_{\smallN_j})$ on
$\mathcal B_{h_n}^{d,g}(\bar{\mathcal X}_{ij}^{\phi}(d,g\mid d',g'))$. We let $\ell_{ij}^\phi$ be the pieces of the
joint boundary $\bar{\mathcal X}_{ij}^{\phi}(d,g\mid d',g')$ of codimension $\bar s_\phi$. With change of variables $\bm x_{F_{\ell_{ij}}}=h_n\bm u$ gives, for each such pair,
\[
\begin{split}
\frac{1}{h_n^{\bar s_\phi}}
\int_{\mathcal B_{h_n}^{d,g}(\bar{\mathcal X}_{ij}^{\phi})}
K^2 r_p r_p^{T}\,\gamma_{d,g,d,g}f_{ij}\,d\bm x_{ij}
=&\sum_{\ell_{ij}}\int_{\mathcal R_{\ell_{ij}}}\int_{N^{\bar s_\phi}}
K(\|\bm u\|)^2 r_p(\|\bm u\|)r_p(\|\bm u\|)^{T}\\
&\quad\times\big(\gamma_{d,g,d,g}f_{ij}\big)(h_n\bm u,\bm x_{R_{\ell_{ij}}};\phi)
\,d\bm u\,d\bm x_{R_{\ell_{ij}}}
+\frac{1}{h_n^{\bar s_\phi}}I_{ij,h_n},
\end{split}
\]
where \(
\mathcal R_{\ell_{ij}}:=\prod_{z\in R_{\ell_{ij}}}I_z(\ell_{ij})
\) is the rectangular domain of the free coordinates $\bm x_{R_{\ell_{ij}}}$, $R_{\ell_{ij}}:=(\widetilde{\mathcal S}_i\cup\widetilde{\mathcal S}_j)\setminus
F_{\ell_{ij}}$, and $I_{ij,h_n}=O(h_n^{\bar s_\phi+1})$ collects the overlap terms, as in Lemma~\ref{lemma:convergencematricesgamma}.
By Assumption~\ref{ass:covariances} the collections $\{\kappa_{ij}^{a,b}(\cdot;\phi)\}$
and $\{f_{ij}(\cdot;\phi)\}$ are uniformly bounded and equicontinuous within each
joint score dimension $q\in\mathcal Q_\phi$, and $m_{d,g}(\cdot;\phi)$ is continuous
and bounded by Assumptions~\ref{ass:cont} and~\ref{ass:bounded}, so
$\gamma_{d,g}f_{ij}$ inherits both properties within each $q$. Since $q_{ij}$ takes
finitely many values and there are finitely many boundary pieces, the integrand may
be replaced by its value at $\bm u=0$ uniformly over dependent pairs, with an error that is $o(1)$ uniformly in 
$(i,j)$. Averaging over the 
$O(n_\phi)$ pairs preserves the 
error order $o(1)$ by Assumption~\ref{ass:localdependence}. Hence,
\[
\begin{split}
T_2=&\Psi_p^{\bar s_\phi}\,\underbrace{\frac{1}{n_\phi}\sum_{i\in\mathcal V_\phi}
\sum_{j\in\mathcal K_i^\phi}\int_{\bar{\mathcal X}_{ij}^{\phi}(d,g\,|\,d',g')}
\gamma_{d,g,d,g}(\bm z;\phi)f_{ij}(\bm z;\phi)\,d\mathcal H^{q_{ij}-\bar s_\phi}(\bm z)}_{\gamma_{d,g,d,g,n}(0;\phi) \to \gamma_{d,g,d,g}(0;\phi)}
+o(1).
\end{split}
\]

Similar arguments obtain the limits for
$n_\phi h_n^{\bar{s}_\phi}\V[\tminus[*]]$ and  $n_\phi h_n^{\bar{s}_\phi}\C[\tplus[*],\tminus[*]].$
}
\end{proof}


\phantomsection
\addcontentsline{toc}{subsubsection}
  {Theorem~\ref{th:asynormaltau}: Asymptotic normality}

\begin{theorem}\label{th:asynormaltau}
Suppose Assumptions~\ref{ass:consistency}--\ref{ass:localdependence},
\ref{ass:regularity} and~\ref{ass:kernel} hold.
If $n h_n^{\bar{s}} \to \infty$, $h_n \to 0$, and 
$h_n = O\!\left(n^{-1/(2p + 2 + \bar{s})}\right)$, then
\[
\sqrt{n h_n^{\bar{s}}}\,
\frac{(\estimatorapp - \boundeff) - h_n^{p+1}\bias}
{\sqrt{V^\phi_{d,g,d',g',p}(h_n)}}
\xrightarrow{d} \mathcal{N}(0,1),
\]
where $\bias = B^\phi_{d,g,d',g',p, p+1} + o(1)$ and $V^\phi_{d,g,d',g'p}(h_n) = V^\phi_{d,g,d',g',p} + o(1)$ as defined in \eqref{eq:asybias} and \eqref{eq:asyvariance}.
\end{theorem}
\begin{proof}[Proof of Theorem \ref{th:asynormaltau}]
Define for generic $(d,g)$
\begin{equation*}
\begin{split}
\Zplus &= e_1^T \Gammaplusbc{,p}{h_n}{-1}\;
           \frac{1}{n_\phi h_n^{\bar{s}_\phi}}\,\Iplus \Kh \rr \nu_{i,d,g}
\end{split}
\end{equation*}
and set $\Z = \Zplus - \Zminus$. Then
\[
\sum_i \Z = e_1^T\Gammaplusbc{p,}{h_n}{-1}\tplus[*] - e_1^T\Gammaminusbc{p,}{h_n}{-1}\tminus[*]
\]

\medskip
\noindent\textbf{Variance.}
Note $n_\phi h_n^{\bar{s}_\phi}\,\V\!\Big[\sum_i \Z\Big] 
= V^\phi_{d,g,d',g',p}(h_n)$ given in \eqref{eq:variancefix}. By Lemmas~\ref{lemma:convergencematricesgamma} and \ref{lemma:variances},
\[
\begin{split}
V^\phi_{d,g,d',g',p}(h_n) =&   \frac{\omegaplus+\omegaminus + \bar \gamma_{d,g,d',g'}(0; \phi)}{\limdensity}\,
  (\Gamma_p^{\bar{s}_\phi})^{-1}\Psi_p^{\bar{s}}(\Gamma_p^{\bar{s}_\phi})^{-1} + o(1) \\
  =& V^\phi_{d,g,d',g',p} + o(1)
  \end{split}
\]

\medskip
\noindent\textbf{CLT via dependency graph.}
The collection $\{\Z\}_i$ admits dependency graph 
$\bm W_n$. Let $\tilde\sigma_n^2 = V^\phi_{d,g,d',g',p}(h_n)/n_\phi h_n^{\bar{s}_\phi}$, the variance of $\Z$, and set 
$\xi_n = \sum_i \Z/ \tilde\sigma_n$.  
Applying the Wasserstein bound of \citet[Theorem A.4]{leung20},
\begin{equation}\label{eq:distfirst}
\begin{split}
\Delta(\xi_n,\mathcal{N}(0,1)) 
&\leq \tfrac{1}{\tilde\sigma_n^3}\max_i\E[|\Z|^3]\sum_i|\bm N_i|^2 \\
&\quad+ \tfrac{1}{\tilde\sigma_n^2}
\Big(\tfrac{4}{\pi}\max_i\E[\Z^4]\Big)^{1/2}
\Big(4\sum_i|\bm N_i|^3 + 3\sum_{i,j}(\bm W^3)_{ij}\Big)^{1/2}.
\end{split}
\end{equation}

\medskip
\noindent\textbf{Moment bounds.}
Using the $c_r$-inequality,
\[
\E[|Z_i|^3] \leq 4(\E[|\Zplus|^3]+\E[|\Zminus|^3]) 
= O(n_\phi^{-3}h_n^{-2\bar{s}_\phi}),
\]
and similarly,
\[
\E[Z_i^4] \leq 8(\E[\Zplus^4]+\E[\Zminus^4]) 
= O(n_\phi^{-4}h_n^{-3\bar{s}_\phi}).
\]
using boundedness of the third and fourth conditional moment of the outcome (Assumption \ref{ass:fourthmoment}) and Assumptions \ref{ass:cont}.

\medskip
\noindent\textbf{Rate of convergence.}
Plugging these bounds into \eqref{eq:distfirst} and using 
Assumption~\ref{ass:localdependence}
\[
\Delta(\xi_n,\mathcal{N}(0,1)) = O((n_\phi h_n^{\bar{s}_\phi})^{-1/2}) = o(1).
\]
Hence $\xi_n \xrightarrow{d} \mathcal{N}(0,1)$. 

\medskip
\noindent\textbf{Conclusion.}
By Lemma~\ref{lemma:convergencematricesgamma},
$
\sqrt{n_\phi h_n^{\bar{s}_\phi}}\cdot \tfrac{\estimatorapp[*]}{\sqrt{V^\phi_{d,g,d',g',p}(h_n)}} = \xi_n + o_p(1)
$, so $\estimatorapp[*]$ (rescaled) has the same asymptotic distribution as $\xi_n$. Moreover, by Lemmas~\ref{lemma:convergencematricesgamma} and \ref{lemma:convergencematricestheta}, $h_n^{p+1}\bias = O(h_n^{p+1})$ and $T_{d,g,d'g'}(h_n) = O(h_n^{p+2})$. 
Then Eq.\eqref{eq:decompositiontau} and $h_n = O(n_\phi^{-1/(2p+2 + {\bar{s}_\phi})})$ yield the conclusion.
\end{proof}

\phantomsection
\addcontentsline{toc}{subsubsection}
  {Proof of Theorem~\ref{th:asynormal}}

\begin{proof}[Proof of Theorem \ref{th:asynormal}]
The result follows from Theorem~\ref{th:asynormaltau} by setting $p = 1$ and writing $B^\phi = B^\phi_{d,g,d',g',1,2}$ and $V^\phi = V^\phi_{d,g,d',g',1}$.
\end{proof}

\subsection{Variance Estimation}

Define 
\begin{equation}\label{eq:varestimatorappendix}
\begin{split}
    \hat{\Psi}_{d,g,p}(h_n; \phi) &= \Mplus \bm{W} \Mplus^T / n_\phi, \\
    \hat{\Psi}_{d',g',p}(h_n; \phi) &= \Mminus \bm{W} \Mminus^T / n_\phi, \\
    \hat{\Psi}_{d,g,d',g',p}(h_n; \phi) &= \Mplus \bm{W} \Mminus^T / n_\phi,
\end{split}
\end{equation}
where $\bm{W}$ is the dependency graph and
\begin{equation*}
\begin{split}
    \Mplus^T &= \Big[\Iplus[1]\Kh[1]\rr[1]\hat{u}_{1,d,g} \;\; \cdots \;\; \Iplus[n]\Kh[n]\rr[n]\hat{u}_{n,d,g}\Big], 
    \end{split}
\end{equation*}
for a generic $(d,g)$ with $\hat{u}_{i,d,g}$ and $\hat{u}_{i,d',g'}$ denoting the regression residuals.

\medskip
Let
\[
\varianceconstant = \frac{V^\phi_{d,g,d',g'}(h_n)}{n_\phi h_n^{\bar{s}_\phi}}
\]

An estimator of $\varianceconstant$ is
\begin{equation}\label{eq:varestimator}
    \hat{\mathcal{V}}^\phi_{d,g,d',g',p}(h_n)
    = \frac{1}{n_\phi}\, e_1^T\!\left[
        \hat{V}_{d,g,p}(h_n; \phi) + \hat{V}_{d',g',p}(h_n; \phi) - 2\,\hat{V}_{d,g,d',g',p}(h_n; \phi)
      \right]e_1,
\end{equation}
where
\begin{equation*}
\begin{split}
    \hat{V}_{d,g,p}(h_n; \phi) &= \Gammaplus^{-1}\,\hat{\Psi}_{d,g}(h_n; \phi)\,\Gammaplus^{-1}, \\
    \hat{V}_{d',g',p}(h_n; \phi) &= \Gammaminus^{-1}\,\hat{\Psi}_{d',g'}(h_n; \phi)\,\Gammaminus^{-1}, \\
    \hat{V}_{d,g,d',g',p}(h_n; \phi) &= \Gammaplus^{-1}\,\hat{\Psi}_{d,g,d',g'}(h_n; \phi)\,\Gammaminus^{-1}.
\end{split}
\end{equation*}
We show that this estimator is consistent for $\varianceconstant$. Convergence of this estimator to the asymptotic variance $(n_\phi h_n^{\bar{s}_\phi})^{-1}V^\phi_{d,g,d',g',p}$ follows since $V^\phi_{d,g,d',g',p}(h_n) = V^\phi_{d,g,d',g',p} + o(1)$ for $h_n \to 0$ by Lemma \ref{lemma:convergencematricesgamma} and \ref{lemma:variances}.

\phantomsection
\addcontentsline{toc}{subsubsection}
  {Lemma~\ref{lemma:psiconsistency}: Variance-matrix consistency}

\begin{lemma}\label{lemma:psiconsistency}
Suppose Assumptions~\ref{ass:consistency}--\ref{ass:localdependence},
\ref{ass:regularity}, and~\ref{ass:kernel} hold.
If $n_\phi h_n^{\bar{s}_\phi} \to \infty$ and $h_n < \kappa_{\phi}$, 
\begin{equation*}
\begin{split}
    \hat{\Psi}_{d,g,p}(h_n; \phi) =& \Psiplusbc{p}{}{h_n}  + o_p(1)\\
    \hat{\Psi}_{d',g',p}(h_n; \phi) =& \Psiminusbc{p}{}{h_n}  + o_p(1)\\
    \hat{\Psi}_{d,g,d',g',p}(h_n; \phi) =& \bar{\Psi}_{d,g,d',g',p}(h_n; \phi) +  o_p(1)
\end{split}
\end{equation*}
%

\begin{proof}
The generic element of $\hat{\Psi}_{d,g,p}(h_n; \phi)$ is 
\begin{equation}\label{eq:varitem}
    \frac{1}{n_\phi h_n^{\bar{s}_\phi}}\sum_i\sum_{j \in \bm{N}_i}
    \hat{u}_{i,d,g}\hat{u}_{j,d,g}\,
    \Iplus[i]\Iplus[j]\Kh[i]\Kh[j]\,
    \left(\Xh[i]\right)^l\left(\Xh[j]\right)^v,
    \qquad l,v=0,\dots,p.
\end{equation}
We prove the claim for $l,v=0$. The proof for the other elements is analogous. We have
\[
\hat{u}_{i,d,g} = ({\beta}_{d,g,p}(h_n; \phi)-\hat{\beta}_{d,g,p}(h_n; \phi))^Tr_p(\tilde{X}_i) + \nu_{i,d,g}, \quad  i = 1,..., n
\]
Expanding the product $\hat{u}_{i,d,g}\hat{u}_{j,d,g}$, expression \eqref{eq:varitem} becomes
\begin{align}\label{eq:expansion_uiuj}
\eqref{eq:varitem} 
&= \frac{1}{n_\phi h_n^{\bar{s}_\phi}}\sum_i\sum_{j \in \bm{N}_i}\Big[ \nonumber\\
&\quad ({\beta}_{d,g,p}(h_n; \phi) - \betaplus)^Tr_p(\tilde{X}_i)({\beta}_{d,g,p}(h_n; \phi) - \betaplus)^Tr_p(\tilde{X}_j) \tag{I}\nonumber\\
&\quad + \nu_{j,d,g}({\beta}_{d,g,p}(h_n; \phi) - \betaplus)^Tr_p(\tilde{X}_i) \tag{II}\nonumber\\
&\quad + \nu_{i,d,g}({\beta}_{d,g,p}(h_n; \phi) - \betaplus)^Tr_p(\tilde{X}_j) \tag{III}\nonumber\\
&\quad + \nu_{i,d,g} \nu_{j,d,g} \tag{IV}\Big]\Iplus[i]\Iplus[j]\Kh[i]\Kh[j].
\end{align}
\noindent\textbf{Leading term (IV).}
For (IV), we have $\E[(IV)] = e_1\V[\tplus[*]]e_1 =  e_1^T\Psiplusbc{p}{}{h_n} e_1$ by Lemma \ref{lemma:variances}. We show concentration of variance. For compactness, let 
\(
\widetilde{\nu}_{i,d,g}
:=
\nu_{i,d,g} I_{i,d,g,\phi}.
\)
Expanding the second moment:
\begin{equation}\label{eq:secondmomentsubgraph}
\begin{split}
\E\big[(IV)^2\big]
&=
\frac{1}{n_\phi^2 h_n^{2\bar{s}_\phi}}
\sum_{i\neq j}
\sum_{k\in\bm N_i}
\sum_{l\in\bm N_j}
\E\left[
\widetilde{\nu}_{i,d,g}
\widetilde{\nu}_{j,d,g}
\widetilde{\nu}_{k,d,g}
\widetilde{\nu}_{l,d,g}
\Kh\Kh[j]\Kh[k]\Kh[l]
\right]
\\
&\quad+
\frac{1}{n_\phi^2 h_n^{2\bar{s}_\phi}}
\sum_i\sum_{j\in\bm N_i}
\E\left[
\widetilde{\nu}_{i,d,g}^2
\widetilde{\nu}_{j,d,g}^2
\Kh^2\Kh[j]^2
\right]
\\
&\quad+
\frac{1}{n_\phi^2 h_n^{2\bar{s}_\phi}}
\sum_i\sum_{j\in\bm N_i}
\sum_{k\in\bm N_i\setminus\{j\}}
\E\left[
\widetilde{\nu}_{i,d,g}^2
\widetilde{\nu}_{j,d,g}
\widetilde{\nu}_{k,d,g}
\Kh^2\Kh[j]\Kh[k]
\right]
\end{split}
\end{equation}
By arithmetic-geometric mean inequality, the last two terms in \eqref{eq:secondmomentsubgraph} are bounded by
\[
\frac{1}{n_\phi^2 h_n^{2\bar{s}_\phi}}\,
   \max_i \E\left[\tilde{\nu}^4_{i,d,g}\Kh^4\right]\,
   \sum_i \big(|\bm{N}_{i,n}| + |\bm{N}_{i,n}|^2\big),
\]
which is $O((n_\phi h_n^{\bar{s}_\phi})^{-1})$ by Assumption \ref{ass:localdependence} and since $\E\left[
\widetilde{\nu}_{i,d,g}^{\,4}\Kh^4
\right]
= O(h_n^{\bar{s}_\phi})$ by Assumptions~\ref{ass:fourthmoment}, \ref{ass:cont},
\ref{ass:densitycont}, and~\ref{ass:kernel}.
The first term is bounded by
{
\begin{equation}\label{eq:bigeqsubgraph}
\begin{split}
&\E\big[(IV)\big]^2
+
\frac{1}{n_\phi^2h_n^{2\bar{s}_\phi}}
\sum_i
\sum_{j\in\bm{N}_i}
\sum_{k\in\bm{N}_i\cup\bm{N}_j}
\sum_{l\in\bm{N}_i\cup\bm{N}_j\cup\bm{N}_k}
\Bigg\{
\\
&\qquad
\E\left[
\widetilde{\nu}_{i,d,g}
\widetilde{\nu}_{j,d,g}
\widetilde{\nu}_{k,d,g}
\widetilde{\nu}_{l,d,g}
\Kh\Kh[j]\Kh[k]\Kh[l]
\right]
\\
&\qquad
-
\E\left[
\widetilde{\nu}_{i,d,g}\widetilde{\nu}_{k,d,g}
\Kh\Kh[k]
\right]
\E\left[
\widetilde{\nu}_{j,d,g}\widetilde{\nu}_{l,d,g}
\Kh[j]\Kh[l]
\right]
\Bigg\}\\
& \leq \E\big[(IV)\big]^2
+
\frac{1}{n_\phi^2h_n^{2\bar{s}_\phi}}
\max_i
\E\left[
\widetilde{\nu}_{i,d,g}^{\,4}\Kh^4
\right]
\sum_i\sum_{j\in\bm{N}_i}
\sum_{k\in\bm{N}_i\cup\bm{N}_j}
\left(
|\bm{N}_i\cup\bm{N}_j|
+
|\bm{N}_k|
\right)
\\
&\leq
\E\big[(IV)\big]^2
+
\frac{1}{n_\phi h_n^{2\bar{s}_\phi}}
\max_i
\E\left[
\widetilde{\nu}_{i,d,g}^{\,4}\Kh^4
\right]
\frac{
\sum_i|\bm{N}_i|^3
+
3\sum_{i,j}(\bm{W}_n^3)_{ij}
}{n_\phi}.
\end{split}
\end{equation}
}
where the first inequality follow by the arithmetic-geometric mean inequality, and the second follows from the dependency-graph counting argument in equation~(35) of \citet{leung20}.
The second summand is
$O((n_\phi h_n^{\bar{s}_\phi})^{-1})$, again by Assumption \ref{ass:localdependence}, ~\ref{ass:fourthmoment}, \ref{ass:cont},
\ref{ass:densitycont}, and~\ref{ass:kernel}.
Hence, since
\(
\V[(IV)]
=
\E[(IV)^2]-\E[(IV)]^2,
\)
we obtain
\[
\begin{split}
\V[(IV)]
&\leq
\frac{1}{n_\phi^2h_n^{2\bar{s}_\phi}}
\max_i
\E\left[
\widetilde{\nu}_{i,d,g}^{\,4}\Kh^4
\right]
\Bigg(
\sum_i\left(
|\bm{N}_{i,n}|+|\bm{N}_{i,n}|^2
\right)
+
\sum_i|\bm{N}_{i,n}|^3
+
3\sum_{i,j}(\bm{W}_n^3)_{ij}
\Bigg)
\\
&=
O\left(
(n_\phi h_n^{\bar{s}_\phi})^{-1}
\right)
\to0.
\end{split}
\]
Hence,
\[
(IV)
=
e_1^\top\Psiplusbc{p}{}{h_n}e_1
+o_p(1).
\]

\noindent\textbf{Remaining terms (I)--(III).}
Define
\[
M(h_n) =
\frac{1}{n_\phi h_n^{\bar{s}_\phi}}
\sum_i\sum_{j\in\bm N_i}
\rr \rr[j]^T
\Iplus[i]\Iplus[j]\Kh[i]\Kh[j].
\]
Term (I) is equal to
\[
(I) =
\big[H^{-1}(h_n)
(\beta_{d,g,p}(h_n;\phi)-\betaplus)\big]^T
M(h_n)
\big[H^{-1}(h_n)
(\beta_{d,g,p}(h_n;\phi)-\betaplus)\big].
\]
By the arithmetic--geometric mean inequality for some $C >0$ 
\[
\E[\|M(h_n)\|]
\leq
\frac{C}{n_\phi h_n^{\bar s_\phi}}
\sum_i |\bm N_i|\,
\E\!\left[\Iplus[i]\Kh[i]^2\right]
=O(1),
\]
where $\E[\Iplus[i]\Kh[i]^2]=O(h_n^{\bar s_\phi})$ by
Assumptions~\ref{ass:pos}, \ref{ass:densitycont}, and
\ref{ass:kernel}, and $\tfrac{\sum_i |\bm N_i|}{n_\phi} = O(1)$ by Assumption~\ref{ass:localdependence}. Hence, by Markov's inequality,
$M(h_n)=O_p(1)$.

Furthermore, $H^{-1}(h_n) ({\beta}_{d,g,p}(h_n; \phi) - \betaplus) = o_p(1)$ by Lemma \ref{lemma:convergencematricesgamma} and \ref{lemma:convergencematricest} as $n_\phi h_n^{\bar{s}\phi} \to \infty$. Thus, $(\text{I}) = o_p(1)$.
Convergence in probability to zero of $(II)$ and $(III)$ can be proved similarly by factorizing $H^{-1}(h_n) ({\beta}{d,g,p}(h_n; \phi) - \betaplus)$.




The proofs for $\hat{\Psi}_{d',g',p}(h_n; \phi)$ and $\hat{\Psi}_{d,g,d',g',p}(h_n; \phi)$ are analogous.
\end{proof}
\end{lemma}

\phantomsection
\addcontentsline{toc}{subsubsection}
  {Theorem~\ref{th:variance}: Variance estimator consistency}
\begin{theorem}\label{th:variance}
Suppose Assumptions~\ref{ass:consistency}--\ref{ass:localdependence},
\ref{ass:regularity}, 
and~\ref{ass:kernel} hold.
If $n_\phi h_n^{\bar{s}_\phi} \to \infty$ and $h_n \to 0$ 
\begin{equation*}
    n_\phi h_n^{\bar{s}_\phi}\hat{\mathcal{V}}^\phi_{d,g,d',g',p}(h_n) \overset{p}{\to} V^\phi_{d,g,d',g',p}(h_n) 
\end{equation*}
where $V^\phi_{d,g,d',g',p}(h_n) =  V^\phi_{d,g,d',g',p} + o(1) $ if $h_n \to 0$.
\end{theorem}

\begin{proof}
The proof follows from Lemma \ref{lemma:convergencematricesgamma} and \ref{lemma:psiconsistency}.  
\end{proof}

\subsection{Differentiability of $\muplus$ and $\muminus$}\label{appendix:differentiability}
The bias properties of $\estimatorapp$ hinge on Assumption~\ref{ass:smoothnessmain}, which imposes smoothness of $\muplus$ and $\muminus$. Continuity of $m_{d,g}(\cdot; \phi)$ and $f(\cdot; \phi)$ is sufficient for consistency of the estimator (Lemma~\ref{lemma:consistency}). However, obtaining the quadratic rate of bias decay, $h_n^{2+2p}$, requires differentiability of $\muplus$ and $\muminus$.
We verify this assumption starting from differentiability of outcome regression functions and density.

\begin{assumption}\label{ass:primitivedifferentiability}

$m_{d,g}(x_i, \bm x_{\smallN_i}; \phi)$, $m_{d',g'}(x_i, \bm x_{\smallN_i}; \phi)$ and $\densityphi$ are $S \geq p + 2$ continuously differentiable.
\end{assumption}
\noindent Under Assumption~\ref{ass:primitivedifferentiability},
Assumption~\ref{ass:smoothnessmain} follows. The argument uses the
same parametrization of the boundary pieces as in
Theorem~\ref{ass:identification}. Let $\partial\eballboundright$
denote the set of points in $\eballboundright$ at distance
$\varepsilon$ from $\Frontierphi$. Then
\[
\muplus[\epsilon]
=
\frac{
\int_{\partial\eballboundright}
m_{d,g}(x_i,\bm x_{\smallN_i};\phi)\densityphi
\,d\mathcal H^{S_\phi}
}{
\int_{\partial\eballboundright}
\densityphi
\,d\mathcal H^{S_\phi}
}.
\]

Recenter the cutoff at zero, and recall that $\Frontierphi$ is a finite
union of the coordinate-aligned pieces $\curve$. Also recall that
$\cindex$ and $\findex$ denote, respectively, the coordinates fixed at
the cutoff and the free coordinates on $\curve$, with
$s_{\curve}=|\cindex|$, and $\freeregion$ is the rectangular domain of
$\bm x_{\findex}$.
Consider a generic $S$-times continuously differentiable function $h$.
Define the $\varepsilon$-shell of $\curve$ as
$\partial\mathcal{B}_\epsilon^{d,g}(\curve)$, the set of points at
distance $\varepsilon$ from $\curve$. 
A point
on the $\varepsilon$-shell of $\curve$ belongs to
$\partial\eballboundright$ only if no other boundary piece is at
distance smaller than $\varepsilon$. Hence, by finite
inclusion--exclusion,
\begin{equation}\label{eq:fullshell}
\int_{\partial\eballboundright} h
=
\sum_{\curve}
\sum_{A\subseteq\{\curve':\curve'\neq\curve\}}
(-1)^{|A|}Q_{\curve,A,h}(\varepsilon),
\end{equation}
where $Q_{\curve,A,h}(\varepsilon)$ is the integral
\begin{equation}\label{eq:overlapshell}
\begin{split}
Q_{\curve,A,h}(\varepsilon)
={}&
\varepsilon^{s_{\curve,A}-1}
\int_{\mathcal R_{\curve,A}}
\int_{D_{\curve,A}}
h(\varepsilon\bm u,\bm x_{R_{\curve,A}})
\,d\mathcal H^{s_{\curve,A}-1}(\bm u)
\,d\bm x_{R_{\curve,A}}.
\end{split}
\end{equation}
The domains of integration are defined as follows. For a given pair
$(\curve,A)$,
\(
F_{\curve,A}
:=
\cindex\cup
\bigcup_{\curve'\in A}F_{\curve'},
\)
$s_{\curve,A}:=|F_{\curve,A}|$, and $R_{\curve,A}$ denotes the
remaining coordinates. The domain $D_{\curve,A}$ is
\(
D_{\curve,A}
:=
\left\{
\bm u:
\|\bm u_{\cindex}\|=1,\;
\|\bm u_{F_{\curve'}}\|<1
\ \text{for all }\curve'\in A
\right\},
\)
with the orthant restrictions understood and with weak inequalities
used whenever required by the tie-breaking rule.
Both $D_{\curve,A}$ and $\mathcal R_{\curve,A}$ are independent of
$\varepsilon$. Therefore, the integral in 
$Q_{\curve,A,h}(\varepsilon)$ is $S$-times continuously
differentiable in $\varepsilon$ by differentiation under the integral sign and
continuous differentiability of $h$

Since $s_{\curve,A}\geq s_{\curve}\geq\bar s_\phi$, define
\[
A_h(\varepsilon)
:=
\sum_{\curve}
\sum_{A\subseteq\{\curve':\curve'\neq\curve\}}
(-1)^{|A|}
\varepsilon^{s_{\curve,A}-\bar s_\phi}
\int_{\mathcal R_{\curve,A}}
\int_{D_{\curve,A}}
h(\varepsilon\bm u,\bm x_{R_{\curve,A}})
\,d\mathcal H^{s_{\curve,A}-1}(\bm u)
\,d\bm x_{R_{\curve,A}}.
\]
Then,
\begin{equation}\label{eq:shellfactor}
\int_{\partial\eballboundright} h
=
\varepsilon^{\bar s_\phi-1}A_h(\varepsilon).
\end{equation}
Since $s_{\curve,A}-\bar s_\phi$ is a nonnegative integer and the
domains of integration are fixed, $A_h(\varepsilon)$ is $S$-times
continuously differentiable in $\varepsilon$ and at $\varepsilon = 0$.
Applying this result to
$h=m_{d,g}(\cdot;\phi)f(\cdot;\phi)$ and to
$h=f(\cdot;\phi)$ gives
\begin{equation}
\muplus[\varepsilon]
=
\frac{A_{m_{d,g}f}(\varepsilon)}
{A_f(\varepsilon)}.
\end{equation}
Since $A_f(0)>0$, $\muplus[\varepsilon]$ is $S$-times continuously
differentiable on a right-neighborhood of zero, including
at $\varepsilon=0$. The argument for $\muminus$ is identical.

\section{Bias Correction }\label{appendix3:biascorrection}
\subsection{Notation and Bias-corrected Expansion}
We establish the asymptotic normality of the bias-corrected estimator $\estimatorbcapp$ of $\boundeff$ 
\begin{equation*}
    \begin{split}
        \estimatorbcapp =& \estimatorapp - \biasbcapp, \\
        \biasbcapp = &h_n^{p+1}e_{p+1}^T\big[\betaminusbc \hat{B}_{d',g',p,p+1}(h_n; \phi) -\betaplusbc\hat{B}_{d,g,p,p+1}(h_n; \phi)\big],
    \end{split}
\end{equation*}
where $\betaplusbc$ and $\betaminusbc$ are the $(p+1)$-order local polynomial estimators with bandwidth $b_n$, $e_{p+1}$ is the $p+2$ -dimensional vector with the $p+2$-th entry equal to 1 and all other zeros, and $\hat{B}_{d,g,l,q}(h_n) = e_1^T{\Gamma}^{-1}_{d,g, l}(h_n; \phi)\theta_{d,g,l, q}(h_n)$ for generic $(d,g)$ and integers $l,q$.
(see Subsection \ref{subsec:notation} for the related definitions).

Since most passages are similar to the derivations for the asymptotic normality of $\estimator$ without bias correction, we only sketch the proof. We consider the centered estimators. 
\[
\estimatorbcapp[*]= \estimatorapp[*] - \biasbcapp[*] 
\]
where $\estimatorapp[*]$ is defined in Subsection \ref{subsec:notation}, and $\biasbcapp[*]$ uses the centered versions of 
$\betaplusbc$ and $\betaminusbc$, obtained by replacing $Y_i$ in their expressions with 
\[
\nu_{i, \bullet, p+1}(b_n)
= Y_i - \mathbf r_{p+1}(\tilde X_i)^T\,\beta_{\bullet,p+1}(b_n),\quad \bullet \in \{(d,g), (d'g')\}
\] 
with $\beta_{\bullet,p+1}(b) = \argmin\limits_{\bm{b} \in \mathbbm{R}^{p+2}}\E\Big[\sum_iI_{i, \bullet, \phi} (Y_i - \bm r_{p+1}(\distvari)^T\bm{b})^2\cdot K_{b}(\distvari)\Big]$.

Under Assumption \ref{ass:smoothness}, if $\max\{h_n,b_n\}<\kappa$, a Taylor expansion yields
\begin{equation}\label{eq:decompositiontaubc}
    \begin{split}
       \estimatorbcapp = & \boundeff + h_n^{p+1}\bias + T_{d,g,d'g'}(h_n; \phi)\\
       - & h_n^{p+1}\biasfinite{d,g,d',g'}{p}{p+1}{h_n; \phi} -  h_n^{p+1}b_n\bigbiasbc + T^{bc}_{d,g,d',g'}(h_n, b_n; \phi) \\
       + & \estimatorbcapp[*]
    \end{split}
\end{equation}
\sloppy
where $\bias$ is defined in Subsection \ref{sec:asyquantities}, 
$\biasfinite{d,g,d',g'}{p}{p+1}{h_n} = \tfrac{\muplusderiv[(p+1)]}{(p+1)!}\hat{B}_{d,g,p,p+1}(h_n)- \tfrac{\muminusderiv[(p+1)]}{(p+1)!}\hat{B}_{d',g',p,p+1}(h_n)$, and $\bigbiasbc$ is equal to 
\begin{equation*}
    \frac{\muplusderiv[(p+2)]}{(p+2)!}\biasfinite{d,g}{p+1}{p+2}{b_n}\frac{\biasplus}{(p+1)!} - \frac{\muminusderiv[(p+2)]}{(p+2)!}\biasfinite{d',g'}{p+1}{p+2}{b_n}\frac{\biasminus}{(p+1)!}
\end{equation*}
%

\medskip
We denote the variance and covariance of $t_{d,g,p+1}^*(b_n; \phi)$ and $t_{d',g',p+1}^*(b_n; \phi)$, multiplied by $nb_n^{\bar{s}_\phi}$, by $\Psiplusbc{p+1}{}{b_n}$, $\Psiminusbc{p+1}{}{b_n}$ and $\Psiplusminusbc{p+1}{}{b_n}$, and we let for generic $(d,g)$ 
\begin{equation*}
\begin{split}
    \Psiplusbc{p,}{p+1}{h_n,b_n} & = \frac{n_\phi(h_nb_n)^{\bar{s}_\phi}}{m_n^{\bar{s}_\phi}}\C[t_{d,g,p}^*(h_n; \phi), t_{d,g,p+1}^*(b_n; \phi)]     
\end{split}
\end{equation*}
\sloppy
and $\Psiplusminusbc{p,}{p+1}{h_n,b_n} = \tfrac{n_\phi(h_nb_n)^{\bar{s}_\phi}}{m_n^{\bar{s}_\phi}}\C[t_{d,g,p}^*(h_n; \phi), t_{d',g',p+1}^*(b_n; \phi)]$
where $m_n$ is the minimum between $h_n$ and $b_n$. Further define $\variancebc = \variancebcplus + \variancebcminus - 2\covarbc$ where for generic $(d,g)$
\begin{equation*}
\begin{split}
    \variancebcplus
    & =  \frac{1}{n_\phi h_n^{\bar{s}_\phi}}e_1^T\Gammaplusbc{,p}{h_n}{-1}\Psiplus\Gammaplusbc{,p}{h_n}{-1}e_1 \\
    &+ \frac{1}{nb_n^{\bar{s}_\phi}}\frac{h_n^{2p + 2}}{b_n^{2+2p}}e_{p+1}^T\Gammaplusbc{,p+1}{b_n}{-1}\Psiplusbc{p+1}{}{b_n}\Gammaplusbc{,p+1}{b_n}{-1}e_{p+1}) \\
    &\times(e_1^T\Gammaplusbc{,p+1}{b_n}{-1}\thetaplusbc{p}{p+1}{h_n})^2 \\
    &-2\frac{m_n^{\bar{s}_\phi}}{n_\phi(h_nb_n)^{\bar{s}_\phi}}\frac{h_n^{p+1}}{b_n^{p+1}}(e_1^T\Gammaplusbc{,p}{h_n}{-1}\Psiplusbc{p}{,p+1}{h_n, b_n}\Gammaplusbc{,p+1}{b_n}{-1}e_{p+1}) \\
    &\times(e_1^T\Gammaplusbc{,p+1}{h_n}{-1}\thetaplusbc{p}{p+1}{h_n})
\end{split}    
\end{equation*}
The first two terms in $\variancebc$ represent the asymptotic variance, for $n \min\{h_n^{\bar s}, b_n^{\bar s}\} \to \infty$ of the (centered) bias-corrected estimator to the right and left of the cutoff respectively, which includes the variance of the original centered estimator, the variance of the centered bias correction, and their covariance.
\noindent$\covarbc$ is given by 
\begin{equation*}
    \begin{split}
        \covarbc &= \covarest{p}{h_n; \phi} + h_n^{2(p+1)}\covarest{p+1}{b_n; \phi}\biasplusbc\biasminusbc \\
    & - h_n^{p+1}\covarest{p+1, p,p+1}{h_n,b_n; \phi}B_{d,g,p,p+1}(b_n; \phi) \\
    & - h_n^{p+1}\covarest{p+1,p,p+1}{b_n,h_n; \phi}\biasplusbc 
    \end{split}
\end{equation*}
where 
\begin{equation*}
    \begin{split}
         \covarest{p}{h_n; \phi} &=  \frac{1}{n_\phi h_n^{\bar{s}_\phi}}e_1^T\Gammaplusbc{,p}{h_n}{-1}\Psiplusminusbc{p}{}{h_n}\Gammaminusbc{,p}{h_n}{-1}e_1 \\
         \covarest{p+1,p,p+1}{h_n,b_n; \phi} & =  \frac{m_n^{\bar{s}_\phi}}{n_\phi(h_nb_n)^{\bar{s}_\phi}}\frac{1}{b_n^{p+1}}(p+1)!e_1^T\Gammaplusbc{,p}{h_n}{-1}\Psiplusminusbc{p}{,p+1}{h_n,b_n} \Gammaplusbc{,p+1}{b_n}{-1}e_{p+1}\\
         \covarest{p+1,p,p+1}{b_n,h_n; \phi} & =  \frac{m_n^{\bar{s}}}{n_\phi(h_nb_n)^{\bar{s}_\phi}}\frac{1}{b_n^{p+1}}(p+1)!e_{p+1}^T\Gammaplusbc{,p+1}{b_n}{-1}\Psiplusminusbc{p}{,p+1}{b_n, h_n} \Gammaminusbc{,p}{h_n}{-1}e_{1}
    \end{split}
\end{equation*}

\medskip
\noindent By similar arguments as Lemma~\ref{lemma:variances}, for $n\min\{h_n^{\bar{s}},b_n^{\bar{s}}\}\to \infty$
\[
\Psiplus,\Psiminus = O\!\left(\frac{1}{n_\phi h_n^{\bar{s}_\phi}}\right), 
\qquad 
\Psiplus[b_n],\Psiminus[b_n] = O\!\left(\frac{1}{n_\phi b_n^{\bar{s}_\phi}}\right).
\]
and the terms $\Psiplusbc{p}{,q}{h_n,b_n},\;\Psiminusbc{p}{,q}{h_n,b_n}$ and $\Psiplusminusbc{p}{,q}{h_n,b_n}$ are $O\!\left(\frac{m_n^{\bar{s}_\phi}}{n_\phi(h_nb_n)^{\bar{s}_\phi}}\right),
$

\subsection{Asymptotic Normality — Theorem \ref{th:asybiascorretiontau}}
\begin{theorem}\label{th:asybiascorretiontau}
Suppose Assumptions~\ref{ass:consistency}--\ref{ass:localdependence},
\ref{ass:regularity}, and~\ref{ass:kernel} hold. If $n_\phi\min\{h_n^{\bar{s}_\phi},b_n^{\bar{s}_\phi}\}\to \infty$, $n_\phi\min\{h_n^{2p+2+\bar{s}_\phi},b_n^{2p+2+\bar{s}_\phi}\}\max\{h_n^2,b_n^{2}\} \to 0$:
\begin{equation*}
    \frac{\estimatorbcapp - \boundeff}{\sqrt{\variancebc}}\xrightarrow{d} \mathcal{N}(0, 1)
\end{equation*}
\end{theorem}

\begin{proof}
Consider $\Zbc = \Zplusbc - \Zminusbc$, where $\Zplus = \Zplusbcone -\Zplusbctwo$ and for generic $(d,g)$
\begin{equation*}
\begin{split}
     \Zplusbcone =&  e_1^T\Gammaplusbc{,p}{h_n}{-1}\Iplus\Kh\rr\nu_{i,d,g,p}(h_n)/(nh_n^{\bar{s}}), \\
     \Zplusbctwo = & \frac{h_n^{p+1}b_n^{-p-1}}{n_\phi b_n^{\bar{s}_\phi }}\biasplus 
    \times e_{p+1}^T\bar{\Gamma}_{d,g,p+1}^{-1}(b_n)\Iplus K\Big(\frac{\tilde{X}_i}{b_n}\Big)r_{p+1}\Big(\frac{\tilde{X}_i}{b_n}\Big)\nu_{i,d,g,p+1}(b_n)
\end{split}  
\end{equation*}
Here, $\nu_{i,d,g, r}(h) = Y_i - \bm r_r(\distvari)\beta_{d,g,r}(h; \phi)$, is the residual obtained from the outcome linear approximation using the population linear projection coefficient $\beta_{\bullet,r}(h; \phi)$. Then
\[
\V\!\Big[\sum_i \Zbc\Big] \;=\; \variancebc
= O_p\!\Big(\frac{1}{n_\phi h_n^{\bar{s}_\phi}}\Big)
+ O_p\!\Big(\frac{h_n^{2(p+1)}}{n_\phi b_n^{2(p+1)+\bar{s}_\phi}}\Big).
\]
Let $\xi_n = \sum_i \Zbc / \sigma_n$ with $\sigma_n^2 = \variancebc$.  
$\{\Zbc\}_{i=1}^n$ has dependency graph $\bm W_n$.  
Stein’s method and \citet[Theorem~A.4]{leung20} then yield
\begin{equation}\label{eq:distbc}
    \begin{split}
        \Delta({\xi}_{n}, \mathcal{N}(0,1)) & \leq \frac{1}{\sigma_{n}^3}\E[|\Zbc|^3]\sum_i|\bm{N}_{i,n}|^2 \\
  & + \frac{1}{\sigma_{n}^2}\Big(\frac{4}{\pi}\E[\Zbc^4]\Big)^{1/2}\Big(4\sum_i|\bm{N}_{i,n}|^3 + 3\sum_{i,j}(\bm{W}_n^3)_{ij}\Big)^{1/2}
    \end{split}
\end{equation}
\medskip
By the $c_r$-inequality,
\[
\E[|\Zbc|^3] \;\leq\; 4\big(\E[|\Zplusbc|^3]+\E[|\Zminusbc|^3]\big).
\]
where by decomposing $\Zplusbc$ 
\[
\E[|\Zplusbc|^3]
\;\leq\; 4\big(\E[|\Zplusbcone|^3]+\E[|\Zplusbctwo|^3]\big)
= O\!\Big(\frac{1}{n_\phi^3 h_n^{2\bar{s}_\phi}}\Big)
+ O\!\Big(\frac{h_n^{3(p+1)}}{n_\phi^3 b_n^{3(p+1)+2\bar{s}_\phi}}\Big),
\]
and $\E[|\Zminusbc|^3]$ has a bound of the same order.  
Similarly,
\[
\E[\Zbc^4]
= O\!\Big(\frac{1}{n_\phi^4 h_n^{3\bar{s}_\phi}}\Big)
+ O\!\Big(\frac{h_n^{4(p+1)}}{n_\phi^4 b_n^{4(p+1)+3\bar{s}_\phi}}\Big).
\]
Let $\rho_n=h_n/b_n$. By Assumption \ref{ass:localdependence}, the first and second terms in \eqref{eq:distbc} are
\[
O\!\Big((n_\phi h_n^{\bar{s}})^{-1/2}\min\{1,\rho_n^{-3(p+1)-\tfrac{3}{2}\bar{s}}\}\Big)
+ O\!\Big((nb_n^{\bar{s}})^{-1/2}\min\{1,\rho_n^{\,3(p+1)+\tfrac{3}{2}\bar{s}}\}\Big)
= o(1),
\]
and
\[
O_p\!\Big((n_\phi h_n^{\bar{s}_\phi})^{-1/2}\min\{1,\rho_n^{-2(p+1)-\bar{s}_\phi}\}\Big)
+ O_p\!\Big((n_\phi b_n^{\bar{s}_\phi})^{-1/2}\min\{1,\rho_n^{\,2(p+1)+\bar{s}_\phi}\}\Big)
= o(1).
\]
\sloppy
Hence $\xi_n \xrightarrow{d} \mathcal N(0,1)$.  
The condition $n_\phi \min\{h_n^{\bar{s}_\phi}, b_n^{\bar{s}_\phi}\} \to \infty$ and Lemmas~\ref{lemma:convergencematricesgamma} and ~\ref{lemma:convergencematricestheta} gives
\(\tfrac{\estimatorbcapp[*]}{\sqrt{\variancebc}} = \xi_n + o_p(1)\),
By $n_\phi \min\{h_n^{\bar{s}_\phi}, b_n^{\bar{s}_\phi}\} \to \infty$ and Lemmas~\ref{lemma:convergencematricesgamma},  \ref{lemma:convergencematricestheta} we have $T_{d,g,d'g'}(h_n; \phi) = O(h_n^{p+2})$, $h_n^{p+1}(\bias - \biasfinite{d,g,d',g'}{p}{p+1}{h_n; \phi}) = o_p(h_n^{p+1})$, $h_n^{p+1}b_n\bigbiasbc = O_p(h_n^{p+1}b_n)$ and $T^{bc}_{d,g,d',g'}(h_n, b_n; \phi) = o_p(h_n^{p+1}b_n)$. The conclusion follows from decomposition~\eqref{eq:decompositiontaubc} and $n_\phi\min\{h_n^{2p+2+\bar{s}_\phi},b_n^{2p+2+\bar{s}_\phi}\}\max\{h_n^2,b_n^{2}\} \to 0$.
\end{proof}

An estimator of the asymptotic variance of $\estimatorbcapp$ can be obtained as follows by plugging the estimators
\begin{equation*}
\begin{split}
    \hat{\tilde{\Psi}}_{d,g,p,q}(h_n, b_n) = \Mplus\bm{W}_n\mathcal{M}_{d,g,q}(b_n)^T/n\\
    \hat{\tilde{\Psi}}_{d',g',p,q}(h_n, b_n) = \Mminus\bm{W}_n\mathcal{M}_{d',g',q}(b_n)^T/n\\
    \hat{\tilde{\Psi}}_{d,g,d',g',p,q}(h_n, b_n) = \mathcal{M}_{d,g,p}(h_n)\bm{W}_n\mathcal{M}_{d',g',q}(b_n)^T/n
\end{split}
\end{equation*}
and the estimators defined in \eqref{eq:varestimatorappendix} for the corresponding quantities in $\variancebc$.

\section{Pooled Estimation}\label{appendix:heterogeneous}

\subsection{Notation, Assumptions and Asymptotic Expansion}
We show asymptotic normality of the pooled estimator $\estimatorpooled$. Define the density integrated over the points that are at distance $\tilde{x}$ from the boundary for stratum $\phi$:
\begin{equation*}
\fdistance = \int_{\partial \mathcal{B}_{\tilde{x}}(\Frontier)} 
\densityphi d\mathcal{H}^{S_\phi + 1 - \bar{s}_\phi}
\end{equation*}
where $\partial\mathcal{B}_{\tilde{x}}(\Frontier) = \{\ppoint \in \scorespace : l_{d,g|d',g'}^{min}(x_i, \bm{x}_{\smallN_i}) = \tilde{x}\}$. We recall $S_\phi$ and $\bar{s}_\phi$ are the interference set size and codimension within stratum $\phi$. Since $\phi(\cdot)$ is a function of the network, we index the set $\Phi_n$ by $n$. Also index by $n$  the stratum proportion $\pi_{\phi, n} = n_\phi/n$. 

%
\begin{assumption}\label{ass:boundaryhet}
The stratum set is bounded $|\Phi_n| < \infty$, and  $\pi_{\phi,n} \to \bar{\pi}_{\phi}$ as $n \to \infty$. \end{assumption}
\vspace{-1em}
{\color{black}
\begin{assumption}\label{ass:smoothnesshet}
For some $\kappa > 0$ and for $\ppoint \in \eballbound[\kappa]$, $\fdistance$ is at least $(p + 2)$-times continuously differentiable.
\end{assumption}}
\textcolor{black}{\noindent Differently from Section~\ref{subsec:asytheory}, we will require Assumption \ref{ass:regularity} to hold for all $\phi$. Assumption \ref{ass:smoothnesshet} further requires $\fdistance$ to be differentiable.}
Moreover, we require the number of strata $|\Phi_n|$ to be asymptotically bounded, and the stratum-specific proportions to converge to a specific limit. Throughout we let $\bar{s}= \min_{\phi \in \Phi_n}\bar{s}_\phi$ be the minimal codimension across strata.
The local polynomial pooled estimator of order $p$ is given by 
\begin{align*}
    &\hat{\tau}_{d,g|d',g',p}(h_n) = \hat{\beta}_{d,g,p}(h_n) - \hat{\beta}_{d',g',p}(h_n) \\  
    &\hat{\beta}_{d,g,p}(h_n) =  \argmin\limits_{\bm{b} \in \mathbbm{R}^{p+1}}\sum_i\mathbbm{1}(D_i = d, G_i = g)(Y_i - \rp[\tilde{X}_{i}]^T\bm{b})^2\cdot K_{h_n^{\bar s}}(\tilde{X}_{i}) \\
    &\hat{\beta}_{d',g',p}(h_n) = \argmin\limits_{\bm{b}\in \mathbbm{R}^{p+1}}\sum_i\mathbbm{1}(D_i = d', G_i = g')(Y_i - \rp[\tilde{X}_{i}]^T\bm{b})^2\cdot K_{h_n^{\bar s}}(\tilde{X}_{i})
\end{align*}
Sample quantities are defined similar to before but now using the whole sample. In particular, we redefine the weighting matrix for generic $(d,g)$ as 
\begin{align*}
        & W_{d,g}(h)= \text{diag}(I_{1,d,g}\cdot K_{h^{\bar s}}(\distvari[1]),...,I_{n,d,g} \cdot K_{h}(\distvari[n]))
\end{align*}
with $I_{i,d,g} = \mathbbm{1}(D_i = d, G_i = g)$. We define $\Gamma_{d,g, p}(h)$, $\theta_{d,g, p}(h)$, and $t_{d,g, p }(h)$ analogously as in  Section \ref{subsec:notation}, but using $W_{d,g}(h)$ and $W_{d',g'}(h)$.
For instance, 
\[
\Gamma_{d,g, p}(h) = \tilde{X}_p(h)^T W_{d,g}(h)\tilde{X}_p(h)/n
\]
whereas previously $\Gamma_{d,g,p}(h; \phi) = \tilde{X}_p(h)^T W_{d,g}(h; \phi)\tilde{X}_p(h)/n_\phi$.
Then we have:
\begin{align*}
&\betapluspooled = H(h_n)\Gamma^{-1}_{d,g,p}(h_n)t_{d,g,p}(h_n), \quad 
\betaminuspooled = H(h_n)\Gamma^{-1}_{d',g',p}(h_n)t_{d',g',p}(h_n)
\end{align*}
and associated population linear projection coefficients:
\begin{equation*}
\beta_{d,g,p}(h_n) = H(h_n)\bar{\Gamma}^{-1}_{d,g,p}(h_n)\bar{t}_{d,g,p}(h_n), \quad
\beta_{d',g',p}(h_n) = H(h_n)\bar{\Gamma}^{-1}_{d,g,p}(h_n)\bar{t}_{d,g,p}(h_n)
\end{equation*}
where we recall the bar symbol $\bar{\cdot}$ denotes expectation of the sample matrix.
Finally, we define the following weighted averages of distance-based outcome regression functions:
\begin{equation*}
\mupluspooledbar = \frac{\sum_{\phi} \muplus \fdistance \cdot \pi_{\phi,n}}{\sum_\phi \fdistance  \cdot \pi_{\phi,n}}, \qquad \muminuspooledbar = \frac{\sum_{\phi} \muminus \fdistance \cdot \pi_{\phi,n}}{\sum_\phi \fdistance  \cdot \pi_{\phi,n}}
\end{equation*}
and the average density over the offset curve of the boundary 
\[
\densitypooledbar = \sum_{\phi} \fdistance \cdot \pi_{\phi,n}
\]
and the limit quantity $\densitypooledbar[0] = \sum_{\phi} \limdensity\cdot \bar{\pi}_{\phi}$

Under Assumption \ref{ass:smoothness} and \ref{ass:smoothnesshet} holding for all $\phi$, we decompose $\betaplus$ as follows:
\begin{equation}
        \hat{\beta}_{d,g,p}(h_n) =  \beta_{d,g,p}(h_n) + \underbrace{H(h_n)\Gammaplushet[-1]\Xp\Wplushet\bm \nu_{d,g}}_{\hat{\beta}^*_{d,g,p}(h_n)} 
\end{equation}
Now note that 
\begin{equation}
\begin{split}
 \bar{t}_{d,g,p}(h_n) =& \frac{1}{n}\sum_i \int_0^{h_n} K\left(\frac{\tilde{x}}{h_n}\right)r_p\left( \frac{\tilde{x}}{h_n}\right)\mu_{d,g}(\tilde{x}; \phi_i)f_{d,g|d',g'}(\tilde{x}; \phi_i) d\tilde{x} \\
 = & \frac{1}{n}\sum_\phi n_\phi \cdot \int_0^{h_n} K\left(\frac{\tilde{x}}{h_n}\right)r_p\left( \frac{\tilde{x}}{h_n}\right)\muplus \fdistance d\tilde{x} \\
 = & \int_0^{h_n} K\left(\frac{\tilde{x}}{h_n}\right)r_p\left( \frac{\tilde{x}}{h_n}\right) \Big(\sum_\phi \muplus \fdistance \cdot \pi_{\phi,n} \Big) d\tilde{x} \\
 = & \int_0^{h_n} K\left(\frac{\tilde{x}}{h_n}\right)r_p\left( \frac{\tilde{x}}{h_n}\right) \mupluspooledbar \densitypooledbar d\tilde{x}
 \end{split}
\end{equation}
Under Assumption \ref{ass:smoothnesshet} a $(p+1)$-order Taylor expansion of $\mupluspooledbar$ to the right of the cutoff yields
\begin{equation*}
\begin{split}
    {\beta}_{d,g,p}(h_n) = & \beta_{d,g,p}(0) + h_n^{p+1}\frac{\bar{\mu}_{d,g}^{(p+1)}(0)}{(p+1)!}H(h_n)\Gammaplusbchet{,p}{h_n}{-1}\thetaplusbchet{p}{p+1}{h_n} + T_{d,g}(h_n)    
\end{split}
\end{equation*}
where $\beta_{d,g,p}(0) = [\bar{\mu}_{d,g}(0), \;\bar{\mu}^{(1)}_{d,g}(0),\,...,\, \bar{\mu}_{d,g}^{(p)}(0)/p!]^T$, and $\bar{\mu}^{(v)}_{d,g}(0)$ denotes the $v$-order derivative of $\mupluspooledbar$. An analogous expansion holds for $\beta_{d',g',p}(h_n)$. Combining the two, we obtain
\begin{equation}\label{eq:decompositiontauhet}
    \begin{split}
        \hat{\tau}_{d,g,d',g',p}(h_n) = & \boundeffpooled + h_n^{p+1}B_{d,g,d',g', p, p+1}(h_n) + T_{d,g,d'g'}(h_n)\\
        + & \hat{\tau}^*_{d,g,d',g',p}(h_n) 
    \end{split}
\end{equation}
where $\hat{\tau}^*_{d,g,d',g',p}(h_n)$ is the centered pooled estimator and $B_{d,g,d',g', p, p+1}(h_n)  = \frac{\bar{\mu}^{(p+1)}_{d,g}(0)}{(p+1)!}B_{d,g, p, p+1}(h_n) - \frac{\bar{\mu}^{(p+1)}_{d',g'}(0)}{(p+1)!}B_{d',g', p, p+1}(h_n)$, with $B_{d,g, p, p+1}(h_n)= e_1^T\Gamma^{-1}_{d,g,p}(h_n)\theta^{-1}_{d,g,p}(h_n)$ for generic $(d,g)$, and $T_{d,g,d'g'}(h_n) = e_1^T\big(T_{d,g}(h_n) - T_{d',g'}(h_n)\big)$. 

\subsection{Consistency}

\phantomsection
\addcontentsline{toc}{subsubsection}
  {Lemma~\ref{lemma:convergencebiashet} and Theorem~\ref{lemma:consistencypooled}: common codimension}

\begin{lemma}[Sample matrices convergence]\label{lemma:convergencebiashet}
{Suppose Assuptions~\ref{ass:consistency}--\ref{ass:networkheterogeneity}, \ref{ass:localdependence}, \ref{ass:boundaryhet}, \ref{ass:kernel} hold. Let Assumption~\ref{ass:identification} and $\ref{ass:variances}$ hold for all $\phi$,} and $\bar{s}_{\phi}=\bar{s}$ for all $\phi$. If $n h_n^{\bar{s}} \to \infty$ and $h_n \to 0 $
then for all $d,g$
\begin{align*}
\Gammaplushet= \densitypooledbar[0]\cdot\Gamma_p^{\bar{s}} + o_p(1), \quad \thetaplushet= \densitypooledbar[0]\cdot\theta_{p, p+1}^{\bar{s}} + o_p(1)  
\end{align*}
and
\[
\tplushet= \left(\sum_\phi \bar{\pi}_\phi \int_{\Frontier}  m_{d,g}(x_i, \bm x_{\smallN_i}; \phi)\densityphi\right)\cdot \theta_{p,0}^{\bar{s}} + o_p(1)
\]
\end{lemma}

\begin{proof}

\noindent. 
Note that $\Gammaplushet = \sum_\phi \pi_{\phi, n}\cdot \Gammaplus$, and
by Lemma \ref{lemma:convergencematricestheta}, $
\Gammaplus = \Gammaplusbc{}{h_n}{} + o_p(1)$ and $
\Gammaplusbc{}{h_n}{} = \limdensity \Gamma_p^{\bar{s}} + o(1)$ as $n_\phi h_n \to \infty$ and $h_n \to 0$. Since $|\Phi_n| < \infty$ and $\pi_{\phi, n} \to \bar{\pi}_\phi$ by Assumption \ref{ass:boundaryhet}, $\Gammaplushet = \sum_\phi \bar{\pi}_{\phi,n} \cdot\Gammaminusbc{}{h_n}{} + o_p(1)$ which yields  
\[
\Gammaplushet = \underbrace{\big(\sum_\phi \bar{\pi}_{\phi}\cdot \limdensity\big)}_{\densitypooledbar[0]} \Gamma_p^{\bar{s}} + o_p(1)
\]
The conclusions for $\thetaplushet$ and $\tplushet$ follow analogously from Lemma \ref{lemma:convergencematricestheta} and \ref{lemma:convergencematricest}, respectively.
\end{proof}


\begin{theorem}[Consistency]\label{lemma:consistencypooled}
\textcolor{black}{Suppose Assumptions~\ref{ass:consistency}--\ref{ass:networkheterogeneity}, \ref{ass:localdependence}, \ref{ass:boundaryhet}, \ref{ass:kernel} hold. Let Assumption~\ref{ass:identification} and $\ref{ass:variances}$ hold for every $\phi$,} and $\bar{s}_{\phi}=\bar{s}$ for all $\phi$.
If $n h_n^{\bar{s}} \to \infty$ and $h_n \to 0 $
then
\[
\estimatorpooled \xrightarrow{p} \boundeffpooled
\]
\end{theorem}
\begin{proof}
The proof follows readily from Lemma~\ref{lemma:convergencebiashet}, and using the identity $e_1^T(\Gamma_p^{\bar{s}})^{-1}\theta^{\bar{s}}_{p, 0} = 1.$
\end{proof}

\phantomsection
\addcontentsline{toc}{subsubsection}
  {Theorem~\ref{lemma:convergencelowcod}: heterogeneous codimension}
\begin{theorem}[Heterogeneous codimensions]\label{lemma:convergencelowcod}
\textcolor{black}{Suppose Assumptions~\ref{ass:consistency}--\ref{ass:networkheterogeneity}, \ref{ass:localdependence}, \ref{ass:boundaryhet}, \ref{ass:kernel} hold. Let Assumption~\ref{ass:identification} and $\ref{ass:variances}$ hold for all $\phi$,} and let $\bar{s} = \min_\phi \bar{s}_\phi$.
If $n h_n^{\bar{s}} \to \infty$ and $h_n \to 0 $
then
\[
\estimatorpooled \xrightarrow{p} \frac{\sum_{\phi : \bar{s}_\phi = \bar{s}} \boundeff \, \cdot \limdensity \cdot \bar{\pi}_{\phi}}
     {\sum_{\phi : \bar{s}_\phi = \bar{s}} \limdensity \cdot \bar{\pi}_\phi}
\]
\end{theorem}
\begin{proof}
Expanding $\Gammaplushet$
\begin{equation*}
   \Gammaplushet = \sum_\phi \pi_{\phi,n} \frac{1}{h_n^{\bar{s}}}\sum_i \Iplus \Kh r_p\left(\Xh\right) 
    = \sum_\phi \pi_{\phi,n}\cdot h_n^{\bar{s}_\phi - \bar{s}}\Gammaplus 
\end{equation*}
Hence, since by Lemma \ref{lemma:convergencematricesgamma} $\Gammaplus = O_p(1)$, Lemma \ref{lemma:convergencematricesgamma} and Assumption \ref{ass:boundaryhet}, yield $\Gammaplushet = \big(\sum_{\phi : \bar{s}_\phi = \bar{s}} \bar{\pi}_{\phi}\cdot \limdensity\big)\Gamma_p^{\bar{s}} + o_p(1)$, for generic $d,g$. Similarly, using Lemma~\ref{lemma:convergencematricest}: 
\[
\tplushet = \sum_{\phi : \bar{s}_\phi = \bar{s}} \bar{\pi}_\phi \int_{\Frontier}  m_{d,g}(x_i, \bm x_{\smallN_i}; \phi)\densityphi\cdot \theta_{p,0}^{\bar{s}} + o_p(1) 
\]
Combining and using the identity $e_1^T(\Gamma_p^{\bar{s}})^{-1}\theta^{\bar{s}}_{p, 0} = 1.$ yields the result.
\end{proof}

\subsection{Variance and Asymptotic Normality}
{\color{black}
For characterizing the limiting variance we make the following assumption extending \ref{ass:covariances}.
\red{Define the conditional covariance across strata for $\phi, \phi' \in \Psi_n$ for $i \in \mathcal{V}_\phi$ and $j \in \mathcal{V}_{\phi'}$}:
\[
\kappa_{ij}^{a,b}(\bm x_{ij}; \phi, \phi') = \C[Y_i(a), Y_j(b)\, | \bm X_{ij} = \bm x_{ij}, \phi, \phi']
\]
with corresponding joint density $f_{ij}(\cdot; \phi, \phi')$, recalling $\bm X_{ij} = \bm X_{\tilde{S}_i \cup \tilde S_j}.$ If $\phi = \phi'$ notation reduces to one index $\phi$. We also define the boundary intersection for dependent units across strata:
\[
\mathcal X_{ij}^{\phi,\phi'}(d,g\mid d',g') = \{\bm x_{\tilde{S}_i \cup \tilde S_j} : \bm x_{\tilde{S}_i} \in \Frontierphi, \, \bm x_{\tilde S_j} \in \mathcal{\bar X}_{j\smallN_j}^{\phi'}(d,g\,|\,d',g')\}
\]
and the set 
\(
\mathcal K_i^{\phi,\phi'}
:=
\left\{
j\in(\bm N_i\setminus\{i\})\cap\mathcal V_{\phi'}:
\operatorname{codim}\!\left(
\mathcal X_{ij}^{\phi,\phi'}(d,g\mid d',g')
\right)=\bar s
\right\},
\). 
}

{\color{black}
\begin{assumption}[Pooled covariance regularity]
\label{ass:pooledcovariance}
For each $\phi,\phi'\in\Phi_n$, for each $q\in\mathcal Q_{\phi, \phi'}$, with $Q_{\phi, \phi'} = \{\max\{S_\phi, S_{\phi'}\}+1, \ldots, S_\phi + S_{\phi'} + 2\}$, the collections
\(\{\kappa_{ij}^{a,b}(\cdot;\phi, \phi')\}_{ij\in\mathcal V_\phi, j\in\mathcal V_{\phi'}:\,q_{ij}=q}\)
 and
\(\{f_{ij}(\cdot;\phi)\}_{ij\in\mathcal V_\phi, j\in\mathcal V_{\phi'}:\,q_{ij}=q}\)
are uniformly bounded and equicontinuous, for
$a,b\in\{(d,g),(d',g')\}$. Moreover, 
\[
\gamma_{a,b,n}(0;\phi,\phi')
:=
\frac{1}{n_\phi}
\sum_{i\in\mathcal V_\phi}
\sum_{j\in\mathcal K_i^{\phi,\phi'}}
\gamma_{a,b, ij}(x_{ij}; \phi, \phi')f_{ij}(x_ij; \phi, \phi')
\,d\mathcal H^{q_{ij}-\bar s}(x)
\to
\gamma_{a,b}(0;\phi,\phi'),
\]
with $ \gamma_{a,b, ij}(x_{ij}; \phi, \phi') = \kappa_{{ij}}^{a,b}(x_{ij};\phi,\phi') + \eta_a(x_i, \bm x_{\smallN_i};\phi) \eta_b(x_j, \bm x_{\smallN_j};\phi') $
\end{assumption}
}

\begin{lemma}[Variance of $t_{d,g,d',g'}^*(h_n)$]\label{lemma:varianceshe}
\textcolor{black}{Suppose Assumptions~\ref{ass:consistency}--\ref{ass:networkheterogeneity}, \ref{ass:localdependence}, \ref{ass:boundaryhet}, \ref{ass:pooledcovariance} \ref{ass:kernel} hold. Further let Assumption~\ref{ass:identification} and \ref{ass:variances} hold for all $\phi$,} and $\bar{s}_{\phi}=\bar{s}$ for all $\phi$.
If $n h_n^{\bar{s}} \to \infty$ and $h_n \to 0 $
\[
n h_n^{\bar{s}}\V[t_{d,g,d',g'}^*(h_n)]
= (\omegaplushet+\omegaminushet + \bar \gamma_{d,g, d',g'}(0))\Psi_p^{\bar{s}} + o(1).
\]
where $\bar{\omega}_{d,g}(0) = \sum_\phi \bar \omega_{d,g}(0; \phi)\cdot \bar{\pi}_\phi$ and $\bar \gamma_{d,g, d',g'}(0) = \sum_{\phi, \phi'} \bar \gamma_{d,g, d',g'}(0; \phi, \phi')\cdot \bar{\pi}_\phi $, with $\bar \gamma_{d,g, d',g'}(0; \phi, \phi') = \gamma_{d,g, d',g'}(0; \phi, \phi') + \gamma_{d,g, d',g'}(0; \phi, \phi') - 2\gamma_{d,g, d',g'}(0; \phi, \phi')$
\end{lemma}

\begin{proof}
We start from
\[
\V[t_{d,g,d',g'}^*(h_n)]
= \V[t_{d,g,}^*(h_n)] + \V[t_{d',g'}^*(h_n)] - 2\C[t_{d,g}^*(h_n), t_{d',g'}^*(h_n)].
\]
We establish the result for $\V[\tplus[*]]$; the arguments for $\V[\tminus[*]]$ and $\C[\tplus[*], \tminus[*])$ are analogous.

\medskip
\noindent\textbf{Decomposition of the variance} We expand the variance as in Lemma \ref{lemma:variances}.
\begin{equation*}
\begin{split}
n h_n^{\bar{s}} \V[\tplushet[*]]
&= \underbrace{\frac{1}{n h_n^{\bar{s}}}\sum_i
\E\!\big[I_{i,d,g} \Kh^2\, \rr\rr^T\, \nu_{i,d,g}^2\big]}_{T_1}\\
&\quad+ \underbrace{\frac{1}{n h_n^{\bar{s}}}\sum_i \sum_{j\in\bm N_i\setminus\{i\}}
\E\!\big[I_{i,d,g} I_{j,d,g}\, \Kh \Kh[j]\, \rr\, r_p\left(\Xh[j]\right)^T\, \nu_{i,d,g} \nu_{j,d,'g'}\big]}_{T_2}.
\end{split}
\end{equation*}

\medskip
\noindent\textbf{Limit of $T_1$.}
The terms $T_1$ can be written as:
\[
T_1 = \sum_\phi \pi_{\phi,n} \underbrace{\frac{1}{n_\phi h_n^{\bar{s}}}
\sum_i \E\!\big[\Iplus \Kh^2\, \rr\rr^T\, \nu_{i,d,g}^2\big]}_{I'_\phi}
\]
From the proof of Lemma \ref{lemma:variances} $I'_\phi \to \omegaplus \Psi_{p}^{\bar{s}}$ for all $\phi$. Hence, since $|\Phi_n| < \infty$
$
T_1 \to \big(\sum_\phi \bar{\pi}_\phi \cdot \omegaplus\big)\Psi_{p}^{\bar{s}} = \omegaplushet \Psi_{p}^{\bar{s}}.
$

\medskip
\noindent\textbf{Limit of $T_2$.}
The terms $T_2$ can be written as:
\[
T_2 = \sum_{\phi, \phi'} \pi_{\phi,n} \underbrace{\frac{1}{n_\phi h_n^{\bar{s}}}\sum_i \sum_{j\in\bm N_i\setminus\{i\}}
\E\!\big[I_{i,d,g, \phi} I_{j,d,g, \phi'}\, \Kh \Kh[j]\, \rr\, r_p\left(\Xh[j]\right)^T\, \nu_{i,d,g} \nu_{j,d,'g'}\big]}_{I^{''}_{\phi, \phi'}}
\]
Analogously to the proof of Lemma \ref{lemma:variances} $I^{''}_{\phi, \phi'} \to \gamma_{d,g}(0; \phi, \phi') \Psi_{p}^{\bar{s}}$ for all $\phi$. Hence, since $|\Phi_n| < \infty$
$
T_2 \to \big(\sum_{\phi, \phi'} \bar{\pi}_\phi \cdot \gamma_{d,g}(0; \phi, \phi')\big)\Psi_{p}^{\bar{s}} = \gamma_{d,g}(0) \Psi_{p}^{\bar{s}}.
$

Analogous arguments for $\V[t_{d',g'}^*(h_n)]$ and $ 2\C[t_{d,g}^*(h_n), t_{d',g'}^*(h_n)]$ yield the conclusion.
\end{proof}

\begin{theorem}\label{th:heterogeneousconvergence}
\textcolor{black}{Suppose Assumptions \ref{ass:consistency}--\ref{ass:networkheterogeneity}, \ref{ass:boundaryhet}, \ref{ass:pooledcovariance}, \ref{ass:kernel} hold, and let \ref{ass:cont}, \ref{ass:regularity}.i and \ref{ass:smoothnesshet} hold for all $\phi$.} Let $\bar{s}_{\phi}=\bar{s}$ for all $\phi$.
If $n h_n^{\bar{s}} \to \infty$, $h_n \to 0$, and 
$h_n = O\!\left(n^{-1/(2p + 2 + \bar{s})}\right)$, then
\[
\sqrt{n h_n^{\bar{s}}}\,
\frac{(\estimatorapp - \boundeffpooled) - h_n^{p+1}\bias}
{\sqrt{V_{d,g,d',g',p}(h_n)}}
\xrightarrow{d} \mathcal{N}(0,1),
\]
where $\bias = B_{d,g,d',g',p, p+1} + o(1)$ and $V_{d,g,d',g'p}(h_n) = V_{d,g,d',g',p} + o(1)$,
\begin{equation}\label{eq:biaspooled}
B_{d,g,d',g',p, p+ 1} 
= \Big(\frac{\bar{\mu}^{(p+1)}_{d,g}(0)-\bar{\mu}^{(p+1)}_{d',g'}(0)}{(p+1)!}\Big)e_1^\top (\Gamma_p^{\bar{s}})^{-1}\theta_{p,p+1}^{\bar{s}},
\end{equation}
and 
\begin{equation}\label{eq:varpooled}
    V_{d,g,d',g',p}
= \Big(\frac{\omegaplushet + \omegaminushet + \bar \gamma_{d,g,d',g'}(0)}{\densitypooledbar[0]^2}\Big)e_1^\top (\Gamma_p^{\bar{s}})^{-1}\Psi_p^{\bar{s}}(\Gamma_p^{\bar{s}})^{-1}
e_1.
\end{equation}
\end{theorem}

\begin{proof}[Proof of Theorem \ref{th:heterogeneousconvergence}]
The proof follows analogously as in Theorem \ref{th:asynormaltau}, but accounting for different strata, so we only report main steps. Define
\begin{equation*}
\begin{split}
\Zplus &= e_1^T \bar{\Gamma}^{-1}_{d,g,p}(h_n)\;
           \frac{1}{n h_n^{\bar{s}}}\,I_{i,d,g} \Kh \rr \nu_{i,d,g},\\
\end{split}
\end{equation*}
for generic $d,g$ and set $\Z = \Zplus - \Zminus$. Then
\[
\sum_i \Z = e_1^T\bar{\Gamma}^{-1}_{d,g,p}(h_n)\tplushet[*] - e_1^T\bar{\Gamma}^{-1}_{d',g',p}(h_n)\tminushet[*]
\]
Denote $nh_n^{\bar{s}}\,\V\!\Big[\sum_i \Z\Big] 
= V_{d,g,d',g',p}(h_n)$. By Lemmas~\ref{lemma:convergencebiashet} and \ref{lemma:varianceshe},
\[
V_{d,g,d',g',p}(h_n) =  \frac{\omegaplushet+\omegaminushet + \bar \gamma_{d,g,d',g'}(0)}{\densitypooledbar[0]^2}\,
  (\Gamma_p^{\bar{s}})^{-1}\Psi_p^{\bar{s}}(\Gamma_p^{\bar{s}})^{-1} + o(1).
\]

\medskip
\noindent\textbf{CLT via dependency graph.}
The collection $\{\Z\}_i$ admits dependency graph 
$\bm W_n$. Let $\tilde\sigma_n^2 = V_{d,g,d',g',p}(h_n)/nh_n^{\bar s}$, and set 
$\xi_n = \sum_i \Z/ \tilde\sigma_n$.  
Applying the Wasserstein bound of \citet[Theorem A.4]{leung20},
\begin{equation}\label{eq:dist2}
\begin{split}
\Delta(\xi_n,\mathcal{N}(0,1)) 
&\leq \tfrac{1}{\tilde\sigma_n^3}\max_i\E[|\Z|^3]\sum_i|\bm N_i|^2 \\
&\quad+ \tfrac{1}{\tilde\sigma_n^2}
\Big(\tfrac{4}{\pi}\max_i\E[\Z^4]\Big)^{1/2}
\Big(4\sum_i|\bm N_i|^3 + 3\sum_{i,j}(\bm W^3)_{ij}\Big)^{1/2}\\
& = O(nh_n^{\bar{s}})^{-1/2}) 
\end{split}
\end{equation}
using Assumption \ref{ass:fourthmoment} for all $\phi$, and the fact that the domain of integration has volume proportional to $h_n^{\bar{s}}$, analogously as Theorem \ref{th:asynormaltau}. Hence $\xi_n \xrightarrow{d} \mathcal{N}(0,1)$. 

\medskip
\noindent\textbf{Conclusion.}
By Lemma~\ref{lemma:convergencebiashet},
$
\sqrt{n h_n^{\bar{s}}}\cdot \tfrac{\estimatorapp[*]}{\sqrt{V_{d,g,d',g',p}(h_n)}} = \xi_n + o_p(1)
$, so $\estimatorapp[*]$ (rescaled) has the same asymptotic distribution as $\xi_n$. Moreover, by Lemmas~\ref{lemma:convergencebiashet},
\[
h_n^{p+1}B_{d,g,d',g', p, p+1} = O(h_n^{p+1})
\] and $T_{d,g,d'g'}(h_n) = O(h_n^{p+2})$. Then Eq.\eqref{eq:decompositiontauhet} and $h_n = O(n^{-1/(2p+2 + {\bar{s}})})$ yield the conclusion.
\end{proof}

\section{Estimation of Boundary Overall Direct Effects}\label{appendix:estimatoroverall}

We provide the asymptotic theory for the estimator $\overallest$ and $\hat{\tau}_{1|0}(h_n)$ defined in section~\ref{sec:mainoverall}.  
The target outcome regression functions for $d \in \{0,1\}$ are $\tilde{\mu}_1\left(x_i ; \phi\right)=\mathbb{E}\left[Y_i \mid X_i=x_i ; \phi\right]$ for $x_i \geq 0$, and $\tilde{\mu}_0\left(x_i ; \phi\right)=\mathbb{E}\left[Y_i \mid X_i=x_i ; \phi\right]$ for $x_i<0$, and the pooled regression functions
\begin{equation*}
\widetilde{\mu}_d(x)
=
\frac{
\sum_{\phi\in\Phi}
\widetilde{\mu}_d(x;\phi)
f_{X_i}(x;\phi)\pi_{\phi}
}{
f_{X_i}(x)
}.
\end{equation*}
where $f_{X_i}(x_i; \phi)$ is the marginal density of $X_i$ and $f_{X_i}(x) =
\sum_{\phi\in\Phi}
f_{X_i}(x;\phi)\pi_{\phi}$.
Let $\overallestapp$ and $\hat{\tau}_{1|0,p}(h_n)$ denote the generic $p$th-order local polynomial estimator. 
The asymptotic analysis follows similar steps as in Section~\ref{appendix:estimator} and \ref{appendix:heterogeneous}.  
For brevity, we only sketch the arguments.
%
For $d \in \{0,1\}$ define  
\begin{align*}
\hat{\beta}_{d,p}(h_n;\phi)
&=
\argmin_{b \in \mathbb{R}^{p+1}}\sum_i
\mathbbm{1}(D_i = d)\mathbbm{1}(\phi_i=\phi)(Y_i - r_p(X_i)b)^2K_{h_n}(X_i),\\
\hat{\beta}_{d,p}(h_n)
&=
\argmin_{b \in \mathbb{R}^{p+1}}\sum_i
\mathbbm{1}(D_i = d)
(Y_i - r_p(X_i)b)^2K_{h_n}(X_i).
\end{align*}
The estimators are given by 
\[
\hat{\tau}_{1|0,p}(h_n;\phi)
=
e_1^\top\!\left [
\hat{\beta}_{1}(h_n;\phi)-\hat{\beta}_{0}(h_n;\phi)
\right ],
\qquad
\hat{\tau}_{1|0,p}(h_n)
=
e_1^\top\!\left [
\hat{\beta}_{1}(h_n)-\hat{\beta}_{0}(h_n)
\right].
\]

%
We set $\mathbbm{1}(D_i =  0, \phi_i = \phi)=I_{i, d, \phi}$, $\mathbbm{1}(D_i = d) = I_{i, d}$, and $\bm{Y}=(Y_1,\dots,Y_n)^T$, and we define $\Xpov = [r(X_1/h_n),\dots,r(X_n/h_n)]^T$ and $S_p(h_n) = [(X_1/h_n)^p,\dots,(X_n/h_n)^p]^T$. We define
\begin{align*}
    W_{d}(h_n; \phi) &= \mathrm{diag}(I_{1,d,\phi}\cdot K_h(X_1),\dots,I_{n,d,\phi}\cdot K_h(X_n)), & 
    W_{d}(h_n) &= \mathrm{diag}(I_{1,d}\cdot K_h(X_1),\dots,I_{n,t}\cdot K_h(X_n)) \\
    \Gamma_{d,p}(h_n; \phi) &= \Xpov^T W_d(h_n; \phi) \Xpov / n_\phi, & 
    \Gamma_{d,p}(h_n) &= \Xpov^T W_d(h_n) \Xpov / n_\phi\\
    t_{d,p}(h_n; \phi)&= \Xpov^T W_d(h_n; \phi) \bm{Y} / n_\phi
    &
    t_{d,p}(h_n)&= \Xpov^T W_d(h_n) \bm{Y} / n_\phi
\end{align*}
The estimators $\hat{\beta}_{d,p}(h_n; \phi)$ and $\hat{\beta}_{d,p}(h_n)$ and linear projection coefficients are:
\begin{align*}
    \hat{\beta}_{d,p}(h_n; \phi) =& H(h_n)\Gamma_{d,p}(h_n; \phi)^{-1}t_{d,p}(h_n; \phi), \quad
    &\hat{\beta}_{d,p}(h_n) =  H(h_n)\Gamma_{d,p}(h_n)^{-1}t_{d,p}(h_n). \\
\beta_{d,p}(h_n; \phi) =& H(h_n)\bar{\Gamma}_{d,p}(h_n; \phi)^{-1}\bar{t}_{d,p}(h_n; \phi), \qquad
& \beta_{d,p}(h_n) = H(h_n)\bar{\Gamma}_{d,p}(h_n)^{-1}\bar{t}_{d,p}(h_n)
\end{align*}
and define the centered quantities $\tplusov[*],\; \tminusov[*],\; \betaoverallplus[*],\; \betaoverallminus[*]$, and $\hat{\tau}^*_{1|0,p}(h_n; \phi)$ obtained by replacing $Y_i$ with $\nu_{i,d}= Y_i-r_p(X_i)\beta_{d,p}(h_n; \phi)$.


\subsection{Asymptotic Theory}
Under Assumptions \ref{ass:consistency}-\ref{ass:identification} for $d \in \{0,1\}$
\[
\tilde{\mu}_d(x_i; \phi) = \sum_{g \in \mathcal{G}_i} \int_{\mathcal{X}^\phi_{\smallN_i}(g)} m_{d, g}(x_i, \bm x_{\smallN_i}; \phi) \cdot\frac{f(x_i, \bm{x}_{\smallN_i}; \phi)}{f_{X_i}(x_i; \phi)}d\bm{x}_{\smallN_i}
\]
are continuous in $x_i = 0$, where $\mathcal{X}^\phi_{\smallN_i}(g)$ is the set of elements $\bm{x}_{\smallN_i}$ such that $e(\bm x_{\smallN_i}) = g$. We further adopt the assumptions from Section~\ref{appendix:estimator}, replacing \ref{ass:smoothness} with:

\begin{assumption}\label{ass:smoothnessov} 
$\muoverallplus[x_i]$ and $\muoverallminus[x_i]$ are $(p+2)$-times continuously differentiable.
\end{assumption}

\noindent Under Assumption~\ref{ass:smoothnessov}, we obtain the decomposition 
\begin{equation}\label{eq:decompositiontauov}
    \begin{split}
        \overallestapp = & \overall(\phi) + h_n^{p+1}\underbrace{\Big(
       \frac{\muoverallderivplus[p+1]}{(p+1)!}\biasplusov
      -\frac{\muoverallderivminus[p+1]}{(p+1)!}\biasminusov\Big)}_{\biasov(h_n; \phi)} + T_{1,0}(h_n; \phi) \\
    + &  \hat{\tau}^*_{1|0,p}(h_n; \phi)
    \end{split}
\end{equation}
\sloppy
with $B_{d, p, p+1}(h_n; \phi)= e_1^T\bar{\Gamma}_{d,p}^{-1}(h_n; \phi)\bar{\theta}_{d,p,p+1}(h_n; \phi)$,  \[\bar{\theta}_{d,p,q}(h_n; \phi)= \E\big[X_p(h_n)^TW_{d}(h_n; \phi)\,S_q(h)/n_\phi \big]
\]
and $T_{1,0}(h_n; \phi)$ depends on the Taylor reminders. 
Because the estimator uses only observations with $|X_i|\leq h_n$, we have
\begin{align*}
    \E\Bigg[\frac{1}{n_\phi h_n}\sum_i \Iplusov K\!\left(\frac{X_i}{h_n}\right)
          \left(\frac{X_i}{h_n}\right)^k\Bigg]
    &=  \int_0^1 K(u)u^k 
       \int_{\pscore} f(uh_n,\bm{x}_{\smallN_i}; \phi)\,d\bm{x}_{\smallN_i}\,du \\
    &= \int_0^1 K(u)u^k f_{X_i}(uh_n; \phi)\,du,
\end{align*}
for nonnegative integer $k$, and analogously for $ \E\Bigg[\frac{1}{nh_n}\sum_i \Iminusov K\!\left(\frac{X_i}{h_n}\right)\Big(\tfrac{X_i}{h_n}\Big)^k\Bigg]$. Moreover, by similar arguments to Lemma~\ref{lemma:convergencematricesgamma}, one obtains 
\[
\V\!\left[\frac{1}{n_\phi h_n}\sum_i I_{i,d, \phi}K\!\Big(\tfrac{X_i}{h_n}\Big)\Big(\tfrac{X_i}{h_n}\Big)^k\right] 
= O\!\Big(\tfrac{1}{n_\phi h_n}\Big).
\]
for $d \in\{0,1\}$, from which for $n_\phi h_n \to \infty$ and $h_n \to 0$, $\Gamma_{d, p}(h_n; \phi) = \limdensityov\,\Gamma_p^{1} + o_p(1)$ and 

\[
t_{d, p}(h_n) =  \theta^1_{p,0} \cdot \sum_{g \in \mathcal{G}_i} \int_{\mathcal{X}^\phi_{\smallN_i}(g)}\hspace{-0.5em}m_{d, g}(x_i, \bm x_{\smallN_i}; \phi) f(x_i, \bm{x}_{\smallN_i}; \phi)d\bm{x}_{\smallN_i}+ o_p(1),
\]
with
\[
\Gamma_p^{1} = \int_0^1 K(u)r_p(u)\,du,
\qquad
\theta_{p,q}^{1} = \int_0^1 K(u)u^{p+q}\,du,
\]
Additionally, $\bar{\theta}_{d,p,q}  = \limdensityov\theta_{p,q}^{1} + o(1)$. Finally, $n_\phi h_n\V\!\big[t_{1,p}^*(h_n; \phi) - t_{1,0,p}^*(h_n; \phi)\big]
= \big(\limvarplusov + \limvarminusov\big)\Psi_p^1 + o(1)$
where for $d \in \{0,1\}$
\[
\bar{\omega}_{d}(0) = \sum_{g \in \mathcal{G}_i}\int_{\mathcal{X}^\phi_{\smallN_i}(g)}\big[\sigma_{d,g}(0, \bm{x}_{\smallN_i}) + (m_{d,g}(0, \bm x_{\smallN_i}; \phi) - \tilde{\mu}_{d}(0; \phi))^2\big] f(0, \bm{x}_{\smallN_i}; \phi)d\bm{x}_{\smallN_i}
\]
Here, only the variance term $\bar{\omega}_{d}(0;\phi)$ enters the asymptotic variance. For $d\in\{0,1\}$, write
\(
n_\phi h_n\operatorname{Var}\!\left(t_{d,p}^{*}(h_n;\phi)\right)
=
T_{1,d}(h_n;\phi)+T_{2,d}(h_n;\phi),
\)
where
\[
T_{1,d}(h_n;\phi)
=
\frac{1}{n_\phi h_n}
\sum_{i\in V_\phi}
\mathbb{E}\left[
I_{i,d,\phi}
K^2\!\left(\frac{X_i}{h_n}\right)
r_p\!\left(\frac{X_i}{h_n}\right)
r_p\!\left(\frac{X_i}{h_n}\right)^\top
\nu_{i,d}^2
\right],
\]
with $\nu_{i, d}=Y_i-r_p\left(X_i\right)^{\top} \beta_{d, p}\left(h_n ; \phi\right)$, and
\[
T_{2,d}(h_n;\phi)
=
\frac{1}{n_\phi h_n}
\sum_{i\in V_\phi}
\sum_{j\in N_i\setminus\{i\}}
\mathbb{E}\left[
I_{i,d,\phi}I_{j,d,\phi}
K\!\left(\frac{X_i}{h_n}\right)
K\!\left(\frac{X_j}{h_n}\right)
r_p\!\left(\frac{X_i}{h_n}\right)
r_p\!\left(\frac{X_j}{h_n}\right)^\top
\nu_{i,d}\nu_{j,d}
\right].
\]
By the same arguments as in Section~S.4,
\(
T_{1,d}(h_n;\phi)
=
\bar{\omega}_{d}(0;\phi)\Psi_p^1+o(1).
\)
For $T_{2,d}(h_n; \phi)$ we have:
\begin{align*}
T_{2,d}(h_n;\phi)
={}&
\frac{1}{n_\phi h_n}
\sum_{i\in V_\phi}
\sum_{j\in N_i\setminus\{i\}}
\sum_{g, g'}
\int_{\mathcal B_{ij}^{g,g'}(h_n;\phi)}
K\!\left(\frac{x_i}{h_n}\right)
K\!\left(\frac{x_j}{h_n}\right)
\\
&\qquad\times
r_p\!\left(\frac{x_i}{h_n}\right)
r_p\!\left(\frac{x_j}{h_n}\right)^\top
\mathbb E\!\left[\nu_{i,d}\nu_{j,d}\mid \bm X_{ij}= \bm x_{ij};\phi\right]
f_{ij}(\bm x_{ij};\phi),d\bm x_{ij},
\end{align*}
where
\[
\mathcal B_{ij}^{g,g'}(h_n;\phi)
=
B_{h_n}^d\!\left(
\overline{\mathcal X}_{iS_i}^{\phi}(1,g\mid0,g)
\right)
\cap
B_{h_n}^d\!\left(
\overline{\mathcal X}_{jS_j}^{\phi}(1,g'\mid0,g')
\right),
\]
Since the boundary intersection $\overline{\mathcal X}_{iS_i}^{\phi}(1,g\mid0,g)
\cap
\overline{\mathcal X}_{jS_j}^{\phi}(1,g'\mid0,g')
$ has codimension $2$ for all $i\neq j$, then
\(
\operatorname{Vol}\!\left(\mathcal B_{ij}^{g,g'}(h_n;\phi)\right)
=
O(h_n^2),
\)
following the arguments in Theorem~\ref{th:variance}, from which under Assumption \ref{ass:localdependence}
\(
T_{2,d}(h_n;\phi)
= O(h_n)\frac{1}{n_\phi}\sum_{i\in V_\phi}|N_i| =
o(1).
\)
We define the asymptotic bias and variance.
For the stratum-specific estimator: 
\begin{equation}\label{eq:bias-overall-phi}
B_{1,0,p,p+1}(h_n;\phi)
=
\frac{
\widetilde{\mu}_1^{(p+1)}(0;\phi)
-
\widetilde{\mu}_0^{(p+1)}(0;\phi)
}{
(p+1)!
}
e_1^\top
(\Gamma_p^1)^{-1}
\theta_{p,p+1}^1
+o(1),
\end{equation}
and 
\begin{equation}\label{eq:variance-overall-phi}
V_{1,0,p}(h_n;\phi)
=
\frac{
\bar{\omega}_1(0;\phi)
+
\bar{\omega}_0(0;\phi)
}{
f_{X_i}(0;\phi)^2
}
e_1^\top
(\Gamma_p^1)^{-1}
\Psi_p^1
(\Gamma_p^1)^{-1}
e_1
+o(1),
\end{equation}
For the pooled estimator,
\begin{equation}\label{eq:bias-overall}
B_{1,0,p,p+1}(h_n)
=
\frac{
\widetilde{\mu}_1^{(p+1)}(0)
-
\widetilde{\mu}_0^{(p+1)}(0)
}{
(p+1)!
}
e_1^\top
(\Gamma_p^1)^{-1}
\theta_{p,p+1}^1
+o(1).
\end{equation}
and
\begin{equation}\label{eq:variance-overall}
V_{1,0,p}(h_n)
=
\frac{
\bar{\omega}_1(0)
+
\bar{\omega}_0(0)
}{
f_{X_i}(0)^2
}
e_1^\top
(\Gamma_p^1)^{-1}
\Psi_p^1
(\Gamma_p^1)^{-1}
e_1
+o(1).
\end{equation}
where
\(
\bar{\omega}_d(0)
=
\sum_{\phi\in\Phi}
\bar{\pi}_\phi\,
\bar{\omega}_d(0;\phi),
\)
ans
\(
\bar f_{X_i}(0)
=
\sum_{\phi\in\Phi}
\bar{\pi}_\phi\,
f_{X_i}(0;\phi),
\)

\begin{theorem}\label{th:asynormaltauov}
Suppose Assumptions~\ref{ass:consistency}--\ref{ass:networkheterogeneity}, \ref{ass:localdependence}, and~\ref{ass:kernel} hold. 
\begin{enumerate}[(i)]
\item For a given $\phi$, suppose that
$\widetilde{\mu}_1(\cdot;\phi)$ and
$\widetilde{\mu}_0(\cdot;\phi)$ are $(p+2)$-times continuously
differentiable, and Assumption  \ref{ass:identification}, \ref{ass:fourthmoment}---\ref{ass:variances} hold. If $n_\phi h_n\to\infty$, $h_n\to0$, and
$h_n=O(n_\phi^{-1/(2p+3)})$, then
\[
\sqrt{n_\phi h_n}
\frac{
\hat\tau_{1\mid0,p}(h_n;\phi)
-\tau_{1\mid0}(\phi)
-h_n^{p+1}B_{1,0,p,p+1}(h_n;\phi)
}{
\sqrt{V_{1,0,p}(h_n;\phi)}
}
\xrightarrow{d}\mathcal N(0,1).
\]

\item For the pooled estimator, suppose that the conditions in part~(i)
hold for all $\phi$, Assumption~\ref{ass:boundaryhet} holds,
and $f_{X_i}(\cdot;\phi)$ is $(p+2)$-times continuously differentiable
for all $\phi$. If $nh_n\to\infty$, $h_n\to0$, and
$h_n=O(n^{-1/(2p+3)})$, then
\[
\sqrt{nh_n}
\frac{
\hat\tau_{1\mid0,p}(h_n)
-\tau_{1\mid0}
-h_n^{p+1}B_{1,0,p,p+1}(h_n)
}{
\sqrt{V_{1,0,p}(h_n)}
}
\xrightarrow{d}\mathcal N(0,1).
\]
\end{enumerate}
\end{theorem}

\begin{proof}
For part~(i), the argument follows Theorem~\ref{th:asynormaltau} with codimension one. 
Define
\[
Z_{i, d, p}(h_n) 
=  e_1^T\bar{\Gamma}_{d,p}^{-1}(h_n; \phi)I_{d,i,\phi}
K\!\Big(\tfrac{X_i}{h_n}\Big) r_p\!\Big(\tfrac{X_i}{h_n}\Big)\nu_{i,d}/n_\phi,
\]
and set $\Zov = \Zplusov - \Zminusov$.  
Then, $n_\phi h_n\V\!\big[\sum_i \Zov\big] = V_{1,0,p}(h_n) = (\limvarplusov+\limvarminusov)\Psi_p^{1}/\limdensityov^2 + o(1)$.  Denote $\sigma_n^2 = V_{1,0,p}(h_n; \phi)/n_\phi h_n$ and $\xi_n = \sum_i \Zov/\sigma_n$.  
By Assumption~\ref{ass:localdependence} and Stein’s method \citep[Theorem~A.4]{leung20}, one obtains
$
\Delta(\xi_n,\mathcal{N}(0,1)) = O\big((n_\phi h_n)^{-1}\big) \to 0.
$
Thus $\xi_n \xrightarrow{d} \mathcal{N}(0,1)$.  
Since $\tfrac{\overallestapp^*}{\sqrt{\sigma^2_n}} = \xi_n + o_p(1)$, the decomposition \eqref{eq:decompositiontauov} and convergence of $\bar{\Gamma}_{d,p}(h_n; \phi)$ and $\bar{\theta}_{d,p,p+1}(h_n; \phi)$, to $f(0; \phi)\Gamma_p^1$ and $f(0; \phi)\theta^{1}_{p, p+1}$ gives the conclusion.

Part~(ii) follows by the same argument as in the pooled analysis of Section~\ref{appendix:heterogeneous} using the full-sample
quantities and replacing $n_\phi$ by $n$.
\end{proof}

\subsection{Variance Estimation and Bias Correction}

Define for $d \in \{0,1\}$
\begin{equation}
\begin{split}
    \mathcal{M}_{d,p}(h_n; \phi)^T 
    &= \Big[\Iplusov[1]\Kh[1]\rr[1]\hat{\xi}_{1,d}\;\cdots\; \Iplusov[n]\Kh[n]\rr[n]\hat{\xi}_{n,d}\Big]
\end{split}
\end{equation}
where $\hat{\xi}_{i,d}$ denotes the regression residuals. The estimator of the asymptotic variance of $\overallestapp$ is
\begin{equation}\label{eq:varestimatorov}
    \hat{\mathcal{V}}_{1,0,p}(h_n; \phi) 
    = \frac{1}{n_\phi}\, e_1^T\Big(\hat{V}_{1,p}(h_n; \phi) 
      + \hat{V}_{0,p}(h_n; \phi) 
      - 2\hat{V}_{1,0,p}(h_n; \phi)\Big)e_1,
\end{equation}
with
\begin{align*}
    \hat{V}_{d,p}(h_n; \phi) 
    &= \Gammaplusov[-1]\hat{\tilde{\Psi}}_{1,p}(h_n; \phi)\Gammaplusov[-1], & \hat{\tilde{\Psi}}_{d,p}(h_n; \phi) 
    &= \mathcal{M}_{d,p}(h_n; \phi)\bm{W}_n\mathcal{M}_{d,p}(h_n; \phi)^T/n_\phi, \\
    \hat{V}_{1,0,p}(h_n; \phi) 
    &= \Gammaplusov[-1]\hat{\tilde{\Psi}}_{1,0,p}(h_n; \phi)\Gammaminusov[-1], & \hat{\tilde{\Psi}}_{1,0,p}(h_n; \phi) 
    &= \mathcal{M}_{1,p}(h_n; \phi)\bm{W}_n\mathcal{M}_{0,p}(h_n; \phi)^T/n_\phi
\end{align*}

\noindent Finally, we give the expression for the bias-corrected estimator:
\begin{equation*}
    \overallestbcapp = \overallestapp - \biasbcappov
\end{equation*}
where $\biasbcappov$ uses the estimator $\hat{\beta}_{d, p+1}(b_n; \phi)$ with bandwidth $b_n$.

\stopcontents[supp]

\putbib[biblio]
\end{bibunit}

\end{document}